\pdfoutput=1

\documentclass[aps,prb,notitlepage,showkeys,showpacs,superscriptaddress,nofootinbib,longbibliography,preprintnumbers,11pt]{revtex4-2}

\usepackage[utf8]{inputenc}

\usepackage{here}
\usepackage{ulem}

\usepackage{url}

\usepackage{dsfont}

\newcounter{n}

\usepackage[top=30truemm,bottom=30truemm,left=25truemm,right=25truemm]{geometry}

\usepackage{spectralsequences}
\usepackage{tikz}
\usetikzlibrary{intersections}
\usetikzlibrary{calc}

\usepackage{xcolor}

\newcommand{\abs}[1]{\left|#1\right|}

\usepackage{tikz-cd}

\usepackage[colorlinks=true,linktoc=page,citecolor=red,linkcolor=blue]{hyperref}	
\usepackage{slashed}

\usepackage{amsthm}
\usepackage{thmtools}
\usepackage{blindtext}
\usepackage{braket}

\declaretheoremstyle[
       shaded={bgcolor=\color{rgb}{0.9,0.9,0.9}}  
]{theorem}

\declaretheoremstyle[
       shaded={bgcolor=\color{rgb}{0.9,0.9,0.9}}
]{question}

\declaretheoremstyle[
       shaded={bgcolor=\color{rgb}{0.9,0.9,0.9}}  
]{remark}

\declaretheoremstyle[
       shaded={bgcolor=\color{rgb}{0.9,0.9,0.9}}  
]{proposition}

\declaretheoremstyle[
       shaded={bgcolor=\color{rgb}{0.9,0.9,0.9}}  
]{definition}

\declaretheoremstyle[
       shaded={bgcolor=\color{rgb}{0.9,0.9,0.9}}  
]{assumption}

\declaretheoremstyle[
       shaded={bgcolor=\color{rgb}{0.9,0.9,0.9}}  
]{conjecture}

\declaretheoremstyle[
       shaded={bgcolor=\color{rgb}{0.9,0.9,0.9}}  
]{corrorary}

\declaretheoremstyle[
       shaded={bgcolor=\color{rgb}{0.9,0.9,0.9}}  
]{axiom}

\declaretheoremstyle[
       shaded={bgcolor=\color{rgb}{0.9,0.9,0.9}}  
]{lemma}

\def\id{{\mathrm{id}}}

\def\>{{\geq }}
\def\<{{\leq }}

\newcommand{\Hom}{\text{Hom}}
\newcommand{\End}{\text{End}}
\newcommand{\Aut}{\text{Aut}}
\newcommand{\Vect}{\text{Vec}}
\newcommand{\Rep}{\text{Rep}}
\newcommand{\Mod}{\text{Mod}}
\newcommand{\Fun}{\text{Fun}}

\newcommand{\Mat}{\text{Mat}}
\def\1{\mathds{1}}

\newcommand{\TY}{\text{TY}}

\usepackage{mathrsfs}
\usepackage{amsmath,amssymb}

\usepackage{mathtools}

\usepackage{esvect}

\usepackage{adjustbox}

\numberwithin{equation}{section}

\usepackage[status=draft,inline,nomargin]{fixme}
\fxusetheme{color}
\FXRegisterAuthor{ki}{aki}{KI}
\FXRegisterAuthor{shu}{shuenv}{\color{red}Shu}

\usetikzlibrary{decorations.pathmorphing,shapes}
\newcounter{sarrow}

\usepackage{xurl}

\begin{document}

\title{Generalized cluster states in 2+1d: non-invertible symmetries, interfaces, and parameterized families}

\author{Kansei Inamura}
\email{kansei.inamura@physics.ox.ac.uk}
\affiliation{Mathematical Institute, University of Oxford, Andrew Wiles Building, Woodstock Road, Oxford, OX2 6GG, UK}
\affiliation{Rudolf Peierls Centre for Theoretical Physics, University of Oxford, Parks Road, Oxford, OX1 3PU, UK}

\author{Shuhei Ohyama}
\email{shuhei.oyama@univie.ac.at}
\affiliation{University of Vienna, Faculty of Physics, Boltzmanngasse 5, A-1090 Vienna, Austria}

\begin{abstract}
We construct 2+1-dimensional lattice models of symmetry-protected topological (SPT) phases with non-invertible symmetries and investigate their properties using tensor networks.
These models, which we refer to as generalized cluster models, are constructed by gauging a subgroup symmetry $H \subset G$ in models with a finite group 0-form symmetry $G$.
By construction, these models have a non-invertible symmetry described by the group-theoretical fusion 2-category $\mathcal{C}(G; H)$.
After identifying the tensor network representations of the symmetry operators, we study the symmetry acting on the interface between two generalized cluster states.
In particular, we will see that the symmetry at the interface is described by a multifusion category known as the strip 2-algebra.
By studying possible interface modes allowed by this symmetry, we show that the interface between generalized cluster states in different SPT phases must be degenerate.
This result generalizes the ordinary bulk-boundary correspondence.
Furthermore, we construct parameterized families of generalized cluster states and study the topological charge pumping phenomena, known as the generalized Thouless pump.
We exemplify our construction with several concrete cases, and compare them with known phases, such as SPT phases with $2\mathrm{Rep}((\mathbb{Z}_{2}^{[1]}\times\mathbb{Z}_{2}^{[1]})\rtimes\mathbb{Z}_{2}^{[0]})$ symmetry.
\end{abstract}

\maketitle

\setcounter{tocdepth}{3}
\tableofcontents

\section{Introduction and summary}
\label{sec: Introduction}

\subsection{Introduction}
Symmetry-protected topological (SPT) phases are the simplest yet non-trivial phases of quantum many-body systems \cite{Gu_2009, Pollmann:2009mhk, Pollmann_2010, Chen:2010zpc, Chen_2011_complete, Chen:2011bcp, Chen:2011pg, Schuch_2011}.
Despite their simplicity, they exhibit a surprisingly rich structure due to the symmetry constraints.
A defining feature of non-trivial SPT phases is the existence of protected edge modes. 
The connection between SPT non-triviality and the presence of edge modes---often referred to as the bulk-boundary correspondence---has been established for various symmetries and spatial dimensions.
It is also well established that adiabatic and periodic driving of an SPT state can induce topological transport, known as the generalized Thouless pump, which has also been demonstrated for various symmetries and spatial dimensions
\cite{Thouless83,PhysRevB.82.115120,Kitaev2011SCGP,Kitaev2013SCGP,Kitaev2015IPAM,Kapustin:2020mkl,PhysRevResearch.2.042024,Tantivasadakarn:2021wdv,Shiozaki:2021weu,Hermele2021CMSA,Wen:2021gwc,Bachmann:2022bhx,Spodyneiko:2023vsw,Ohyama:2022cib,Inamura:2024jke,Jones:2025khc,Shiozaki:2025pyo,Li:2025wes}.

In a related but distinct context, recent studies have explored generalizations of symmetry~\cite{Gaiotto:2014kfa}.
As established in a seminal work by Wigner \cite{Wigner1959}, conventional symmetries are represented by unitary or anti-unitary operators, which are invertible and form a group.
By contrast, within the framework of generalized symmetries, the invariance under the action of certain non-invertible operators is also regarded as a symmetry.
Correspondingly, the algebraic structure of such a symmetry, called non-invertible symmetry, is modeled by an appropriate (higher) category.
A notable example of a non-invertible symmetry is the Kramers-Wannier self-duality of the critical Ising model \cite{KW1941, Frohlich:2004ef, Chang:2018iay}.\footnote{More precisely, the critical Ising model on the lattice is invariant under the Kramers-Wannier transformation followed by the lattice translation by half a site \cite{Seiberg:2023cdc, Seiberg:2024gek}.}
See \cite{Cordova:2022ruw, McGreevy:2022oyu, Schafer-Nameki:2023jdn, Brennan:2023mmt, Bhardwaj:2023kri, Luo:2023ive, Shao:2023gho, Carqueville:2023jhb, Iqbal:2024pee, Costa:2024wks} for recent reviews of non-invertible symmetries.
Given this generalization, it is natural to investigate the properties of SPT phases in the presence of non-invertible symmetries.
There are various recent studies along this line \cite{Thorngren:2019iar,Inamura:2021wuo,Inamura:2021szw,Garre-Rubio:2022uum,Fechisin:2023odt,Seifnashri:2024dsd,Choi:2024rjm, Jia:2024bng, Li:2024fhy, Inamura:2024jke, Pace:2024acq, Meng:2024nxx, Cao:2025qhg, Aksoy:2025rmg, Maeda:2025rxc, Lu:2025rwd, Furukawa:2025flp, ParayilMana:2025nxw, Inamura:2025cum, You:2025uxo, Lu:2025yru}.

Tensor networks provide an efficient representation of quantum many-body states, particularly those in gapped phases, including SPT phases; see \cite{Cirac:2020obd} for a review.
Beyond this original motivation, tensor networks have increasingly been recognized as powerful tools for giving microscopic descriptions of non-invertible symmetries \cite{Molnar:2022nmh, Garre-Rubio:2022uum, Lootens:2021tet, Lootens:2022avn, Gorantla:2024ocs}. 
In particular, in 1+1-dimensional systems, matrix product states (MPS) and matrix product operators (MPO) provide a systematic framework for constructing topological invariants that distinguish different SPT phases with non-invertible symmetries \cite{Garre-Rubio:2022uum, Inamura:2024jke}.
Tensor networks have also been employed to investigate the bulk-boundary correspondence and the generalized Thouless pumps for non-invertible SPT phases in 1+1 dimensions \cite{Inamura:2024jke, Li:2025wes}.
Given the success of tensor network methods in 1+1d, it is natural to expect that they are also useful in studying non-invertible SPT phases in higher dimensions.

Non-invertible symmetries in higher dimensions are now being actively studied in both mathematics and physics.
In 1+1 dimensions, finite non-invertible symmetries are described by fusion 1-categories \cite{Bhardwaj:2017xup, Chang:2018iay, Thorngren:2019iar}, which have been extensively studied for a long time; see \cite{EGNO2015} for a review.
Similarly, in 2+1 dimensions, finite non-invertible symmetries are expected to be described by fusion $2$-categories, which were defined in \cite{Douglas:2018qfz} and classified in \cite{Decoppet:2024htz} (see also \cite{Bhardwaj:2024qiv, Bullimore:2024khm}).
Recent studies have revealed various realizations of fusion 2-category symmetries in 2+1-dimensional systems, both on the lattice and in the continuum \cite{Kaidi:2021xfk, Bhardwaj:2022lsg, Bhardwaj:2022maz, Bartsch:2022mpm, Bartsch:2022ytj, Delcamp:2023kew, Choi:2024rjm, Inamura:2023qzl, Hsin:2024aqb, Bhardwaj:2024qiv, Bullimore:2024khm, Cordova:2024jlk, Cordova:2024mqg, Cao:2025qhg, Bhardwaj:2025piv, Bhardwaj:2025piv, Eck:2025ldx, Vancraeynest-DeCuiper:2025wkh, Hsin:2025ria, Furukawa:2025flp, Inamura:2025cum, KNBalasubramanian:2025vum}.
In particular, in~\cite{Choi:2024rjm}, Choi, Sanghavi, Shao, and Zheng constructed two gapped lattice Hamiltonians with a certain fusion 2-category symmetry and showed that their ground states are in distinct SPT phases protected by the fusion 2-category symmetry.
More recently, lattice models for more general SPT phases were constructed in \cite{Cao:2025qhg, Inamura:2025cum} using the gauging of finite group symmetries \cite{Delcamp:2023kew}.
On the mathematical side, in \cite{Decoppet:2023bay}, D\'{e}coppet and Yu established the classification of fusion $2$-categories admitting a fiber $2$-functor, as well as the classification of fiber $2$-functors of such fusion 2-categories.
While the result of \cite{Decoppet:2023bay} is purely mathematical, from a physical point of view, it can be regarded as a categorical formulation of the classification of SPT phases with fusion 2-category symmetries.

Although there has been progress in various lattice realizations of SPT phases with fusion 2-category symmetry, there has not yet been a systematic understanding of bulk-boundary correspondence and the generalized Thouless pump for such SPT phases.
As in the case of 1+1 dimensions, it is expected that tensor network techniques allow a systematic study of edge modes (or more generally, interface modes) and topological pumping for 2+1d SPT phases with fusion 2-category symmetries.

In this paper, we construct Hamiltonians for 2+1d SPT phases protected by group-theoretical fusion 2-category symmetry $\mathcal{C}(G;H)$ using gauging techniques in lattice systems following the construction in \cite{Cao:2025qhg,Inamura:2025cum}.\footnote{We note that the group-theoretical fusion 2-category $\mathcal{C}(G;H)$ includes $2\Vect_{\mathbb{G}}$ and $2\Rep(\mathbb{G})$ for any split 2-group $\mathbb{G}$~\cite{Decoppet:2023bay}.}
We further provide tensor network representations of the ground states and the symmetry operators.
Since the ground states exhibit a structure similar to that of the cluster state, we refer to them as generalized cluster states.
In addition, we identify the symmetry acting on interfaces---namely, the strip 2-algebra\footnote{The strip 2-algebra is a natural categorification of the strip algebra~\cite{Cordova:2024iti}. Strip algebras are known by various names in the literature, such as module annular algebras~\cite{Bridgeman:2022gdx}, boundary tube algebras \cite{Jia:2024rzr, Choi:2024tri, Choi:2024wfm, Jia:2024zdp, Jia:2025yph}, and interface algebras \cite{Inamura:2024jke, Lu:2025rwd}. Recent applications of strip algebras are also discussed in, e.g., \cite{Choi:2023xjw, Cordova:2024vsq, Copetti:2024onh, Copetti:2024dcz, Bhardwaj:2024igy, Cordova:2024nux, Heymann:2024vvf, AliAhmad:2025bnd, Benini:2025lav}. See also \cite{Kitaev:2011dxc, Lan:2013wia, Bridgeman:2018jdv, Bridgeman:2019axg, Bridgeman:2019wyu, Barter_2022} for earlier works on their relation to gapped excitations on gapped boundaries of 2+1d topological orders. \label{foot: strip}} \cite{Gagliano:2025gwr}---and analyze its representations.
This allows us to systematically study interface modes between states belonging to different phases as well as within the same phase. 
A more detailed summary of the paper is provided below.

\subsection{Outline and summary}
We begin with several remarks before presenting the outline and summary. 
Since we develop the theory in a general setting, it is inevitable to introduce various notations and technical machinery, thereby rendering the paper long.
To improve readability, we include, as the final subsection of each section, an instantiation of the general theory.
These subsections focus on the case of the $G$-cluster state, which is a simple example of an SPT state with $2\mathrm{Rep}(G)\boxtimes 2\Vect_{G}$ symmetry.
Since most of the notation is defined in the main text rather than in the example subsections, we also provide a list of notation at the beginning of the paper for quick reference.
Accordingly, for readers who are not primarily interested in technical details or the general theory, the paper is organized so that it can be read in an almost self-contained manner by referring to the example subsections at the end of each section, i.e., Sections~\ref{sec: Example: G-cluster state}, \ref{sec: Example: symmetry of the G-cluster state}, \ref{sec: Example: gapped boundary of the G-cluster state}, and \ref{sec: Example: parameterized families of the G-cluster states}.

The structure of the paper is as follows.
In Section \ref{sec: Preliminaries}, we provide some mathematical concepts, including fusion 2-categories and fiber 2-functors.
We also review a subset of the results proved in~\cite{Decoppet:2023bay}.
After that, we present the four main aspects of this paper, which we briefly summarize below.

\vspace*{\baselineskip}
\noindent{\bf Models.}
In Section \ref{sec: 2+1d generalized cluster states with non-invertible symmetries},
we introduce generalized cluster models in 2+1 dimensions, which realize SPT phases with fusion 2-category symmetries.
This model is constructed by gauging a subgroup symmetry $H\subset G$ in a system with finite group $G$ symmetry.
The gauged model realizes an SPT phase when the model before gauging has no topological order and spontaneously breaks the symmetry $G$ down to $K$ satisfying $G=HK$ and $H \cap K = \{e\}$.
After constructing the Hamiltonian, we identify the ground state and provide its tensor network representation.
This section can also be read as a review of gauging finite group symmetries in lattice systems.
The general construction is illustrated with an example of the $G$-cluster model in Section~\ref{sec: Example: G-cluster state}.

\vspace*{\baselineskip}
\noindent{\bf Symmetries.}
In Section \ref{sec: Non-invertible symmetries of 2+1d generalized cluster states}, we identify the symmetry structure of the generalized cluster models.
As we mentioned above, the symmetry of the model is described by a group-theoretical fusion 2-category $\mathcal{C}(G; H)$.
Some of the symmetry operators are constructed as $\mathsf{D}_H U_g \overline{\mathsf{D}}_H$, where $\mathsf{D}_H$ is the gauging operator, $\overline{\mathsf{D}}_H$ is the ungauging operator, and $U_g$ is the on-site symmetry operator of the ungauged model.
The other symmetry operators are obtained by condensing some line operators on the above symmetry operators.
By construction, these operators have a natural tensor network representation.
In addition, we also compute the partial action of the symmetry operators $\mathsf{D}_H U_g \overline{\mathsf{D}}_H$ on a subregion of the system, and reveal the fractionalization of the symmetry operators on the boundary of the region.
In the subsequent section, the fractionalized symmetry operators will serve as building blocks of the symmetry operators at the interface of generalized cluster states.
The general computation is illustrated with an example of the $G$-cluster model in Section~\ref{sec: Example: symmetry of the G-cluster state}.

Along the way, we observe that the bond dimensions of the symmetry operators $\mathsf{D}_H U_g \overline{\mathsf{D}}_H$ can be reduced by using the fact that the symmetry is non-anomalous.
This may be viewed as the ``onsiteability" of non-anomalous non-invertible symmetries in 2+1 dimensions in the same vein as~\cite{Meng:2024nxx}.

\vspace*{\baselineskip}
\noindent{\bf Interfaces.}
In Section~\ref{sec: Interfaces of 2+1d generalized cluster states},
we study the symmetry structure at the interface---the strip 2-algebra---between two generalized cluster states.
We first define the strip 2-algebra as an endomorphism category of an object of a suitable 2-category.
As we will see, in general, the strip 2-algebra is not a fusion category but a multifusion category.
We then analyze the 2-category of modules over the strip 2-algebra, which physically corresponds to the 2-category of gapped interface modes.
We provide a (partly conjectural) description of the 2-category of modules as a functor category, and establish formulas for computing the category of interface modes.
We also give tensor network representations of the symmetry operators at the interface and compute their fusion rule, which agrees with that of the strip 2-algebra.
Using this fusion rule, we show that the interface between the generalized cluster states in different SPT phases must be degenerate, establishing a generalized version of the bulk-boundary correspondence.
This degeneracy cannot be lifted unless the symmetry is explicitly broken, even if the interface is gapless.
This is in contrast to the case of invertible symmetries, where the edge modes can be non-degenerate when they are gapless.
We also exemplify the general result with the $G$-cluster model; see Section~\ref{Example: gapped boundary of the G-cluster state CT} for a category-theoretical point of view and Section~\ref{sec: Example: gapped boundary of the G-cluster state} for a tensor network point of view.

\vspace*{\baselineskip}
\noindent{\bf Parameterized families.}
In Section \ref{sec: Parameterized families},
we construct parameterized families of the generalized cluster states and study the generalized Thouless pumps.
To construct the families, we employ the method of symmetry interpolation, which produces an $S^{1}$-parameterized family that preserves the non-invertible symmetry of the generalized cluster state.
More concretely, we first construct $S^1$-parameterized families of $G/K$ symmetry-breaking states and then gauge the subgroup $H$ to obtain $S^1$-families of generalized cluster states.
We then perform an adiabatic evolution to show that our parameterized families are non-trivial.
More specifically, we show that the adiabatic evolution along the parameter space $S^1$ pumps a non-trivial self-interface mode, which is in a non-trivial representation of the strip 2-algebra constructed in Section \ref{sec: Interfaces of 2+1d generalized cluster states}.
As a simple example, we also discuss the case of the $G$-cluster model in detail in Section~\ref{sec: Example: parameterized families of the G-cluster states}.

\vspace*{\baselineskip}
To illustrate the above general results, we finally present two classes of concrete examples in Sections \ref{sec: Gx2RepG SPT phases} and \ref{sec: Tambara-Yamagami SPT phases}.
The first class consists of general SPT phases that preserve the $2\mathrm{Rep}(G) \boxtimes 2\Vect_{G}$ symmetry; the $G$-cluster state mentioned above is an example of this class.
The second class consists of SPT phases that preserve the Tambara-Yamagami fusion 2-category symmetry. 
For each class, we explicitly construct the tensor network representation of the ground state for each SPT phase and analyze the interface symmetries and parameterized families.

Various technical details are relegated to the appendices.
In Appendix \ref{sec: Remarks on the choice of representatives}, 
we discuss the dependence of the generalized cluster states on the choice of gauge fixing.
In Appendix \ref{sec: Condensation},
we discuss condensation surfaces and the condensation of line operators on them.
In Appendix \ref{sec: Derivations},
we provide detailed derivations of several technical results used in the main text.
In Appendix \ref{sec: Uniqueness of action tensors},
we prove the uniqueness of the action tensors derived in Section \ref{sec: Non-invertible symmetries of 2+1d generalized cluster states}.
In Appendix \ref{sec: Multifusion categories},
we briefly review the basics of multifusion categories.
In Appendix \ref{sec: Tambara-Yamagami cluster model in another gauge},
we provide an alternative gauge fixing for the Tambara-Yamagami cluster model discussed in Section \ref{sec: Tambara-Yamagami SPT phases}.
In Appendix~\ref{sec: cluster prime model}, we discuss the relation between the SPT phases of the models in \cite{Choi:2024rjm} and those of the generalized cluster models.

\subsection{Notations and conventions}
Throughout the paper, we consider lattice models defined on a square lattice with periodic boundary conditions.
The sets of plaquettes, edges, and vertices of the square lattice are denoted by $P$, $E$, and $V$, respectively.
In all tensor network diagrams in the subsequent sections, the physical legs are supposed to be oriented upwards.

The following is a list of notations that we will employ in this paper.
\begin{itemize}
\item $G$: a finite group
\item $H$: a subgroup of $G$
\item $K$: a complement of $H$ in $G$, i.e., a subgroup of $G$ that satisfies $HK = G$ and $H \cap K = \{e\}$. \\
We note that every $g \in G$ can be uniquely decomposed as
\begin{equation}
g = hk
\label{eq: unique decomposition}
\end{equation}
for $h \in H$ and $k \in K$.
\item $S_{H \backslash G}$: a set of representatives of right $H$-cosets in $G$
\item $\mathfrak{r}[g] \in S_{H \backslash G}$: the representative of the right $H$-coset that contains $g \in G$ \footnote{For $g = hk$, we have $\mathfrak{r}[g] = \mathfrak{r}[k]$. The map $\mathfrak{r}|_K: K \to S_{H \backslash G}$ is bijective due to the decomposition~\eqref{eq: unique decomposition}.}
\item $h[k] \in H$: the unique element of $H$ that satisfies $\mathfrak{r}[k] = h[k]^{-1}k$
\item $U_{g; K}(h) \in H$: the unique element of $H$ that satisfies $gh^{-1} = U_{g; K}(h)^{-1} k$ for some $k \in K$
\item $\widetilde{U}_{g; S_{H \backslash G}}(h) \in H$: the unique element of $H$ that satisfies $gh = \widetilde{U}_{g; S_{H \backslash G}}(h) r$ for some $r \in S_{H \backslash G}$
\item $\overrightarrow{X}_g$, $\overleftarrow{X}_g$: the left and right multiplications of $g \in G$ \footnote{In \cite{Fechisin:2023odt}, $\overleftarrow{X}_g$ denotes the right multiplication of $g^{-1}$.}
\item $U_g$: the on-site symmetry operator for $g \in G$ (i.e., the tensor product of $\overrightarrow{X}_g$'s on all sites)
\item $\mathsf{D}_H$, $\overline{\mathsf{D}}_H$: the gauging and ungauging operators for a subgroup symmetry $H \subset G$
\item $\mathcal{C}(G; H)$: the 2-category of $(\Vect_H, \Vect_H)$-bimodules in $2\Vect_G$ (i.e., a group-theoretical fusion 2-category obtained by gauging $H$ in $G$)
\item $V^g$: a simple object of $2\Vect_G$
\item $V_H^g = \Vect_H \boxtimes V^g \boxtimes \Vect_H$: a simple object of $\mathcal{C}(G; H)$ 
\item $\mathsf{D}[V_H^g] = \mathsf{D}_H U_g \overline{\mathsf{D}}_H$: the symmetry operator labeled by $V_H^g \in \mathcal{C}(G; H)$
\item $\ket{\text{cluster}(G; H; K)}$: a generalized cluster state with symmetry $\mathcal{C}(G; H)$
\item $\mathcal{C}_{K_1, K_2}$: the symmetry category at the interface of $\ket{\text{cluster}(G; H; K_1)}$ and $\ket{\text{cluster}(G; H; K_2)}$ (i.e., the strip 2-algebra)
\item $\phi_K[V_H^g]$, $\overline{\phi_K}[V_H^g]$: the action tensors for $\mathsf{D}[V_H^g]$ acting on $\ket{\text{cluster}(G; H; K)}$
\item $\{P_{s}\}_{s=1}^{N}$: for a unitary operator $M$ satisfying $M^{N} = \id$, $P_{s}[M]$ is the projector onto the eigenspace with eigenvalue $e^{i\frac{2\pi s}{N}}$, i.e., $P_{s}[M]:=\frac{1}{N}\sum_{n=0}^{N-1} e^{-i\frac{2\pi s}{N}n} M^{n}$.
\item $\log M$: the logarithm of an operator $M$, i.e., an operator satisfying $e^{\log M}=M$. Such an operator is not unique. However, for a unitary operator $M$ with $M^{N}=\mathrm{id}$, we always take
      \begin{equation}
       \log M = \sum_{s=0}^{N-1} i \frac{2\pi s}{N} P_{s}[M].
      \end{equation}
\end{itemize}

\section{Preliminaries}
\label{sec: Preliminaries}
In this section, we briefly recall the basics of fusion 2-categories and fiber 2-functors.
All 2-categories and 1-categories discussed in this paper are finite, semisimple, and $\mathbb{C}$-linear.

\subsection{Fusion 2-categories}
A fusion 2-category consists of various data that capture the algebraic structure of topological defects in 2+1 dimensions.
In what follows, we will briefly describe these data.
See \cite{Douglas:2018qfz} for a precise definition of a fusion 2-category.

\begin{figure}[t]
\centering
\adjincludegraphics[trim={10, 10, 10, 10}, scale=1.1]{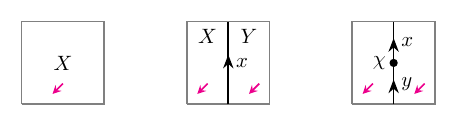}
\caption{An object $X$ corresponds to a topological surface. A 1-morphism $x: Y \to X$ corresponds to a topological line between $X$ and $Y$. A 2-morphism $\chi: y \Rightarrow x$ corresponds to a topological point operator between $x$ and $y$. The magenta arrows perpendicular to the surfaces represent their coorientation.}
\label{fig: objects and morphisms}
\end{figure}

Basic ingredients of a fusion 2-category are objects, 1-morphisms between objects, and 2-morphisms between 1-morphisms.
Physically, objects, 1-morphisms, and 2-morphisms correspond to topological surfaces, topological lines, and topological point operators in (2+1)-dimensional spacetime, cf. Figure~\ref{fig: objects and morphisms}.
A 1-morphism $x$ is said to be simple if its endomorphism space $\End(x)$ is one-dimensional. Similarly, an object $X$ is said to be simple if its identity 1-morphism $\id_X$ is simple. Physically, this means that topological surfaces and topological lines corresponding to simple objects and simple 1-morphisms do not support non-trivial topological point operators.
Any objects and 1-morphisms are assumed to be isomorphic to finite direct sums of simple ones.
Physically, simple objects and simple 1-morphisms correspond to indecomposable topological surfaces and topological lines, respectively.

The fusion of topological surfaces is given by the tensor product of the corresponding objects. 
For any pair of simple objects $X$ and $Y$, their tensor product $X \square Y$ can be decomposed into a finite direct sum of simple objects as
\begin{equation}
X \square Y \cong \bigoplus_{Z: \text{simples}} N_{XY}^Z Z,
\end{equation}
where $N_{XY}^Z$ is a non-negative integer.
The summation on the right-hand side is taken over all (isomorphism classes of) simple objects.
The unit object for this tensor product is denoted by $I$, which is supposed to be simple.
Physically, the unit object $I$ corresponds to the identity surface.

\begin{figure}[t]
\centering
\adjincludegraphics[trim={10, 10, 10, 10}, scale=1.1]{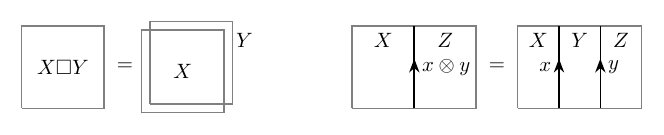}
\caption{The fusion of topological surfaces corresponds to the tensor product of objects, while the fusion of topological lines corresponds to the composition of 1-morphisms.}
\label{fig: fusion}
\end{figure}

Similarly, the fusion of topological lines is given by the composition of the corresponding 1-morphisms.
For any pair of composable 1-morphisms $x: Y \to X$ and $y: Z \to Y$, their composition $x \otimes y: Z \to X$ can be decomposed into a finite direct sum of simple 1-morphisms.
We note that the fusion of 1-endomorphisms of $X$ is again a 1-endomorphism of $X$.
Accordingly, 1-endomorphisms of $X$ (and the 2-morphisms between them) form a multifusion category $\End(X)$.\footnote{A multifusion category is a finite semisimple $\mathbb{C}$-linear rigid monoidal category \cite{EGNO2015}. A fusion category is a multifusion category whose unit object is simple.}
By definition, $\End(X)$ is fusion if and only if $X$ is simple.
See Figure~\ref{fig: fusion} for the diagrammatic representation of the fusion of surfaces and lines.
The fusion of topological point operators is also defined similarly.

As in the case of fusion 1-categories, the associativity of the tensor product is captured by a natural family of 1-isomorphisms called associators:
\begin{equation}
\alpha_{X, Y, Z}: (X \square Y) \square Z \to X \square (Y \square Z).
\end{equation}
The associators satisfy the pentagon equation up to a natural family of 2-isomorphisms called pentagonators:
\begin{equation}
\adjincludegraphics[valign = c, width = 0.75\linewidth]{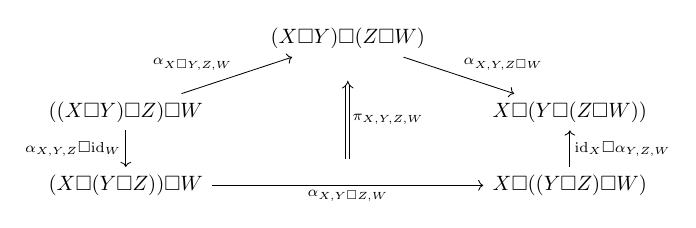}
\end{equation}
The pentagonators must satisfy appropriate coherence conditions, which are represented by the commutativity of the Stasheff polytope $K_5$ \cite{KV1994, gordon1995coherence, Gurski2007}.
The pentagonators are encoded in the collection of complex numbers called 10-j symbols \cite{Douglas:2018qfz}, which are higher-dimensional generalizations of $F$-symbols in fusion 1-categories.
Throughout most of this paper, we will not explicitly use the data of the associators and pentagonators.

\vspace*{\baselineskip}
\noindent{\bf Connected components.}
Simple objects $X$ and $Y$ are said to be connected to each other if and only if there exists a non-zero 1-morphism between them.
We note that $X$ and $Y$ can be connected even if they are not isomorphic to each other.
As a 2-category, a fusion 2-category $\mathcal{C}$ can be decomposed into a finite direct sum of connected components as
\begin{equation}
\mathcal{C} \cong \bigoplus_{i = 1, \cdots, n} \mathcal{C}_i, \qquad n \in \mathbb{Z}_{\geq 1}.
\end{equation}
All simple objects in the same component are connected to each other, while any simple objects in different components are not connected.
Physically, two simple objects are connected if and only if they are related by condensation \cite{Kong:2014qka, Gaiotto:2019xmp, Johnson-Freyd:2020ivj, Kong:2024ykr}.
That is, for any pair of simple objects $X$ and $Y$ in the same connected component, the topological surface corresponding to $X$ is obtained by condensing some line operators on the surface corresponding to $Y$, and vice versa.

\vspace*{\baselineskip}
\noindent{\bf Examples.}
Let us provide some examples of fusion 2-categories that are relevant to later discussions. 
Further examples will be discussed in Sections \ref{sec: Group-theoretical fusion 2-categories} and \ref{sec: Example: Tambara-Yamagami fusion 2-categories}.
\begin{itemize}
\item {\bf Non-anomalous invertible 0-form symmetry $2\Vect_G$.} The simplest non-trivial example of a fusion 2-category is the 2-category $2\Vect_G$ of finite semisimple $G$-graded 1-categories, where $G$ is a finite group.
This 2-category describes a non-anomalous invertible 0-form symmetry $G$ in 2+1 dimensions.\footnote{More precisely, $2\Vect_G$ describes the product of the 0-form symmetry $G$ and a 2-form symmetry $\mathrm{U}(1)$ generated by the identity point operator. The 2-form symmetry $\mathrm{U}(1)$ exists in any quantum system in 2+1d.}
Objects of $2\Vect_G$ are finite semisimple $G$-graded categories, 1-morphisms are functors that preserve the $G$-grading, and 2-morphisms are natural transformations between them.
Simple objects of $2\Vect_G$ are labeled by elements of $G$ and are denoted by $V^g$ for $g \in G$.
Here, $V^g$ is a $G$-graded semisimple category defined by
\begin{equation}
V^g = \bigoplus_{g^{\prime} \in G} (V^g)_{g^{\prime}}, \qquad
(V^g)_{g^{\prime}} = \delta_{g, g^{\prime}} \Vect,
\end{equation}
where $(V^g)_{g^{\prime}}$ is the componet of $V^g$ with grading $g^{\prime} \in G$.
The fusion rules of these simple objects are given by
\begin{equation}
V^g \boxtimes V^{g^{\prime}} \cong V^{gg^{\prime}},
\label{eq: fusion 2VecG}
\end{equation}
where $\boxtimes$ denotes the Deligne tensor product.
The simple objects $V^g$ and $V^{g^{\prime}}$ are connected to each other if and only if $g = g^{\prime}$.
The 1-category of 1-endomorphisms of $V^g$ is equivalent to $\Vect$, whose unique simple object is given by the identity 1-morphism of $V^g$.
Namely, we have an equivalence of 1-categories
\begin{equation}
\Hom_{2\Vect_G}(V^g, V^{g^{\prime}}) \cong \delta_{g, g^{\prime}} \Vect.
\label{eq: 2VecG Hom cat}
\end{equation}
The 10-j symbols of $2\Vect_G$ are trivial.

\item {\bf Anomalous 0-form symmetry $2\Vect_G^{\omega}$.} A slightly more general example of a fusion 2-category is $2\Vect_G^{\omega}$, where $\omega \in Z^4(G, \mathrm{U}(1))$ is a 4-cocycle on $G$.
This fusion 2-category describes an anomalous 0-form symmetry $G$ with an anomaly $[\omega] \in H^4(G, \mathrm{U}(1))$.
Simple objects $V^g \in 2\Vect_G^{\omega}$ are labeled by elements of $G$ and obey the fusion rules~\eqref{eq: fusion 2VecG}.
As in the case of $2\Vect_G$, the 1-category of 1-morphisms between $V^g$ and $V^{g^{\prime}}$ is given by \eqref{eq: 2VecG Hom cat}.
The 10-j symbols of $2\Vect_G^{\omega}$ are given by the 4-cocycle $\omega$.
When $\omega$ is trivial, $2\Vect_G^{\omega}$ reduces to $2\Vect_G$.

\item {\bf Non-anomalous non-invertible 1-form symmetry $2\Rep(G)$.}
Another example is the fusion 2-category $2\Rep(G)$ that describes the condensation completion of a non-anomalous non-invertible 1-form symmetry $\Rep(G)$ \cite{Douglas:2018qfz, Gaiotto:2019xmp}.\footnote{More precisely, $2\Rep(G)$ is defined as the 2-category of 2-representations of $G$ and is monoidally equivalent to the 2-category of module 1-categories over $\Rep(G)$ \cite[Lemma 1.3.8]{Decoppet2023Morita}. The latter 2-category is also equivalent to the condensation completion of $\Rep(G)$ \cite{Douglas:2018qfz}.}
Here, $\Rep(G)$ denotes the category of finite dimensional representations of $G$.
The 1-form symmetry defects correspond to 1-endomorphisms of the unit object $I \in 2\Rep(G)$ and form a symmetric fusion 1-category
\begin{equation}
\End_{2\Rep(G)}(I) \cong \Rep(G).
\end{equation}
Every simple object of $2\Rep(G)$ is obtained from the unit object $I$ by condensing a set of line operators labeled by objects of $\Rep(G)$.
In particular, any simple object of $2\Rep(G)$ is connected to $I$.
The condensed line operators form a special symmetric Frobenius algebra $A \in \Rep(G)$, and its condensation gives rise to an object denoted by $I[A] \in 2\Rep(G)$.
The object $I[A]$ is simple if $A$ is indecomposable as an algebra in $\Rep(G)$. The tensor product of two objects is given by
\begin{equation}
I[A_1] \square I[A_2] = I[A_1 \otimes A_2],
\end{equation}
where $\otimes$ on the right-hand side denotes the tensor product in $\Rep(G)$.
We note that $A_1 \otimes A_2$ has a natural algebra structure because $\Rep(G)$ is braided.
The fusion rules of $2\Rep(G)$ are computed more explicitly in \cite{Greenough:0911.4979}.

\end{itemize}

\subsection{Group-theoretical fusion 2-categories}
\label{sec: Group-theoretical fusion 2-categories}
Basic examples of fusion 2-categories are those that arise from gauging subgroups of (0-form) finite group symmetries in 2+1 dimensions.
Such fusion 2-categories are called group-theoretical fusion 2-categories.
In this subsection, we provide a brief review of this class of fusion 2-categories following \cite{Decoppet:2023bay}.
We refer the reader to \cite{Bartsch:2022ytj, Bhardwaj:2022maz, Delcamp:2023kew} for more physical descriptions.\footnote{See also \cite{Tachikawa:2017gyf} for the symmetries arising from the gauging of 0-form symmetries in general dimensions.}
In later sections, we will study SPT phases with these symmetries.

Let us first introduce the notion of group-theoretical fusion 2-categories from a physics perspective.
To this end, we start from a 2+1d system with a finite group symmetry $G$ that possibly has an anomaly $\omega \in Z^4(G, \mathrm{U}(1))$.
We then consider gauging a non-anomalous subgroup $H$ with a discrete torsion, which is given by a 3-cochain $\nu \in C^3(H, \mathrm{U}(1))$ such that $d\nu = \omega|_H^{-1}$.
Here, $\omega|_H$ denotes the restriction of $\omega$ to $H$.
The symmetry 2-category of the gauged model is described by a group-theoretical fusion 2-category, which is denoted by $\mathcal{C}(G, \omega; H, \nu)$.
When both $\omega$ and $\nu$ are trivial, this 2-category will be denoted by
\begin{equation}
\mathcal{C}(G; H) := \mathcal{C}(G, 1; H, 1).
\end{equation}
In the rest of the paper, we will focus on the case where $\omega$ and $\nu$ are trivial.

The symmetry category $\mathcal{C}(G; H)$ admits a particularly simple description for certain choices of $G$ and $H$.
For example, when $G$ is a semidirect product $A \rtimes_{\rho} G_0$ where $A$ is an abelian group and $\rho: G_0 \to \Aut(A)$ is a group homomorphism, we have \cite{Bartsch:2022mpm, Delcamp:2023kew, Decoppet:2023bay}
\begin{equation}
\mathcal{C}(A \rtimes_{\rho} G_0; A) \cong 2\Vect_{\mathbb{G}}, \qquad
\mathcal{C}(A \rtimes_{\rho} G_0; G_0) \cong 2\Rep(\mathbb{G}).
\end{equation}
Here, $\mathbb{G}$ is a split 2-group with $\pi_1(\mathbb{G}) = G_0$, $\pi_2(\mathbb{G}) = A$, and the $\pi_1(\mathbb{G})$-action on $\pi_2(\mathbb{G})$ given by $\rho$.
In what follows, we will give a categorical description of $\mathcal{C}(G; H)$ for more general $G$ and $H$.

Mathematically, $\mathcal{C}(G; H)$ is defined as the 2-category of $(\Vect_H, \Vect_H)$-bimodules in $2\Vect_G$.
Here, $\Vect_H$ is the 1-category of finite dimensional $H$-graded vector spaces.
We note that $\Vect_H$ is an algebra object in $2\Vect_G$, and hence $(\Vect_H, \Vect_H)$-bimodules in $2\Vect_G$ make sense.
In the form of an equality, we have\footnote{In general, the 2-category of $(A, B)$-bimodules in a fusion 2-category $\mathcal{C}$ is denoted by ${}_A \mathcal{C}_B$, where $A$ and $B$ are algebras in $\mathcal{C}$.}
\begin{equation}
\mathcal{C}(G; H) = {}_{\Vect_H} (2\Vect_G)_{\Vect_H}.
\end{equation}
Objects, 1-morphisms, and 2-morphisms of $\mathcal{C}(G, H)$ are finite semisimple $G$-graded $(\Vect_H, \Vect_H)$-bimodule 1-categories, $(\Vect_H, \Vect_H)$-bimodule functors preserving the $G$-grading, and $(\Vect_H, \Vect_H)$-bimodule natural transformations.
The tensor product in $\mathcal{C}(G, H)$ is defined by the relative tensor product over $\Vect_H$, which is denoted by $\boxtimes_{\Vect_H}$.
The unit object is given by the regular bimodule $\Vect_H$.
The connected components of $\mathcal{C}(G, H)$ are in one-to-one correspondence with double $H$-cosets in $G$.
Namely, $\mathcal{C}(G; H)$ can be decomposed into a direct sum of connected components as
\begin{equation}
\mathcal{C}(G; H) \cong \bigoplus_{x \in H \backslash G / H} \mathcal{C}(G; H)_x.
\end{equation}
The connected component $\mathcal{C}(G; H)_x$ for $x \in H \backslash G / H$ contains simple objects of the form 
\begin{equation}
V_H^g := \Vect_H \boxtimes V^g \boxtimes \Vect_H, \quad \forall g \in x.
\label{eq: simple obj CGH}
\end{equation}
We note that $V^g_H \cong V^{g^{\prime}}_H$ for all $g, g^{\prime} \in x$ because $\Vect_H \boxtimes V^h \cong V^h \boxtimes \Vect_H \cong \Vect_H$ for all $h \in H$.
The fusion rules of the simple objects $V^g_H$ and $V^{g^{\prime}}_H$  can be computed as
\begin{equation}
V_H^g \boxtimes_{\Vect_H} V_H^{g^{\prime}} \cong \bigoplus_{g^{\prime \prime} \in gHg^{\prime}} V_H^{g^{\prime \prime}},
\label{eq: fusion rules VHg}
\end{equation}
See, e.g., \cite{ENO2010, Greenough:0911.4979, Douglas:1406.4204} for the relative tensor product of module categories over a monoidal category.

By definition, any simple object in $\mathcal{C}(G; H)_x$ is connected to $V^g_H$ for $g \in x$.
Physically, this means that the topological surface operator labeled by any simple object of $\mathcal{C}(G; H)_x$ can be obtained by condensing some line operators on the surface operator labeled by $V^g_H$.
The line operators on the surface labeled by $V^g_H$ are labeled by 1-endomorphisms of $V^g_H$, which form a fusion 1-category \cite{Decoppet:2023bay}\footnote{More precisely, in \cite{Decoppet:2023bay}, it was shown that $\End_{\mathcal{C}(G; H)}(V^g_H)$ is monoidally equivalent to $\Vect_{H \cap gHg^{-1}}$. This is also monoidally equivalent to $\Vect_{H \cap g^{-1}Hg}$ because $H \cap gHg^{-1}$ is isomorphic to $H \cap g^{-1}Hg$.}
\begin{equation}
\End_{\mathcal{C}(G; H)}(V^g_H) \cong \Vect_{H \cap g^{-1}Hg}.
\end{equation}
For example, the line operators on the condensation defect labeled by $V_H^e$ form $\Vect_H$.
Condensing an algebra object $\mathbb{C}[H] \in \Vect_H$ on this condensation defect gives us back the identity surface, which corresponds to the unit object of $\mathcal{C}(G; H)$.

\vspace*{\baselineskip}
\noindent{\bf Example: $2\Rep(G)$ as a group-theoretical fusion 2-category.}
The fusion 2-category $2\Rep(G)$ introduced in the previous subsection is an example of a group-theoretical fusion 2-category.
More specifically, there is an equivalence of fusion 2-categories \cite[Example 3.1.2]{Decoppet:2023bay} (see also \cite{Delcamp:2021szr})
\begin{equation}
2\Rep(G) \cong \mathcal{C}(G; G).
\end{equation}
Physically, this equivalence implies that the $2\Rep(G)$ symmetry is obtained by starting from a $G$-symmetric model in 2+1d and gauging the full symmetry $G$ \cite{Bhardwaj:2022lsg, Bartsch:2022mpm}.\footnote{The invertible part of $2\Rep(G)$ symmetry was studied also in \cite{Barkeshli:2022edm}.}
In this context, the non-invertible 1-form symmetry $\Rep(G) \cong \End_{2\Rep(G)}(I)$ can be understood as the symmetry generated by the Wilson lines in the gauged theory.
The simple objects of the form $V^g_H$ for $g \in G$ are all isomorphic to each other because all $g \in G$ are in the same double $G$-coset in $G$.
The corresponding symmetry operator is obtained by condensing the Wilson line for the regular representation $\mathbb{C}[G]^* \in \Rep(G)$, which is equipped with the structure of the dual group algebra.

\subsection{Fiber 2-functors}
\label{sec: Fiber 2-functors}
A fiber 2-functor of a fusion 2-category $\mathcal{C}$ is a tensor 2-fucntor\footnote{Here, a 2-functor means a weak 2-functor (i.e., a pseudofunctor), just as a 2-category in this paper means a weak 2-category (i.e., a bicategory).}
\begin{equation}
F: \mathcal{C} \to 2\Vect,
\end{equation}
where $2\Vect$ is the 2-category of finite semisimple 1-categories.
In \cite{Decoppet:2023bay}, it was shown that a fusion 2-category $\mathcal{C}$ admits a fiber 2-functor only when $\mathcal{C}$ is group-theoretical, i.e., $\mathcal{C} = \mathcal{C}(G, \omega; H, \nu)$ for some $G$, $\omega$, $H$, and $\nu$.
The classification of fiber 2-functors of $\mathcal{C}(G, \omega; H, \nu)$ was given in \cite[Corollary 4.1.4]{Decoppet:2023bay}.
In particular, when $\omega$ and $\nu$ are trivial, a fiber 2-functor of $\mathcal{C}(G; H)$ is labeled by a pair $(K, \lambda)$, where $K$ is a subgroup of $G$ that satisfies
\begin{equation}
HK = G, \quad H \cap K = \{e\},
\label{eq: ZS product}
\end{equation}
and $\lambda \in Z^3(K, \mathrm{U}(1))$ is a 3-cocycle on $K$.
A subgroup $K$ that satisfies~\eqref{eq: ZS product} is called a complement of $H$ in $G$.
Fiber 2-functors labeled by $(K, \lambda)$ and $(K^{\prime}, \lambda^{\prime})$ are equivalent to each other if and only if there exists an element $g \in G$ such that $K^{\prime} = gKg^{-1}$ and $\lambda(k_1, k_2, k_3)/\lambda^{\prime}(gk_1g^{-1}, gk_2g^{-1}, gk_3g^{-1})$ is a 3-coboundary on $K$.
The classification of fiber 2-functors for general $\omega$ and $\nu$ is similar; see \cite[Corollary 4.1.4]{Decoppet:2023bay} for more details.

\vspace*{\baselineskip}
\noindent{\bf Classification of SPT phases.}
In 1+1 dimensions, it is known that bosonic SPT phases with fusion 1-category symmetry $\mathcal{D}$ are in one-to-one correspondence with (isomorphism classes of) fiber functors of $\mathcal{D}$ \cite{Thorngren:2019iar,Komargodski:2020mxz}.
Similarly, it is natural to expect that 2+1d bosonic SPT phases with fusion 2-category symmetry $\mathcal{C}$ are in one-to-one correspondence with (isomorphism classes of) fiber 2-functors of $\mathcal{C}$.
Here, an SPT phase is an invertible gapped phase that can be continuously deformed into the trivial phase if we break the symmetry explicitly.
All SPT phases defined this way are non-chiral, i.e., they admit gapped boundaries, which may or may not preserve the symmetry.

To argue the correspondence between fiber 2-functors and SPT phases, we postulate that the symmetry $\mathcal{C}$ of a topological field theory $\mathcal{T}$, which is the low-energy limit of a gapped phase, is implemented by a tensor 2-functor from $\mathcal{C}$ to the 2-category $\mathcal{C}_{\mathcal{T}}$ of topological defects in $\mathcal{T}$.
When $\mathcal{T}$ is invertible, the 2-category $\mathcal{C}_{\mathcal{T}}$ is equivalent to $2\Vect$.\footnote{More generally, when $\mathcal{T}$ is the Turaev-Viro-Barrett-Westbury TFT constructed from a fusion 1-category $\mathcal{D}$ \cite{Turaev:1992hq, Barrett:1993ab}, $\mathcal{C}_{\mathcal{T}}$ is equivalent to the 2-category $\mathrm{Bimod}(\mathcal{D}, \mathcal{D})$ of $(\mathcal{D}, \mathcal{D})$-bimodule categories \cite{Fuchs:2012dt, Kitaev:2011dxc, Meusburger:2022zul}. The invertible TFT corresponds to the case of $\mathcal{D} = \Vect$.}
Therefore, the symmetry $\mathcal{C}$ in the invertible gapped phase should be implemented by a fiber 2-functor of $\mathcal{C}$.
We expect that non-isomorphic fiber 2-functors correspond to different SPT phases because there is no continuous deformation between such fiber 2-functors.
Assuming in addition that isomorphic fiber 2-functors correspond to the same phase, we conclude that 2+1d SPT phases with symmetry $\mathcal{C}$ are in one-to-one correspondence with isomorphism classes of fiber 2-functors of $\mathcal{C}$.

We emphasize that the finite non-invertible symmetry of a 2+1d SPT phase must be group-theoretical because a fusion 2-category admits a fiber 2-functor only when it is group-theoretical.\footnote{The converse is not true: not every group-theoretical fusion 2-category admits an SPT phase. The condition for the group-theoretical non-invertible symmetries to admit an SPT phase in general dimension was studied in \cite{Hsin:2025ria}.}
Therefore, to study SPT phases in 2+1d, it suffices to consider group-theoretical fusion 2-categories $\mathcal{C}(G, \omega; H, \nu)$.
In this paper, we restrict ourselves to the case where $\omega$ and $\nu$ are trivial.

According to the classification of fiber 2-functors, 2+1d SPT phases with symmetry $\mathcal{C}(G; H)$ should be labeled by the equivalence classes of pairs $(K, \lambda)$.
Physically, these SPT phases are obtained by starting from a $G$-symmetric gapped phase that spontaneously breaks $G$ down to $K$ and gauging the subgroup symmetry $H$.
The 3-cocycle $\lambda$ specifies the SPT class for the unbroken symmetry $K$ before gauging.
More precisely, $K$ is the unbroken symmetry of one of the ground states. The other ground states generally have different unbroken symmetries that are conjugate to $K$. Thus, the phase depends only on the conjugacy class of $(K, \lambda)$, which agrees with the equivalence class of fiber 2-functors.
In later sections, we will study lattice models for all of these SPT phases with trivial $\lambda$.

\vspace*{\baselineskip}
\noindent{\bf Example: SPT phases with non-invertible 1-form symmetry.}
A non-invertible 1-form symmetry in 2+1d is generally described by a braided fusion 1-category $\mathcal{B}$.
Its condensation completion is $\Mod(\mathcal{B})$, the 2-category of $\mathcal{B}$-module categories.
It is known that a fusion 2-category $\Mod(\mathcal{B})$ is group-theoretical if and only if $\mathcal{B} = \Rep(G)$ for some $G$ \cite[Theorem 3.1.3]{Decoppet:2023bay}, i.e.,
\begin{equation}
\Mod(\mathcal{B}) = \Mod(\Rep(G)) \cong 2\Rep(G).
\end{equation}
The fiber 2-functor of $2\Rep(G)$ is unique and is given by the forgetful functor, which corresponds to $(K, \lambda) = (\{e\}, 1)$.
Physically, this means that there is no non-trivial SPT phase with only a 1-form symmetry.

\subsection{Example: Tambara-Yamagami 2-categories}
\label{sec: Example: Tambara-Yamagami fusion 2-categories}
Let us consider simple examples of group-theoretical fusion 2-categories known as Tambara-Yamagami 2-categories \cite{Decoppet:2023bay}.
These are 2-categorical analogues of Tambara-Yamagami 1-categories \cite{TY1998}.
We recall that a Tambara-Yamagami 1-category is a $\mathbb{Z}_2$-graded fusion 1-category $\mathcal{D} = \mathcal{D}_0 \oplus \mathcal{D}_1$, where $\mathcal{D}_0 = \Vect_A$ for a finite abelian group $A$ and $\mathcal{D}_1 = \Vect$.
Similarly, a Tambara-Yamagami 2-category is defined as a $\mathbb{Z}_2$-graded fusion 2-category $\mathcal{C} = \mathcal{C}_0 \oplus \mathcal{C}_1$, where $\mathcal{C}_0 = 2\Rep(A) \boxtimes 2\Vect_A$ for a finite abelian group $A$ and $\mathcal{C}_1 = 2\Vect$.
It was shown in \cite{Decoppet:2023bay} that any Tambara-Yamagami 2-category is group-theoretical and is of the form
\begin{equation}
2\TY(A, \zeta) := \mathcal{C}(A \wr \mathbb{Z}_2, \zeta; A_L, \nu).
\end{equation}
Here, $A \wr \mathbb{Z}_2 := (A \times A) \rtimes \mathbb{Z}_2$ is the wreath product of $A$ by $\mathbb{Z}_2$, $\zeta$ is a 4-cocycle on $A \wr \mathbb{Z}_2$ that is trivializable on $A \times A \times \{0\} \subset A \wr \mathbb{Z}_2$, and $A_L$ is a subgroup of $A \wr \mathbb{Z}_2$ defined by 
\begin{equation}
A_L := \{(a, e, 0) \mid a \in A\},
\label{eq: AL}
\end{equation}
where $e \in A$ and $0 \in \mathbb{Z}_2$ denote the unit elements of $A$ and $\mathbb{Z}_2$, respectively.
The above 2-category $2\TY(A, \zeta)$ does not depend on the choice of a 3-cochain $\nu \in C^3(A_L, \mathrm{U}(1))$ that satisfies $d\nu = \zeta|_{A_L}^{-1}$.
The group structure on $A \wr \mathbb{Z}_2$ is given by
\begin{equation}
(a_l, a_r, n) \cdot (b_l, b_r, m) =
\begin{cases}
(a_l b_l, a_r b_r, n + m \bmod 2) \quad & \text{when $n = 0$},\\
(a_l b_r, a_r b_l, n + m \bmod 2) \quad & \text{when $n = 1$}.
\end{cases}
\label{eq: wreath prod}
\end{equation}
The Tambara-Yamagami 2-category $2\TY(A, \zeta)$ is indeed a $\mathbb{Z}_2$-graded fusion 2-category
\begin{equation}
2\TY(A, \zeta) = 2\TY(A, \zeta)_0 \oplus 2\TY(A, \zeta)_1,
\end{equation}
where $2\TY(A, \zeta)_0 \cong 2\Rep(A) \boxtimes 2\Vect_A$ and $2\TY(A, \zeta)_1 \cong 2\Vect$.
The unique simple object with a non-trivial grading is given by $D := V_{A_L}^{(e, e, 1)}$.
Due to \eqref{eq: fusion rules VHg}, $D$ obeys the fusion rule \cite{Decoppet:2023bay}
\begin{equation}
D \boxtimes_{\Vect_{A_L}} D \cong \bigoplus_{a \in A} V_{A_L}^{(e, a, 0)} \cong \bigoplus_{a \in A} V_A^e \boxtimes V^a,
\end{equation}
where $V_A^e \boxtimes V^a \in 2\Rep(A) \boxtimes 2\Vect_A \cong 2\TY(A, \zeta)_0$.
The above fusion rule is reminiscent of the fusion rule of the non-invertible object of a Tambara-Yamagami 1-category \cite{TY1998}.

Physically, $2\TY(A, \zeta)$ describes the symmetry of a 2+1d system that is invariant under gauging the 2-group symmetry $A^{[0]} \times A^{[1]}$, where the superscripts specify the form degrees.
In Section~\ref{sec: Tambara-Yamagami SPT phases}, we will study 2+1d SPT phases with symmetry $2\TY(A, 1)$, i.e., the Tambara-Yamagami symmetry with trivial $\zeta$.
See Section~\ref{sec: Classification TY} for the classification of fiber 2-functors of these fusion 2-categories.

\vspace*{\baselineskip}
\noindent{\bf Example: $A = \mathbb{Z}_2$.}
When $A = \mathbb{Z}_2$, the wreath product $A \wr\mathbb{Z}_2$ is isomorphic to $D_8$, the dihedral group of order 8.
Thus, the Tambara-Yamagami 2-category for $A = \mathbb{Z}_2$ with trivial $\zeta$ can be written as
\begin{equation}
2\TY(\mathbb{Z}_2, 1) = \mathcal{C}(D_8; \mathbb{Z}_2),
\end{equation}
where $\mathbb{Z}_2$ on the right-hand side is a non-normal subgroup of $D_8$.
This Tambara-Yamagami 2-category is monoidally equivalent to the 2-category of 2-representations of a 2-group $(\mathbb{Z}_2 ^{[1]} \times \mathbb{Z}_2^{[1]}) \rtimes \mathbb{Z}_2^{[0]}$, i.e., we have
\begin{equation}
2\TY(\mathbb{Z}_2, 1) \cong 2\Rep((\mathbb{Z}_2 ^{[1]} \times \mathbb{Z}_2^{[1]}) \rtimes \mathbb{Z}_2^{[0]}).
\end{equation}
This symmetry has been studied in both continuum field theories \cite{Kaidi:2021xfk, Bartsch:2022mpm, Bhardwaj:2022maz, Bartsch:2022ytj} and lattice models \cite{Choi:2024rjm, Cao:2025qhg, Furukawa:2025flp}.
We note that not every Tambara-Yamagami 2-category is monoidally equivalent to the 2-category of 2-representations of a finite 2-group.
In particular, when the order of $A$ is odd, $2\TY(A, 1)$ cannot be written as the 2-category of 2-representations of any finite 2-group \cite{Decoppet:2023bay}.

\section{Generalized cluster states in 2+1d}
\label{sec: 2+1d generalized cluster states with non-invertible symmetries}

In this section, we define concrete lattice models for 2+1d SPT phases with group-theoretical non-invertible symmetry $\mathcal{C}(G; H)$.
As discussed in Section~\ref{sec: Fiber 2-functors}, such SPT phases are expected to be classified by equivalence classes of pairs $(K, \lambda)$, where $K$ is a complement of $H$ in $G$ (i.e., a subgroup of $G$ that satisfies \eqref{eq: ZS product}) and $\lambda \in Z^3(K, \mathrm{U}(1))$ is a 3-cocycle on $K$.
Our lattice models realize all of these SPT phases with trivial $\lambda$.
We refer the reader to~\cite{Inamura:2025cum} for more general gapped lattice models with more general fusion 2-category symmetries.\footnote{The lattice models constructed in \cite{Inamura:2025cum} include those that realize all SPT phases with arbitrary group-theoretical fusion 2-category symmetry $\mathcal{C}(G, \omega; H, \nu)$. However, the properties of these phases were not studied in detail. When all cocycle data, including $\lambda$, are trivial, the tensor network representation of the ground states given in \cite[Section 7.3.3]{Inamura:2025cum} simplifies significantly, which makes the analyses in the later sections much easier.}
The ground states of our models are natural generalizations of the cluster state, and hence we call them generalized cluster states.
We will provide explicit tensor network representations of these states.

\subsection{Generalized cluster models}
\label{sec: Generalized cluster models}
\begin{figure}[t]
\centering
\adjincludegraphics[trim={10, 10, 10, 10}]{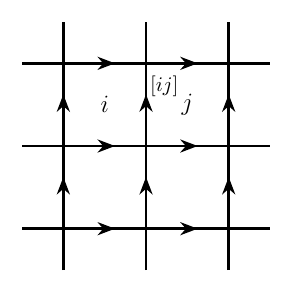}
\caption{The square lattice on which our model is defined. We impose periodic boundary conditions in both directions. The orientation of the edges is fixed as shown above.}
\label{fig: lattice}
\end{figure}

Our lattice model is defined by gauging a subgroup $H$ in a $G$-symmetric model that spontaneously breaks the symmetry $G$ down to $K$.
By construction, the gauged model realizes a $\mathcal{C}(G; H)$-symmetric SPT phase labeled by $(K, 1)$ if $K$ satisfies \eqref{eq: ZS product}.
For concreteness, we will consider the model on a square lattice with periodic boundary conditions and fix the orientation of the edges as shown in Figure~\ref{fig: lattice}.
In what follows, $P$, $E$, and $V$ denote the set of plaquettes, the set of edges, and the set of vertices of the square lattice.
The edge between the neighboring plaquettes $i, j \in P$ is denoted by $[ij] \in E$, where $i$ is on the left side of $[ij]$ (cf. Figure~\ref{fig: lattice}).

\subsubsection{$G$-symmetric model}
Let us first define the $G$-symmetric model whose symmetry is spontaneously broken down to $K$.
The state space of the model is given by
\begin{equation}
\mathcal{H} = \bigotimes_{i \in P} \mathbb{C}^{|G|}.
\end{equation}
Namely, we have a $|G|$-dimensional qudit on each plaquette.
The basis of the local state space on each plaquette $i$ is denoted by $\{\ket{g}_i \mid g \in G\}$.
We will use the following shorthand notation for the basis of the total state space:
\begin{equation}
\ket{\{g_i\}} := \bigotimes_{i \in P} \ket{g_i}_i.
\end{equation}
The Hamiltonian of the model is given by the sum
\begin{equation}
\hat{H} = -\sum_{i \in P} \hat{\mathsf{h}}_i - \sum_{[ij] \in E} \hat{\mathsf{h}}_{ij}.
\label{eq: Ham G/K}
\end{equation}
The plaquette term $\hat{\mathsf{h}}_i$ and the edge term $\hat{\mathsf{h}}_{ij}$ are defined by
\begin{equation}
\hat{\mathsf{h}}_i \ket{g_i}_i = \frac{1}{|K|} \sum_{k \in K} \ket{g_i k}_i,
\qquad
\hat{\mathsf{h}}_{ij} \ket{g_i}_i \otimes \ket{g_j}_j = \delta_{g_i^{-1} g_j \in K} \ket{g_i}_i \otimes \ket{g_j}_j,
\end{equation}
where $\delta_{g_i^{-1} g_j \in K}$ is one if $g_i^{-1} g_j \in K$ and zero otherwise.
We note that $\hat{\mathsf{h}}_i$ acts non-trivially only on the plaquette $i$, while $\hat{\mathsf{h}}_{ij}$ acts non-trivially only on the neighboring plaquettes $i$ and $j$.
The above model has a $G$ symmetry, whose symmetry operators are defined by
\begin{equation}
U_g \ket{\{g_i\}} = \ket{\{g g_i\}}, \quad \forall g \in G.
\end{equation}
It is straightforward to check that $\hat{H}$ commutes with $U_g$ for all $g \in G$.

The Hamiltonian~\eqref{eq: Ham G/K} consists of local commuting projectors, i.e., each local term satisfies
\begin{equation}
[\hat{\mathsf{h}}_i, \hat{\mathsf{h}}_j] = [\hat{\mathsf{h}}_i, \hat{\mathsf{h}}_{jk}] = [\hat{\mathsf{h}}_{ij}, \hat{\mathsf{h}}_{kl}] = 0, \qquad
\hat{\mathsf{h}}_i^2 = \hat{\mathsf{h}}_{ij}^2 = 1.
\end{equation}
Thus, the model is eactly solvable.
The ground states are obtained by acting with the projector $\prod_{i \in P} \hat{\mathsf{h}}_i \prod_{[ij] \in E} \hat{\mathsf{h}}_{ij}$ on generic states in $\mathcal{H}$.
One can immediately see that the (normalized) ground states are explicitly given by
\begin{equation}
\ket{\text{GS}; \mu} := \bigotimes_{i \in P} \left( \frac{1}{\sqrt{|K|}} \sum_{g \in x_{\mu}} \ket{g}_i \right), \qquad \mu = 1, 2, \cdots, |G/K|,
\label{eq: GS G/K}
\end{equation}
where $x_{\mu}$ is a left $K$-coset in $G$, i.e., $x_{\mu} = \{r_{\mu} k \mid k \in K\}$ with $r_{\mu} \in G$ a representative of $x_{\mu}$.

The above ground states spontaneously break the symmetry $G$ down to $K$.
More precisely, the symmetry action on the ground states~\eqref{eq: GS G/K} is given by
\begin{equation}
U_g \ket{\text{GS}; \mu} = \ket{\text{GS}; \nu},
\end{equation}
where $g = g_{\nu} g_{\mu}^{-1}$ for any $g_{\mu} \in x_{\mu}$ and $g_{\nu} \in x_{\nu}$.
In particular, each vacuum $\ket{\text{GS}; \mu}$ preserves the subgroup symmetry $g_{\mu} K g_{\mu}^{-1}$, which is isomorphic to $K$.

\subsubsection{Gauged model}
Now, we gauge a subgroup $H$ in the above $G$-symmetric model.
For the gauging procedure on the lattice and the corresponding gauging map, see, e.g., \cite{Haegeman:2014maa, Williamson:2017uzx, Tantivasadakarn:2022hgp, Delcamp:2023kew}.

\vspace*{\baselineskip}
\noindent{\bf State space.}
The state space of the gauged model is given by
\begin{equation}
\mathcal{H}_{\text{gauged}} = \hat{\pi}_{\text{Gauss}} \left[ \left(\bigotimes_{i \in P} \mathbb{C}^{|G|}\right) \otimes \hat{\pi}_{\text{flat}} \left(\bigotimes_{[ij] \in E} \mathbb{C}^{|H|}\right) \right],
\end{equation}
where $\hat{\pi}_{\text{Gauss}}$ and $\hat{\pi}_{\text{flat}}$ are the projectors that implement the Gauss law constraint and the flatness condition, respectively.
The details of these projectors will be explained shortly.
The dynamical degrees of freedom are $|G|$-dimensional qudits on the plaquettes and $|H|$-dimensional qudits on the edges.
The qudits on the plaquettes are called matter fields, while the qudits on the edges are called gauge fields.
The projector $\hat{\pi}_{\text{Gauss}}$ is given by the product of local commuting projectors
\begin{equation}
\hat{\pi}_{\text{Gauss}} = \prod_{i \in P} \hat{G}_i.
\end{equation}
Each projector $\hat{G}_i$ is the Gauss law operator defined by
\begin{equation}
\hat{G}_i = \frac{1}{|H|} \sum_{h \in H} \overrightarrow{X}_h^{(i)} L_h^{(\partial i)},
\end{equation}
where $\overrightarrow{X}_h^{(i)}$ and $L_h^{(\partial i)}$ are defined by
\begin{equation}
\overrightarrow{X}_h^{(i)} \ket{g}_i = \ket{hg}_i, \quad
L_h^{(\partial i)} \Ket{\adjincludegraphics[valign = c, width = 0.15\linewidth, trim={10, 10, 10, 10}]{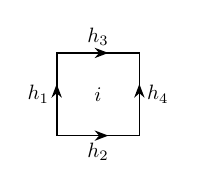}} = \Ket{\adjincludegraphics[valign = c, width = 0.15\linewidth, trim={10, 10, 10, 10}]{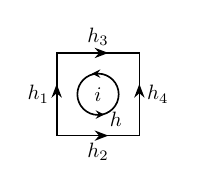}} = \Ket{\adjincludegraphics[valign = c, width = 0.185\linewidth, trim={10, 10, 10, 10}]{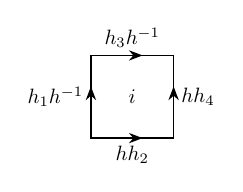}}.
\label{eq: loop op}
\end{equation}
These operators act non-trivially only on the plaquette $i$ and the edges on the boundary of $i$, respectively.
We note that the product $\overrightarrow{X}_h^{(i)} L_h^{(\partial i)}$ acts as a gauge transformation around the plaquette $i \in P$.
Gauge equivalent configurations of the dynamical variables are identified in the image of $\hat{\pi}_{\text{Gauss}}$ because
\begin{equation}
\hat{\pi}_{\text{Gauss}} = \hat{\pi}_{\text{Gauss}} \overrightarrow{X}_h^{(i)} L_h^{(\partial i)}, \quad \forall h \in H, ~ \forall i \in P.
\end{equation}
Thus, in the projected subspace, the matter fields and the gauge fields satisfy the Gauss law constraint.
Similarly, the projector $\hat{\pi}_{\text{flat}}$ is given by the product of local commuting projectors
\begin{equation}
\hat{\pi}_{\text{flat}} = \prod_{v \in V} \hat{F}_v,
\label{eq: pi flat}
\end{equation}
where each projector $F_v$ imposes the flatness condition on the gauge fields around the vertex $v \in V$:
\begin{equation}
\hat{F}_v \Ket{\adjincludegraphics[valign = c, trim={10, 10, 10, 10}]{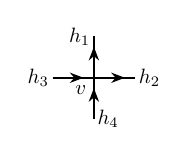}} = \delta_{h_1h_2, h_3h_4} \Ket{\adjincludegraphics[valign = c, trim={10, 10, 10, 10}]{tikz/out/vertex.pdf}}.
\end{equation}
Thus, in the image of $\hat{\pi}_{\text{flat}}$, the gauge fields satisfy the flatness condition everywhere on the square lattice.

\vspace*{\baselineskip}
\noindent{\bf Hamiltonian.}
The Hamiltonian of the gauged model is obtained by introducing the minimal coupling between the matter fields and the gauge fields.
More specifically, the gauged Hamiltonian is given by
\begin{equation}
\hat{H}_{\text{gauged}} = -\sum_{i \in P} \hat{\mathsf{h}}_i^{\text{gauged}} - \sum_{[ij] \in E} \hat{\mathsf{h}}_{ij}^{\text{gauged}},
\label{eq: gauged Ham}
\end{equation}
where $\hat{\mathsf{h}}_{i}^{\text{gauged}}$ and $\hat{\mathsf{h}}_{ij}^{\text{gauged}}$ are defined by
\begin{align}
\hat{\mathsf{h}}_i^{\text{gauged}} \ket{g_i}_i & = \frac{1}{|K|} \sum_{k \in K} \ket{g_ik}_i, \\
\hat{\mathsf{h}}_{ij}^{\text{gauged}} \ket{g_i}_i \otimes \ket{h_{ij}}_{ij} \otimes \ket{g_j}_j & = \delta_{g_i^{-1}h_{ij}g_j \in K} \ket{g_i}_i \otimes \ket{h_{ij}}_{ij} \otimes \ket{g_j}_j.
\end{align}
We note that the above operators commute with the projectors $\hat{\pi}_{\text{Gauss}}$ and $\hat{\pi}_{\text{flat}}$, i.e.,
\begin{equation}
[\hat{\mathsf{h}}_i^{\text{gauged}}, \hat{\pi}_{\text{Gauss}}] = [\hat{\mathsf{h}}_i^{\text{gauged}}, \hat{\pi}_{\text{flat}}] = [\hat{\mathsf{h}}_{ij}^{\text{gauged}}, \hat{\pi}_{\text{Gauss}}] = [\hat{\mathsf{h}}_{ij}^{\text{gauged}}, \hat{\pi}_{\text{flat}}] = 0.
\end{equation}
This implies that the Hamiltonian~\eqref{eq: gauged Ham} maps any state in $\mathcal{H}_{\text{gauged}}$ to another state in $\mathcal{H}_{\text{gauged}}$ because
\begin{equation}
\begin{aligned}
\ket{\psi} \in \mathcal{H}_{\text{gauged}} ~ &\Leftrightarrow ~ \ket{\psi} = \hat{\pi}_{\text{Gauss}} \hat{\pi}_{\text{flat}} \ket{\psi} \\
~ &\Rightarrow ~ H_{\text{gauged}} \ket{\psi} = \hat{\pi}_{\text{Gauss}} \hat{\pi}_{\text{flat}} H \ket{\psi}
~ \Leftrightarrow ~ H \ket{\psi} \in \mathcal{H}_{\text{gauged}}.
\end{aligned}
\end{equation}
Therefore, the gauged Hamiltonian is well-defined as a linear map from $\mathcal{H}_{\text{gauged}}$ to itself.

\subsubsection{Gauge fixing}
Let us fix the gauge in order to solve the model explicitly.
To this end, we first recall that the matter fields on the plaquettes can be shifted by any element of $H$ by the gauge transformation:
\begin{equation}
\Ket{\adjincludegraphics[valign = c, width = 0.15\linewidth, trim={10, 10, 10, 10}]{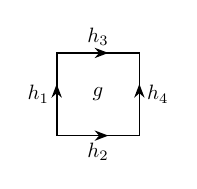}} \sim \Ket{\adjincludegraphics[valign = c, width = 0.19\linewidth, trim={10, 10, 10, 10}]{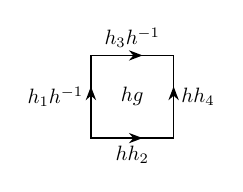}}.
\end{equation}
Here, $\ket{\psi} \sim \ket{\psi^{\prime}}$ means that $\ket{\psi}$ and $\ket{\psi^{\prime}}$ are gauge equivalent.
In other words, they are identified with each other upon applying the Gauss law projector:
\begin{equation}
\ket{\psi} \sim \ket{\psi^{\prime}} ~ \overset{\text{def}}{\Longleftrightarrow} ~ \hat{\pi}_{\text{Gauss}} \ket{\psi} = \hat{\pi}_{\text{Gauss}} \ket{\psi^{\prime}}.
\end{equation}
Using this gauge transformation, we can always shift the matter field $g_i \in G$ to a representative of the right $H$-coset $Hg_i = \{hg_i \mid h \in H\}$.
This operation is referred to as gauge fixing.
The state space after the gauge fixing can be written as
\begin{equation}
\mathcal{H}_{\text{g.f.}} = \left(\bigotimes_{i \in P} \mathbb{C}^{|S_{H \backslash G}|}\right) \otimes \hat{\pi}_{\text{flat}} \left( \bigotimes_{[ij] \in E} \mathbb{C}^{|H|} \right),
\label{eq: state space gf}
\end{equation}
where $S_{H \backslash G}$ is the set of representatives of right $H$-cosets in $G$.
The matter fields after the gauge fixing are labeled by elements of $S_{H \backslash G}$.
We note that $|S_{H \backslash G}| = |G|/|H|$.

We emphasize that no information is lost by the gauge fixing because the state spaces before and after the gauge fixing are isomorphic to each other:
\begin{equation}
\mathcal{H}_{\text{g.f.}} \cong \mathcal{H}_{\text{gauged}} .
\label{eq: gf isom}
\end{equation}
An isomorphism $\mathcal{U}_{\text{g.f.}}: \mathcal{H}_{\text{gauged}} \to \mathcal{H}_{\text{g.f.}}$ is given by
\begin{equation}
\mathcal{U}_{\text{g.f.}} (\hat{\pi}_{\text{Gauss}} \hat{\pi}_{\text{flat}} \ket{\{g_i; h_{ij}\}}) = \hat{\pi}_{\text{flat}} \ket{\{r_i; h_i^{-1}h_{ij}h_j\}},
\label{eq: gauge fixing}
\end{equation}
where $r_i \in S_{H \backslash G}$ is a representative of the right $H$-coset $Hg_i$, and $h_i \in H$ is the unique element that satisfies $g_i = h_i r_i$.
Diagrammatically, the above isomorphism can be expressed as
\begin{equation}
\mathcal{U}_{\text{g.f.}}\Ket{\adjincludegraphics[valign = c, width = 0.18\linewidth, trim={10, 10, 10, 10}]{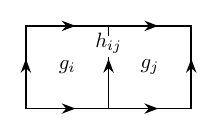}} =
\Ket{\adjincludegraphics[valign = c, width = 0.18\linewidth, trim={10, 10, 10, 10}]{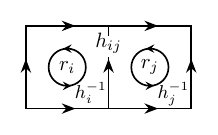}} = \Ket{\adjincludegraphics[valign = c, width = 0.18\linewidth, trim={10, 10, 10, 10}]{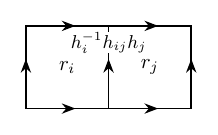}},
\end{equation}
where $g_i = h_i r_i$ and $g_j = h_j r_j$ with $h_i, h_j \in H$ and $r_i, r_j \in S_{H \backslash G}$.
The inverse $\mathcal{U}_{\text{g.f.}}^{-1}: \mathcal{H}_{\text{g.f.}} \to \mathcal{H}_{\text{gauged}}$ is given by
\begin{equation}
\mathcal{U}_{\text{g.f.}}^{-1} (\hat{\pi}_{\text{flat}} \ket{\{r_i; h_{ij}\}}) = \hat{\pi}_{\text{Gauss}} \hat{\pi}_{\text{flat}} \ket{\{r_i; h_{ij}\}}.
\end{equation}
By a direct computation, one can vertify that $\mathcal{U}_{\text{g.f.}}^{-1} \mathcal{U}_{\text{g.f.}} = \id_{\mathcal{H}_{\text{gauged}}}$ and $\mathcal{U}_{\text{g.f.}} \mathcal{U}_{\text{g.f.}}^{-1} = \id_{\mathcal{H}_{\text{g.f.}}}$.

The Hamiltonian after the gauge fixing is then given by
\begin{equation}
\hat{H}_{\text{g.f.}} = \mathcal{U}_{\text{g.f.}} \hat{H}_{\text{gauged}} \mathcal{U}_{\text{g.f.}}^{-1} = -\sum_{i \in P} \hat{\mathsf{h}}_i^{\text{g.f.}} - \sum_{[ij] \in E} \hat{\mathsf{h}}_{ij}^{\text{g.f.}}.
\label{eq: H gf}
\end{equation}
The local terms $\hat{\mathsf{h}}_i^{\text{g.f.}} := \mathcal{U}_{\text{g.f.}} \hat{\mathsf{h}}_i^{\text{gauged}} \mathcal{U}_{\text{g.f.}}^{-1}$ and $\hat{\mathsf{h}}_{ij}^{\text{g.f.}} := \mathcal{U}_{\text{g.f.}} \hat{\mathsf{h}}_{ij}^{\text{gauged}} \mathcal{U}_{\text{g.f.}}^{-1}$ can be computed as
\begin{align}
\hat{\mathsf{h}}_i^{\text{g.f.}} \Ket{\adjincludegraphics[valign = c, width = 0.1\linewidth, trim={10, 10, 10, 10}]{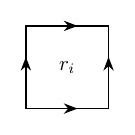}} &= \frac{1}{|K|} \sum_{k \in K} \Ket{\adjincludegraphics[valign = c, width = 0.1\linewidth, trim={10, 10, 10, 10}]{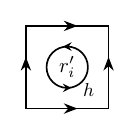}}, \label{eq: hi gf}\\
\hat{\mathsf{h}}_{ij}^{\text{g.f.}} \ket{r_i}_i \otimes \ket{h_{ij}}_{ij} \otimes \ket{r_j}_j &= \delta_{r_i^{-1}h_{ij}r_j \in K} \ket{r_i}_i \otimes \ket{h_{ij}}_{ij} \otimes \ket{r_j}_j, \label{eq: hij gf}
\end{align}
where $r_i^{\prime} \in S_{H \backslash G}$ and $h \in H$ in \eqref{eq: hi gf} are uniquely determined by
\begin{equation}
r_ik = h^{-1} r_i^{\prime}
\label{eq: r prime}
\end{equation}
for given $r_i \in S_{H \backslash G}$ and $k \in K$.

By construction, the gauged model after the gauge fixing does not depend on the choice of representatives up to isomorphism.
Nevertheless, the Hamiltonian~\eqref{eq: H gf} after the gauge fixing does depend on $S_{H \backslash G}$ because the isomorphism $\mathcal{U}_{\text{g.f.}}$ does.
Accordingly, the ground states of \eqref{eq: H gf} also depend on $S_{H \backslash G}$.
The relation between the models with different choices of representatives will be discussed in more detail in Appendix~\ref{sec: Remarks on the choice of representatives}.

\subsubsection{Generalized cluster model}
The gauged model defined above realizes an SPT phase when $K$ is a complement of $H$ in $G$, i.e., when $K$ is a subgroup of $G$ that satisfies \eqref{eq: ZS product}.
The model for such a choice of $K$ will be called a generalized cluster model, for the reason that will become clear in the next subsection.
Below, we will write down the Hamiltonian of this model.

When $K$ is a complement of $H$, any element $g \in G$ can be uniquely decomposed into the product
\begin{equation}
g = hk, \qquad h \in H, \quad k \in K.
\label{eq: g=hk}
\end{equation}
This decomposition implies that any right $H$-coset in $G$ can be written as
\begin{equation}
x[k] := Hk.
\end{equation}
Furthermore, the uniqueness of the decomposition~\eqref{eq: g=hk} implies that $x[k] = x[k^{\prime}]$ if and only if $k = k^{\prime}$.
The representative of $x[k]$ is denoted by $\mathfrak{r}[k] \in S_{H \backslash G}$, which can also be uniquely decomposed as
\begin{equation}
\mathfrak{r}[k] = h[k]^{-1} k, \qquad h[k] \in H.
\end{equation}
Using the above decomposition, one can solve \eqref{eq: r prime} explicitly as follows:
\begin{equation}
\mathfrak{r}[k_i] k = h^{-1} \mathfrak{r}[k_i^{\prime}] ~ \Leftrightarrow ~
\begin{cases}
k_i^{\prime} = k_i k, \\
h = h[k_i^{\prime}]^{-1} h[k_i].
\end{cases}
\end{equation}
This enables us to slightly simplify the expressions \eqref{eq: hi gf} and \eqref{eq: hij gf}.

We can now write down the Hamiltonian of the generalized cluster model as
\begin{equation}
\hat{H}_{\text{g.f.}} = -\sum_{i \in P} \hat{\mathsf{h}}_i^{\text{g.f.}} - \sum_{[ij] \in E} \hat{\mathsf{h}}_{ij}^{\text{g.f.}},
\label{eq: generalized cluster ham}
\end{equation}
where the local terms $\hat{\mathsf{h}}_i^{\text{g.f.}}$ and $\hat{\mathsf{h}}_{ij}^{\text{g.f.}}$ are given by
\begin{align}
\hat{\mathsf{h}}_i^{\text{g.f.}} \Ket{\adjincludegraphics[valign = c, width = 0.1\linewidth, trim={10, 10, 10, 10}]{tikz/out/hi_gf1.pdf}} & = \frac{1}{|K|} \sum_{k_i^{\prime} \in K} \Ket{\adjincludegraphics[valign = c, width = 0.11\linewidth, trim={10, 10, 10, 10}]{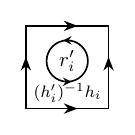}}, \label{eq: hi gf 2}\\
\hat{\mathsf{h}}_{ij}^{\text{g.f.}} \ket{r_i}_i \otimes \ket{h_{ij}}_{ij} \otimes \ket{r_j}_j & = \delta_{h_{ij}, h_i^{-1}h_j} \ket{r_i}_i \otimes \ket{h_{ij}}_{ij} \otimes \ket{r_j}_j. \label{eq: hij gf 2}
\end{align}
Here, we employed the following shorthand notations:
\begin{equation}
r_i = \mathfrak{r}[k_i], \quad r_i^{\prime} = \mathfrak{r}[k_i^{\prime}], \quad h_i = h[k_i], \quad h_i^{\prime} = h[k_i^{\prime}], \qquad \forall k_i, k_i^{\prime} \in K.
\label{eq: shorthand}
\end{equation}
The Kronecker delta in \eqref{eq: hij gf 2} comes from the equality
\begin{equation}
\delta_{r_i^{-1}h_{ij}r_j \in K} = \delta_{h_i h_{ij} h_j^{-1} \in K} = \delta_{h_{ij}, h_i^{-1}h_j},
\end{equation}
where the last equality follows from $H \cap K = \{e\}$.

\subsection{Generalized cluster states}
\label{sec: Generalized cluster states}
In this subsection, we write down the ground state of the generalized cluster model using tensor networks.
The ground state turns out to be a natural generalization of the ordinary cluster state.
Thus, we will call this state a generalized cluster state.

\vspace*{\baselineskip}
\noindent{\bf Generalized cluster states.}
To find a ground state of the generalized cluster model, we note that the Hamiltonian~\eqref{eq: generalized cluster ham} is a sum of local commuting projectors:
\begin{equation}
[\hat{\mathsf{h}}_i^{\text{g.f.}}, \hat{\mathsf{h}}_j^{\text{g.f.}}] = [\hat{\mathsf{h}}_i^{\text{g.f.}}, \hat{\mathsf{h}}_{jk}^{\text{g.f.}}] = [\hat{\mathsf{h}}_{ij}^{\text{g.f.}}, \hat{\mathsf{h}}_{kl}^{\text{g.f.}}] = 0, \qquad
(\hat{\mathsf{h}}_i^{\text{g.f.}})^2 = (\hat{\mathsf{h}}_{ij}^{\text{g.f.}})^2 = 1.
\end{equation}
Therefore, a ground state is obtained by applying the product of all local projectors to a generic state in $\mathcal{H}_{\text{g.f.}}$. i.e., we have
\begin{equation}
\ket{\text{GS}} = \prod_{i \in P} \hat{\mathsf{h}}_i^{\text{g.f.}} \prod_{[ij] \in E} \hat{\mathsf{h}}_{ij}^{\text{g.f.}} \ket{\psi}, \qquad \ket{\psi} \in \mathcal{H}_{\text{g.f.}}.
\label{eq: GS projector}
\end{equation}
We can show that the above state does not depend on $\ket{\psi}$ up to normalization, which means that the ground state is unique.
To see this, we first notice that in the state $\prod_{[ij]} \hat{\mathsf{h}}_{ij}^{\text{g.f.}} \ket{\psi}$, the configuration of the gauge fields is uniquely determined by the configuration of the matter fields.
Specifically, when the matter fields on the neighboring plaquettes $i, j \in P$ are given by $r_i, r_j \in S_{H \backslash G}$, the gauge field on the edge $[ij] \in E$ must be given by $h_i^{-1}h_j \in H$.
Here, $r_i$ and $h_i$ are related as in \eqref{eq: shorthand}, that is, $h_i$ is the unique element of $H$ such that $h_i r_i \in K$.
Thus, we find
\begin{equation}
\prod_{[ij] \in E} \hat{\mathsf{h}}_{ij}^{\text{g.f.}} \ket{\psi} = \sum_{\{r_i \in S_{H \backslash G}\}} \psi(\{r_i\}) \ket{\{r_i; h_i^{-1}h_j\}},
\end{equation}
where $\psi(\{r_i\})$ is a complex number that depends only on the configuration of the matter fields.
By applying the projector $\prod_{i} \hat{\mathsf{h}}_i^{\text{g.f.}}$ to the above state, we obtain
\begin{equation}
\begin{aligned}
\ket{\text{GS}} &= \frac{1}{|K|^{|P|}} \sum_{\{r_i \in S_{H \backslash G}\}} \psi(\{r_i\}) \sum_{\{r_i^{\prime} \in S_{H \backslash G}\}} \ket{\{r_i^{\prime}; (h_i^{\prime})^{-1} h_j^{\prime}\}} \\
& \propto \sum_{\{r_i \in S_{H \backslash G}\}} \ket{\{r_i; h_i^{-1}h_j\}},
\end{aligned}
\end{equation}
where $|P|$ is the number of plaquettes.
The last expression of the above equation does not depend on $\ket{\psi}$, and hence it is the unique ground state of \eqref{eq: generalized cluster ham}.
This ground state will be referred to as the generalized cluster state and denoted by
\begin{equation}
\begin{aligned}
\ket{\text{cluster}(G; H; K)} &:= \frac{1}{|K|^{|P|/2}} \sum_{\{r_i \in S_{H \backslash G}\}} \ket{\{r_i; h_i^{-1}h_j\}} \\
&~= \frac{1}{|K|^{|P|/2}} \sum_{\{k_i \in K\}} \ket{\{\mathfrak{r}[k_i]; h[k_i]^{-1}h[k_j]\}}.
\end{aligned}
\label{eq: generalized cluster}
\end{equation}
Here, $1/|K|^{|P|/2}$ is a normalization factor.
In the second equality, we used \eqref{eq: shorthand} to rewrite the summation over $r_i \in S_{H \backslash G}$ by the summation over $k_i \in K$.
We note that the state $\ket{\text{cluster}(G; H; K)}$ depends on the choice of representatives $S_{H \backslash G}$ because $h[k_i]$ does.\footnote{In particular, $\ket{\text{cluster}(G; H; K)}$ becomes the trivial product state when $S_{H \backslash G} = K$.}
We will discuss this dependency in more detail in Appendix~\ref{sec: Remarks on the choice of representatives}.

\vspace*{\baselineskip}
\noindent{\bf PEPS representation.}
The generalized cluster state~\eqref{eq: generalized cluster} can be represented by the following tensor network:
\begin{equation}
\ket{\text{cluster}(G; H; K)} = \adjincludegraphics[valign = c, trim={10, 10, 10, 10}]{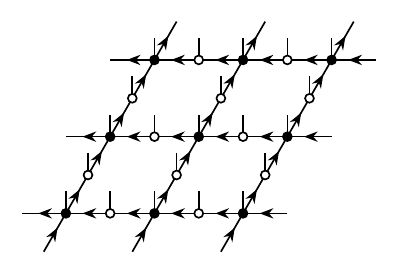}
\label{eq: generalized cluster PEPS}
\end{equation}
A tensor network state of the above form is known as a projected-entangled pair state (PEPS)~\cite{Cirac:2020obd}.
The black and white dots in \eqref{eq: generalized cluster PEPS} are located on the plaquettes and edges of the underlying square lattice.
Accordingly, the physical legs of the black and white tensors are labeled by elements on $S_{H \backslash G}$ and $H$, respectively.
On the other hand, the virtual bonds are all labeled by elements of $H$.
In particular, the bond dimension is $|H|$.
The components of the local tensors in \eqref{eq: generalized cluster PEPS} are defined by
\begin{equation}
\adjincludegraphics[valign = c, trim={10, 10, 10, 10}]{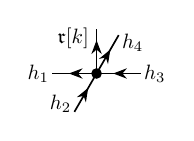} = \frac{1}{\sqrt{|K|}} \prod_{i = 1, 2, 3, 4} \delta_{h_i, h[k]}, \quad
\adjincludegraphics[valign = c, trim={10, 10, 10, 10}]{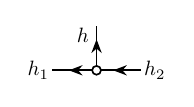} = \delta_{h, h_1^{-1}h_2}.
\label{eq: local tensors}
\end{equation}
One can easily check that the tensor network in~\eqref{eq: generalized cluster PEPS} correctly reproduces the generalized cluster state~\eqref{eq: generalized cluster}.
Indeed, the first equality of \eqref{eq: local tensors} implies that when the physical leg on the plaquette $i \in P$ is labeled by $r_i = \mathfrak{r}[k_i]$ for $k_i \in K$, the virtual bonds around it are all labeled by $h_i = h[k_i]$.
In particular, the virtual bonds on the left side and the right side of the edge $[ij] \in E$ are labeled by $h_i$ and $h_j$, respectively.
Then, the second equality of \eqref{eq: local tensors} implies that the physical leg on the edge $[ij]$ is labeled by $h_i^{-1} h_j$.
This shows that the tensor network state~\eqref{eq: generalized cluster PEPS} is an equal-weight superposition of the states $\ket{\{r_i; h_i^{-1}h_j\}}$ for all configurations of the matter fields $\{r_i \in S_{H \backslash G} \mid i \in P\}$.
Therefore, the tensor network state~\eqref{eq: generalized cluster PEPS} agrees with the generalized cluster state~\eqref{eq: generalized cluster}.

We note that the black tensor on each plaquette can be thought of as a generalization of the copy tensor, which copies a part of the matter field to the nearby virtual bonds.
Specifically, when the matter field on the plaquette $i \in P$ is given by $r_i = h_i^{-1} k_i$, the black tensor copies $h_i$ to the virtual bonds around $i$.
On the other hand, the white tensor on each edge is the multiplication tensor.

\subsection{Example: $G$-cluster state}
\label{sec: Example: G-cluster state}
As a simple example, we consider the generalized cluster state whose input data are given by
\begin{equation}
G = G_0 \times G_0, \quad
H = G_0^{\text{left}} := \{(g, e) \mid g \in G_0\}, \quad
K = G_0^{\text{diag}} := \{(g, g) \mid g \in G_0\},
\end{equation}
where $G_0$ is an arbitrary finite group.
We choose the set of representatives of right $H$-cosets in $G$ to be
\begin{equation}
S_{H \backslash G} = G_0^{\text{right}} := \{(e, g) \mid g \in G_0\}.
\end{equation}
The generalized cluster state obtained from the above input data is denoted by
\begin{equation}
\ket{\text{$G_0$-cluster}} := \ket{\text{cluster}(G_0 \times G_0; G_0^{\text{left}}; G_0^{\text{diag}})}.
\label{eq: G-cluster state}
\end{equation}
This state will be called the $G_0$-cluster state, which is a 2+1d analogue of the 1+1d $G_0$-cluster state studied in~\cite{Brell_2015, Fechisin:2023odt}.
The corresponding model will be referred to as the $G_0$-cluster model.
The above input data implies that, as in 1+1d \cite{Fechisin:2023odt}, the $G_0$-cluster model is obtained by gauging the $G_0^{\text{left}}$ symmetry in the $G_0 \times G_0$-symmetric model that spontaneously breaks the symmetry down to $G_0^{\text{diag}}$.
Accordingly, the symmetry of this model is described by $2\Rep(G_0) \boxtimes 2\Vect_{G_0}$; see Section~\ref{sec: Example: symmetry of the G-cluster state} for more details on this symmetry.
As we will see below, the $G_0$-cluster model reduces to the ordinary cluster model when $G_0 = \mathbb{Z}_2$.

In what follows, we will write down the $G_0$-cluster model and its ground state explicitly based on the general construction in the previous subsection.
More details of this model are investigated in our companion paper \cite{Ohyama:2026oay}.

\vspace*{\baselineskip}
\noindent{\bf The $G_0$-cluster model.}
The state space of the $G_0$-cluster model on a square lattice is given by
\begin{equation}
\mathcal{H}_{\text{g.f.}} = \left(\bigotimes_{i \in P} \mathbb{C}^{|G_0|}\right) \otimes \hat{\pi}_{\text{flat}} \left(\bigotimes_{[ij] \in E} \mathbb{C}^{|G_0|} \right).
\label{eq: G-cluster state space}
\end{equation}
The matter field on plauquette $i \in P$ and the gauge field on edge $[ij] \in E$ are labeled by $g_i \in G_0$ and $g_{ij} \in G_0$.\footnote{According to the general construction in Section~\ref{sec: Generalized cluster models}, the matter field and the gauge filed are labeled by elements of $G_0^{\text{right}}$ and $G_0^{\text{left}}$, respectively. These elements are identified with elements of $G_0$ in this subsection.}
The corresponding states are denoted by $\ket{g_i}_i$ and $\ket{g_{ij}}_{ij}$, respectively.
The projector $\hat{\pi}_{\text{flat}}$ in \eqref{eq: G-cluster state space} imposes the flatness condition on the gauge fields.
The Hamiltonian of the model is
\begin{equation}
\hat{H}_{\text{g.f.}} = -\sum_{i \in P} \hat{\mathsf{h}}^{\text{g.f.}}_i - \sum_{[ij] \in E} \hat{\mathsf{h}}^{\text{g.f.}}_{ij},
\label{eq: G-cluster ham}
\end{equation}
where $\hat{\mathsf{h}}^{\text{g.f.}}_i$ and $\hat{\mathsf{h}}^{\text{g.f.}}_{ij}$ are the local commuting projectors defined by
\begin{align}
\hat{\mathsf{h}}^{\text{g.f.}}_i \Ket{\adjincludegraphics[valign = c, trim={10, 10, 10, 10}]{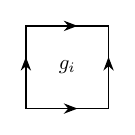}} &= \frac{1}{|G_0|} \sum_{g \in G_0} \Ket{\adjincludegraphics[valign = c, trim={10, 10, 10, 10}]{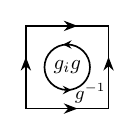}},
\label{eq: G-cluster ham plaquette} \\
\hat{\mathsf{h}}^{\text{g.f.}}_{ij} \ket{g_i}_i \otimes \ket{g_{ij}}_{ij} \otimes \ket{g_j}_j &= \delta_{g_{ij}, g_i^{-1}g_j} \ket{g_i}_i \otimes \ket{g_{ij}}_{ij} \otimes \ket{g_j}_j.
\label{eq: G-cluster ham edge}
\end{align}
Here, the loop on the right-hand side of \eqref{eq: G-cluster ham plaquette} represents the loop operator acting on the gauge fields, cf.~\eqref{eq: loop op}.
When $G_0 = \mathbb{Z}_2$, the above local terms become $\hat{\mathsf{h}}_i^{\text{g.f.}} = \frac{1}{2}(1+X^{(i)} \prod_{[ij] \in \partial i} X^{(ij)})$ and $\hat{\mathsf{h}}_{ij}^{\text{g.f.}} = \frac{1}{2}(1+Z^{(i)}Z^{(ij)}Z^{(j)})$, where $X$ and $Z$ are the Pauli-X and Pauli-Z operators.
Thus, in this case, the $G_0$-cluster model reduces to the ordinary cluster model \cite{Briegel:0004051, PhysRevLett.86.5188, Raussendorf:0301052}, which realizes an SPT phase with $2\Rep(\mathbb{Z}_2) \boxtimes 2\Vect_{\mathbb{Z}_2}$ symmetry \cite{Yoshida:2015cia}.
In general, the $G_0$-cluster model realizes an SPT phase with $2\Rep(G_0) \boxtimes 2\Vect_{G_0}$ symmetry.
The symmetry structure of this model will be discussed in more detail in Section~\ref{sec: Example: symmetry of the G-cluster state}.

\vspace*{\baselineskip}
\noindent{\bf The $G_0$-cluster state.}
The Hamiltonian~\eqref{eq: G-cluster ham} is solvable because it is a sum of local commuting projectors.
The ground state of this model can be written explicitly as (cf. \eqref{eq: generalized cluster})
\begin{equation}
\ket{\text{$G_0$-cluster}} = \frac{1}{|G_0|^{|P|/2}} \sum_{\{g_i \in G_0\}} \ket{\{g_i; g_i^{-1}g_j\}}.
\label{eq: G cluster}
\end{equation}
This state can be represented by a PEPS of the form
\begin{equation}
\ket{\text{$G_0$-cluster}} = \adjincludegraphics[valign = c, trim={10, 10, 10, 10}]{tikz/out/cluster_PEPS.pdf},
\label{eq: G cluster PEPS}
\end{equation}
where both the physical legs and the virtual legs take values in $G_0$.
The black dot and the white dot in~\eqref{eq: G cluster PEPS} are the copy tensor and the multiplication tensor, respectively, i.e.,
\begin{equation}
\adjincludegraphics[valign = c, trim={10, 10, 10, 10}]{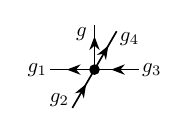} = \frac{1}{\sqrt{|G_0|}} \prod_{i = 1, 2, 3, 4} \delta_{g, g_i}, \qquad
\adjincludegraphics[valign = c, trim={10, 10, 10, 10}]{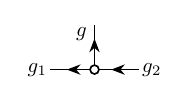} = \delta_{g, g_1^{-1}g_2}.
\label{eq: local tensor G-cluster}
\end{equation}
When $G_0 = \mathbb{Z}_2$, equation \eqref{eq: G cluster PEPS} agrees with the PEPS representation of the ordinary cluster state.

\section{Non-invertible symmetries of 2+1d generalized cluster states}
\label{sec: Non-invertible symmetries of 2+1d generalized cluster states}
By construction, the generalized cluster model defined in Section~\ref{sec: 2+1d generalized cluster states with non-invertible symmetries} has a group-theoretical fusion 2-category symmetry $\mathcal{C}(G; H)$.
In this section, we will write down the symmetry operators explicitly using tensor networks.
In addition, we study the symmetry action on the generalized cluster states~\eqref{eq: generalized cluster}.
We will see that the non-invertible symmetry operators act fractionally on the virtual legs of the tensor network representation~\eqref{eq: generalized cluster PEPS}.

\subsection{Tensor network representations of symmetry operators}
\label{sec: Tensor network representations of symmetry operators}

\begin{figure}[t]
\centering
\adjincludegraphics{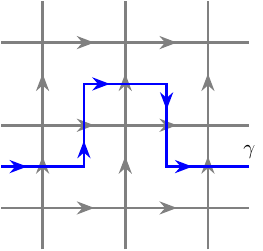}
\caption{A loop $\gamma$ on the dual lattice.}
\label{fig: Wilson line}
\end{figure}

\noindent{\bf 1-form symmetry.}
The 1-form symmetry operators are labeled by 1-endomorphisms of the unit object $I \in \mathcal{C}(G; H)$.
These 1-morphisms form a symmetric fusion category
\begin{equation}
\End_{\mathcal{C}(G; H)}(I) \cong \Rep(H).
\end{equation}
The corresponding symmetry operators are the Wilson lines for the $H$ gauge fields.
The action of the Wilson line $W_{\rho}$ labeled by a representation $\rho \in \Rep(H)$ is defined by
\begin{equation}
W_{\rho}(\gamma) \hat{\pi}_{\text{flat}} \ket{\{r_i; h_{ij}\}} = \mathrm{tr}_{V_{\rho}}(\rho(\text{hol}_{\gamma}(\{h_{ij}\}))) \hat{\pi}_{\text{flat}} \ket{\{r_i; h_{ij}\}},
\end{equation}
where $V_{\rho}$ is the representation space of $\rho$, $\gamma$ is an oriented closed loop on the dual lattice as shown in Figure~\ref{fig: Wilson line}, and $\text{hol}_{\gamma}(\{h_{ij}\})$ is the holonomy of the $H$ gauge field along $\gamma$.
More specifically, the holonomy $\text{hol}_{\gamma}(\{h_{ij}\})$ is defined by
\begin{equation}
\text{hol}_{\gamma}(\{h_{ij}\}) = \prod_{[ij] \in E_{\gamma}} h_{ij}^{s_{ij}},
\end{equation}
where $\prod_{[ij] \in E_{\gamma}}$ is the path-ordered product over all edges that intersect $\gamma$,\footnote{Precisely, we need to specify the base point of $\gamma$ so that the path-ordered product is well-defined. Nevertheless, the Wilson line $W_{\rho}$ is independent of the base point because we take the trace over $V_{\rho}$.} and $s_{ij}$ is the sign defined by
\begin{equation}
s_{ij} = +1 \quad \text{for } \adjincludegraphics[valign = c, trim={10, 10, 10, 10}]{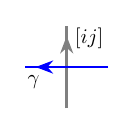}, \qquad
s_{ij} = -1 \quad \text{for } \adjincludegraphics[valign = c, trim={10, 10, 10, 10}]{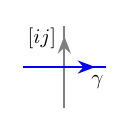}.
\label{eq: sij}
\end{equation}
We note that the Wilson line $W_{\rho}$ is topological in spatial directions due to the flatness condition on the gauge fields.
The generalized cluster state~\eqref{eq: generalized cluster} preserves this 1-form symmetry because
\begin{equation}
W_{\rho}(\gamma) \ket{\text{cluster}(G; H; K)} = \dim(\rho) \ket{\text{cluster}(G; H; K)}
\end{equation}
for any loop $\gamma$ and any representation $\rho \in \Rep(H)$.

The Wilson line $W_{\rho}(\gamma)$ can be written as a matrix product operator (MPO) along $\gamma$, which is expressed diagrammatically as
\begin{equation}
\adjincludegraphics[valign = c, scale=0.9, trim={10, 10, 10, 10}]{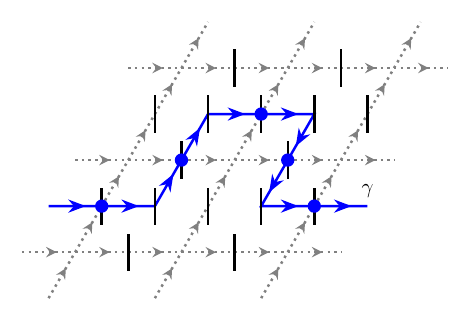}.
\label{eq: W MPO}
\end{equation}
Here, the gray dotted lines represent the underlying square lattice, and the vertical black segments represent the physical legs.
The bond Hilbert space of the MPO is the representation space $V_{\rho}$, and the local MPO tensors represented by the blue dots are defined by
\begin{equation}
\adjincludegraphics[valign = c, trim={10, 10, 10, 10}]{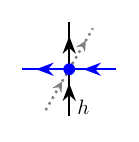} = \adjincludegraphics[valign = c, trim={10, 10, 10, 10}]{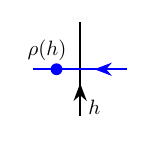}, \qquad
\adjincludegraphics[valign = c, trim={10, 10, 10, 10}]{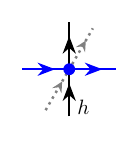} = \adjincludegraphics[valign = c, trim={10, 10, 10, 10}]{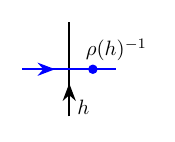}.
\label{eq: W MPO tensor}
\end{equation}
We note that the MPO tensors depend on the relative orientation of $\gamma$ and the edge of the square lattice.
The intersections without the blue dots in \eqref{eq: W MPO} and \eqref{eq: W MPO tensor} represent the trivial braiding.

\vspace*{\baselineskip}
\noindent{\bf 0-form symmetry.}
The 0-form symmetry operators are labeled by objects of $\mathcal{C}(G; H)$.
As we reviewed in Section~\ref{sec: Group-theoretical fusion 2-categories}, all simple objects of $\mathcal{C}(G; H)$ are connected to the simple objects of the following form \cite{Decoppet:2023bay}:
\begin{equation}
V_H^g = \Vect_H \boxtimes V^g \boxtimes \Vect_H, \quad g \in G.
\label{eq: VHg}
\end{equation}
Here, we recall that $V^g$ is a simple object of $2\Vect_G$, and $\Vect_H = \bigoplus_{h \in H} V^h$ is an algebra object of $2\Vect_G$.
The symmetry operator corresponding to~\eqref{eq: VHg} is given by \cite{Inamura:2025cum}
\begin{equation}
\mathsf{D}[V_H^g] := \mathsf{D}_H U_g \overline{\mathsf{D}}_H,
\label{eq: 0-form}
\end{equation}
where $\mathsf{D}_H: \mathcal{H} \to \mathcal{H}_{\text{g.f.}}$ is the gauging operator, $\overline{\mathsf{D}}_{H}: \mathcal{H}_{\text{g.f.}} \to \mathcal{H}$ is the ungauging operator, and $U_g: \mathcal{H} \to \mathcal{H}$ is a symmetry operator of the original $G$-symmetric model.
See below for more details of the gauging and ungauging operators.
We note that $V_H^g$ is isomorphic to $V_H^{g^{\prime}}$ if and only if $g$ and $g^{\prime}$ are in the same double $H$-coset in $G$.
Correspondingly, the symmetry operators~\eqref{eq: 0-form} satisfy
\begin{equation}
\mathsf{D}[V_H^g] = \mathsf{D}[V_H^{g^{\prime}}] ~ \Leftrightarrow ~ HgH = Hg^{\prime}H,
\label{eq: same operator}
\end{equation}
which imediately follows from $\mathsf{D}_H U_h = \mathsf{D}_H$ and $U_h \overline{\mathsf{D}}_H = \overline{\mathsf{D}}_H$ for all $h \in H$.

In what follows, we will write down the 0-form symmetry operators~\eqref{eq: 0-form} using tensor networks.
Furthermore, we will also obtain tensor network representations of all the other symmetry operators by condensing line operators on the symmetry operators~\eqref{eq: 0-form}.

\subsubsection{Gauging operator}
\label{sec: Gauging operator}
Let us first write down the gauging operator $\mathsf{D}_H$, which serves as a building block of the symmetry operators~\eqref{eq: 0-form}.
The gauging operator $\mathsf{D}_H: \mathcal{H} \to \mathcal{H}_{\text{g.f.}}$ is an operator that implements the gauging of the subgroup symmetry $H \subset G$ in the sense that
\begin{equation}
\mathsf{D}_H \hat{H} = \hat{H}_{\text{g.f.}} \mathsf{D}_H,
\label{eq: intertwiner}
\end{equation}
where $\hat{H}$ is the Hamiltonian of the original $G$-symmetric model and $\hat{H}_{\text{g.f.}}$ is the Hamiltonian of the gauged model after the gauge fixing.\footnote{The gauging operator $\mathsf{D}_H$ is defined so that \eqref{eq: intertwiner} holds for any $G$-symmetric Hamiltonian $\hat{H}$ and the corresponding gauged Hamiltonian $\hat{H}_{\text{g.f.}}$.}
Such a gauging operator is given by the composition
\begin{equation}
\mathsf{D}_H = \mathcal{U}_{\text{g.f.}} \mathsf{D}_H^{\prime},
\label{eq: Dgf def}
\end{equation}
where $\mathcal{U}_{\text{g.f.}}: \mathcal{H}_{\text{gauged}} \to \mathcal{H}_{\text{g.f.}}$ is the gauge fixing operator~\eqref{eq: gauge fixing} and $\mathsf{D}_H^{\prime}: \mathcal{H} \to \mathcal{H}_{\text{gauged}}$ is the gauging operator before the gauge fixing.
The explicit form of $\mathsf{D}_H^{\prime}$ is given by \cite{Haegeman:2014maa}
\begin{equation}
\mathsf{D}_H^{\prime} \ket{\{g_i\}} = \hat{\pi}_{\text{Gauss}} \ket{\{g_i; e\}},
\label{eq: D gauged}
\end{equation}
where $\ket{\{g_i; e\}}$ is a state in which the gauge fields are trivial on all edges.
By construction, the above gauging operator satisfies
\begin{equation}
\mathsf{D}_H^{\prime} \hat{H} = \hat{H}_{\text{gauged}} \mathsf{D}_H^{\prime},
\end{equation}
which in turn imples \eqref{eq: intertwiner}.
Combining \eqref{eq: gauge fixing}, \eqref{eq: Dgf def}, and \eqref{eq: D gauged}, we find \cite{Tantivasadakarn:2022hgp, Inamura:2025cum, Delcamp:2023kew}
\begin{equation}
\mathsf{D}_H \ket{\{g_i\}} = \mathcal{U}_{\text{g.f.}} \hat{\pi}_{\text{Gauss}} \ket{\{g_i; e\}} = \ket{\{r_i; h_i^{-1}h_j\}},
\label{eq: D gf}
\end{equation}
where $r_i \in S_{H \backslash G}$ and $h_i \in H$ are uniquely determined by $g_i = h_i r_i$.

We note that the gauging operator $\mathsf{D}_H$ is neither injective nor surjective.
It is not injective because it satisfies
\begin{equation}
\mathsf{D}_H U_h = \mathsf{D}_H, \quad \forall h \in H,
\end{equation}
which implies that any state charged non-trivially under the $H$ symmetry is in the kernel of $\mathsf{D}_H$.
Similarly, it is not surjective because it satisfies
\begin{equation}
W_{\rho}(\gamma) \mathsf{D}_H = \dim(\rho) \mathsf{D}_H, \quad \forall \rho \in \Rep(H), ~ \forall \gamma: \text{loop},
\end{equation}
which implies that any state charged non-trivially under the $\Rep(H)$ symmetry are not in the image of $\mathsf{D}_H$.
As we will see later, $\mathsf{D}_H$ is invertible on the symmetric subspace of $\mathcal{H}$, even though it is non-invertible on the entire $\mathcal{H}$.

By construction, the gauging operator $\mathsf{D}_H$ maps the ground states of the $G$-symmetric model to the ground states of the gauged model.
In particular, the ground states of the $G/K$-SSB Hamiltonian~\eqref{eq: Ham G/K} are all mapped to the generalized cluster state~\eqref{eq: generalized cluster}, i.e., 
\begin{equation}
\mathsf{D}_H \ket{\text{GS}; \mu} = \ket{\text{cluster}(G; H; K)}, \quad \mu = 1, 2, \cdots, |G/K|,
\label{eq: D gf GS}
\end{equation}
where $\ket{\text{GS}; \mu}$ is defined by \eqref{eq: GS G/K}.

\vspace*{\baselineskip}
\noindent{\bf PEPO representation.}
The gauging operator $\mathsf{D}_H$ defined by \eqref{eq: D gf} can be represented by the following tensor network:
\begin{equation}
\mathsf{D}_H = \adjincludegraphics[valign = c, trim={10, 10, 10, 10}]{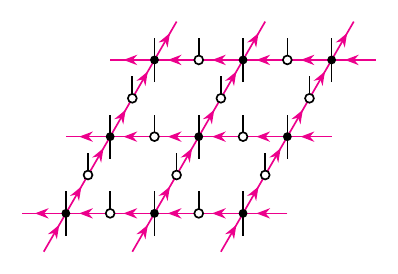}.
\label{eq: gauging PEPO}
\end{equation}
A tensor network operator of this form is called a projected-entangled pair operator (PEPO)~\cite{Cirac:2020obd}.
As in the PEPS representation~\eqref{eq: generalized cluster PEPS} of the generalized cluster state, the black and white dots in \eqref{eq: gauging PEPO} are located on the plaquettes and edges of the underlying square lattice.
The bottom and top legs of each black dot are the physical legs associated with the matter fields of the original model and the gauged model.
Accordingly, they are labeled by elements of $G$ and $S_{H \backslash G}$, respectively.
Similarly, the top leg of each white dot is the physical leg associated with the gauge fields of the gauged model, and hence it is labeled by an element of $H$.
Each virtual bond, written in magenta, is also labeled by an element of $H$, meaning that the bond dimension is $|H|$.
The local tensors in \eqref{eq: gauging PEPO} are defined by
\begin{equation}
\adjincludegraphics[valign = c, trim={10, 10, 10, 10}]{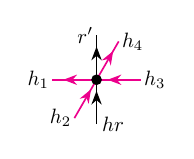} = \delta_{r, r^{\prime}} \prod_{i = 1, 2, 3, 4} \delta_{h, h_i}, \quad
\adjincludegraphics[valign = c, trim={10, 10, 10, 10}]{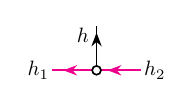} = \delta_{h, h_1^{-1}h_2}.
\label{eq: DH tensors}
\end{equation}
It is straightforward to check that the action of the PEPO~\eqref{eq: gauging PEPO} on a state $\ket{\{g_i\}}$ agrees with \eqref{eq: D gf}.

\subsubsection{Ungauging operator}
\label{sec: Ungauging operator}
The ungauging operator $\overline{\mathsf{D}}_H: \mathcal{H}_{\text{g.f.}} \to \mathcal{H}$ is an operator that implements the ungauging of the symmetry $H$ in the sense that
\begin{equation}
\overline{\mathsf{D}}_H \hat{H}_{\text{g.f.}} = \hat{H} \overline{\mathsf{D}}_H.
\label{eq: ungauging intertwiner}
\end{equation}
The explicit form of $\overline{\mathsf{D}}_H$ is given by
\begin{equation}
\overline{\mathsf{D}}_H \ket{\{r_i; h_{ij}\}} = \sum_{\{h_i \in H\}} \left(\prod_{[ij] \in E} \delta_{h_{ij}, h_i^{-1}h_j}\right) \ket{\{h_i r_i\}}.
\label{eq: D gf bar}
\end{equation}
We note that the gauging and ungauging operators are related by
\begin{equation}
\overline{\mathsf{D}}_H = \mathsf{D}_H^{\dagger},
\end{equation}
where the Hermitian conjugate is defined with respect to the following Hermitian inner product:
\begin{equation}
\braket{\{g_i\}| \{g_i^{\prime}\}} = \prod_{i \in P} \delta_{g_i, g_i^{\prime}}, \quad
\braket{\{r_i; h_{ij}\} | \{r_i^{\prime}; h_{ij}^{\prime}\}} = \prod_{i \in P} \delta_{r_i, r_i^{\prime}} \prod_{[ij] \in E} \delta_{h_{ij}, h_{ij}^{\prime}}.
\end{equation}
Given that the Hamiltonians are Hermitian, equation \eqref{eq: ungauging intertwiner} automatically follows from \eqref{eq: intertwiner}.\footnote{In \cite{Inamura:2025cum}, the ungauging operator~\eqref{eq: D gf bar} was derived without assuming the Hermiticity of the Hamiltonians.}

A direct computation shows that the gauging and ungauging operators satisfy
\begin{equation}
\overline{\mathsf{D}}_H \mathsf{D}_H = |H| \hat{\pi}_{\text{sym}}, \quad 
\mathsf{D}_H \overline{\mathsf{D}}_H = |H| \hat{\pi}_{\text{hol}},
\label{eq: DD bar}
\end{equation}
where $\hat{\pi}_{\text{sym}} := \frac{1}{|H|} \sum_{h \in H} U_h$ is the projection to the symmetric subspace of the original state space $\mathcal{H}$, while $\hat{\pi}_{\text{hol}}$ is the projection to the subspace of $\mathcal{H}_{\text{g.f.}}$ in which the gauge fields have the trivial holonomy along any loop (i.e., the gauge fields have the trivial charge under the $\Rep(H)$ symmetry).
The first equality of \eqref{eq: DD bar} holds because
\begin{equation*}
\overline{\mathsf{D}}_H \mathsf{D}_H \ket{\{g_i\}} \overset{\eqref{eq: D gf}}{=} \overline{\mathsf{D}}_H \ket{\{r_i; h_i^{-1} h_j\}}
\overset{\eqref{eq: D gf bar}}{=} \sum_{\{h_i^{\prime} \in H\}} \left(\prod_{[ij] \in E} \delta_{h_i^{-1}h_j, (h_i^{\prime})^{-1}h_j^{\prime}}\right) \ket{\{h_i^{\prime}r_i\}} = \sum_{h \in H} \ket{\{hh_ir_i\}}.
\end{equation*}
In the last equality, we used the fact that the summand on the left-hand side is non-zero only when there exists $h \in H$ such that $h_i^{\prime} = hh_i$ for all $i \in P$.
Similarly, the second equality of \eqref{eq: DD bar} follows from
\begin{equation}
\begin{aligned}
\mathsf{D}_H \overline{\mathsf{D}}_H \ket{\{r_i; h_{ij}\}}
& \overset{\eqref{eq: D gf bar}}{=} \mathsf{D}_H \sum_{\{h_i \in H\}} \left(\prod_{[ij] \in E} \delta_{h_{ij}, h_i^{-1}h_j}\right) \ket{\{h_ir_i\}} 
\overset{\eqref{eq: D gf}}{=} \sum_{\{h_i \in H\}} \left(\prod_{[ij] \in E} \delta_{h_{ij}, h_i^{-1}h_j}\right) \ket{\{r_i; h_i^{-1}h_j\}} \\
& ~~ \, = \sum_{\{h_i \in H\}} \left(\prod_{[ij] \in E} \delta_{h_{ij}, h_i^{-1}h_j} \right) \ket{\{r_i; h_{ij}\}} 
= |H| \delta_{\text{hol}}(\{h_{ij}\}) \ket{\{r_i; h_{ij}\}},
\end{aligned}
\label{eq: pi hol}
\end{equation}
where $\delta_{\text{hol}}(\{h_{ij}\}) = 1$ if the holonomy of the gauge fields $\{h_{ij} \mid [ij] \in E\}$ is trivial along any loop and $\delta_{\text{hol}}(\{h_{ij}\}) = 0$ otherwise.
Equation~\eqref{eq: DD bar} implies that $\mathsf{D}_H$ and $\overline{\mathsf{D}}_H$ are mutually inverse bijections (up to normalization) on the symmetric subspaces of $\mathcal{H}$ and $\mathcal{H}_{\text{g.f.}}$.

As with the gauging operator, the ungauging operator $\overline{\mathsf{D}}_H$ maps the ground states of the gauged model to the ground states of the original $G$-symmetric model.
In particular, the generalized cluster state~\eqref{eq: generalized cluster} is mapped to the symmetric linear combination of the symmetry-breaking ground states~\eqref{eq: GS G/K}, i.e., 
\begin{equation}
\overline{\mathsf{D}}_H \ket{\text{cluster}(G; H; K)} = \sum_{\mu = 1, 2, \cdots, |G/K|} \ket{\text{GS}; \mu}.
\label{eq: ungauging GS}
\end{equation}
This result immediately follows from \eqref{eq: D gf GS} and \eqref{eq: DD bar}.

\vspace*{\baselineskip}
\noindent{\bf PEPO representation.}
The tensor network representation of $\overline{\mathsf{D}}_H$ is obtained by turning the tensor network for $\mathsf{D}_H$ upside down as follows:
\begin{equation}
\overline{\mathsf{D}}_H = \adjincludegraphics[valign = c, trim={10, 10, 10, 10}]{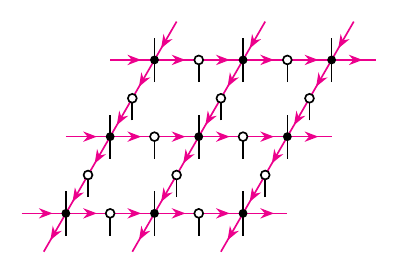}.
\label{eq: ungauging PEPO}
\end{equation}
We note that the orientations of the virtual bonds are opposite to those of \eqref{eq: gauging PEPO}.\footnote{As always, the physical legs are still oriented upwards.}
Furthermore, as opposed to \eqref{eq: gauging PEPO}, the bottom legs of the black tensors are now labeled by elements of $S_{H \backslash G}$, while the top legs are labeled by elements of $G$.
The physical legs of the white tensors, as well as the virtual bonds, are still labeled by elements of $H$.
The local tensors in \eqref{eq: ungauging PEPO} are defined by
\begin{equation}
\adjincludegraphics[valign = c, trim={10, 10, 10, 10}]{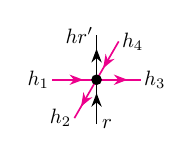} = \delta_{r, r^{\prime}} \prod_{i = 1, 2, 3, 4} \delta_{h, h_i}, \quad \adjincludegraphics[valign = c, trim={10, 10, 10, 10}]{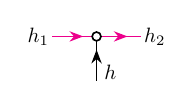}  = \delta_{h, h_1^{-1}h_2}.
\label{eq: DH bar tensors}
\end{equation}
It is straightforward to check that the PEPO~\eqref{eq: ungauging PEPO} acting on $\ket{\{r_i; h_{ij}\}}$ agrees with \eqref{eq: D gf bar}.

\subsubsection{Symmetry operators}
\label{sec: Symmetry operators}
Using the PEPO representations~\eqref{eq: gauging PEPO} and \eqref{eq: ungauging PEPO} of the gauging and ungauging operators, we can write down the PEPO representation of the symmetry operator~\eqref{eq: 0-form} as follows:
\begin{equation}
\mathsf{D}[V_H^g] = \adjincludegraphics[valign = c, width = 0.45\linewidth, trim={10, 10, 10, 10}]{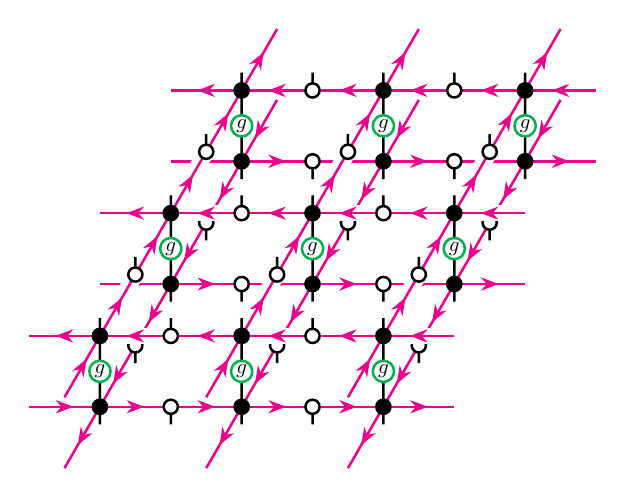}.
\label{eq: symmetry PEPO}
\end{equation}
The local tensors on the top and bottom layers are defined by \eqref{eq: DH tensors} and \eqref{eq: DH bar tensors}, respectively.
The green dots on the middle layer represent the on-site symmetry operator $U_g$, which is given by the left multiplication of $g$.
The composite tensor on each plaquette of the square lattice can be computed as
\begin{equation}
\adjincludegraphics[valign = c, trim={10, 10, 10, 10}]{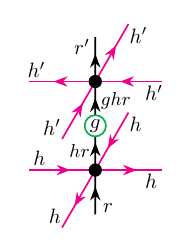} = \delta_{h^{\prime}, ghr \mathfrak{r}[ghr]^{-1}} \delta_{r^{\prime}, \mathfrak{r}[ghr]},
\label{eq: composite}
\end{equation}
where $\mathfrak{r}[ghr] \in S_{H \backslash G}$ is the representative of the right $H$-coset $Hghr$, i.e.,
\begin{equation}
ghr \in H \mathfrak{r}[ghr].
\end{equation}
We note that $ghr \mathfrak{r}[ghr]^{-1}$ is an element of $H$ by definition.
The total bond dimension of the PEPO~\eqref{eq: symmetry PEPO} is $|H|^2$ because both the top and bottom layers have bond dimension $|H|$.
As we will see below, this bond dimension can be reduced to $|H|$, which turns out to be crucial for later computations.

Before proceeding, we mention that the symmetry operator $\mathsf{D}[V_H^g]$ correctly obeys the same fusion rule as $V_H^g$.
One can verify this by computing the fusion rule of $\mathsf{D}[V_H^g]$ explicitly as
\begin{equation}
\mathsf{D}[V_H^g] \mathsf{D}[V_H^{g^{\prime}}]
= \mathsf{D}_H U_g \overline{\mathsf{D}}_{H} \mathsf{D}_H U_{g^{\prime}} \mathsf{D}_H
= \sum_{h \in H} \mathsf{D}_H U_{ghg^{\prime}} \overline{\mathsf{D}}_H
= \sum_{g^{\prime \prime} \in gHg^{\prime}} \mathsf{D}[V_H^{g^{\prime \prime}}],
\end{equation}
where the second equality follows from \eqref{eq: DD bar}.
The above fusion rule agrees with \eqref{eq: fusion rules VHg}, which justifies the claim that $\mathsf{D}[V_H^g]$ is the symmetry operator corresponding to $V_H^g \in \mathcal{C}(G; H)$.

\vspace*{\baselineskip}
\noindent{\bf Reduction of bond dimension.}
One can reduce the bond dimension of the PEPO~\eqref{eq: symmetry PEPO} to $|H|$ when the symmetry $\mathcal{C}(G; H)$ is non-anomalous.\footnote{This may be viewed as an analogue of the ``onsiteability" of non-anomalous fusion category symmetries in 1+1d~\cite{Meng:2024nxx, Evans:2025msy}. In \cite{Meng:2024nxx}, symmetry operators are said to be on-site if they are represented by MPOs whose bond dimensions are equal to the quantum dimensions of the corresponding objects in the fusion category. See also~\cite{Seifnashri:2025vhf, Feng:2025yge} for the relation between onsiteability and anomalies for invertible symmetries.}
To reduce the bond dimension, we choose the representatives of right $H$-cosets in $G$ so that they form a subgroup of $G$.
In this case, the set $S_{H \backslash G}$ becomes a complement of $H$ in $G$ because it satisfies
\begin{equation}
HS_{H \backslash G} = G, \quad H \cap S_{H \backslash G} = \{e\}.
\end{equation}
Such a choice of representatives is possible because $\mathcal{C}(G; H)$ is supposed to be non-anomalous.
For this choice of $S_{H \backslash G}$, the representative $\mathfrak{r}[ghk]$ satisfies
\begin{equation}
\mathfrak{r}[ghr] = \mathfrak{r}[gh] r.
\end{equation}
Therefore, the right-hand side of \eqref{eq: composite} reduces to
\begin{equation}
\delta_{h^{\prime}, ghr \mathfrak{r}[ghr]^{-1}} \delta_{r^{\prime}, \mathfrak{r}[ghr]} = \delta_{h^{\prime}, gh\mathfrak{r}[gh]^{-1}} \delta_{r^{\prime}, \mathfrak{r}[gh] r}.
\label{eq: h prime r prime 2}
\end{equation}
Namely, the composite tensor \eqref{eq: composite} is non-zero only when $h^{\prime} = gh\mathfrak{r}[gh]^{-1}$ and $r^{\prime} = \mathfrak{r}[gh] r$.
We note that $h^{\prime} = gh \mathfrak{r}[gh]^{-1}$ does not depend on $r$.
This property allows us to decompose the composite tensor \eqref{eq: composite} as follows:
\begin{equation}
\adjincludegraphics[valign = c, trim={10, 10, 10, 10}]{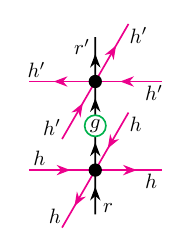} = \adjincludegraphics[valign = c, trim={10, 10, 10, 10}]{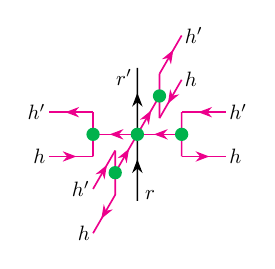}.
\label{eq: decomposition}
\end{equation}
The local tensors on the right-hand side are given by
\begin{equation}
\adjincludegraphics[valign = c, trim={10, 10, 10, 10}]{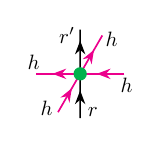} = \delta_{r^{\prime}, \mathfrak{r}[gh]r}, \quad
\adjincludegraphics[valign = c, trim={10, 10, 10, 10}]{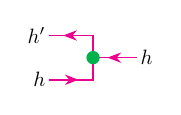}  = \adjincludegraphics[valign = c, trim={10, 10, 10, 10}]{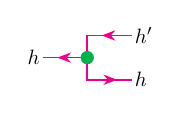}  = \delta_{h^{\prime}, gh\mathfrak{r}[gh]^{-1}}.
\label{eq: filled}
\end{equation}
Using the decomposition~\eqref{eq: decomposition}, we can deform the PEPO representation~\eqref{eq: symmetry PEPO} as follows:
\begin{equation}
\mathsf{D}[V_H^g] = \adjincludegraphics[valign = c, trim={10, 10, 10, 10}]{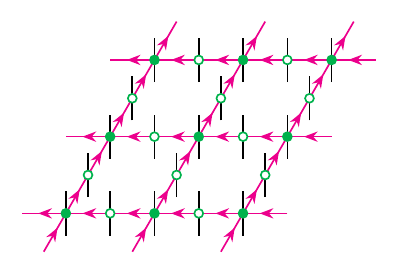}.
\label{eq: reduced PEPO}
\end{equation}
The tensors represented by the filled green dots are defined by the first equality of \eqref{eq: filled}, while the tensors represented by the unfilled dots are defined by
\begin{equation}
\adjincludegraphics[valign = c, trim={10, 10, 10, 10}]{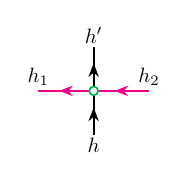} := \adjincludegraphics[valign = c, trim={10, 10, 10, 10}]{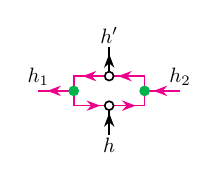} = \delta_{h, h_1^{-1}h_2} \delta_{h^{\prime}, \mathfrak{r}[gh_1]h\mathfrak{r}[gh_2]^{-1}}.
\label{eq: unfilled}
\end{equation}
We emphasize that the bond dimension of the PEPO representation~\eqref{eq: reduced PEPO} is $|H|$ rather than $|H|^2$.

\vspace*{\baselineskip}
\noindent{\bf Line operators on $\mathsf{D}[V_H^g]$.}
Line operators on the surface operator $\mathsf{D}[V_H^g]$ are labeled by 1-endomorphisms of $V_H^g$, which form a fusion 1-category \cite{Decoppet:2023bay}
\begin{equation}
\End_{\mathcal{C}(G; H)}(V_H^g) \cong \Vect_{H \cap g^{-1}Hg}.
\end{equation}
These line operators will be denoted by $\{\mathsf{L}_h \mid h \in H \cap g^{-1}Hg\}$.
The surface operator in the presence of a line operator $\mathsf{L}_h$ can be represented by the following tensor network:
\begin{equation}
\adjincludegraphics[valign = c, trim={10, 10, 10, 10}]{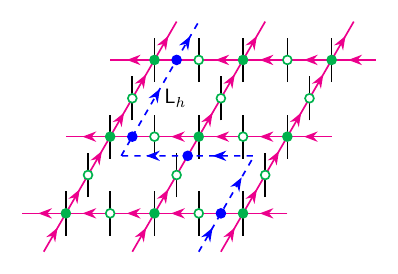}.
\label{eq: line on surface}
\end{equation}
The bond dimension of the line operator $\mathsf{L}_h$ is one, which means that $\mathsf{L}_h$ is the tensor product of on-site operators along the line.
Nevertheless, as a visual aid, we draw a dashed line as in \eqref{eq: line on surface} to represent the line operator $\mathsf{L}_h$.
Each on-site operator along the line is given by the left multiplication of $h$ or $h^{-1}$, depending on the orientation:
\begin{equation}
\adjincludegraphics[valign = c, trim={10, 10, 10, 10}]{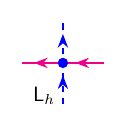} := \adjincludegraphics[valign = c, trim={10, 0, 10, 10}]{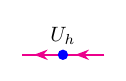}, \qquad
\adjincludegraphics[valign = c, trim={10, 10, 10, 10}]{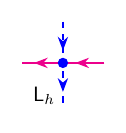} := \adjincludegraphics[valign = c, trim={10, 0, 10, 10}]{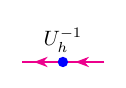}.
\end{equation}
The line operator $\mathsf{L}_h$ defined above is topological in spatial directions because it satisfies the following pulling-through equations:
\begin{equation}
\adjincludegraphics[valign = c, trim={10, 10, 10, 10}]{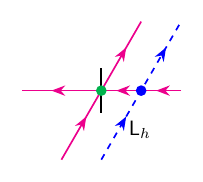} = \adjincludegraphics[valign = c, trim={10, 10, 10, 10}]{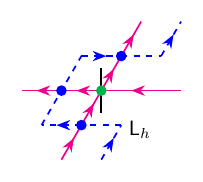}, \qquad
\adjincludegraphics[valign = c, trim={10, 10, 10, 10}]{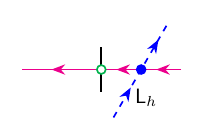} = \adjincludegraphics[valign = c, trim={10, 10, 10, 10}]{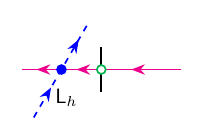}.
\label{eq: pulling through}
\end{equation}
To show the pulling-through equations, we first notice that these equations are equivalent to
\begin{align}
\adjincludegraphics[valign = c, trim={10, 10, 10, 10}]{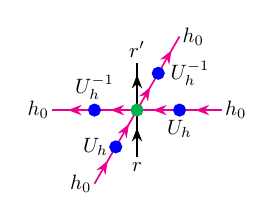} = \adjincludegraphics[valign = c, trim={10, 10, 10, 10}]{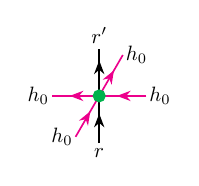}, \qquad
\adjincludegraphics[valign = c, trim={10, 10, 10, 10}]{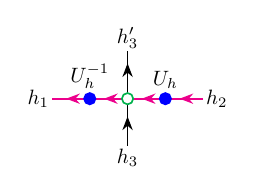} = \adjincludegraphics[valign = c, trim={10, 10, 10, 10}]{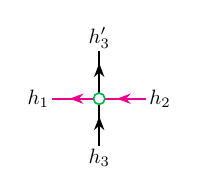}.
\end{align}
Due to \eqref{eq: filled} and \eqref{eq: unfilled}, these equations can also be written more explicitly as
\begin{equation}
\delta_{r^{\prime}, \mathfrak{r}[ghh_0]r} = \delta_{r^{\prime}, \mathfrak{r}[gh_0]r}, \qquad \delta_{h_3, h_1^{-1}h_2} \delta_{h_3^{\prime}, \mathfrak{r}[ghh_1]h_3\mathfrak{r}[ghh_2]^{-1}} = \delta_{h_3, h_1^{-1}h_2} \delta_{h_3^{\prime}, \mathfrak{r}[gh_1]h_3\mathfrak{r}[gh_2]^{-1}}.
\label{eq: pulling through 2}
\end{equation}
The first equality holds if and only if $\mathfrak{r}[ghh_0] = \mathfrak{r}[gh_0]$ for all $h_0 \in H$, that is, if and only if $ghh_0$ and $gh_0$ are in the same right $H$-coset in $G$.
This condition is equivalent to
\begin{equation}
Hgh = Hg ~\Leftrightarrow~ h \in g^{-1}Hg.
\end{equation}
The second equality of \eqref{eq: pulling through 2} also holds if and only if $h \in g^{-1}Hg$.
Therefore, we find that the line operator $\mathsf{L}_h$ labeled by $h \in H \cap g^{-1}Hg$ satisfies the pulling-through equations~\eqref{eq: pulling through}.

\vspace*{\baselineskip}
\noindent{\bf Other symmetry operators.}
Since all simple objects of $\mathcal{C}(G; H)$ are connected to $V_H^g$ for some $g \in G$, all 0-form symmetry operators can be obtained by condensing line operators on the corresponding surface operator $\mathsf{D}[V_H^g]$.
The line operators condensed on $\mathsf{D}[V_H^g]$ form a ($\Delta$-separable symmetric) Frobenius algebra in $\End_{\mathcal{C}(G; H)}(V_H^g) \cong \Vect_{H \cap g^{-1}Hg}$.
A Frobesnius algebra object of $\Vect_{H \cap g^{-1}Hg}$ is labeled by a pair $(H^{\prime}, \kappa)$, where $H^{\prime}$ is a subgroup of $H \cap g^{-1}Hg$ and $\kappa \in Z^2(H^{\prime}, \mathrm{U}(1))$ is a 2-cocycle on $H^{\prime}$ \cite{EGNO2015}.
The algebra labeled by $(H^{\prime}, \kappa)$ can be chosen to be the twisted group algebra $\mathbb{C}[H^{\prime}]^{\kappa}$.\footnote{More generally, an algebra object labeled by $(H^{\prime}, \kappa)$ can be any algebra that is Morita equivalent to $\mathbb{C}[H^{\prime}]^{\kappa}$ in $\Vect_{H \cap g^{-1}Hg}$. The surface operator obtained by the condensation depends only on the Morita class of the algebra.}
The corresponding line operator is $\bigoplus_{h \in H^{\prime}} \mathsf{L}_h$.
The multiplication $m$ and the comultiplication $\Delta$ for the Frobenius algebra structure on $\mathbb{C}[H^{\prime}]^{\kappa}$ are given by
\begin{equation}
m(v_{h_1} \otimes v_{h_2}) = \kappa(h_1, h_2) v_{h_1 h_2}, \quad 
\Delta(v_{h_1}) = \frac{1}{|H^{\prime}|} \sum_{h_2 \in H^{\prime}} \kappa(h_1h_2^{-1}, h_2)^{-1} v_{h_1 h_2^{-1}} \otimes v_{h_2},
\end{equation}
where $\{v_h \mid h \in H^{\prime}\}$ is a basis of $\mathbb{C}[H^{\prime}]^{\kappa}$.
One can check that $\mathbb{C}[H^{\prime}]^{\kappa}$ equipped with the above multiplication and comultiplication is indeed a $\Delta$-separable symmetric Frobenius algebra, whose unit $u: \mathbb{C} \to \mathbb{C}[H^{\prime}]^{\kappa}$ and counit $\epsilon: \mathbb{C}[H^{\prime}]^{\kappa} \to \mathbb{C}$ are given by
\begin{equation}
u(1) = v_e, \qquad
\epsilon(v_h) = |H^{\prime}| \delta_{h, e},
\end{equation}
where $e$ denotes the unit element of $H$.
Here, without loss of generality, we supposed that $\kappa$ is normalized so that $\kappa(e, h) = \kappa(h, e) = 1$ for any $h \in H^{\prime}$.

In the context of tensor networks, the condensation of an algebra object $\mathbb{C}[H^{\prime}]^{\kappa}$ on $\mathsf{D}[V_H^g]$ is defined by\footnote{One may define the condensation by a finer mesh of an algebra object. However, one can always reduce the mesh using the Frobenius relation and the $\Delta$-separability of $\mathbb{C}[H^{\prime}]^{\kappa}$.}
\begin{equation}
\mathsf{D}[V_H^g; \mathbb{C}[H^{\prime}]^{\kappa}] = \adjincludegraphics[valign = c, trim={10, 10, 10, 10}]{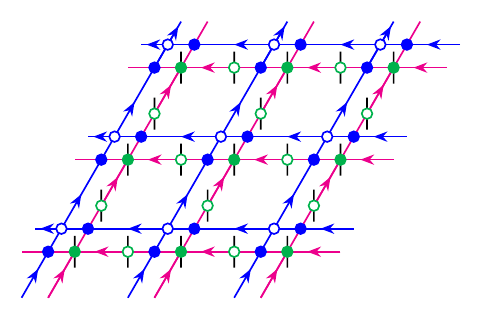}.
\label{eq: sym op condensation}
\end{equation}
The virtual Hilbert space associated with the blue bond is $\mathbb{C}[H^{\prime}]^{\kappa}$, and the local tensors represented by the filled and unfilled blue dots are given by
\begin{equation}
\adjincludegraphics[valign = c, trim={10, 10, 10, 10}]{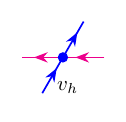} = \adjincludegraphics[valign = c, trim={10, 10, 10, 10}]{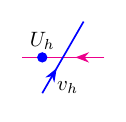}, \qquad
\adjincludegraphics[valign = c, trim={10, 10, 10, 10}]{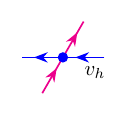} = \adjincludegraphics[valign = c, trim={10, 10, 10, 10}]{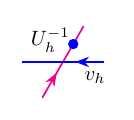}, \qquad
\adjincludegraphics[valign = c, trim={10, 10, 10, 10}]{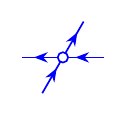} = \adjincludegraphics[valign = c, trim={10, 10, 10, 10}]{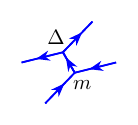}.
\end{equation}
We note that the PEPO representation~\eqref{eq: sym op condensation} has the bond dimension $|H||H^{\prime}|$.
As we will show in Appendix~\ref{sec: Condensation on a condensation surface}, one can recover the identity surface by condensing the algebra object $\mathbb{C}[H]$ on a condensation surface $\mathsf{D}[V_H^e]$, i.e., we have $\mathsf{D}[V_H^e; \mathbb{C}[H]] = \id$.

One can coarsen the blue mesh in \eqref{eq: sym op condensation} by using the pulling-through equations~\eqref{eq: pulling through} and the $\Delta$-separable Frobenius algebra structure on $\mathbb{C}[H^{\prime}]^{\kappa}$.
In particular, for periodic boundary conditions, we can reduce the number of vertical lines and horizontal lines to one:
\begin{equation}
\mathsf{D}[V_H^g; \mathbb{C}[H^{\prime}]^{\kappa}] =  \adjincludegraphics[valign = c, trim={10, 10, 10, 10}]{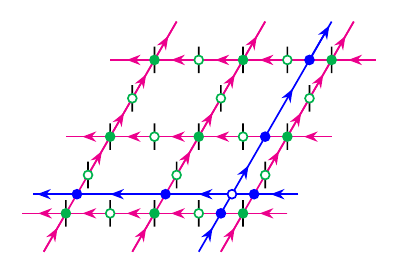}.
\label{eq: sym op condensation reduced}
\end{equation}
This expression makes it manifest that $\mathsf{D}[V_H^g; \mathbb{C}[H^{\prime}]^{\kappa}]$ commutes with the Hamiltonian~\eqref{eq: generalized cluster ham}, i.e., 
\begin{equation}
[\mathsf{D}[V_H^g; \mathbb{C}[H^{\prime}]^{\kappa}], \hat{H}_{\text{g.f.}}] = 0.
\label{eq: commutativity}
\end{equation}
This equation automatically follows from the commutativity of the local tensors of $\mathsf{D}[V_H^g]$ and the local terms of $\hat{H}_{\text{g.f.}}$, because one can freely move the line operators in \eqref{eq: sym op condensation reduced} so that they are far away from the local operators involved in the commutation relation.
Therefore, the surface operator~\eqref{eq: sym op condensation reduced} is indeed a symmetry operator of the generalized cluster model.

\subsection{Fractionalized symmetries}
\label{sec: MPO representations of fractionalized symmetries}
In this subsection, we study the symmetry action on the generalized cluster state~\eqref{eq: generalized cluster}.
In particular, we will see that the symmetry operator $\mathsf{D}[V_H^g]$ acting on the physical degrees of freedom is ``fractionalized'' into operators acting on the virtual degrees of freedom of the PEPS~\eqref{eq: generalized cluster PEPS}.
In what follows, we suppose that $S_{H \backslash G}$ is a complement of $H$ in $G$ so that the symmetry operator $\mathsf{D}[V_H^g]$ is represented by the PEPO~\eqref{eq: reduced PEPO} with bond dimension $|H|$.

The generalized cluster state~\eqref{eq: generalized cluster} must be invariant (up to scalar) under the symmetry action because it is the unique ground state of the symmetric Hamiltonian~\eqref{eq: generalized cluster ham}.
Indeed, due to \eqref{eq: D gf GS} and \eqref{eq: ungauging GS}, one can compute the global action of $\mathsf{D}[V_H^g]$ on the generalized cluster state as
\begin{equation}
\mathsf{D}[V_H^g] \ket{\text{cluster}(G; H; K)} = |H|\ket{\text{cluster}(G; H; K)}.
\label{eq: global action}
\end{equation}
On the other hand, at the level of the local tensors, $\mathsf{D}[V_H^g]$ acts on the generalized cluster PEPS~\eqref{eq: generalized cluster PEPS} as
\begin{equation}
\adjincludegraphics[valign = c, trim={10, 10, 10, 10}]{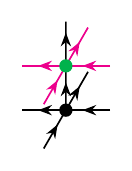} ~ = ~ \adjincludegraphics[valign = c, trim={10, 10, 10, 10}]{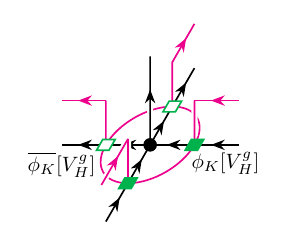}\,, \qquad
\adjincludegraphics[valign = c, trim={10, 10, 10, 10}]{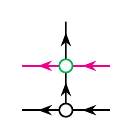} ~ = ~ \adjincludegraphics[valign = c, trim={10, 10, 10, 10}]{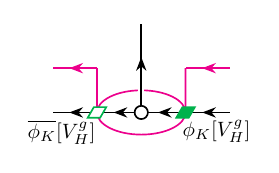}\,.
\label{eq: fractionalization}
\end{equation}
See Appendix~\ref{sec: Derivation of fractionalized symmetry operators} for a derivation.
All the virtual bonds in the above equation take values in $H$, and the tensors $\phi_K[V_H^g]$ and $\overline{\phi_K}[V_H^g]$ are defined by
\begin{equation}
\adjincludegraphics[valign = c, trim={10, 10, 10, 10}]{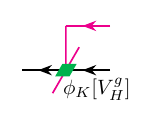} := \adjincludegraphics[valign = c, trim={10, 10, 10, 10}]{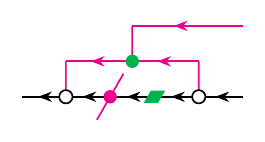}\,, \qquad
\adjincludegraphics[valign = c, trim={10, 10, 10, 10}]{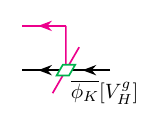} := \adjincludegraphics[valign = c, trim={10, 10, 10, 10}]{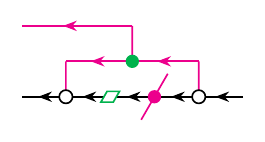}\,,
\label{eq: fractionalization MPO}
\end{equation}
where the local tensors on the right-hand sides are given by
\begin{align}
\adjincludegraphics[valign = c, trim={10, 10, 10, 10}]{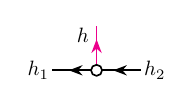} &= \adjincludegraphics[valign = c, trim={10, 10, 10, 10}]{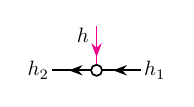} = \delta_{h, h_1^{-1}h_2}, \qquad
\adjincludegraphics[valign = c, trim={10, 10, 10, 10}]{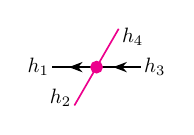} = \prod_{i, j = 1, 2, 3, 4} \delta_{h_i, h_j}, \label{eq: phiK 1} \\
\adjincludegraphics[valign = c, trim={10, 10, 10, 10}]{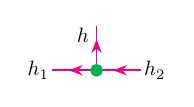} &= \adjincludegraphics[valign = c, trim={10, 10, 10, 10}]{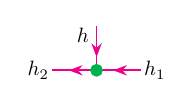} = \delta_{h_1, h} \delta_{h_2, gh\mathfrak{r}[gh]^{-1}}, \quad
\adjincludegraphics[valign = c, trim={10, 10, 10, 10}]{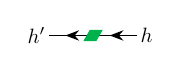} = \adjincludegraphics[valign = c, trim={10, 10, 10, 10}]{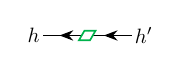} = \delta_{h^{\prime}, U_{g; K}(h)}. \label{eq: phiK 2}
\end{align}
Here, $U_{g; K}(h)$ is the unique element of $H$ that satisfies
\begin{equation}
U_{g; K}(h)gh^{-1} \in K.
\label{eq: UgK}
\end{equation}
We note that the white and magenta dots in \eqref{eq: phiK 1} are the multiplication and copy tensors, while the three-leg tensors in \eqref{eq: phiK 2} are the same as those in \eqref{eq: filled}.

\vspace*{\baselineskip}
\noindent{\bf Action tensors.}
For later use, we decompose $\phi_K[V_H^g]$ and $\overline{\phi_K}[V_H^g]$ into the components labeled by $h \in H$ as follows:
\begin{equation}
\phi_K[V_H^g]_h := \adjincludegraphics[valign = c, trim={10, 10, 10, 10}]{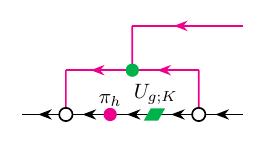}, \qquad \overline{\phi_K}[V_H^g]_h := \adjincludegraphics[valign = c, trim={10, 10, 10, 10}]{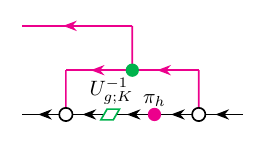}.
\label{eq: action tensors}
\end{equation}
Here, $\pi_h$ is the projection to the virtual state labeled by $h \in H$, and $U_{g; K}: H \to H$ is the map defined by \eqref{eq: UgK}.\footnote{We note that $U_{g; K}: H \to H$ is bijective and hence invertible. A proof goes as follows. By defintion, for any $h_1, h_2 \in H$, there exist $k_1, k_2 \in K$ such that $U_{g; K}(h_1)gh_1^{-1}=k_1$ and $U_{g; K}(h_2)gh_2^{-1} = k_2$. If $U_{g; K}(h_1)=U_{g; K}(h_2)$, we have $k_1h_1 = k_2h_2$, which implies $h_1 = h_2$ because $K$ is a complement of $H$. Therefore, $U_{g; K}$ is injective. Furthermore, since the source and target of $U_{g; K}$ are the same, the injectivity implies the surjectivity. Thus, $U_{g; K}$ is bijective. More explicitly, the inverse of $U_{g; K}$ is given by $U_{g; K}^{-1}(h) = hg^{-1}U_{g; K}(h)^{-1}hg$. \label{fn: UgK inverse}}
The tensors $\phi_K[V_H^g]_h$ and $\overline{\phi_K}[V_H^g]_h$ are called action tensors.
By using the action tensors, the local action~\eqref{eq: fractionalization} of the symmetry operator $\mathsf{D}[V_H^g]$ on the generalized cluster state can be written as
\begin{equation}
\adjincludegraphics[valign = c, trim={10, 10, 10, 10}]{tikz/out/local_action1.pdf} ~ = \sum_{h \in H} \adjincludegraphics[valign = c, trim={10, 10, 10, 10}]{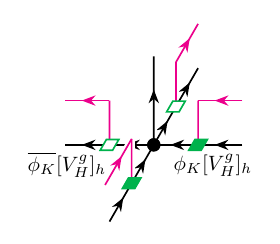}, \qquad
\adjincludegraphics[valign = c, trim={10, 10, 10, 10}]{tikz/out/local_action3.pdf} ~ = \sum_{h \in H} \adjincludegraphics[valign = c, trim={10, 10, 10, 10}]{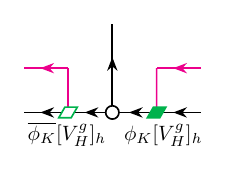}.
\label{eq: decomposed fractionalization}
\end{equation}
The action tensors defined by \eqref{eq: action tensors} satisfy the following orthogonality and completeness relations:
\begin{equation}
\adjincludegraphics[valign = c, trim={10, 10, 10, 10}]{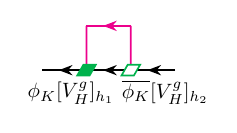} = \delta_{h_1, h_2} \adjincludegraphics[valign = c, trim={10, 10, 10, 10}]{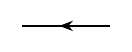}, \qquad
\sum_{h \in H} \adjincludegraphics[valign = c, trim={10, 10, 10, 10}]{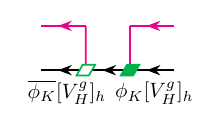} = \adjincludegraphics[valign = c, trim={10, 10, 10, 10}]{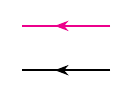}.
\label{eq: orthogonality and completeness}
\end{equation}
A derivation of the above equations will be given in Appendix~\ref{sec: Orthogonality and completeness relations}.\footnote{As shown in Appendix~\ref{sec: Orthogonality and completeness relations}, the completeness relation automatically follows from the orthogonality relation.}
In Appendix~\ref{sec: Uniqueness of action tensors}, we will show that the action tensors satisfying the orthogonality and completeness relations are essentially unique.
Due to the orthogonality relation, one can freely reconnect the fractionalized symmetry operators as 
\begin{equation}
\adjincludegraphics[valign = c, trim={10, 10, 10, 10}]{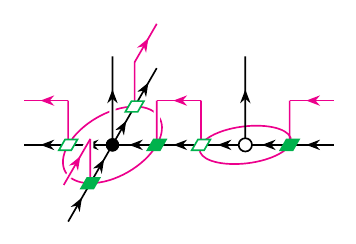} ~ = ~ \adjincludegraphics[valign = c, trim={10, 10, 10, 10}]{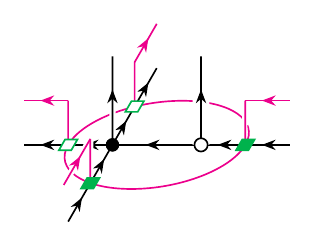}.
\label{eq: reconnection}
\end{equation}
Using the local symmetry action~\eqref{eq: fractionalization} and the reconnection~\eqref{eq: reconnection}, one can recover the global symmetry action~\eqref{eq: global action}.

\subsection{Example: symmetry of the $G$-cluster state}
\label{sec: Example: symmetry of the G-cluster state}
As a simple example, let us consider the symmetry of the $G_0$-cluster model and its action on the $G_0$-cluster state~\eqref{eq: G cluster}, where $G_0$ is any finite group.
To this end, we first recall that the $G_0$-cluster model is obtained by gauging the $G_0^{\text{left}}$ symmetry in a model with $G_0 \times G_0$ symmetry as mentioned in Section~\ref{sec: Example: G-cluster state}.
Thus, by construction, this model has a group-theoretical fusion 2-category symmetry
\begin{equation}
\mathcal{C}(G_0 \times G_0; G_0^{\text{left}}) \cong 2\Rep(G_0) \boxtimes 2\Vect_{G_0}.
\end{equation}
In what follows, we will write down the tensor network representations of the (non-invertible) 0-form symmetry operators and compute their local actions on the $G_0$-cluster state.

\vspace*{\baselineskip}
\noindent{\bf Symmetry operators.}
As discussed in Section~\ref{sec: Tensor network representations of symmetry operators}, all the 0-form symmetry operators up to condensation are of the form
\begin{equation}
\mathsf{D}[V_{G_0^{\text{left}}}^{(g_{\ell}, g_r)}] \coloneq \mathsf{D}_{G_0^{\text{left}}} U_{(g_{\ell}, g_r)} \overline{\mathsf{D}}_{G_0^{\text{left}}}, \qquad
\forall g_{\ell}, g_r \in G_0,
\label{eq: G-cluster sym op}
\end{equation}
where $\mathsf{D}_{G_0^{\text{left}}}$ and $\overline{\mathsf{D}}_{G_0^{\text{left}}}$ are the gauging and ungauging operators for the $G_0^{\text{left}}$ symmetry, and $U_{(g_{\ell}, g_r)}$ is the invertible symmetry operator of the ungauged model.
More explicitly, the operators on the right-hand side are given by\footnote{The general definitions of the gauging and ungauging operators are given in \eqref{eq: D gf} and \eqref{eq: D gf bar}, respectively.}
\begin{align}
\mathsf{D}_{G_0^{\text{left}}} \ket{\{g_i, g_i^{\prime}\}} = \ket{\{g_i^{\prime}; g_i^{-1}g_j\}}, \quad
\overline{\mathsf{D}}_{G_0^{\text{left}}} = \mathsf{D}_{G_0^{\text{left}}}^{\dagger}, \quad
U_{(g_{\ell}, g_r)} \ket{\{g_i, g_i^{\prime}\}} = \ket{\{g_{\ell}g_i, g_rg_i^{\prime}\}},
\label{eq: D G0left}
\end{align}
where $\{\ket{\{g_i, g_i^{\prime}\}} \mid g_i, g_i^{\prime} \in G_0\}$ is a basis of the state space of the ungauged model.
Any other 0-form symmetry operators can be obtained from \eqref{eq: G-cluster sym op} by condensing line operators.
Due to the general result in Section~\ref{sec: Symmetry operators}, the symmetry operator~\eqref{eq: G-cluster sym op} can be represented by a tensor network of the form
\begin{equation}
\mathsf{D}[V_{G_0^{\text{left}}}^{(g_{\ell}, g_r)}] = \adjincludegraphics[valign = c, trim={10, 10, 10, 10}]{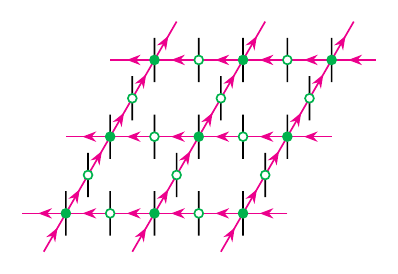},
\label{eq: G-cluster sym op PEPO}
\end{equation}
where both the physical legs and the virtual legs take values in $G_0$.
The filled and unfilled green dots on the right-hand side are defined as in \eqref{eq: filled} and \eqref{eq: unfilled}, i.e.,
\begin{equation}
\adjincludegraphics[valign = c, trim={10, 10, 10, 10}]{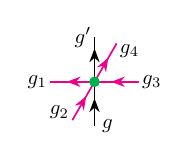} = \delta_{g^{\prime}, g_rg} \prod_{i, j = 1, 2, 3, 4} \delta_{g_i, g_j}, \qquad
\adjincludegraphics[valign = c, trim={10, 10, 10, 10}]{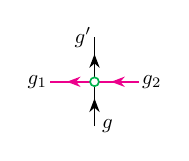} = \delta_{g, g_1^{-1}g_2} \delta_{g^{\prime}, g}.
\end{equation}
From the $(g_{\ell}, g_r)$-dependence of the above tensors, we find
\begin{equation}
\mathsf{D}[V_{G_0^{\text{left}}}^{(g_{\ell}, g_r)}] = U_{g_r} \mathsf{D}[V_{G_0^{\text{left}}}^{(e, e)}],
\label{eq: G-cluster sym op2}
\end{equation}
where $U_{g_r}$ is the left multiplication of $g_r$ acting on the matter fields:
\begin{equation}
U_{g_r} \coloneq \bigotimes_{i \in P} \overrightarrow{X}_{g_r}^{(i)}.
\end{equation}
We note that $\mathsf{D}[V_{G_0^{\text{left}}}^{(g_{\ell}, g_r)}]$ in \eqref{eq: G-cluster sym op2} does not depend on $g_{\ell} \in G_0$.
This is an immediate consequence of the fact that the symmetry operator for $G_0^{\text{left}}$ is absorbed by the gauging operator $\mathsf{D}_{G_0^{\text{left}}}$ in \eqref{eq: G-cluster sym op}.

\vspace*{\baselineskip}
\noindent{\bf Action tensors.}
Due to the general discussions in Section~\ref{sec: MPO representations of fractionalized symmetries}, the symmetry operators of the form~\eqref{eq: G-cluster sym op PEPO} act on the local tensors of the $G_0$-cluster state~\eqref{eq: G cluster PEPS} as
\begin{equation}
\adjincludegraphics[valign = c, trim={10, 10, 10, 10}]{tikz/out/local_action1.pdf} ~ = \sum_{g \in G_0} \adjincludegraphics[valign = c, trim={10, 10, 10, 10}]{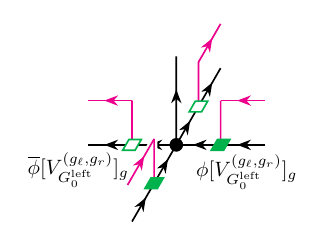}, \quad
\adjincludegraphics[valign = c, trim={10, 10, 10, 10}]{tikz/out/local_action3.pdf} ~ = \sum_{g \in G_0} \adjincludegraphics[valign = c, trim={10, 10, 10, 10}]{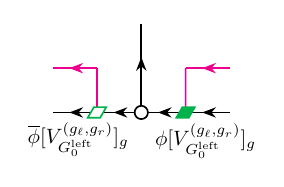},
\end{equation}
where the action tensors $\phi[V_{G_0^{\text{left}}}^{(g_{\ell}, g_r)}]_g$ and $\overline{\phi}[V_{G_0^{\text{left}}}^{(g_{\ell}, g_r)}]_g$ are defined as in \eqref{eq: action tensors}.
Based on the general expression \eqref{eq: action tensors}, we can compute the components of these action tensors explicitly as\footnote{Although these action tensors depend on $g_{\ell}$, the set of the action tensors $\{\phi[V_{G_0^{\text{left}}}^{(g_{\ell}, g_r)}]_g \mid g \in G_0\}$ does not because $g_{\ell}$ only shifts the subscript $g$.
This is consistent with the fact that the symmetry operator is independent of $g_{\ell}$.}
\begin{equation}
\adjincludegraphics[valign = c, trim={10, 10, 10, 10}]{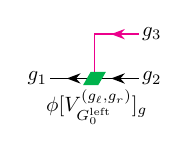} = \adjincludegraphics[valign = c, trim={10, 10, 10, 10}]{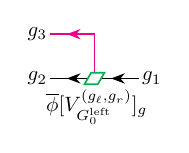} = \delta_{g_1g_3^{-1}, gg_{\ell}} \delta_{g_2g_3^{-1}, g_r^{-1}gg_{\ell}}.
\label{eq: G-cluster action tensor}
\end{equation}
These action tensors are the building blocks of the symmetry operators on the boundary of the $G_0$-cluster state, which will be discussed in Section~\ref{sec: Example: gapped boundary of the G-cluster state}.
See also Section~\ref{Example: gapped boundary of the G-cluster state CT} for a category-theoretical description of the symmetry on the boundary of the $G_0$-cluster state.

\section{Interfaces of 2+1d generalized cluster states}
\label{sec: Interfaces of 2+1d generalized cluster states}
In this section, we discuss the interface of 2+1d generalized cluster states with group-theoretical fusion 2-category symmetry $\mathcal{C}(G; H)$.
Specifically, we determine the symmetry structure at the interface and investigate gapped interfaces allowed by the symmetry.
We provide both a category-theoretical description and a tensor network description of the interfaces in subsections~\ref{sec: Category theoretical description} and \ref{sec: Tensor network description}, respectively.
The readers interested only in the tensor network description may skip subsection~\ref{sec: Category theoretical description}.

\subsection{Category theoretical description}
\label{sec: Category theoretical description}
We first provide a purely categorical description of the interface between the SPT phases realized by the generalized cluster states.
Our discussion does not rely on the concrete lattice model.
As such, this subsection can be read independently of Sections~\ref{sec: 2+1d generalized cluster states with non-invertible symmetries} and \ref{sec: Non-invertible symmetries of 2+1d generalized cluster states}.

\subsubsection{Interface symmetries: strip 2-algebras}
\label{sec: Interface symmetries CT}
As we reviewed in Section~\ref{sec: Fiber 2-functors}, fiber 2-functors of fusion 2-category $\mathcal{C}(G; H)$ are labeled by pairs $(K, \lambda)$, where $K$ is a complement of $H$ in $G$ (i.e., a subgroup of $G$ satisfying \eqref{eq: ZS product}) and $\lambda \in Z^3(K, \mathrm{U}(1))$ is a 3-cocycle on $K$.
In what follows, we will restrict our attention to fiber 2-functors with trivial $\lambda$, in which case a fiber 2-functor is labeled only by $K$.
The fiber 2-functor labeled by $K$ can be described as a left $\mathcal{C}(G; H)$-module 2-category \cite{Decoppet:2023bay}\footnote{We recall that a fiber 2-functor is equivalently defined as a module 2-category that is equivalent to $2\Vect$ as a semisimple 2-category.}
\begin{equation}
\mathcal{M}(K) := {}_{\Vect_H} (2\Vect_G)_{\Vect_K},
\label{eq: MK}
\end{equation}
which is equivalent to $2\Vect$ as a semisimple 2-category because $K$ is a complement of $H$ in $G$~\cite{Decoppet:2023bay}.
The left action of $\mathcal{C}(G; H)$ on $\mathcal{M}(K)$ is given by the relative tensor product over $\Vect_H$.
Physically, this fiber 2-functor corresponds to the SPT phase realized by the generalized cluster state~\eqref{eq: generalized cluster}.
This SPT phase will be denoted by $\mathsf{SPT}_{K}$.

Now, let us consider an interface between $\mathsf{SPT}_{K_1}$ and $\mathsf{SPT}_{K_2}$.
A symmetry operator acting on the interface can be depicted as
\begin{equation}
\mathsf{L}(X; m_1, m_2) := \adjincludegraphics[valign = c, trim={10, 10, 10, 10}]{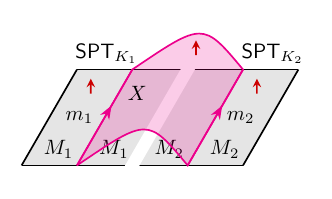},
\label{eq: interface line op}
\end{equation}
where $X \in \mathcal{C}(G; H)$ is an object of $\mathcal{C}(G; H)$, $M_1 \in \mathcal{M}(K_1)$ and $M_2 \in \mathcal{M}(K_2)$ are the unique simple objects defined by
\begin{equation}
M_1 = \Vect_H \boxtimes \Vect_{K_1}, \qquad
M_2 = \Vect_H \boxtimes \Vect_{K_2},
\end{equation}
and $m_1 \in \Hom_{\mathcal{M}(K_1)}(X \boxtimes_{\Vect_H} M_1, M_1)$ and $m_2 \in \Hom_{\mathcal{M}(K_2)}(M_2, X \boxtimes_{\Vect_H} M_2)$ are 1-morphisms of $\mathcal{M}(K_1)$ and $\mathcal{M}(K_2)$.
Physically, $X$ corresponds to a symmetry operator in the bulk, $M_1$ and $M_2$ correspond to the ground states of $\mathsf{SPT}_{K_1}$ and $\mathsf{SPT}_{K_2}$, and $m_1$ and $m_2$ correspond to topological lines on which the bulk symmetry operator ends.\footnote{From the tensor network point of view, the 1-morphisms $m_1$ and $m_2$ correspond to the fractionalized symmetry operators defined in Section~\ref{sec: MPO representations of fractionalized symmetries}. See Section \ref{sec: Tensor network description} for more details on this point.}
The small red arrows in \eqref{eq: interface line op} specify the coorientations of the surfaces, which will be omitted when no confusion can arise.

We interpret $\mathsf{L}(X; m_1, m_2)$ as a line operator acting on the interface between $\mathsf{SPT}_{K_1}$ and $\mathsf{SPT}_{K_2}$.
Physically, its action is defined by shrinking the red surface in \eqref{eq: interface line op} onto the middle interface.
In \eqref{eq: interface line op}, the interface is left blank to indicate that we do not specify the interface degrees of freedom.
In general, the interface degrees of freedom form a module category over the symmetry category at the interface.
Such module categories will be discussed in Section~\ref{sec: Gapped interface modes CT}, while here, we focus only on the symmetry category itself.

Similarly, a topological point operator between $\mathsf{L}(X; m_1, m_2)$ and $\mathsf{L}(X^{\prime}; m_1^{\prime}, m_2^{\prime})$ can be depicted as
\begin{equation}
\adjincludegraphics[valign = c, trim={10, 10, 10, 10}]{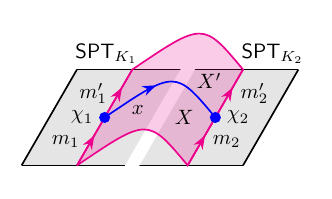},
\label{eq: interface point op}
\end{equation}
where $x \in \Hom_{\mathcal{C}(G; H)} (X, X^{\prime})$ is a 1-morphism in $\mathcal{C}(G; H)$, and $\chi_1 \in \Hom_{\mathcal{M}(K_1)}(m_1, m_1^{\prime} \circ (x \boxtimes_{\Vect_H} \id_{M_1}))$ and $\chi_2 \in \Hom_{\mathcal{M}(K_2)}((x \boxtimes_{\Vect_H} \id_{M_2}) \circ m_2, m_2^{\prime})$ are 2-morphisms in $\mathcal{M}(K_1)$ and $\mathcal{M}(K_2)$.
The line operator $\mathsf{L}(X; m_1, m_2)$ is not necessarily simple even if $X$, $m_1$, and $m_2$ are simple because there may exist a non-trivial topological point operator on it.
To obtain a simple summand, one may condense an idempotent of the algebra of topological point operators on $\mathsf{L}(X; m_1, m_2)$. Fortunately, as we will see later, the line operator $\mathsf{L}(X; m_1, m_2)$ becomes simple if we choose $X$, $m_1$, and $m_2$ appropriately, and any simple line can be obtained in that way.
Thus, we do not discuss the condensation of an idempotent on the line operator.

If we fold the picture in \eqref{eq: interface line op} along the interface, we obtain
\begin{equation}
\mathsf{L}(X; m_1, m_2) = \adjincludegraphics[valign = c, trim={10, 10, 10, 10}]{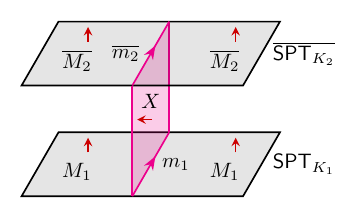}.
\label{eq: strip}
\end{equation}
The interface is on the right side of the diagram, which we left blank as in \eqref{eq: interface line op}.
Here, $\overline{\mathsf{SPT}_{K_2}}$ denotes the orientation-reversal of $\mathsf{SPT}_{K_2}$.
This SPT phase is labeled by a right $\mathcal{C}(G; H)$-module 2-category\footnote{The orientation-reversal of $\mathsf{SPT}_{K_2}$ has a symmetry $\mathcal{C}(G; H)^{\text{mop}}$, which is the fusion 2-category obtained by reversing the order of the tensor product in $\mathcal{C}(G; H)$. Therefore, $\overline{\mathsf{SPT}_{K_2}}$ is labeled by a left $\mathcal{C}(G; H)^{\text{mop}}$-module 2-category, or equivalently, a right $\mathcal{C}(G; H)$-module 2-category.}
\begin{equation}
\overline{\mathcal{M}(K_2)} = {}_{\Vect_{K_2}}(2\Vect_G)_{\Vect_H}.
\end{equation}
The unique simple object of this 2-category is given by $\overline{M_2} = \Vect_{K_2} \boxtimes \Vect_H$.
The figure in \eqref{eq: strip} suggests that the symmetry operators acting on the interface can be identified with 1-endomorphisms of $\overline{M}_2 \boxtimes_{\Vect_H} M_1$ in the 2-category ${}_{\Vect_{K_2}}(2\Vect_G)_{\Vect_{K_1}}$.
Similarly, topological point operators between these symmetry operators can be identified with 2-morphisms in the same 2-category.
Therefore, the symmetry category at the interface of $\mathsf{SPT}_{K_1}$ and $\mathsf{SPT}_{K_2}$ should be
\begin{equation}
\begin{aligned}
\mathcal{C}_{K_1, K_2} & := \End_{{}_{\Vect_{K_2}}(2\Vect_G)_{\Vect_{K_1}}} (\overline{M_2} \boxtimes_{\Vect_H} M_1) \\
&~ \cong \End_{{}_{\Vect_{K_2}}(2\Vect_G)_{\Vect_{K_1}}} (\Vect_{K_2} \boxtimes \Vect_H \boxtimes \Vect_{K_1}),
\label{eq: strip 2-algebra}
\end{aligned}
\end{equation}
which is a multifusion 1-category, whose monoidal structure is given by the composition of 1-morphisms in ${}_{\Vect_{K_2}}(2\Vect_G)_{\Vect_{K_1}}$.
The above category will be called a strip 2-algebra following~\cite{Cordova:2024iti, Gagliano:2025gwr}.\footnote{In this context, a 2-algebra means an algebra in the 2-category $2\Vect$, i.e., a monoidal 1-category.}

\vspace*{\baselineskip}
\noindent{\bf Strip 2-algebra $\mathcal{C}_{K_1, K_2}$.}
Let us describe the multifusion category structure on $\mathcal{C}_{K_1, K_2}$ in more detail.
To this end, we first enumerate all simple objects of $\mathcal{C}_{K_1, K_2}$.
Pictorially, simple objects of the category on the second line of \eqref{eq: strip 2-algebra} can be represented by the following diagram:
\begin{equation}
\mathsf{L}(h_1, h_2, k_1, k_2) := 
\adjincludegraphics[valign = c, trim={10, 10, 10, 10}]{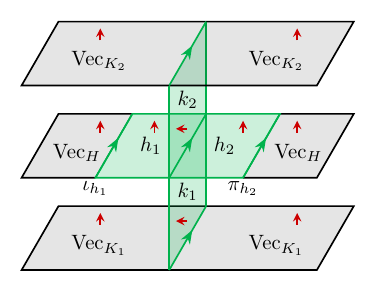}.
\label{eq: simples of strip}
\end{equation}
The relation to the objects of the form \eqref{eq: strip} will be explained later.
The green surface labeled by $h_1, h_2 \in H$, $k_1 \in K_1$, and $k_2 \in K_2$ represent the simple objects $V^{h_1}, V^{h_2}, V^{k_1}, V^{k_2} \in 2\Vect_G$.
These objects must satisfy
\begin{equation}
h_1 k_1 = k_2 h_2
\label{eq: h1k1 = k2h2}
\end{equation}
so that the configuration of the surfaces is compatible with the fusion rules.
We note that $h_2$ and $k_2$ are uniquely determined by $h_1$ and $k_1$ because $K_2$ is a complement of $H$ in $G$.
The 1-morphisms $\iota_{h_1}: \Vect_H \to V^{h_1}$ and $\pi_{h_2}: V^{h_2} \to \Vect_H$ in \eqref{eq: simples of strip} are the inclusion and projection 1-morphisms.
The simple 1-morphism at the junction of the vertical surface and the middle surface is unique because of the following equivalence of semisimple 1-categories:
\begin{equation}
\Hom_{2\Vect_G} (V^{h_1} \boxtimes V^{k_1}, V^{k_2} \boxtimes V^{h_2}) \cong \delta_{h_1k_1, k_2h_2} \Vect.
\end{equation}
Similarly, the simple 1-morphisms on the top and bottom surfaces where the vertical surface ends are also unique because
\begin{equation}
\Hom_{(2\Vect_G)_{\Vect_{K_1}}} (V^{k_1} \boxtimes \Vect_{K_1}, \Vect_{K_1}) \cong \Hom_{{}_{\Vect_{K_2}}(2\Vect_G)} (\Vect_{K_2}, \Vect_{K_2} \boxtimes V^{k_2}) \cong \Vect.
\end{equation}
We note that $\mathsf{L}(h_1, h_2, k_1, k_2)$ is a simple object because we cannot put any non-trivial topological line on the vertical surface in \eqref{eq: simples of strip}, meaning that $\mathsf{L}(h_1, h_2, k_1, k_2)$ cannot have a non-trivial topological point operator on it.
Furthermore, the same argument shows that there exists a topological point operator between $\mathsf{L}(h_1, h_2, k_1, k_2)$ and $\mathsf{L}(h_1^{\prime}, h_2^{\prime}, k_1^{\prime}, k_2^{\prime})$ if and only if $(h_1, h_2, k_1, k_2) = (h_1^{\prime}, h_2^{\prime}, k_1^{\prime}, k_2^{\prime})$, which implies that
\begin{equation}
\mathsf{L}(h_1, h_2, k_1, k_2) \cong \mathsf{L}(h_1^{\prime}, h_2^{\prime}, k_1^{\prime}, k_2^{\prime}) ~\Leftrightarrow~ (h_1, h_2, k_1, k_2) = (h_1^{\prime}, h_2^{\prime}, k_1^{\prime}, k_2^{\prime}).
\label{eq: isom class of L}
\end{equation}
Since $\mathsf{L}(h_1, h_2, k_1, k_2)$ is uniquely determined by the pair $(h_1, k_1)$, the number of the isomorphism classes of simple objects is $|H| |K_1| = |G|$.\footnote{This is consistent with the equivalence of semisimple 1-categories $\mathcal{C}_{K_1, K_2} \cong \Hom_{2\Vect_G}(\Vect_H, \Vect_{K_2} \boxtimes \Vect_H \boxtimes \Vect_{K_1}) \cong \Vect^{\oplus |G|}$, where the first equivalence follows from \cite[Lemma 3.2.2]{Decoppet:2023bay}, and $\Vect^{\oplus |G|}$ denotes the direct sum of $|G|$ copies of $\Vect$. We emphasize that the equivalence $\mathcal{C}_{K_1, K_2} \cong \Vect^{\oplus |G|}$ is not a monoidal equivalence.}

The tensor product of the objects \eqref{eq: simples of strip} is given by the horizontal concatenation
\begin{equation}
\mathsf{L}(h_1, h_2, k_1, k_2) \otimes \mathsf{L}(h_1^{\prime}, h_2^{\prime}, k_1^{\prime}, k_2^{\prime}) = \adjincludegraphics[scale=0.95, valign = c, trim={10, 10, 10, 10}]{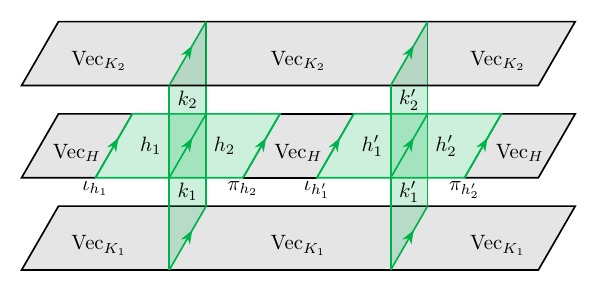}.
\end{equation}
Since $\pi$ and $\iota$ on the right-hand side satisfy $\pi_{h_2} \circ \iota_{h_1^{\prime}} = \delta_{h_2, h_1^{\prime}} \id_{V^{h_1^{\prime}}}$, it follows that the simple objects obey the following fusion rules:
\begin{equation}
\mathsf{L}(h_1, h_2, k_1, k_2) \otimes \mathsf{L}(h_1^{\prime}, h_2^{\prime}, k_1^{\prime}, k_2^{\prime}) \cong \delta_{h_2, h_1^{\prime}} \mathsf{L}(h_1, h_2^{\prime}, k_1k_1^{\prime}, k_2k_2^{\prime}).
\label{eq: fusion rules}
\end{equation}
The $F$-symbols of $\mathcal{C}_{K_1, K_2}$ are trivial because we have chosen all the cocycle data to be trivial.
We note that the fusion outcome of two non-zero simple objects can be empty, indicating that $\mathcal{C}_{K_1, K_2}$ is generally a multifusion category rather than a fusion category.\footnote{A multifusion category is a finite semisimple $\mathbb{C}$-linear rigid monoidal category \cite{EGNO2015}. A fusion category is a multifusion category whose unit object is simple. See Appendix~\ref{sec: Multifusion categories} for more details of multifusion categories.}

\vspace*{\baselineskip}
\noindent{\bf Direct sum decomposition of $\mathcal{C}_{K_1, K_2}$.}
We can decompose $\mathcal{C}_{K_1, K_2}$ into a direct sum of indecomposable multifusion categories by using the following monoidal equivalence:
\begin{equation}
\begin{aligned}
\mathcal{C}_{K_1, K_2} & \cong \bigoplus_{h_1 \in H, h_2 \in H} \Hom_{{}_{\Vect_{K_2}}(2\Vect_G)_{\Vect_{K_1}}} (\Vect_{K_2} \boxtimes V^{h_2} \boxtimes \Vect_{K_1}, \Vect_{K_2} \boxtimes V^{h_1} \boxtimes \Vect_{K_1}) \\
& \cong \bigoplus_{x \in K_2 \backslash G / K_1} \bigoplus_{h_1, h_2 \in H \cap x} \Hom_{{}_{\Vect_{K_2}}(2\Vect_G)_{\Vect_{K_1}}} (\Vect_{K_2} \boxtimes V^{h_2} \boxtimes \Vect_{K_1}, \Vect_{K_2} \boxtimes V^{h_1} \boxtimes \Vect_{K_1}) \\ 
& = \bigoplus_{x \in K_2 \backslash G / K_1} \bigoplus_{h_1, h_2 \in H \cap x} (\mathcal{C}_{K_1, K_2}^x)_{h_1, h_2}.
\end{aligned}
\label{eq: monoidal structure}
\end{equation}
Here, we defined
\begin{equation}
(\mathcal{C}_{K_1, K_2}^x)_{h_1, h_2} := \Hom_{{}_{\Vect_{K_2}}(2\Vect_G)_{\Vect_{K_1}}} (\Vect_{K_2} \boxtimes V^{h_2} \boxtimes \Vect_{K_1}, \Vect_{K_2} \boxtimes V^{h_1} \boxtimes \Vect_{K_1})
\end{equation}
for $x \in K_2 \backslash G / K_1$ and $h_1, h_2 \in H \cap x$.
The first equivalence of \eqref{eq: monoidal structure} immediately follows from \eqref{eq: strip 2-algebra}, and the second equivalence holds because the direct sum component of the right-hand side on the first line is non-empty if and only if $h_1$ and $h_2$ are in the same $(K_2, K_1)$-double coset in $G$.
The monoidal structure on the right-hand side of \eqref{eq: monoidal structure} is given by the composition of 1-morphisms in ${}_{\Vect_{K_2}}(2\Vect_G)_{\Vect_{K_1}}$.
We note that for any pair of objects $\mathsf{L} \in (\mathcal{C}^x_{K_1, K_1})_{h_1, h_2}$ and $\mathsf{L}^{\prime} \in (\mathcal{C}^{x^{\prime}}_{K_1, K_2})_{h_1^{\prime}, h_2^{\prime}}$, the tensor product $\mathsf{L} \otimes \mathsf{L}^{\prime}$ is zero if $x \neq x^{\prime}$.
Thus, we find that the category
\begin{equation}
\mathcal{C}^x_{K_1, K_2} := \bigoplus_{h_1, h_2 \in H \cap x} (\mathcal{C}_{K_1, K_2}^x)_{h_1, h_2}
\label{eq: Cx K1 K2}
\end{equation}
is itself a multifusion category for each $x \in K_2 \backslash G / K_1$.
Furthermore, $\mathcal{C}_{K_1, K_2}^x$ is indecomposable because for any object $\mathsf{L} \in \mathcal{C}_{K_1, K_2}^x$, there exists $\mathsf{L}^{\prime} \in \mathcal{C}_{K_1, K_2}^x$ such that $\mathsf{L} \otimes \mathsf{L}^{\prime}$ is non-zero.
Therefore, the strip 2-algebra $\mathcal{C}_{K_1, K_2}$ can be decomposed into a direct sum of indecomposable multifusion categories as
\begin{equation}
\mathcal{C}_{K_1, K_2} \cong \bigoplus_{x \in K_2 \backslash G / K_1} \mathcal{C}_{K_1, K_2}^x,
\label{eq: strip decomposition}
\end{equation}
where $\mathcal{C}_{K_1, K_2}^x$ is the full subcategory that contains simple objects $\mathsf{L}(h_1, h_2, k_1, k_2)$ for all $h_1, h_2 \in H \cap x$.

\vspace*{\baselineskip}
\noindent{\bf Indecomposable multifusion category $\mathcal{C}_{K_1, K_2}^x$.}
Let us unpack the definition of $\mathcal{C}_{K_1, K_2}^x$.
To this end, we first recall that $\mathcal{C}_{K_1, K_2}^x$ can be decomposed into a direct sum~\eqref{eq: Cx K1 K2} as a semisimple category.
Simple objects of each component $(\mathcal{C}_{K_1, K_2}^x)_{h_1, h_2}$ are $\mathsf{L}(h_1, h_2, k_1, k_2)$, where 
\begin{equation}
h_1k_1 = k_2 h_2 ~ \Leftrightarrow ~ k_1 = h_1^{-1} k_2 h_2 \in K_1 \cap h_1^{-1}K_2h_2.
\end{equation}
The set $K_1 \cap h_1^{-1}K_2h_2$ is not empty because $h_1$ and $h_2$ are in the same double $(K_2, K_1)$-coset in $G$.
Furthermore, for each $k_1 \in K_1 \cap h_1^{-1}K_2h_2$, there exists a unique $k_2 \in K_2$ that satisfies the above condition.
Thus, the simple objects of $(\mathcal{C}_{K_1, K_2}^x)_{h_1, h_2}$ are in one-to-one correspondence with elements of $K_1 \cap h_1^{-1}K_2h_2$.
This implies the equivalence of semisimple categories
\begin{equation}
(\mathcal{C}_{K_1, K_2}^x)_{h_1, h_2} \cong \Vect_{K_1 \cap h_1^{-1} K_2 h_2},
\end{equation}
where the right-hand side is the category of finite dimensional vector spaces graded by the set $K_1 \cap h_1^{-1} K_2 h_2$.
Therefore, we find
\begin{equation}
\mathcal{C}_{K_1, K_2}^x \cong \bigoplus_{h_1, h_2 \in H \cap x} \Vect_{K_1 \cap h_1^{-1}K_2h_2}.
\label{eq: Cx K1 K2 semisimple}
\end{equation}
We note that each component $\Vect_{K_1 \cap h_1^{-1} K_2 h_2}$ is not necessarily a monoidal category because $K_1 \cap h_1^{-1} K_2 h_2$ is not necessarily a group.
Nevertheless, the direct sum~\eqref{eq: Cx K1 K2 semisimple} has a monoidal structure, which we describe shortly.

The decomposition~\eqref{eq: Cx K1 K2 semisimple} implies that an object of $\mathcal{C}_{K_1, K_2}^x$ can be identified with a $|H \cap x| \times |H \cap x|$ matrix whose $(h_1, h_2)$-component is an object of $\Vect_{K_1 \cap h_1^{-1}K_2h_2}$.
The tensor product of objects is then given by matrix multiplication, where the sum and the product of components are defined by the direct sum and the tensor product of graded vector spaces.
In particular, a simple object $\mathsf{L}(h_1, h_2, k_1, k_2) \in \mathcal{C}_{K_1, K_2}^x$ corresponds to a matrix $(\mathbb{C}_{k_1})_{h_1, h_2}$ whose $(h_1, h_2)$-component is a one-dimensional vector space $\mathbb{C}_{k_1}$ graded by $k_1 \in K_1 \cap h_1^{-1}K_2h_2$ and other components are empty.
The tensor product of the simple objects corresponds to the matrix multiplication
\begin{equation}
(\mathbb{C}_{k_1})_{h_1, h_2} (\mathbb{C}_{k_1^{\prime}})_{h_1^{\prime}, h_2^{\prime}} = \delta_{h_2, h_1^{\prime}} (\mathbb{C}_{k_1k_1^{\prime}})_{h_1, h_2^{\prime}},
\end{equation}
which agrees with the fusion rule~\eqref{eq: fusion rules}.

More abstractly, the monoidal structure on $\mathcal{C}_{K_1, K_2}^x$ is specified by a $((\mathcal{C}_{K_1, K_2}^x)_{h_1,h_1}, (\mathcal{C}_{K_1, K_2}^x)_{h_2, h_2})$-bimodule category structure on $(\mathcal{C}_{K_1, K_2}^x)_{h_1, h_2}$ for every pair $h_1, h_2 \in H \cap x$.
This bimodule structure is given by the obvious one induced by the group multiplication, i.e.,
\begin{equation}
\mathbb{C}_{k_1} \vartriangleright \mathbb{C}_{k_1^{\prime}} \vartriangleleft \mathbb{C}_{k_1^{\prime \prime}} = \mathbb{C}_{k_1k_1^{\prime}k_1^{\prime \prime}}
\end{equation}
for all $k_1 \in K_1 \cap h_1^{-1}K_2h_1$, $k_1^{\prime} \in K_1 \cap h_1^{-1}K_2h_2$, and $k_1^{\prime \prime} \in K_1 \cap h_2^{-1}K_2h_2$.
Here, $\vartriangleright$ and $\vartriangleleft$ denote the left action of $(\mathcal{C}_{K_1, K_2}^x)_{h_1, h_1} \cong \Vect_{K_1 \cap h_1^{-1}K_2h_1}$ and the right action of $(\mathcal{C}_{K_1, K_2}^x)_{h_2, h_2} \cong \Vect_{K_1 \cap h_2^{-1}K_2h_2}$ on $(\mathcal{C}_{K_1, K_2}^x)_{h_1, h_2} \cong \Vect_{K_1 \cap h_1^{-1}K_2h_2}$.

\vspace*{\baselineskip}
\noindent{\bf Relation between \eqref{eq: strip} and \eqref{eq: simples of strip}.}
The objects of the strip 2-algebra $\mathcal{C}_{K_1, K_2}$ were originally expressed as in \eqref{eq: strip}.
On the other hand, in the above discussion, we used another expression~\eqref{eq: simples of strip}.
We now explain the relation between them.
According to \eqref{eq: strip}, the strip 2-algebra $\mathcal{C}_{K_1, K_2}$ has an object of the form
\begin{equation}
\adjincludegraphics[valign = c, trim={10, 10, 10, 10}]{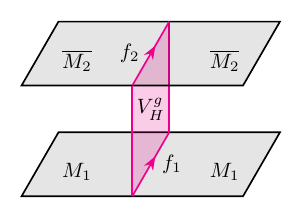},
\label{eq: strip obj}
\end{equation}
where $f_1 \in \Hom_{{}_{\Vect_H}(2\Vect_G)_{\Vect_{K_1}}}(V_H^g \boxtimes_{\Vect_H} M_1, M_1)$ and $f_2 \in \Hom_{{}_{\Vect_{K_2}}(2\Vect_G)_{\Vect_H}}(\overline{M_2}, \overline{M_2} \boxtimes_{\Vect_H} V_H^g)$ are simple 1-morphisms.
Recalling that $V_H^g = \Vect_H \boxtimes V^g \boxtimes \Vect_H$, $M_1 = \Vect_H \boxtimes \Vect_{K_1}$, and $\overline{M_2} = \Vect_{K_2} \boxtimes \Vect_H$, we can write any simple 1-morphisms $f_1$ and $f_2$ as follows:
\begin{equation}
f_1 = f_1(g; h_1) := \adjincludegraphics[valign = c, trim={10, 10, 10, 10}]{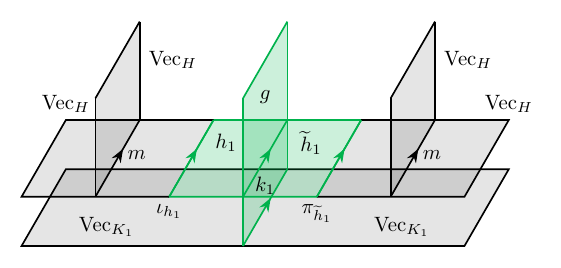},
\label{eq: f1}
\end{equation}
\begin{equation}
f_2 = f_2(g; h_2) = \adjincludegraphics[valign = c, trim={10, 10, 10, 10}]{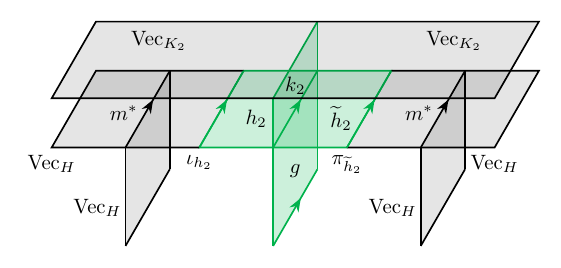}.
\label{eq: f2}
\end{equation}
Here, the elements $h_1, \widetilde{h}_1, h_2, \widetilde{h}_2 \in H$, $k_1 \in K_1$, and $k_2 \in K_2$ must satisfy
\begin{equation}
h_1 k_1 = g \widetilde{h}_1, \qquad h_2 g = k_2 \widetilde{h}_2.
\label{eq: hk = gh}
\end{equation}
We note that $\widetilde{h}_1$ and $k_1$ are uniquely determined by $g$ and $h_1$ because $K_1$ is a complement of $H$ in $G$.
Similarly, $\widetilde{h}_2$ and $k_2$ are uniquely determined by $g$ and $h_2$.
More specifically, $\tilde{h}_1$ and $\tilde{h}_2$ can be written as
\begin{equation}
\widetilde{h}_1 = (U_{g; K_1}^{-1}(h_1^{-1}))^{-1}, \qquad
\widetilde{h}_2 = (U_{g; K_2}^{-1}(h_2^{-1}))^{-1},
\label{eq: h tilde}
\end{equation}
where $U_{g; K}: H \to H$ is defined by \eqref{eq: UgK}.
The 1-morphism $m: \Vect_H \boxtimes \Vect_H \to \Vect_H$ in \eqref{eq: f1} is the multiplication 1-morphism of $\Vect_H \in 2\Vect_G$ (i.e., the tensor product of $\Vect_H$), and $m^*: \Vect_H \to \Vect_H \boxtimes \Vect_H$ in \eqref{eq: f2} is the dual of $m$.
Using~\eqref{eq: f1}, \eqref{eq: f2}, and the equivalence $\Vect_H \boxtimes_{\Vect_H} \Vect_H \cong \Vect_H$, we can express the diagram in~\eqref{eq: strip obj} as
\begin{equation}
\adjincludegraphics[valign = c, trim={10, 10, 10, 10}]{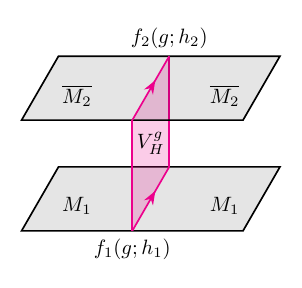} ~ = ~ \adjincludegraphics[valign = c, trim={10, 10, 10, 10}]{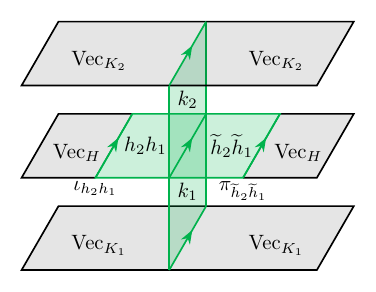}.
\label{eq: strip VHg}
\end{equation}
By comparing this with \eqref{eq: simples of strip}, we find that the object~\eqref{eq: strip obj} with $f_1=f_1(g; h_1)$ and $f_2=f_2(g; h_2)$ can be identified with $\mathsf{L}(h_2h_1, \widetilde{h}_2\widetilde{h}_1, k_1, k_2)$.
In particular, this shows that the object \eqref{eq: strip obj} is simple, and any simple object is of this form.
The isomorphism class of the above object depends only on the pair $(h_2h_1, k_1)$ due to \eqref{eq: isom class of L}. 
Since $k_1$ is the unique element  of $K_1 \cap h^{-1}gH$, we find
\begin{equation}
\adjincludegraphics[valign = c, trim={10, 10, 10, 10}]{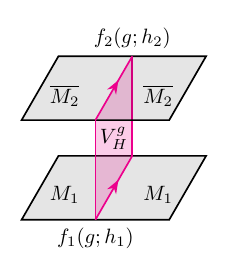} \cong \adjincludegraphics[valign = c, trim={10, 10, 10, 10}]{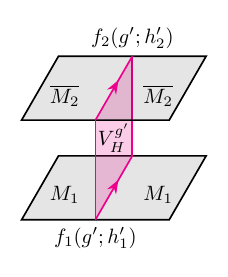} ~\Leftrightarrow~ (h_2h_1, h_1^{-1}gH) = (h_2^{\prime}h_1^{\prime}, (h_1^{\prime})^{-1}g^{\prime}H).
\label{eq: isom of strip obj}
\end{equation}

\vspace*{\baselineskip}
\noindent{\bf Fusion rules revisited.}
For later convenience, we compute the fusion rules of the objects of the form~\eqref{eq: strip VHg}.\footnote{We recall that the fusion rules of the objects~\eqref{eq: simples of strip} are already computed as in~\eqref{eq: fusion rules}.}
The tensor product of the objects is given by the concatenation
\begin{equation}
\adjincludegraphics[valign = c, trim={10, 10, 10, 10}]{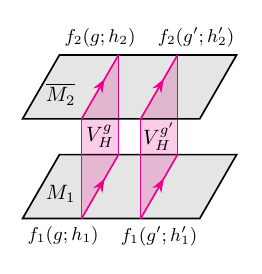}
= \adjincludegraphics[valign = c, trim={10, 10, 10, 10}]{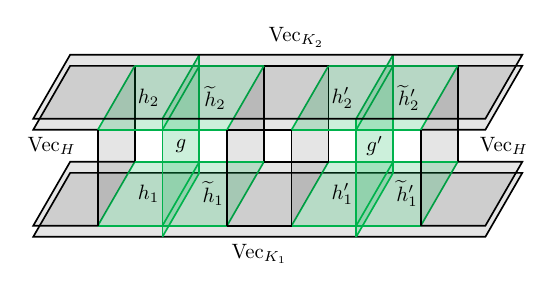},
\end{equation}
where $\widetilde{h}_1$ and $\widetilde{h}_2$ are defined by \eqref{eq: h tilde}, and $\widetilde{h}_1^{\prime}$ and $\widetilde{h}_2^{\prime}$ are also defined similarly.
By fusing the middle gray surfaces on the right-hand side, we end up with
\begin{equation}
\text{RHS} \cong \bigoplus_{h \in H} \delta_{h, \widetilde{h}_1(h_1^{\prime})^{-1}} \delta_{h, \widetilde{h}_2(h_2^{\prime})^{-1}} \adjincludegraphics[valign = c, trim={10, 10, 10, 10}]{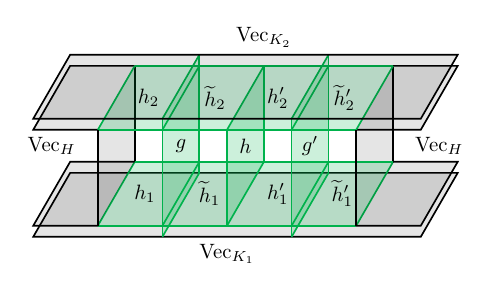}.
\end{equation}
By combining the above two equations, we find
\begin{equation}
\adjincludegraphics[valign = c, trim={10, 10, 10, 10}]{tikz/out/strip_obj_TN_fusion1.pdf} \cong \delta_{\widetilde{h}_1(h_1^{\prime})^{-1}, \widetilde{h}_2(h_2^{\prime})^{-1}} \adjincludegraphics[valign = c, trim={10, 10, 10, 10}]{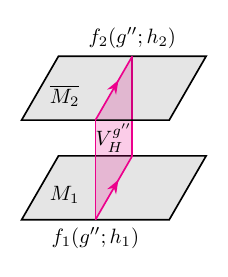},
\label{eq: fusion rules 2}
\end{equation}
where $g^{\prime \prime} = g\widetilde{h}_1(h_1^{\prime})^{-1}g^{\prime}$.

\subsubsection{Gapped interface modes}
\label{sec: Gapped interface modes CT}
Following the classification of 1+1d gapped phases with non-invertible symmetries \cite{Thorngren:2019iar, Komargodski:2020mxz}, we expect that gapped interfaces between $\mathsf{SPT}_{K_1}$ and $\mathsf{SPT}_{K_2}$ are classified by indecomposable $\mathcal{C}_{K_1, K_2}$-module categories.\footnote{In \cite{Thorngren:2019iar, Komargodski:2020mxz}, it was shown that 1+1d gapped phases described by topological field theories with fusion category symmetry $\mathcal{C}$ are classified by indecomposable $\mathcal{C}$-module categories. We expect that the same result holds when $\mathcal{C}$ is a multifusion category.}
The degeneracy of the interface modes should be equal to the number of simple objects of the corresponding module category.
In particular, a non-degenerate interface should correspond to a $\mathcal{C}_{K_1, K_2}$-module category $\Vect$, i.e., a fiber functor of $\mathcal{C}_{K_1, K_2}$.
In what follows, we discuss the necessary and sufficient condition for $\mathcal{C}_{K_1, K_2}$ to admit a fiber functor.
We will also study the 2-category of gapped interfaces and its relation to the Morita dual of $\mathcal{C}(G; H)$.

\vspace*{\baselineskip}
\noindent{\bf Existence of non-degenerate gapped interfaces.}
We show that $\mathcal{C}_{K_1, K_2}$ admits a fiber functor if and only if $\mathsf{SPT}_{K_1}$ and $\mathsf{SPT}_{K_2}$ correspond to isomorphic fiber 2-functors.
In other words, $\mathcal{C}_{K_1, K_2}$ admits a fiber functor if and only if $\mathcal{M}(K_1)$ and $\mathcal{M}(K_2)$ are equivalent as $\mathcal{C}(G; H)$-module 2-categories. Here, $\mathcal{M}(K)$ is defined by \eqref{eq: MK}.
This result implies that the symmetry at the interface of different SPT phases is anomalous in the sense that it does not admit a non-degenerate gapped ground state.

To show the above statement, we first notice that $\mathcal{C}_{K_1, K_2}$ admits a fiber functor if and only if at least one of the direct sum components in \eqref{eq: strip decomposition} does.
Since each component $\mathcal{C}_{K_1, K_2}^x$ is an indecomposable multifusion category, it admits a fiber functor only if it is fusion (see Appendix~\ref{sec: Fiber functors of multifusion categories} for a proof).
By definition, $\mathcal{C}_{K_1, K_2}^x$ is fusion if and only if $|H \cap x| = 1$.
When $\mathcal{C}_{K_1, K_2}^x$ is fusion, it admits a fiber functor because it is monoidally equivalent to $\Vect_{K_1 \cap h^{-1}K_2h}$, where $h \in H \cap x$.
Therefore, $\mathcal{C}_{K_1, K_2}$ admits a fiber functor if and only if there exists a double coset $x \in K_2 \backslash G / K_1$ such that $|H \cap x| = 1$.

Now, we show that the existence of a fiber functor of $\mathcal{C}_{K_1, K_2}$ implies an isomorphism between the fiber 2-functors corresponding to $\mathsf{SPT}_{K_1}$ and $\mathsf{SPT}_{K_2}$.
To this end, we suppose that $\mathcal{C}_{K_1, K_2}$ admits a fiber functor.
In this case, there exists $x \in K_2 \backslash G / K_1$ such that $|H \cap x| = 1$.
Such a double coset $x$ can be written as
\begin{equation}
x = K_2 h K_1 = h K_1,
\label{eq: x=hK1}
\end{equation}
where $h$ is the unique element of $H \cap x$.
The second equality holds because if $K_2 h K_1$ were strictly bigger than $hK_1$, $x$ would contain other elements of $H$.
The above equation implies
\begin{equation}
K_2 h K_1 = h K_1 ~ \Leftrightarrow ~ K_2 h = h K_1 ~ \Leftrightarrow ~ K_2 = hK_1h^{-1}.
\end{equation}
This shows that if $\mathcal{C}_{K_1, K_2}$ admits a fiber functor, there exists $h \in H$ that satisfies $K_2=hK_1h^{-1}$.
Thus, due to the results reviewed in Section~\ref{sec: Fiber 2-functors}, the fiber 2-functors corresponding to $\mathsf{SPT}_{K_1}$ and $\mathsf{SPT}_{K_2}$ are isomorphic to each other.

Conversely, we can also show that an isomorphism between the fiber 2-functors corresponding to $\mathsf{SPT}_{K_1}$ and $\mathsf{SPT}_{K_2}$ implies the existence of a fiber functor of $\mathcal{C}_{K_1, K_2}$.
To see this, we suppose that $\mathsf{SPT}_{K_1}$ and $\mathsf{SPT}_{K_2}$ correspond to isomorphic fiber 2-functors.
Namely, we suppose that there exists $h \in H$ such that $K_2 = hK_1h^{-1}$.\footnote{We recall that fiber 2-functors corresponding to $\mathsf{SPT}_{K_1}$ and $\mathsf{SPT}_{K_2}$ are isomorphic to each other if and only if there exists $g \in G$ such that $K_2 = gK_1g^{-1}$. Since $K_1$ is a complement of $H$ in $G$, there exist $h \in H$ and $k_1 \in K_1$ such that $g = hk_1$, which implies that $gK_1g^{-1} = hK_1h^{-1}$.}
In this case, we have
\begin{equation}
K_2 h K_1 = hK_1h^{-1} hK_1 = hK_1,
\end{equation}
which implies that $x := K_2 h K_1$ satisfies $|H \cap x| = 1$.
This shows that if $\mathsf{SPT}_{K_1}$ and $\mathsf{SPT}_{K_2}$ correspond to isomorphic fiber 2-functors, there exists $x \in K_2 \backslash G / K_1$ such that $|H \cap x| = 1$, and hence $\mathcal{C}_{K_1, K_2}$ admits a fiber functor.

\vspace*{\baselineskip}
\noindent{\bf Set of non-degenerate gapped interfaces.}
As mentioned above, non-degenerate gapped interfaces between $\mathsf{SPT}_{K_1}$ and $\mathsf{SPT}_{K_2}$ should correspond to fiber functors of $\mathcal{C}_{K_1, K_2}$.
The fiber functors of $\mathcal{C}_{K_1, K_2}$ are in one-to-one correspondence with those of the direct sum components of $\mathcal{C}_{K_1, K_2}$.
Each direct sum component $\mathcal{C}_{K_1, K_2}^x$ admits a fiber functor if and only if $x$ satisfies $|H \cap x| = 1$.
Such an $x$ can always be written as $x = K_2 h K_1$, where $h \in H$ satisfies $K_2 = hK_1h^{-1}$.
When $x = K_2hK_1$, there is a monoidal equivalence $\mathcal{C}_{K_1, K_2}^x \cong \Vect_{K_1 \cap h^{-1}K_2h} = \Vect_{K_1}$.
Thus, fiber functors of $\mathcal{C}_{K_1, K_2}^x$ are classified by the second group cohomology $H^2(K_1, \mathrm{U}(1))$.
Taking all direct sum components into account, we find that fiber functors of $\mathcal{C}_{K_1, K_2}$ are in one-to-one correspondence with pairs $(h, [\xi])$, where $h \in H$ satisfies $K_2 = hK_1h^{-1}$ and $[\xi] \in H^2(K_1, \mathrm{U}(1))$ is an element of the second group cohomology of $K_1$.
In other words, we have the following equality of sets:
\begin{equation}
\{\text{Fiber functors of $\mathcal{C}_{K_1, K_2}$}\} = \{(h, [\xi]) \mid h \in H \text{ s.t. } K_2 = hK_1h^{-1}, ~ [\xi] \in H^2(K_1, \mathrm{U}(1))\}.
\label{eq: set of fiber functors}
\end{equation}
Here, the left-hand side is the set of (isomorphism classes of) fiber functors of $\mathcal{C}_{K_1, K_2}$.

\vspace*{\baselineskip}
\noindent{\bf 2-category of gapped interfaces.}
General gapped interfaces between $\mathsf{SPT}_{K_1}$ and $\mathsf{SPT}_{K_2}$ should be classified by module categories over $\mathcal{C}_{K_1, K_2}$.
Similarly, gapped junctions between gapped interfaces would correspond to $\mathcal{C}_{K_1, K_2}$-module functors, and junction-changing operators, which are point-like operators in spacetime, would correspond to module natural transformations.
Thus, we expect that gapped interfaces, gapped junctions, and junction-changing operators form a 2-category $\Mod(\mathcal{C}_{K_1, K_2})$, whose objects, 1-morphisms, and 2-morphisms are $\mathcal{C}_{K_1, K_2}$-module categories, module functors, and module natural transformations.
In what follows, we will show the equivalence of semisimple 2-categories
\begin{equation}
\Mod(\mathcal{C}_{K_1, K_2}) \cong \Fun_{\mathcal{C}(G; H)}(\mathcal{M}(K_2), \mathcal{M}(K_1)),
\label{eq: Mod strip}
\end{equation}
where the objects, 1-morphisms, and 2-morphisms of the right-hand side are $\mathcal{C}(G; H)$-module 2-functors from $\mathcal{M}(K_1)$ to $\mathcal{M}(K_2)$, module natural transformations between them, and module modifications between the module natural transformations.

To show the equivalence~\eqref{eq: Mod strip}, let us first compute the left-hand side of \eqref{eq: Mod strip}.
The direct sum decomposition~\eqref{eq: strip decomposition} implies
\begin{equation}
\Mod(\mathcal{C}_{K_1, K_2}) \cong \bigoplus_{x \in K_2 \backslash G / K_1} \Mod(\mathcal{C}_{K_1, K_2}^{x}),
\end{equation}
where $\mathcal{C}_{K_1, K_2}^x$ is an indecomposable multifusion category~\eqref{eq: Cx K1 K2 semisimple}.
Since $\mathcal{C}_{K_1, K_2}^x$ is indecomposable, $\Mod(\mathcal{C}_{K_1, K_2}^x)$ is equivalent to $\Mod((\mathcal{C}_{K_1, K_2}^x)_{h, h})$ for any diagonal component $(\mathcal{C}_{K_1, K_2}^x)_{h, h}$ \cite[Example 1.4.16]{Douglas:2018qfz}.
Therefore, we find
\begin{equation}
\Mod(\mathcal{C}_{K_1, K_2}) \cong \bigoplus_{x \in K_2 \backslash G / K_1} \Mod(\Vect_{K_1 \cap h_x^{-1}K_2h_x}),
\label{eq: Mod strip LHS}
\end{equation}
where $h_x \in H \cap x$ is arbitrary.
On the other hand, the right-hand side of~\eqref{eq: Mod strip} is equivalent to
\begin{equation}
\begin{aligned}
\Fun_{\mathcal{C}(G; H)}(\mathcal{M}(K_2), \mathcal{M}(K_1)) &\cong \Fun_{2\Vect_G}((2\Vect_G)_{\Vect_{K_2}}, (2\Vect_G)_{\Vect_{K_1}}) \\
&\cong {}_{\Vect_{K_2}}(2\Vect_G)_{\Vect_{K_1}},
\end{aligned}
\end{equation}
where the first equivalence follows from the Morita equivalence between $\mathcal{C}(G; H)$ and $2\Vect_G$ \cite[Theorem 5.4.3]{Decoppet2023Morita} and the second equivalence follows from \cite[Theorem 5.1.2]{Decoppet2023Morita}.
The right-hand side of the above equation is also equivalent to $\bigoplus_{x \in K_2 \backslash G / K_1} \Mod(\Vect_{K_1 \cap h_x^{-1}K_2h_x})$ due to \cite[Proposition 3.2.1]{Decoppet:2023bay}.
Therefore, we find
\begin{equation}
\Fun_{\mathcal{C}(G; H)} (\mathcal{M}(K_2), \mathcal{M}(K_1)) \cong \bigoplus_{x \in K_2 \backslash G / K_1} \Mod(\Vect_{K_1 \cap h_x^{-1}K_2h_x}).
\label{eq: Mod strip RHS}
\end{equation}
Equations \eqref{eq: Mod strip LHS} and \eqref{eq: Mod strip RHS} imply the equivalence \eqref{eq: Mod strip}.

\vspace*{\baselineskip}
\noindent{\bf Relation to the dual 2-category.}
When $K_1 = K_2 =: K$, equation~\eqref{eq: Mod strip} reduces to the equivalence of semisimple 2-categories
\begin{equation}
\Mod(\mathcal{C}_{K, K}) \cong \Fun_{\mathcal{C}(G; H)}(\mathcal{M}(K), \mathcal{M}(K)).
\label{eq: Morita dual}
\end{equation}
The right-hand side is known as the Morita dual of $\mathcal{C}(G; H)$ with respect to $\mathcal{M}(K)$.
This 2-category has a monoidal structure defined by the composition of module functors, and there is a monoidal equivalence \cite[Corollary 4.1.5]{Decoppet:2023bay}
\begin{equation}
\Fun_{\mathcal{C}(G; H)}(\mathcal{M}(K), \mathcal{M}(K)) \cong \mathcal{C}(G; K).
\end{equation}
On the other hand, the left-hand side of \eqref{eq: Morita dual} should also have a monoidal structure, which is physically given by the fusion of gapped self-interfaces of $\mathsf{SPT}_K$ (i.e., gapped excitations within $\mathsf{SPT}_K$).
It is then natural to expect that \eqref{eq: Morita dual} is a monoidal equivalence, not just an equivalence of 2-categories.\footnote{The 1-categorical analogue of this monoidal equivalence is known to hold \cite{Kitaev:2011dxc, Bai:2025zze}.}
Namely, we expect a monoidal equivalence
\begin{equation}
\Mod(\mathcal{C}_{K, K}) \cong \mathcal{C}(G; K).
\label{eq: monoidal equiv conjecture}
\end{equation}
To show this monoidal equivalence, we need to rigorously define a monoidal structure on $\Mod(\mathcal{C}_{K, K})$ by using an additional structure on $\mathcal{C}_{K, K}$.
We will not do this in this paper.\footnote{Nevertheless, we mention that invertible objects of $\mathcal{C}(G; K)$ should correspond to fiber functors of $\mathcal{C}_{K, K}$ under the conjectural monoidal equivalence \eqref{eq: monoidal equiv conjecture}.
In particular, due to \eqref{eq: set of fiber functors}, invertible objects of $\mathcal{C}(G; K)$ should be labeled by pairs $(h, [\xi])$, where $h \in H$ satisfies $K = hKh^{-1}$ and $[\xi] \in H^2(K, \mathrm{U}(1))$.
The object labeled by $(h, [\xi])$ can be obtained by starting from a non-invertible object $V_K^h = \Vect_K \boxtimes V^h \boxtimes \Vect_K$ and condensing the line operators forming the twisted group algebra $\mathbb{C}[H]^{\xi} \in \Vect_H \cong \End_{\mathcal{C}(G; K)}(V_K^h)$.
}

\subsubsection{Example: gapped boundary of the $G$-cluster state}
\label{Example: gapped boundary of the G-cluster state CT}
As a simple example, let us consider the interface between the SPT phases realized by the $G_0$-cluster state~\eqref{eq: G-cluster state} and the trivial product state.
As discussed in Section \ref{sec: Example: G-cluster state}, the SPT phase of the $G_0$-cluster state is obtained by the following choice of the input data:
\begin{equation}
G = G_0 \times G_0, \quad 
H = G_0^{\text{left}}, \quad
K = K_{\text{cluster}} := G_0^{\text{diag}}.
\end{equation}
The symmetry of the $G_0$-cluster state is $2\Rep(G_0) \boxtimes 2\Vect_{G_0}$.
On the other hand, the trivial product state with the same symmetry is obtained by choosing $G$ and $H$ as above and $K$ to be
\begin{equation}
K = K_{\text{triv}} := G_0^{\text{right}}.
\end{equation}

The symmetry at the interface of the above SPT phases is described by the strip 2-algebra~\eqref{eq: strip decomposition} with $K_1 = K_{\text{cluster}}$ and $K_2 = K_{\text{triv}}$.
Since a $(K_{\text{triv}}, K_{\text{cluster}})$-double coset $x$ in $G_0 \times G_0$ is unique, the right-hand side of~\eqref{eq: strip decomposition} consists only of a single component, which implies that the strip 2-algebra $\mathcal{C}_{K_{\text{cluster}}, K_{\text{triv}}}$ is an indecomposable multifusion category.
The unique double coset $x$ is given by $G_0 \times G_0$ itself, and hence it follows that $H \cap x = H = G_0^{\text{left}}$.
Therefore, as a semisimple category, we have
\begin{equation}
\mathcal{C}_{K_{\text{cluster}}, K_{\text{triv}}} \cong \bigoplus_{g_1, g_2 \in G_0^{\text{left}}} \Vect_{G^{\text{diag}}_{0} \cap g_1^{-1} G_0^{\text{right}}g_2}.
\label{eq: G-cluster strip}
\end{equation}
Each component on the right-hand side has a unique simple object up to isomorphism because the set $G_0^{\text{diag}} \cap g_1^{-1}G_0^{\text{right}}g_2$ is a singleton for any pair $g_1, g_2 \in G_0^{\text{left}}$.
If we denote the simple object of each component by $\mathsf{L}(g_1, g_2)$, the fusion rules can be written as
\begin{equation}
\mathsf{L}(g_1, g_2) \otimes \mathsf{L}(g_1^{\prime}, g_2^{\prime}) \cong \delta_{g_2, g_1^{\prime}} \mathsf{L}(g_1, g_2^{\prime}).
\label{eq: G-cluster boundary fusion}
\end{equation}
The above fusion rules imply that there is a monoidal equivalence
\begin{equation}
\mathcal{C}_{K_{\text{cluster}}, K_{\text{triv}}} \cong \Mat_{|G_0|}(\Vect) := \End(\Vect^{\oplus |G_0|}),
\label{eq: strip G-cluster}
\end{equation}
where $\Vect^{\oplus |G|}$ denotes the direct sum of $|G_0|$ copies of $\Vect$.

An indecomposable module category over $\mathcal{C}_{K_{\text{cluster}}, K_{\text{triv}}} \cong \Mat_{|G_0|}(\Vect)$ is unique up to equivalence and is given by $\Vect^{\oplus |G_0|}$.
In particular, $\mathcal{C}_{K_{\text{cluster}}, K_{\text{triv}}}$ does not have a fiber functor.
Therefore, the SPT phase realized by the $G_0$-cluster state does not admit a non-degenerate gapped boundary unless the symmetry is explicitly broken.
This result can be regarded as the bulk-boundary correspondence.

\subsection{Tensor network description}
\label{sec: Tensor network description}
We now describe the interface of the generalized cluster states using the tensor networks.
We will see that the tensor network description is consistent with the category-theoretical description.
Unless otherwise stated, the set $S_{H \backslash G}$ of right $H$-cosets in $G$ is taken to be a complement of $H$ in $G$ as in Section~\ref{sec: MPO representations of fractionalized symmetries}.
This choice of $S_{H \backslash G}$ guarantees that the symmetry operator $\mathsf{D}[V_H^g]$ is represented by the PEPO~\eqref{eq: reduced PEPO}, and the action tensors for it are given by~\eqref{eq: action tensors}.

\subsubsection{Interface symmetries}
\label{sec: Interface symmetries TN}
We first write down the tensor network representations of the symmetry operators acting on the interface.
For simplicity, we will consider the interface represented by the following MPO:\footnote{More generally, one can consider the interface that supports physical degrees of freedom as long as the entire system preserves the symmetry $\mathcal{C}(G; H)$. We will not consider such an interface in this paper.}
\begin{equation}
\ket{\mathcal{I}} := \adjincludegraphics[valign = c, trim={10, 10, 10, 10}]{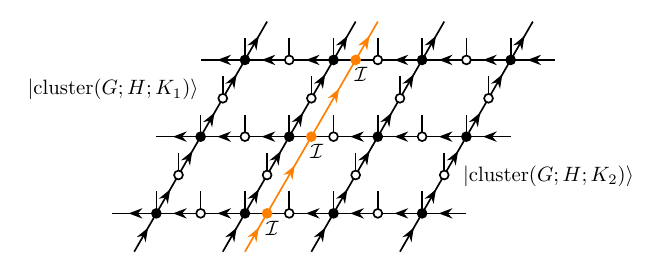}.
\label{eq: interface state}
\end{equation}
Here, the left side of the interface is occupied by $\ket{\text{cluster}(G; H; K_1)}$, while the right side is occupied by $\ket{\text{cluster}(G; H; K_2)}$.
The orange dot represents an MPO tensor $\mathcal{I}$ on the interface.
Using \eqref{eq: decomposed fractionalization} and \eqref{eq: orthogonality and completeness}, one can compute the action of the symmetry operator $\mathsf{D}[V_H^g]$ on $\ket{\mathcal{I}}$ as
\begin{equation}
\mathsf{D}[V_H^g] \ket{\mathcal{I}} = \sum_{h_1, h_2 \in H} \ket{\phi_{K_1}[V_H^g]_{h_1} \cdot \mathcal{I} \cdot \overline{\phi_{K_2}}[V_H^g]_{h_2}},
\end{equation}
where the new MPO tensor on the right-hand side is defined by
\begin{equation}
\phi_{K_1}[V_H^g]_{h_1} \cdot \mathcal{I} \cdot \overline{\phi_{K_2}}[V_H^g]_{h_2} := \adjincludegraphics[valign = c, trim={10, 10, 10, 10}]{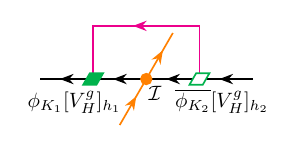}.
\label{eq: interface sym action}
\end{equation}
We recall that $\phi_K[V_H^g]_h$ and $\overline{\phi_K}[V_H^g]_h$ are the action tensors defined by \eqref{eq: action tensors}.
The above equation implies that the symmetry operators on the interface are of the form
\begin{equation}
\hat{\mathsf{L}}(g; h_1, h_2) := \bigotimes_{i: \text{sites}} \hat{\mathcal{O}}_i(g; h_1, h_2).
\label{eq: interface symmetry op}
\end{equation}
The tensor product is taken over all sites on the interface, and each on-site operator $\hat{\mathcal{O}}_i(g; h_1, h_2)$ is defined by
\begin{equation}
\hat{\mathcal{O}}_i(g; h_1, h_2) := \adjincludegraphics[valign = c, trim={10, 10, 10, 10}]{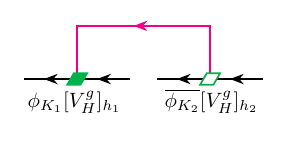}
= \adjincludegraphics[valign = c, trim={10, 10, 10, 10}]{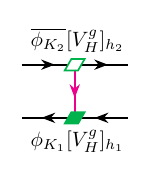}.
\label{eq: interface symmetry op TN}
\end{equation}
In the second equality, we folded the diagram along the interface.
The last diagram represents an operator acting on the two legs on the right.
Since each leg has the bond Hilbert space $\mathbb{C}[H]$, the on-site operator $\hat{\mathcal{O}}_i(g; h_1, h_2)$ acts on $\mathbb{C}[H]^{\otimes 2}$.
Using \eqref{eq: action tensors} and \eqref{eq: interface symmetry op TN}, one can compute the symmetry action on each site explicitly as
\begin{equation}
\begin{aligned}
&\quad \hat{\mathcal{O}}_i(g; h_1, h_2) \ket{h_{\ell}, h_r}_i = \adjincludegraphics[valign = c, trim={10, 10, 10, 10}]{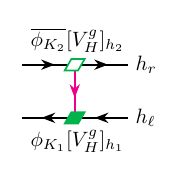} \\
&= \delta_{h_rh_{\ell}^{-1}, U_{g; K_2}^{-1}(h_2) (U_{g; K_1}^{-1}(h_1))^{-1}} \ket{h_1 \widetilde{U}_{g; S_{H \backslash G}}((U_{g; K_1}^{-1}(h_1))^{-1}h_{\ell}), h_2 \widetilde{U}_{g; S_{H \backslash G}}((U_{g; K_2}^{-1}(h_2))^{-1} h_r)}_i,
\end{aligned}
\label{eq: on-site sym action}
\end{equation}
where $U_{g; K}: H \to H$ and $\widetilde{U}_{g; S_{H \backslash G}}: H \to H$ are the invertible maps uniquely determined by
\begin{equation}
U_{g; K}(h) gh^{-1} \in K, \qquad 
(\widetilde{U}_{g; S_{H \backslash G}}(h))^{-1} gh \in S_{H \backslash G},
\qquad \forall h \in H.
\label{eq: U and U tilde}
\end{equation}
A direct computation based on \eqref{eq: on-site sym action} shows that the symmetry operators~\eqref{eq: interface symmetry op} obey the following fusion rules:
\begin{equation}
\hat{\mathsf{L}}(g; h_1, h_2) \hat{\mathsf{L}}(g^{\prime}; h_1^{\prime}, h_2^{\prime}) = \delta_{(h_1^{\prime})^{-1} U^{-1}_{g; K_1}(h_1), (h_2^{\prime})^{-1}U^{-1}_{g; K_2}(h_2)} \hat{\mathsf{L}}(g (U_{g; K_1}^{-1}(h_1))^{-1} h_1^{\prime}g^{\prime}; h_1, h_2).
\label{eq: interface sym fusion rules}
\end{equation}
A detailed derivation will be provided in Appendix~\ref{sec: Derivation of the fusion rules}.
The Kronecker delta on the right-hand side indicates that the symmetry at the interface is described by a multifusion category, which is not necessarily fusion.
In particular, as opposed to the case of a fusion category, the identity operator can be decomposed into a sum of symmetry operators as
\begin{equation}
\id = \sum_{h \in H} \hat{\mathsf{L}}(e; e, h),
\label{eq: id decomp}
\end{equation}
which readily follows from $\id_{\mathbb{C}[H]^{\otimes 2}} = \sum_{h \in H} \hat{\mathcal{O}}_i(e; e, h)$.

\vspace*{\baselineskip}
\noindent{\bf Relation to the category theoretical description.}
The fusion rules~\eqref{eq: interface sym fusion rules} suggest that the symmetry operator $\hat{\mathsf{L}}(g; h_1, h_2)$ corresponds to the following simple object of the strip 2-algebra $\mathcal{C}_{K_1, K_2}$:
\begin{equation}
\hat{\mathsf{L}}(g; h_1, h_2) ~ \longleftrightarrow ~ \adjincludegraphics[valign = c, trim={10, 10, 10, 10}]{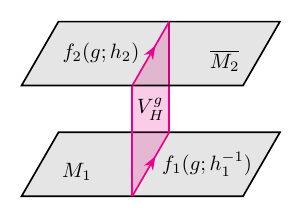}.
\label{eq: strip obj TN}
\end{equation}
Here, $f_1(g; h_1^{-1})$ and $f_2(g; h_2)$ are the simple 1-morphisms defined by \eqref{eq: f1} and \eqref{eq: f2}.
We note that the fusion rules~\eqref{eq: fusion rules 2} of the objects on the right-hand side agree with the fusion rules~\eqref{eq: interface sym fusion rules} of the operators on the left-hand side.
Equation~\eqref{eq: strip obj TN} implies that the action tensors $\phi_{K_1}[V_H^g]_{h_1}$ and $\overline{\phi_{K_2}}[V_H^g]_{h_2}$ correspond to the following simple 1-morphisms:
\begin{equation}
\phi_{K_1}[V_H^g]_{h_1} \leftrightarrow f_1(g; h_1^{-1}), \qquad
\overline{\phi_{K_2}}[V_H^g]_{h_2} \leftrightarrow f_2(g; h_2).
\end{equation}

The symmetry operators $\hat{\mathsf{L}}(g; h_1, h_2)$ and $\hat{\mathsf{L}}(g^{\prime}; h_1^{\prime}, h_2^{\prime})$ are equal to each other if the corresponding objects of $\mathcal{C}_{K_1, K_2}$ are isomorphic to each other.
More specifically, we have (cf. \eqref{eq: isom of strip obj})
\begin{equation}
(h_2h_1^{-1}, h_1gH) = (h_2^{\prime}(h_1^{\prime})^{-1}, h_1^{\prime}g^{\prime}H) ~\Rightarrow~ \hat{\mathsf{L}}(g; h_1, h_2) = \hat{\mathsf{L}}(g^{\prime}; h_1^{\prime}, h_2^{\prime}).
\label{eq: isomorphic sym op}
\end{equation}
See Appendix~\ref{sec: Equality of symmetry operators} for a proof.
On the other hand, the converse is not necessarily true.
Namely, in general, the symmetry operators corresponding to non-isomorphic objects of $\mathcal{C}_{K_1, K_2}$ can coincide with each other.
In other words, the action of the symmetry $\mathcal{C}_{K_1, K_2}$ on the interface is generally unfaithful.\footnote{We may say that the strip 2-algebra $\mathcal{C}_{K_1, K_2}$ acts via a tensor functor from $\mathcal{C}_{K_1, K_2}$ to another multifusion category $\widetilde{\mathcal{C}}_{K_1, K_2}$ that acts faithfully on the interface~\eqref{eq: interface state}. The unfaithfulness of the symmetry action does not affect the bulk-boundary correspondence because the symmetry $\mathcal{C}_{K_1, K_2}$ is (non-)anomalous if and only if its faithful part $\widetilde{\mathcal{C}}_{K_1, K_2}$ is (non-)anomalous.}
Some examples of unfaithful action of $\mathcal{C}_{K_1, K_2}$ will be discussed in Section~\ref{sec: Interfaces TY}.

\subsubsection{Gapped interface modes}
\label{sec: Gapped interface modes TN}
In this subsection, we study the necessary and sufficient conditions for the symmetry at the interface of the generalized cluster states to admit a non-degenerate gapped state.
Here, a state $\ket{\psi}$ at the interface is said to be non-degenerate if it forms a singlet under the symmetry action, that is, it satisfies
\begin{equation}
\hat{\mathsf{L}}(g; h_1, h_2) \ket{\psi} \propto \ket{\psi}, \qquad \forall g \in G, ~ \forall h_1, h_2 \in H.
\label{eq: non-degenerate interface}
\end{equation}
A state $\ket{\psi}$ is said to be gapped if it admits a gapped parent Hamiltonian. In particular, if $\ket{\psi}$ on a 1d interface is written as an MPS with finite bond dimension, then it is gapped \cite{Cirac:2020obd, FNW92, PhysRevB.73.094423, Perez-Garcia:2006nqo,Hastings:2007iok}.\footnote{The result we will show below is not very sensitive to the precise definition of the gappedness. Indeed, the only fact we will use about the gappedness is that the trivial product state is gapped; see the subsequent paragraphs for more precise statements and their proofs.}
We note that the non-degeneracy and gappedness are independent notions; a non-degenerate state may or may not be gapped, and a gapped state may or may not be non-degenerate.

According to the above definition, the non-degeneracy only means that the state $\ket{\psi}$ is symmetric.
In particular, the existence of a non-degenerate interface state is a weaker condition than the existence of a Hamiltonian that has a unique ground state in the presence of an interface.
Nevertheless, our result below shows that the existence of a gapped non-degenerate interface state is equivalent to the condition that the SPT phases on both sides of the interface are the same.

In what follows, we will show that the symmetry at the interface of $\ket{\text{cluster}(G; H; K_1)}$ and $\ket{\text{cluster}(G; H; K_2)}$ admits a non-degenerate gapped state if and only if there exists $h \in H$ such that $K_2 = hK_1h^{-1}$.
More precisely, we will show the following:
\begin{itemize}
\item If there exists $h \in H$ such that $K_2 = hK_1h^{-1}$, the symmetry at the interface admits a non-degenerate product state, which is gapped.
\item If the symmetry at the interface admits a non-degenerate state, which is not necessarily gapped, there exists $h \in H$ such that $K_2 = hK_1h^{-1}$.
\end{itemize}
This result is consistent with the category theoretical description of the interface modes discussed in Section~\ref{sec: Gapped interface modes CT}.\footnote{From the category theoretical point of view, non-degenerate gapped interfaces should correspond to fiber functors of the strip 2-algebra $\mathcal{C}_{K_1, K_2}$. In particular, when $K_1=K_2=K$, they would also correspond to invertible objects of the dual 2-category $\mathcal{C}(G; K)$ under the conjectural monoidal equivalence~\eqref{eq: monoidal equiv conjecture}. This correspondence suggests that the non-degenerate interfaces would be invertible in an appropriate sense. Indeed, the non-degenerate interface that we construct in \eqref{eq: product interface} is invertible as an operator acting on the virtual bonds.}

Let us first show the first bullet point.
To this end, we suppose that there exists $h \in H$ such that $K_2 = hK_1h^{-1}$.
Under this assumption, we show that the following product state is non-degenerate:
\begin{equation}
\ket{\psi(h)} = \bigotimes_{i: \text{sites}} \ket{\psi(h)}_i, \qquad
\ket{\psi(h)}_i := \sum_{h_{\ell} \in H} \ket{h_{\ell}, hh_{\ell}}_i.
\label{eq: product interface}
\end{equation}
We note that the product state $\ket{\psi(h)}$ is gapped because it can be written as an MPO with bond dimension one.
A direct computation shows that the symmetry operators act on $\ket{\psi(h)}$ as
\begin{equation}
\hat{\mathsf{L}}(g; h_1, h_2) \ket{\psi(h)} = \delta_{h, U^{-1}_{g; K_2}(h_2)(U^{-1}_{g; K_1}(h_1))^{-1}} \ket{\psi(h_2h_1^{-1})}.
\label{eq: O psi(h)}
\end{equation}
Due to the definition of $U_{g; K_2}$, there exists $k_2 \in K_2$ such that $h_2g = k_2U^{-1}_{g; K_2}(h_2)$.
Furthermore, due to the assumption that $K_2 = hK_1h^{-1}$, there exists $k_1 \in K_1$ such that $k_2 = hk_1h^{-1}$.
Therefore, we find $h^{-1}h_2g = k_1h^{-1}U^{-1}_{g; K_2}(h_2)$ for some $k_1 \in K_1$.
This implies
\begin{equation}
h^{-1}U^{-1}_{g; K_2}(h_2) = U^{-1}_{g; K_1}(h^{-1}h_2).
\end{equation}
Thus, the Kronecker delta in \eqref{eq: O psi(h)} can be written as
\begin{equation}
\delta_{h, U^{-1}_{g; K_2}(h_2)(U^{-1}_{g; K_1}(h_1))^{-1}} = \delta_{U^{-1}_{g; K_1}(h_1), U^{-1}_{g; K_1}(h^{-1}h_2)} = \delta_{h_1, h^{-1}h_2}.
\end{equation}
Plugging this into \eqref{eq: O psi(h)}, we find
\begin{equation}
\hat{\mathsf{L}}(g; h_1, h_2) \ket{\psi(h)} = \delta_{h, h_2h_1^{-1}} \ket{\psi(h)}.
\end{equation}
This shows that $\ket{\psi(h)}$ is a non-degenerate gapped state at the interface.

Next, we show the second bullet point.
To this end, we first relabel the symmetry operators as
\begin{equation}
\hat{\mathcal{L}}(h_1, h_2, k_1, k_2) := \hat{\mathsf{L}}(k_1; e, h_1), \qquad \forall h_1 \in H, ~ \forall k_1 \in K_1,
\label{eq: sym op relabel}
\end{equation}
where $h_2 \in H$ and $k_2 \in K_2$ are uniquely determined by $h_1 \in H$ and $k_1 \in K_1$ via
\begin{equation}
h_1k_1 = k_2h_2.
\label{eq: k2 h2}
\end{equation}
We note that any symmetry operator is equal to the operator of the form \eqref{eq: sym op relabel} due to \eqref{eq: isomorphic sym op}.
The fusion rules~\eqref{eq: interface sym fusion rules} then reduce to
\begin{equation}
\hat{\mathcal{L}}(h_1, h_2, k_1, k_2) \hat{\mathcal{L}}(h_1^{\prime}, h_2^{\prime}, k_1^{\prime}, k_2^{\prime}) = \delta_{h_2, h_1^{\prime}} \hat{\mathcal{L}}(h_1, h_2^{\prime}, k_1k_1^{\prime}, k_2 k_2^{\prime}),
\label{eq: fusion rules simplified}
\end{equation}
which, due to \eqref{eq: strip obj TN}, immediately follows from the fusion rules \eqref{eq: fusion rules} of the strip 2-algebra $\mathcal{C}_{K_1, K_2}$.\footnote{One can also verify the fusion rules~\eqref{eq: fusion rules simplified} by a direct computation based on \eqref{eq: interface sym fusion rules} and \eqref{eq: sym op relabel}.}
Now, we suppose that the symmetry at the interface admits a (non-zero) non-degenerate state $\ket{\psi}$, which is not necessarily gapped.
Since $\ket{\psi}$ is non-degenerate, it satisfies
\begin{equation}
\hat{\mathcal{L}}(h_1, h_2, k_1, k_2) \ket{\psi} = \alpha(h_1, k_1) \ket{\psi}, \qquad \forall h_1 \in H, ~ \forall k_1 \in K_1,
\end{equation}
where $\alpha(h_1, k_1) \in \mathbb{C}$ is a complex number.
We note that $\alpha$ depends only on $h_1$ and $k_1$ because $h_2$ and $k_2$ are uniquely determined by $h_1$ and $k_1$.
Due to the fusion rules~\eqref{eq: fusion rules simplified}, $\alpha$ satisfies
\begin{equation}
\alpha(h_1, k_1) \alpha(h_2, k_1^{-1}) = \alpha(h_1, e), \qquad \forall k_1 \in K_1,
\end{equation}
where $h_2 \in H$ is determined by \eqref{eq: k2 h2}.
The above equation implies that $\alpha(h_1, k_1)$ is non-zero for all $k_1 \in K_1$ if $\alpha(h_1, e)$ is non-zero.
On the other hand, the decomposition~\eqref{eq: id decomp} of the identity operator implies that $\alpha$ satisfies
\begin{equation}
\sum_{h_1 \in H} \alpha(h_1, e) = 1,
\end{equation}
which in turn implies that there exists $h_1 \in H$ such that $\alpha(h_1, e) \neq 0$.
Thus, we find that there exists $h_1 \in H$ such that $\alpha(h_1, k_1) \neq 0$ for any $k_1 \in K_1$.
When $\alpha(h_1, k_1)$ is non-zero, we have 
\begin{equation}
\alpha(h_1, k_1)^2 = \delta_{h_1, h_2} \alpha(h_1, k_1^2) \neq 0,
\end{equation}
where the first equality follows from the fusion rules~\eqref{eq: fusion rules simplified}.
The above equation shows that $h_1 = h_2$ for all $k_1 \in K_1$.
Recalling that $h_2 \in H$ is defined by \eqref{eq: k2 h2}, we find
\begin{equation}
h_1k_1 h_1^{-1} = h_1k_1h_2^{-1} \in K_2, \qquad \forall k_1 \in K_1,
\end{equation}
which shows that $h_1K_1h_1^{-1} \subset K_2$.
Since $|K_1| = |K_2|$, it then follows that
\begin{equation}
h_1K_1 h_1^{-1} = K_2.
\end{equation}
This concludes the proof.

Notably, our proof of the second part relies only on the fusion rules~\eqref{eq: fusion rules simplified} of the symmetry operators~\eqref{eq: sym op relabel}.
Therefore, as long as the symmetry operators at the interface obey the same fusion rules as \eqref{eq: fusion rules simplified}, the proof can be applied to more general interfaces than~\eqref{eq: interface state}.

\vspace*{\baselineskip}
\noindent{\bf Symmetry-enforced degeneracy.}
As shown above, if there exists no element $h \in H$ such that $K_2 = hK_1h^{-1}$, any state at the interface of $\ket{\text{cluster}(G; H; K_1)}$ and $\ket{\text{cluster}(G; H; K_2)}$ cannot be non-degenerate under the symmetry.
This implies that the ground states in the presence of an interface must be degenerate unless the symmetry is explicitly broken.
In other words, the degeneracy is enforced by the symmetry.
Notably, this statement is irrespective of whether the interface is gapped or gapless.
For example, it follows that the boundary of the ordinary $\mathbb{Z}_2$-cluster model must be degenerate unless the symmetry is explicitly broken.
This is consistent with the observation made in \cite{Yoshida:2015cia} that the boundary of the $\mathbb{Z}_2$-cluster model has a ferromagnetic order.
We emphasize that this type of symmetry-enforced degeneracy is stronger than the usual LSM-type constraint that allows a non-degenerate gapless state.
This stronger constraint originates from the fact that the strip 2-algebra $\mathcal{C}_{K_1, K_2}$ is not a fusion category but a multifusion category.\footnote{We refer the reader to \cite{Chang:2018iay} for the first examples of the LSM-type constraint for fusion category symmetries. To the best of our knowledge, it is still open whether every fusion category can be realized as the symmetry of a 1+1d gapless system with a non-degenerate ground state. Nevertheless, many fusion categories are known to be realizable in non-degenerate gapless systems, including the Haagerup fusion categories \cite{Huang:2021nvb, Vanhove:2021zop} (see also \cite{Liu:2022qwn, Hung:2025gcp, Albert:2025umy, Bottini:2025hri}).}

\subsubsection{Example: gapped boundary of the $G$-cluster state}
\label{sec: Example: gapped boundary of the G-cluster state}
As an example, we consider the gapped interface between the $G_0$-cluster state~\eqref{eq: G cluster} and the trivial product state with $2\Rep(G_0) \boxtimes 2\Vect_{G_0}$ symmetry, where $G_0$ is an arbitrary finite group.
To this end, we first recall that the input data of the $G_0$-cluster state and the trivial product state are given by
\begin{equation}
G = G_0 \times G_0, \qquad
H = G_0^{\text{left}}, \qquad
K_{\text{cluster}} = G_0^{\text{diag}}, \qquad
K_{\text{triv}} = G_0^{\text{right}}, \qquad
S_{H \backslash G} = G_0^{\text{right}}.
\end{equation}
For simplicity, we will focus on the interfaces represented by the tensor network of the form~\eqref{eq: interface state}.
In what follows, an element $(g, e) \in H = G_0^{\text{left}}$ will be written simply as $g \in G_0$.

\vspace*{\baselineskip}
\noindent{\bf Symmetry operators on the boundary.}
According to the general discussions in Section~\ref{sec: Interface symmetries TN}, the symmetry operators at the interface of the $G_0$-cluster state and the trivial product state can be written as
\begin{equation}
\hat{\mathsf{L}}(g_{\ell}, g_r; g_1, g_2) = \bigotimes_{i: \text{sites}} \hat{\mathcal{O}}_i(g_{\ell}, g_r; g_1, g_2), \qquad
\forall g_{\ell}, g_r, g_1, g_2 \in G_0,
\end{equation}
where the tensor product is taken over all sites on the interface, and $\hat{\mathcal{O}}_i(g_{\ell}, g_r; g_1, g_2)$ is defined by
\begin{equation}
\hat{\mathcal{O}}_i(g_{\ell}, g_r; g_1, g_2) \coloneq \adjincludegraphics[valign = c, trim={10, 10, 10, 10}]{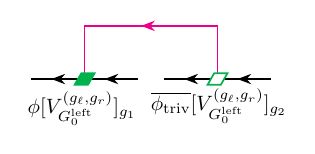}.
\label{eq: G-cluster boundary sym op}
\end{equation}
Here, $\phi[V_{G_0^{\text{left}}}^{(g_{\ell}, g_r)}]_g$ is the action tensor~\eqref{eq: G-cluster action tensor} for the $G_0$-cluster state, and $\phi_{\text{triv}}[V_{G_0^{\text{left}}}^{(g_{\ell}, g_r)}]_g$ is the action tensor for the trivial product state, which is given by\footnote{One can derive \eqref{eq: trivial action tensor} by using the general formula~\eqref{eq: action tensors} for the action tensors.}
\begin{equation}
\adjincludegraphics[valign = c, trim={10, 10, 10, 10}]{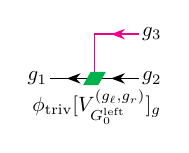} = \adjincludegraphics[valign = c, trim={10, 10, 10, 10}]{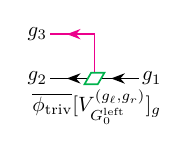} = \delta_{g_1g_3^{-1}, gg_{\ell}} \delta_{g_1, g_2}.
\label{eq: trivial action tensor}
\end{equation}
By plugging \eqref{eq: G-cluster action tensor} and \eqref{eq: trivial action tensor} into \eqref{eq: G-cluster boundary sym op}, we can compute the action of the symmetry operator at the interface as
\begin{equation}
\hat{\mathcal{O}}_i(g_{\ell}, g_r; g_1, g_2) \ket{h_{\ell}, h_r}_i = \adjincludegraphics[valign = c, trim={10, 10, 10, 10}]{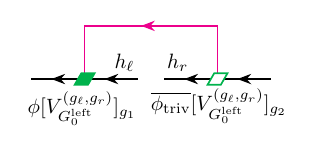} = \delta_{h_rh_{\ell}^{-1}, g_2g_1^{-1}g_r} \ket{g_rh_{\ell}, h_r}_i,
\label{eq: G-cluster boundary sym op2}
\end{equation}
where $h_{\ell}, h_r \in G_0$.
Since the above operator depends only on $g_r$ and $g_2g_1^{-1}$, we can choose $g_{\ell} = g_1 = e$ without loss of generality.
The independent symmetry operators can then be written as
\begin{equation}
\hat{\mathsf{L}}(g, g^{\prime}) \coloneq \hat{\mathsf{L}}(e, g^{-1}g^{\prime}; e, g), \qquad \forall g, g^{\prime} \in G_0.
\label{eq: G-cluster boundary sym op3}
\end{equation}
A direct computation based on \eqref{eq: G-cluster boundary sym op2} shows that the redefined symmetry operators~\eqref{eq: G-cluster boundary sym op3} obey the fusion rule
\begin{equation}
\hat{\mathsf{L}}(g_1, g_2) \hat{\mathsf{L}}(g_1^{\prime}, g_2^{\prime}) = \delta_{g_2, g_1^{\prime}} \hat{\mathsf{L}}(g_1, g_2^{\prime}).
\end{equation}
This fusion rule agrees with the fusion rule~\eqref{eq: G-cluster boundary fusion} derived within the context of category theory.
In particular, the above fusion rule suggests that the symmetry at the interface is described by a multifusion category~\eqref{eq: strip G-cluster}, that is,
\begin{equation}
\mathcal{C}_{K_{\text{cluster}}, K_{\text{triv}}} \cong \Mat_{|G_0|}(\Vect).
\label{eq: G-cluster strip TN}
\end{equation}

\vspace*{\baselineskip}
\noindent{\bf Gapped boundary mode.}
Since the $G_0$-cluster state and the trivial product state are in different SPT phases,\footnote{This is because there is no element $h \in H$ such that $K_\text{triv} = hK_{\text{cluster}}h^{-1}$.} we cannot have a non-degenerate gapped state at the interface without explicitly breaking the symmetry.
Thus, any gapped states at the interface must be degenerate.\footnote{More stringently, as shown in Section \ref{sec: Gapped interface modes TN}, any states at the interface must be degenerate, whether or not they are gapped.}
Indeed, the product states \eqref{eq: product interface} at the interface form a multiplet under the symmetry action, and hence they are degenerate.
More specifically, for each $h \in G_0$, we have the following product state
\begin{equation}
\ket{\psi(h)} = \bigotimes_{i: \text{sites}} \ket{\psi(h)}_i, \qquad
\ket{\psi(h)}_i := \sum_{h_{\ell} \in G_0} \ket{h_{\ell}, hh_{\ell}}_i,
\end{equation}
where the tensor product is taken over all sites on the interface, and the states $\{\ket{\psi(h)} \mid h \in G_0\}$ are permuted by the action of the symmetry operators~\eqref{eq: G-cluster boundary sym op3} as follows:
\begin{equation}
\hat{\mathsf{L}}(g, g^{\prime}) \ket{\psi(h)} = \delta_{g^{\prime}, h} \ket{\psi(g)}.
\label{eq: multiplet}
\end{equation}
From the category-theoretical point of view, equation~\eqref{eq: multiplet} suggests that the product state $\ket{\psi(h)}$ corresponds to the unique simple object of each component of an indecomposable module category $\Vect^{\oplus |G_0|}$ over the symmetry category~\eqref{eq: G-cluster strip TN}.

\section{Parameterized families of 2+1d generalized cluster states}
\label{sec: Parameterized families}
In this section, we construct $S^{1}$-parameterized families of generalized cluster states and study the pumping phenomena of gapped interface modes.
Specifically, for each $h\in H$ satisfying $hKh^{-1}=K$, we construct an $S^1$-parameterized family of $\mathcal{C}(G; H)$-symmetric states that are unitarily equivalent to the generalized cluster state $\ket{\text{cluster}(G; H; K)}$.
We show that the generalized Thouless pump associated with this family is given by the non-degenerate self-interface mode~\eqref{eq: product interface}.
We also briefly discuss the relation to the conjectural classification of $S^1$-parameterized families based on category theory.

\subsection{$S^1$-parameterized families}
\label{sec: S1-parameterized families}

\subsubsection{Families of $G/K$-SSB models}
We first construct $S^1$-parameterized families of $G/K$-SSB models.
The $S^1$-parameterized families of generalized cluster models are obtained by gauging the subgroup symmetry $H$ in the $G/K$-SSB models.

\vspace*{\baselineskip}
\noindent{\bf Symmetry interpolation method.}
In general, there are several ways to construct parameterized families of quantum systems on the lattice~\cite{PhysRevResearch.2.042024,Tantivasadakarn:2021wdv,Shiozaki:2021weu,Wen:2021gwc}.
In our construction, we employ the symmetry interpolation method, which we now explain briefly.
First, suppose that a Hamiltonian $\hat{H}$ possesses a symmetry $U = \prod_i u_i$, that is
\begin{equation}
       [ \hat{H}, U ] = 0.
\end{equation}
Here, $u_i$ is a local unitary operator acting on a finite number of degrees of freedom around a single site $i$.
Since each $u_i$ is unitary, there exists a continuous path $\{ u_i(\theta) \mid 0 \leq \theta \leq 2\pi \}$ connecting it to the identity, that is $u_i(0)=\id$ and $u_i(2\pi)=u_i$.
Using this, we define $U(\theta) = \prod_i u_i(\theta)$ and deform the Hamiltonian by conjugation with $U(\theta)$:\footnote{The Hamiltonian $\hat{H}(\theta)$ after the conjugation is a sum of local terms as long as $H$ is because $U(\theta)$ is a finite-depth local unitary.}
\begin{equation}
    \hat{H}(\theta) = U(\theta) \hat{H} U(\theta)^{\dagger}.
\label{eq: H theta}
\end{equation}
Then, due to the condition $[ \hat{H}, U ] = 0$, the Hamiltonian $\hat{H}(\theta)$ becomes $2\pi$-periodic.
In this way, when the Hamiltonian has a (unitary) symmetry $U$, one can construct an $S^1$-parameterized family by interpolating between $U$ and the identity.
Note that the Hamiltonian $\hat{H}(\theta)$ generally does not possess the symmetry $U$ except for $\theta = 0, 2\pi$.

For our purposes, we take $\hat{H}$ to be the parent Hamiltonian~\eqref{eq: Ham G/K} of the $G/K$-SSB states \eqref{eq: GS G/K}.
Moreover, to construct a $G$-symmetric parameterized family, we require that the interpolated operator $U(\theta)$ commutes with the $G$ symmetry for all $\theta \in [0, 2\pi]$.
We note that $U$ is not necessarily a symmetry operator for the $G$ symmetry under consideration.

\vspace*{\baselineskip}
\noindent{\bf An $S^1$-parameterized family of $\mathbb{Z}_2$-SSB models.}
As a warm-up, we first construct an $S^1$-parameterized family of the models where the symmetry $G = \mathbb{Z}_2 = \{e, \eta\}$ is spontaneously broken down to $K = \{e\}$.
To this end, we consider the following $\mathbb{Z}_2$-SSB Hamiltonian of qubits on a square lattice:
\begin{equation}
\hat{H}_{\text{$\mathbb{Z}_2$-SSB}} = -\sum_{[ij] \in E} Z^{(i)}Z^{(j)}.
\label{eq: Z2 SSB ham}
\end{equation}
The $\mathbb{Z}_2$ symmetry of this model is generated by $X = \bigotimes_{i \in P} X^{(i)}$.
The above Hamiltonian has twofold degenerate ground states $\bigotimes_{i \in P} \ket{e}_i$ and $\bigotimes_{i \in P} \ket{\eta}_i$, where $\ket{e}_i$ and $\ket{\eta}_i$ are the eigenstates of $Z^{(i)}$ with eigenvalues $+1$ and $-1$, respectively.
These two ground states are exchanged by the symmetry operator $X$, and hence the $\mathbb{Z}_2$ symmetry is spontaneously broken.

The symmetry operator $X$ satisfies all the properties required for $U$, i.e., it is a product of local unitaries and commutes with both the Hamiltonian and the symmetry operator.
Thus, one can construct a $\mathbb{Z}_2$-symmetric $S^1$-parameterized family by interpolating between $X$ and the identity operator.
An example of an interpolation preserving the $\mathbb{Z}_2$ symmetry is
\begin{equation}
U(\theta) = \bigotimes_{i \in P} e^{i\frac{\theta}{4}(X^{(i)} - 1)}.
\end{equation}
By conjugating the Hamiltonian~\eqref{eq: Z2 SSB ham} with this $U(\theta)$, we can obtain an $S^1$-parameterized family of $\mathbb{Z}_2$-SSB models.

The $S^1$-parameterized family constructed in this way is non-trivial.
To see this, we consider the Hamiltonian \eqref{eq: Z2 SSB ham} conjugated by $U(\theta)$ only on a subregion $P^{\prime} \subset P$ of the square lattice, i.e., 
\begin{equation}
\hat{H}_{\text{$\mathbb{Z}_2$-SSB}}^{\text{text.}}(\theta) = U^{\prime}(\theta) \hat{H}_{\text{$\mathbb{Z}_2$-SSB}} U^{\prime}(\theta)^{\dagger}, \qquad
U^{\prime}(\theta) = \bigotimes_{i \in P^{\prime}} e^{i\frac{\theta}{4}(X^{(i)}-1)}.
\end{equation}
We will refer to this Hamiltonian as a textured Hamiltonian.
We note that the textured Hamiltonian is $\mathbb{Z}_2$ symmetric for all $\theta \in [0, 2\pi]$, although it is not $2\pi$-periodic.
The ground states of the textured Hamiltonian are obtained by applying $U_{P^{\prime}}(\theta)$ to the ground states of $\hat{H}_{\text{$\mathbb{Z}_2$-SSB}}$.
In particular, at $\theta = 2\pi$, the ground states become the following domain wall configurations:
\begin{align}
\left(\bigotimes_{i \in P \setminus P^{\prime}} \ket{e}_i \right) \otimes \left(\bigotimes_{i \in P^{\prime}} \ket{\eta}_i\right), \qquad
\left(\bigotimes_{i \in P \setminus P^{\prime}} \ket{\eta}_i \right) \otimes \left(\bigotimes_{i \in P^{\prime}} \ket{e}_i\right).
\end{align}
Therefore, adiabatically changing the parameter $\theta$ from $0$ to $2\pi$ pumps a domain wall on the boundary of $P^{\prime}$.
This indicates that the $S^1$-parameterized family obtained above is non-trivial as a family of $\mathbb{Z}_2$-SSB models.

\vspace*{\baselineskip}
\noindent{\bf Domain wall creation operators in $G/K$-SSB model.}
The above example suggests that one can construct a non-trivial $S^1$-parameterized family of $G/K$-SSB models for general $G$ and $K$ by taking $U$ to be an operator that permutes the symmetry-breaking ground states.
Such an operator creates a domain wall when it acts on a ground state.
Thus, we refer to it as a domain wall creation operator.
As in the case of the $\mathbb{Z}_2$-SSB models, we require that the domain wall creation operator commutes with the $G$ symmetry so that the resulting family becomes $G$-symmetric.

To find the domain wall creation operators in a $G/K$-SSB model, we consider the Hamiltonian \eqref{eq: Ham G/K} of $G$-qudits on a square lattice, i.e.,
\begin{equation}
\hat{H} = -\sum_{i \in P} \hat{\mathsf{h}}_i - \sum_{[ij] \in E} \hat{\mathsf{h}}_{ij},
\label{eq: G/K SSB ham}
\end{equation}
where $\hat{\mathsf{h}}_i $ and $\hat{\mathsf{h}}_{ij}$ are defined by
\begin{equation}
\hat{\mathsf{h}}_i \ket{g}_i = \frac{1}{|K|} \sum_{k \in K} \ket{g_ik}_i, \qquad
\hat{\mathsf{h}}_{ij} \ket{g_i}_i \otimes \ket{g_j}_j = \delta_{g_i^{-1}g_j \in K} \ket{g_i}_i \otimes \ket{g_j}_j.
\end{equation}
This model possesses a $G$ symmetry whose action is given by the left multiplication of $G$.
Namely, the Hamiltonian \eqref{eq: G/K SSB ham} commutes with the following symmetry operators:
\begin{equation}
\overrightarrow{X}_g \coloneq \bigotimes_{g \in G} \overrightarrow{X}_g^{(i)}, \qquad \forall g \in G.
\label{eq: left mult g}
\end{equation}
The ground states of the Hamiltonian \eqref{eq: G/K SSB ham} are given by \eqref{eq: GS G/K}, which spontaneously break the symmetry $G$ down to $K$.

One can show that the Hamiltonian $\hat{H}$ commutes with the right multiplication of $h \in H$ satisfying $hKh^{-1} = K$, that is,
\begin{equation}
[\hat{H}, \overleftarrow{X}_h] = 0, \qquad \forall h \in H \text{ s.t. } hKh^{-1} = K,
\end{equation}
where $\overleftarrow{X}_h \coloneq \bigotimes_{i \in P} \overleftarrow{X}_h^{(i)}$.
The operator $\overleftarrow{X}_h$ obviously commutes with the symmetry operators~\eqref{eq: left mult g} because it acts from the right, while the symmetry operators act from the left.
Furthermore, it permutes the $G/K$-SSB ground states \eqref{eq: GS G/K}.
Therefore, the right multiplication $\overleftarrow{X}_h$ for $h \in H$ satisfying $hKh^{-1} = K$ is a domain wall creation operator preserving the $G$ symmetry.

\vspace*{\baselineskip}
\noindent{\bf $S^1$-parameterized families of $G/K$-SSB models.}
Now, we construct an $S^1$-parameterized family of $G/K$-SSB models by interpolating the domain wall creation operator $\overleftarrow{X}_h$ and the identity operator.
An example of an interpolation preserving the $G$ symmetry is
\begin{equation}
U_h(\theta) = \bigotimes_{i \in P} u_h^{(i)}(\theta), \qquad u_h^{(i)}(\theta) \coloneq \exp{\left(\frac{\theta}{2\pi}\log\overleftarrow{X}^{(i)}_{h}\right)}.
\label{eq: Uh theta}
\end{equation}
Here, we choose the principal branch of the logarithm as
\begin{equation}
       \log \overleftarrow{X}_h^{(i)} = \sum_{s=0}^{\abs{H}-1}i\frac{2\pi s}{\abs{H}}P_{s}[\overleftarrow{X}^{(i)}_{h}],
\label{eq: log}
\end{equation}
where $P_s[\overleftarrow{X}_h^{(i)}]$ is the projection onto the eigenspace of $\overleftarrow{X}_h^{(i)}$ with eigenvalue $e^{i\frac{2\pi}{|H|}s}$.\footnote{We note that the eigenvalues of $\overleftarrow{X}_h^{(i)}$ are quantized to the $|H|$-th roots of unity because $(\overleftarrow{X}_h^{(i)})^{|H|} = \id$.}
This projection can be written explicitly as a polynomial
\begin{equation}
P_{s}[\overleftarrow{X}_h^{(i)}]:=\frac{1}{\abs{H}}\sum_{n=0}^{\abs{H}-1} e^{-i\frac{2\pi s}{\abs{H}}n} (\overleftarrow{X}_h^{(i)})^n,
\end{equation}
which obeys the following relations:
\begin{equation}
P_{s}[\overleftarrow{X}_h^{(i)}] P_{s'}[\overleftarrow{X}_h^{(i)}] = \delta_{s,s'}P_{s}[\overleftarrow{X}_h^{(i)}], \qquad
P_{s}[\overleftarrow{X}_h^{(i)}]^{\dagger} = P_{s}[\overleftarrow{X}_h^{(i)}].
\end{equation}
We note that $u_h^{(i)}(\theta)$ in \eqref{eq: Uh theta} acts as $e^{i \frac{\theta}{|H|}s}$ on the eigenspace of $\overleftarrow{X}_h^{(i)}$ with eigenvalue $e^{i \frac{2\pi}{|H|} s}$.
In particular, it satisfies $u_h^{(i)}(0) = \id$ and $u_h^{(i)}(2\pi) = \overleftarrow{X}_h^{(i)}$.
Furthermore, since $u_h^{(i)}(\theta)$ is a polynomial of the right multiplication $\overleftarrow{X}_h^{(i)}$, it automatically commutes with the symmetry operators~\eqref{eq: left mult g} that are given by the left multiplication of $G$.
Therefore, we can obtain an $S^1$-parameterized family of $G/K$-SSB models by conjugating the Hamiltonian~\eqref{eq: G/K SSB ham} with the above $U_h(\theta)$, i.e.,
\begin{equation}
\hat{H}_h(\theta) \coloneq U_h(\theta) \hat{H} U_h(\theta).
\label{eq: Hh theta}
\end{equation}
The ground states of $\hat{H}_h(\theta)$ are obtained by applying $U_h(\theta)$ to the ground states \eqref{eq: GS G/K} of $\hat{H}$.

We can see the non-triviality of the $S^1$-parameterized family~\eqref{eq: Hh theta} by using the following textured Hamiltoian:
\begin{equation}
\hat{H}_h^{\text{text.}}(\theta) \coloneq U_{h}^{\prime}(\theta) \hat{H} U_h^{\prime}(\theta)^{\dagger}, \qquad
U_h^{\prime}(\theta) \coloneq \bigotimes_{i \in P^{\prime}} u_h^{(i)}(\theta).
\label{eq: G/K textured theta}
\end{equation}
Here, $P^{\prime} \subset P$ is a subregion of the square lattice.
By construction, if we change $\theta$ adiabatically from $0$ to $2\pi$ in the textured Hamiltonian~\eqref{eq: G/K textured theta}, the ground states aquire a non-trivial domain wall along the boundary of $P^{\prime}$.
This indicates that the $S^1$-parameterized family constructed above is non-trivial.

\subsubsection{Families of generalized cluster models}
\label{sec: Families of generalized cluster models}
We now construct $S^1$-parameterized families of generalized cluster models with $\mathcal{C}(G; H)$ symmetry.
This is achieved by gauging the subgroup symmetry $H$ in the models defined in the previous subsection.
Unless otherwise stated, in this subsection, the gauged model refers to the model after the gauge fixing.

\vspace*{\baselineskip}
\noindent{\bf Dual of the domain wall creation operator.}
We first define the dual of the domain wall creation operator, which we will use later to write down the Hamiltonian of the gauged model.
The dual of the domain wall creation operator $\overleftarrow{X}_h$ is defined by
\begin{equation}
\left.\overleftarrow{X}_h\right|_{\text{g.f.}} = \bigotimes_{i \in P} \left.\overleftarrow{X}_h^{(i)}\right|_{\text{g.f.}},
\end{equation}
where $\left.\overleftarrow{X}_h^{(i)}\right|_{\text{g.f.}}$ is the local operator obtained by transforming $\overleftarrow{X}_h^{(i)}$ with the gauging operator~\eqref{eq: D gf}:\footnote{In general, for a local operator $\mathcal{O}$, an operator $\mathcal{O}|_{\text{g.f.}}$ satisfying $\mathcal{O}|_{\text{g.f.}} \mathsf{D}_H = \mathsf{D}_H \mathcal{O}$ is not unique: it is unique only up to operators that act as the identity on the image of $\mathsf{D}_H$. However, the {\it local} operator $\mathcal{O}|_{\text{g.f.}}$ satisfying $\mathcal{O}|_{\text{g.f.}} \mathsf{D}_H = \mathsf{D}_H \mathcal{O}$ is unique if it exists, because there is no non-trivial local operator that acts as the identity on the image of $\mathsf{D}_H$.}
\begin{equation}
\left.\overleftarrow{X}_h^{(i)}\right|_{\text{g.f.}} \mathsf{D}_H = \mathsf{D}_H \overleftarrow{X}_h^{(i)}.
\label{eq: dual of Xhi 1}
\end{equation}
By definition, we have $\left.\overleftarrow{X}_h\right|_{\text{g.f.}} \mathsf{D}_H = \mathsf{D}_H \overleftarrow{X}_h$.
As we will see shortly, the explicit form of $\left.\overleftarrow{X}_h^{(i)}\right|_{\text{g.f.}}$ is given by
\begin{equation}
\left.\overleftarrow{X}^{(i)}_{h}\right|_{\text{g.f.}}\Ket{\adjincludegraphics[valign = c, trim={10, 10, 10, 10}]{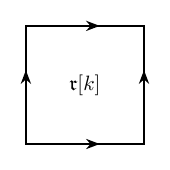}}
= \Ket{\adjincludegraphics[valign = c, trim={10, 10, 10, 10}]{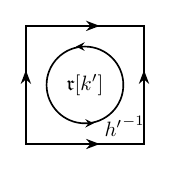}},
\end{equation}
where $k':=h^{-1}kh$ and $h':=h[k]^{-1}hh[k']$.
We note that $k^{\prime}$ is an element of $K$ because $h \in H$ satisfies $K = hKh^{-1}$.
The above operator can be written more concisely as
\begin{equation}
\left.\overleftarrow{X}^{(i)}_{h}\right|_{\text{g.f.}} = \sum_{k \in K} L^{(\partial i)}_{h[h^{-1}kh]^{-1}h^{-1}h[k]} \otimes \ket{\mathfrak{r}[h^{-1}kh]}\bra{\mathfrak{r}[k]},
\label{eq: dual of Xhi 2}
\end{equation}
where $L^{(\partial i)}_{h[h^{-1}kh]^{-1}h^{-1}h[k]}$ is the loop operator defined in \eqref{eq: loop op}.

Let us now verify that $\overleftarrow{X}_h^{(i)}$ in \eqref{eq: dual of Xhi 2} satisfies the defining equation \eqref{eq: dual of Xhi 1}.
To this end, we use the tensor network representation~\eqref{eq: gauging PEPO} of the gauging operator $\mathsf{D}_H$.
To show \eqref{eq: dual of Xhi 1}, it suffices to focus on the local tensors of $\mathsf{D}_H$ around the plaquette $i \in P$.
For the local tensor acting on the matter field, we have
\begin{equation}
\adjincludegraphics[scale=1,trim={10pt 10pt 10pt 10pt},valign = c]{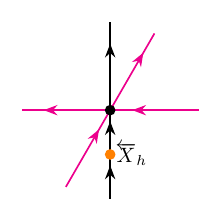}
\;\;=\;\;
\sum_{k \in K}\adjincludegraphics[scale=1,trim={10pt 10pt 10pt 10pt},valign = c]{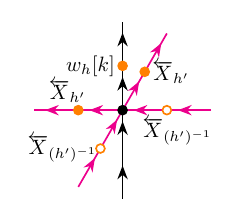},
\label{eq: Xh CR matter}
\end{equation}
where $w_h[k] := \ket{\mathfrak{r}[h^{-1}kh]} \bra{\mathfrak{r}[k]}$  and $h^{\prime} = h[k]^{-1}hh[h^{-1}kh]$.
One can show the above equation by comparing the tensors on both sides while fixing the label of the bottom (physical) leg to $h_0 \mathfrak{r}[k_0]$ for any $h_0 \in H$ and $k_0 \in K$.
On the other hand, for the local tensor acting on the gauge field, we have
\begin{equation}
\adjincludegraphics[scale=1,trim={10pt 10pt 10pt 10pt},valign = c]{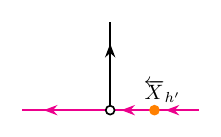}
= \adjincludegraphics[scale=1,trim={10pt 10pt 10pt 10pt},valign = c]{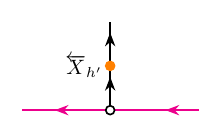}, \qquad
\adjincludegraphics[scale=1,trim={10pt 10pt 10pt 10pt},valign = c]{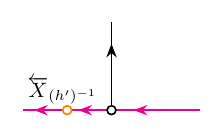}
= \adjincludegraphics[scale=1,trim={10pt 10pt 10pt 10pt},valign = c]{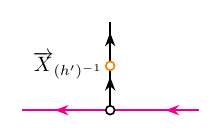},
\label{eq: Xh CR gauge}
\end{equation}
which immediately follows from the definition of the local tensor.
By combining \eqref{eq: Xh CR matter} and \eqref{eq: Xh CR gauge}, we find that $\overleftarrow{X}^{(i)}_{h}$ is transformed into $\left.\overleftarrow{X}^{(i)}_{h}\right|_{\text{g.f.}}$ under the action of the gauging operator as in \eqref{eq: dual of Xhi 1}.

\vspace*{\baselineskip}
\noindent{\bf $S^1$-parameterized families of generalized cluster models.}
An $S^1$-parameterized family of generalized cluster models with $\mathcal{C}(G; H)$ symmetry is obtained by gauging the subgroup symmetry $H$ in the $G/K$-SSB models \eqref{eq: Hh theta}.
The gauged Hamiltonian $\hat{H}^h_{\text{g.f.}}(\theta)$ is related to the Hamiltonian $\hat{H}_h(\theta)$ of the ungauged model as
\begin{equation}
\hat{H}^h_{\text{g.f.}}(\theta) \mathsf{D}_H = \mathsf{D}_H \hat{H}_h(\theta).
\label{eq: Hh theta duality}
\end{equation}
The local Hamiltonian $\hat{H}^h_{\text{g.f.}}(\theta)$ satisfying the above equation is uniquely given by
\begin{equation}
\hat{H}^h_{\text{g.f.}}(\theta) = U^h_{\text{g.f.}}(\theta) \hat{H}_{\text{g.f.}} U^h_{\text{g.f.}}(\theta)^{\dagger},
\label{eq: Hh gf theta}
\end{equation}
where $\hat{H}_{\text{g.f.}}$ is the Hamiltonian \eqref{eq: generalized cluster ham} of the generalized cluster model, and $U^h_{\text{g.f.}}(\theta)$ is the dual of $U_h(\theta)$ satisfying $U^h_{\text{g.f.}}(\theta)\mathsf{D}_H = \mathsf{D}_H U_h(\theta)$.
An explicit form of $U^h_{\text{g.f.}}(\theta)$ is given by
\begin{equation}
U^h_{\text{g.f.}}(\theta) = \prod_{i \in P} \left.u_h^{(i)}(\theta)\right|_{\text{g.f.}}, \qquad
\left.u_h^{(i)}(\theta)\right|_{\text{g.f.}} \coloneq \exp\left(\frac{\theta}{2\pi} \log\left.\overleftarrow{X}_h^{(i)}\right|_{\text{g.f.}}\right).
\label{eq: Uh theta gf}
\end{equation}
Here, we again choose the principal branch of the logarithm as
\begin{equation}
       \log \left.\overleftarrow{X}_h^{(i)}\right|_{\text{g.f.}} = \sum_{s=0}^{\abs{H}-1} i \frac{2\pi s}{\abs{H}} P_s[\left.\overleftarrow{X}_h^{(i)}\right|_{\text{g.f.}}],
\end{equation}
where $P_s[\left.\overleftarrow{X}_h^{(i)}\right|_{\text{g.f.}}]$ is the projection onto the eigenspace of $\left.\overleftarrow{X}_h^{(i)}\right|_{\text{g.f.}}$ with eigenvalue $e^{i\frac{2\pi}{\abs{H}}s}$.\footnote{The eigenvalues of $\left.\overleftarrow{X}_h^{(i)}\right|_{\text{g.f.}}$ are quantized to the $|H|$-th roots of unity because $(\left.\overleftarrow{X}_h^{(i)}\right|_{\text{g.f.}})^{|H|} = \id$.}
The projection $P_s[\left.\overleftarrow{X}_h^{(i)}\right|_{\text{g.f.}}]$ can be written as a polynomial of $\left.\overleftarrow{X}_h^{(i)}\right|_{\text{g.f.}}$ as
\begin{equation}
P_s[\left.\overleftarrow{X}_h^{(i)}\right|_{\text{g.s.}}] = \frac{1}{\abs{H}}\sum_{n=0}^{\abs{H}-1} e^{-i\frac{2\pi s}{\abs{H}}n} (\left.\overleftarrow{X}_h^{(i)}\right|_{\text{g.f.}})^n.
\end{equation}
In particular, $\left.u_h^{(i)}(\theta)\right|_{\text{g.f.}}$ in \eqref{eq: Uh theta gf} is also a polynomial of $\left.\overleftarrow{X}_h^{(i)}\right|_{\text{g.f.}}$, and hence it follows from \eqref{eq: dual of Xhi 1} that
\begin{equation}
\left.u_h^{(i)}(\theta)\right|_{\text{g.f.}} \mathsf{D}_H = \mathsf{D}_H u_h^{(i)}(\theta), \qquad
U^h_{\text{g.f.}}(\theta) \mathsf{D}_H = \mathsf{D}_H U_h(\theta).
\label{eq: Uh theta duality}
\end{equation}
Equations \eqref{eq: Uh theta duality} and \eqref{eq: intertwiner} imply that the gauged Hamiltonian \eqref{eq: Hh gf theta} satisfies \eqref{eq: Hh theta duality}.
We note that $\hat{H}^h_{\text{g.f.}}(\theta)$ is $2\pi$-periodic because $U^h_{\text{g.f.}}(2\pi) = \left.\overleftarrow{X}_h\right|_{\text{g.f.}}$ commutes with $\hat{H}_{\text{g.f.}}$ by construction.

\vspace*{\baselineskip}
\noindent{\bf Tensor network representation of the ground state.}
The ground state of the $\theta$-dependent Hamiltonian \eqref{eq: Hh gf theta} is obtained by applying the unitary operator $U^h_{\text{g.f.}}(\theta)$ to the ground state of $\hat{H}_{\text{g.f.}}$, which is the generalized cluster state~\eqref{eq: generalized cluster}.
Furthermore, as shown in \eqref{eq: D gf GS}, the generalized cluster state is obtained by applying the gauging operator $\mathsf{D}_H$ to a $G/K$-SSB ground state $\bigotimes_{i \in P} \ket{K}_i$, where $\ket{K}_i \coloneq \frac{1}{\sqrt{|K|}} \sum_{k \in K} \ket{k}_i$.
Therefore, the ground state of $\hat{H}^h_{\text{g.f.}}(\theta)$ in \eqref{eq: Hh gf theta} can be written as
\begin{equation}
U^h_{\text{g.f.}}(\theta) \ket{\text{cluster}(G; H; K)} = U^h_{\text{g.f.}}(\theta) \mathsf{D}_H \left(\bigotimes_{i \in P} \ket{K}_i\right) = \mathsf{D}_H \left(\bigotimes_{i \in P} u_h^{(i)}(\theta) \ket{K}_i\right),
\end{equation}
where the second equality follows from the duality relation \eqref{eq: Uh theta duality}.
A tensor network representation of this ground state is given by
\begin{equation}
U^h_{\text{g.f.}}(\theta) \ket{\text{cluster}(G; H; K)} = \adjincludegraphics[scale=1,trim={10pt 10pt 10pt 10pt},valign = c]{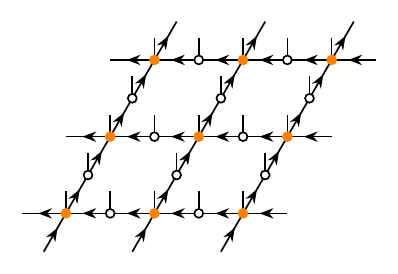},
\end{equation}
where the white dot represents the multiplication tensor defined in \eqref{eq: local tensors}, and the orange dot represents the following tensor:
\begin{equation}
\adjincludegraphics[scale=1,trim={10pt 10pt 10pt 10pt},valign = c]{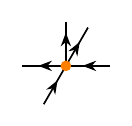}\;\;
\coloneq \adjincludegraphics[scale=1,trim={10pt 10pt 10pt 10pt},valign = c]{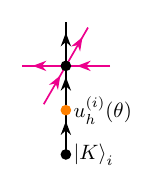}.
\label{eq: local tensor orange}
\end{equation}
Here, the six-leg tensor on the right-hand side is the local tensor of $\mathsf{D}_H$, which is defined in \eqref{eq: DH tensors}.

\subsection{Pumping of gapped interface modes}
\label{sec: Pumping of gapped interface modes}
In this subsection, we study the generalized Thouless pumps associated with the $S^1$-parameterized families of $\mathcal{C}(G; H)$-symmetric models constructed in the previous subsection.
For simplicity, throughout this subsection, we suppose that our lattice models are defined on an infinite plane.\footnote{This is not an essential assumption: one can consider the generalized Thouless pumps on the lattice with periodic boundary conditions in the same way.}

\vspace*{\baselineskip}
\noindent{\bf Generalized Thouless pump.}
We first define the generalized Thouless pump associated with the $S^1$-parameterized family of $\mathcal{C}(G; H)$-symmetric Hamiltonians~\eqref{eq: Hh gf theta}.
To this end, we consider a $\theta$-dependend textured Hamiltonian $\hat{H}_{\text{g.f.}}^{h; \text{text.}}(\theta)$ that satisfy the following properties:
\begin{itemize}
\item On the left half of the plane, the local terms of $\hat{H}_{\text{g.f.}}^{h; \text{text.}}(\theta)$ agree with those of $\hat{H}_{\text{g.f.}}^h(0) = \hat{H}_{\text{g.f.}}$.
\item On the right half of the plane, the local terms of $\hat{H}_{\text{g.f.}}^{h; \text{text.}}(\theta)$ agree with those of $\hat{H}_{\text{g.f.}}^h(\theta)$.
\item Around the middle line at $x=0$, the local terms of $\hat{H}_{\text{g.f.}}^{h; \text{text.}}(\theta)$ are arbitrary as long as the following conditions hold:
\begin{itemize}
\item $\hat{H}_{\text{g.f.}}^{h; \text{text.}}(\theta)$ is continuous in $\theta \in [0, 2\pi]$ with the initial condition $\hat{H}_{\text{g.f.}}^{h; \text{text.}}(0) = \hat{H}_{\text{g.f.}}$.
\item $\hat{H}_{\text{g.f.}}^{h; \text{text.}}(\theta)$ preserves the symmetry $\mathcal{C}(G; H)$ for all $\theta \in [0, 2\pi]$.
\item $\hat{H}_{\text{g.f.}}^{h; \text{text.}}(\theta)$ has a unique gapped ground state for all $\theta \in [0, 2\pi]$.
\end{itemize}
\end{itemize}
We note that $\hat{H}_{\text{g.f.}}^{h; \text{text.}}(\theta)$ is not necessarily $2\pi$-periodic because the local terms around $x=0$ are not.
Now, to define the generalized Thouless pump, we consider changing $\theta$ adiabatically from $0$ to $2\pi$.
Since the textured Hamiltonian $\hat{H}_{\text{g.f.}}^{h; \text{text.}}(\theta)$ is not $2\pi$-periodic, its ground state does not generally go back to the initial state at the end of the adiabatic cycle.
Nevertheless, the ground states at $\theta=0$ and $\theta=2\pi$ differ only around $x=0$ because $\hat{H}_{\text{g.f.}}^{h; \text{text.}}(0)$ and $\hat{H}_{\text{g.f.}}^{h; \text{text.}}(2\pi)$ differ only around $x=0$.
This means that the adiabatic evolution pumps a string-like excitation, which we refer to as the generalized Thouless pump.
The pumped excitation must be non-degenerate under the symmetry action because the ground state of the textured Hamiltonian is unique and hence preserves the symmetry.

In what follows, we take the $\theta$-dependent textured Hamiltonian to be
\begin{equation}
\hat{H}_{\text{g.f.}}^{h; \text{text.}}(\theta) \coloneq U_{\text{g.f.}}^{h; x>0}(\theta) \hat{H}_{\text{g.f.}} U_{\text{g.f.}}^{h; x>0}(\theta)^{\dagger},
\label{eq: text ham theta}
\end{equation}
where $U_{\text{g.f.}}^{h; x>0}(\theta)$ is the unitary operator $U_{\text{g.f.}}^h(\theta)$ restricted to the right half of the plane, i.e.,
\begin{equation}
U_{\text{g.f.}}^{h; x>0}(\theta) \coloneq \prod_{i \in P_{> 0}} \left.u_h^{(i)}(\theta)\right|_{\text{g.f.}}.
\end{equation}
Here, $P_{> 0}$ is the set of the plaquettes whose $x$-coordinates are greater than 0, and $\left.u_h^{(i)}(\theta)\right|_{\text{g.f.}}$ is the local unitary operator defined in \eqref{eq: Uh theta gf}.
We note that the textured Hamiltonian \eqref{eq: text ham theta} satisfies all the properties listed in the previous paragraph.
In particular, this Hamiltonian preserves the symmetry $\mathcal{C}(G; H)$ for all $\theta \in [0, 2\pi]$ because it is obtained by gauging the subgroup symmetry $H$ in the $G$-symmetric textured Hamiltonian~\eqref{eq: G/K textured theta}.\footnote{The subregion $P^{\prime}$ in \eqref{eq: G/K textured theta} is chosen to be $P_{>0}$ here.}
The ground state of the Hamiltonian~\eqref{eq: text ham theta} is obtained by applying $U_{\text{g.f.}}^{h; x>0}(\theta)$ to the ground state of $\hat{H}_{\text{g.f.}}$, that is,
\begin{equation}
U_{\text{g.f.}}^{h; x>0}(\theta) \ket{\text{cluster}(G; H; K)}.
\label{eq: textured GS theta}
\end{equation}
The generalized Thouless pump can then be understood as the gapped excitation localized around $x =0$ in the ground state at $\theta = 2\pi$.
In the remainder of this subsection, we will show that this gapped excitation is given by the non-degenerate self-interface mode~\eqref{eq: product interface}.

\vspace*{\baselineskip}
\noindent{\bf Tensor network representation.}
To explicitly write down the gapped excitation localized around $x=0$, we use the following tensor network representation of the ground state~\eqref{eq: textured GS theta}:
\begin{equation}
U_{\text{g.f.}}^{h; x>0}(\theta) \ket{\text{cluster}(G; H; K)} = \adjincludegraphics[scale=1,trim={10pt 10pt 10pt 10pt},valign = c]{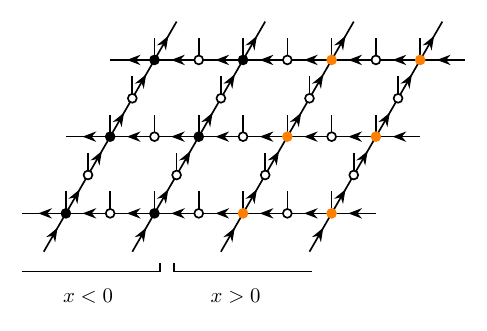}.
\label{eq: textured GS theta TN}
\end{equation}
Here, the local tensors on the right-hand side are defined by \eqref{eq: local tensors} and \eqref{eq: local tensor orange}.
As is clear from the above tensor network representation, the left half of the plane is occupied by the ground state of $\hat{H}_{\text{g.f.}}$, while the right half of the plane is occupied by the ground state of $\hat{H}_{\text{g.f.}}^h(\theta)$.

To identify the pumped excitation, we look into the ground state~\eqref{eq: textured GS theta TN} at $\theta = 2\pi$.
At $\theta = 2\pi$, the local tensor at each vertex on the right half-plane reduces to
\begin{equation}
\adjincludegraphics[scale=1,trim={10pt 10pt 10pt 10pt},valign = c]{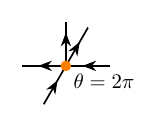} \;=\;
\adjincludegraphics[scale=1,trim={10pt 10pt 10pt 10pt},valign = c]{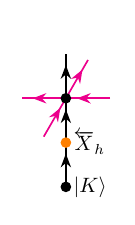} \;=\;
\adjincludegraphics[scale=1,trim={10pt 10pt 10pt 10pt},valign = c]{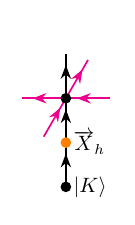} \;=\;
\adjincludegraphics[scale=1,trim={10pt 10pt 10pt 10pt},valign = c]{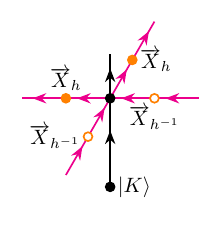} \;=\;
\adjincludegraphics[scale=1,trim={10pt 10pt 10pt 10pt},valign = c]{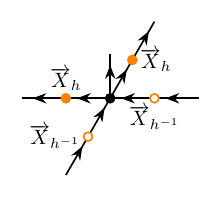}.
\end{equation}
In the second equality, we used $\overleftarrow{X}_h \ket{K} = \overrightarrow{X}_h \ket{K}$, which follows from the condition $K = hKh^{-1}$.
On the other hand, the local tensor on each link (i.e., the multiplication tensor) commutes with the left multiplication $\overrightarrow{X}_h$ acting on the virtual bonds.
Therefore, the PEPS tensor on each unit cell at $\theta = 2\pi$ is related to the one at $\theta = 0$ by the following gauge transformation:
\begin{equation}
 \adjincludegraphics[scale=1,trim={10pt 10pt 10pt 10pt},valign = c]{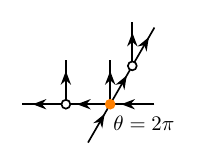} \;\;=\;\;
\adjincludegraphics[scale=1,trim={10pt 10pt 10pt 10pt},valign = c]{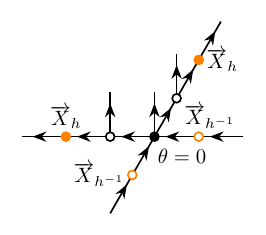}\,.
\end{equation}
By plugging this into \eqref{eq: textured GS theta TN}, we find that changing $\theta$ adiabatically from $0$ to $2\pi$ pumps a gapped excitation along the $y$-axis as follows:
\begin{equation}
\adjincludegraphics[scale=1,trim={10pt 10pt 20pt 10pt},valign = c]{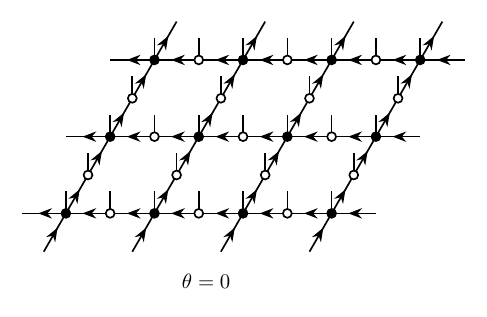}
\longrightarrow\;\;
\adjincludegraphics[scale=1,trim={10pt 10pt 10pt 10pt},valign = c]{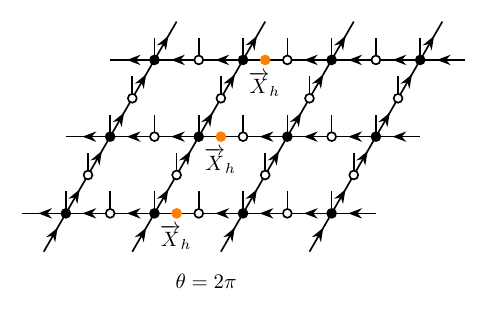}
\nonumber
\end{equation}
The pumped excitation is exactly the non-degenerate gapped self-interface mode $\ket{\psi(h^{-1})}$ defined in \eqref{eq: product interface}.

\vspace*{\baselineskip}
\noindent{\bf Category-theoretical interpretation.}
Let us comment on the relation to the category-theoretical classification of $S^1$-parameterized families.
From the category-theoretical viewpoint, it is expected that $S^1$-parameterized families of 2+1d gapped systems in an SPT phase are classified by monoidal natural automorphisms of the corresponding fiber 2-functor \cite{Inamura:2024jke}.
In particular, for the $\mathcal{C}(G; H)$-symmetric SPT phase labeled by $K$, such monoidal natural automorphisms are in one-to-one correspondence with fiber functors of the strip 2-algebra $\mathcal{C}_{K, K}$ due to the equivalence~\eqref{eq: Morita dual}.
Physically, the fiber functors of $\mathcal{C}_{K, K}$ should be interpreted as the gapped exciations pumped by the adiabatic evolution.
As shown in~\eqref{eq: set of fiber functors}, the isomorphism classes of the fiber functors of $\mathcal{C}_{K, K}$ are classified by
\begin{equation}
\{(h, [\xi]) \mid h \in H \text{ s.t. } K= hKh^{-1}, ~ [\xi] \in H^2(K, \mathrm{U}(1))\}.
\label{eq: S1 family classification}
\end{equation}
On the other hand, in the lattice construction discussed above, we realized the generalized Thouless pump labeled by any $h \in H$ satisfying $K = hKh^{-1}$.
Since the pumped interface mode~\eqref{eq: product interface} is a product state, the corresponding fiber functor labeled by $(h, [\xi])$ should have trivial $[\xi]$.
Thus, we conclude that among the $S^1$-parameterized families classified by~\eqref{eq: S1 family classification}, we constructed all those with trivial $[\xi]$.

\subsection{Example: $S^1$-parameterized families of the $G$-cluster states}
\label{sec: Example: parameterized families of the G-cluster states}
As a simple example, let us consider $S^1$-parameterized families of $2\Rep(G_0) \boxtimes 2\Vect_{G_0}$-symmetric models in the same phase as the $G_{0}$-cluster model~\eqref{eq: G-cluster ham}.
Recall that the $G_{0}$-cluster model is a special case of the generalized cluster model with
\begin{equation}
G=G_{0} \times G_{0}, \quad
H=G_{0}^{\text{left}}, \quad
K=G_0^{\text{diag}}, \quad
S_{H\backslash G}=G_{0}^{\text{right}}.
\end{equation}

\vspace*{\baselineskip}
\noindent{\bf $S^1$-parameterized families.}
As shown in Section~\ref{sec: S1-parameterized families}, we can construct an $S^1$-parameterized family of generalized cluster states for each $h \in H$ satisfying $K = hKh^{-1}$.
In the case of the $G_0$-cluster model, since $H = G_0^{\text{left}}$ and $K = G_0^{\text{diag}}$, the condition $K = hKh^{-1}$ reduces to
\begin{equation}
h \in \{(g, e) \mid g \in Z(G_0)\} = Z(G_0^{\text{left}}),
\end{equation}
where $Z(G_0)$ is the center of $G_0$.
For each $h \in Z(G_0^{\text{left}})$, the Hamiltonian of an $S^1$-parameterized family is obtained by conjugating the $G_0$-cluster Hamilonian~\eqref{eq: G-cluster ham} with the following unitary operator parameterized by $\theta \in [0, 2\pi]$:
\begin{equation}
U_{\text{g.f.}}^h(\theta) \coloneq \prod_{i \in P} \left.u_h^{(i)}(\theta)\right|_{\text{g.f.}}, \qquad
\left.u_h^{(i)}(\theta)\right|_{\text{g.f.}} \coloneq \exp\left(\frac{\theta}{2\pi} \log L_{h^{-1}}^{(\partial i)}\right).
\label{eq: Ugf h theta G-cluster}
\end{equation}
Here, we recall that $L_{h^{-1}}^{(\partial i)}$ is the loop operator defined in \eqref{eq: loop op}, which acts only on the gauge fields around the plaquette $i$.
We choose the branch of the logarithm as 
\begin{equation}
       \log L_{h^{-1}}^{(\partial i)} = \sum_{s = 0}^{\abs{G_{0}}-1} i \frac{2\pi}{|G_{0}|} P_s[L_{h^{-1}}^{(\partial i)}],
\end{equation}
where $P_s[L_{h^{-1}}^{(\partial i)}]$ is the projection operator onto the eigenspace of $L_{h^{-1}}^{(\partial i)}$ with eigenvalue $e^{i\frac{2\pi s}{|G_0|}}$.
The above operator \eqref{eq: Ugf h theta G-cluster} is obtained from the general formula \eqref{eq: Uh theta gf} by using the fact that the dual of the domain wall creation operator on each plaquette $i$ is now given by
\begin{equation}
\left.\overleftarrow{X}_h^{(i)}\right|_{\text{g.f.}} = L_{h^{-1}}^{(\partial i)}.
\end{equation}
We note that $\left.u_h^{(i)}(\theta)\right|_{\text{g.f.}}$ in \eqref{eq: Ugf h theta G-cluster} interpolates between the identity operator and the loop operator $L_{h^{-1}}^{(\partial i)}$.
In particular, on a square lattice with periodic boundary conditions, the unitary operator~\eqref{eq: Ugf h theta G-cluster} is $2\pi$-periodic because\footnote{The $2\pi$-periodicity of $U_{\text{g.f.}}^h(\theta)$ is a special property of the $G_0$-cluster model: in general, the unitary operator $U_{\text{g.f.}}^h(\theta)$ defined by \eqref{eq: Uh theta gf} is not necessarily $2\pi$-periodic. Nevertheless, the conjugated Hamiltonian \eqref{eq: Hh gf theta} is $2\pi$-periodic.}
\begin{equation}
U_{\text{g.f.}}^h(2\pi) = \prod_{i \in P} L_{h^{-1}}^{(\partial i)} = \prod_{[ij] \in E} \overrightarrow{X}_{h^{-1}}^{(ij)} \overleftarrow{X}_h^{(ij)} = \id,
\end{equation}
where the last equality follows from the fact that $h$ is in the center of $G_0^{\text{left}}$.
Therefore, the $G_0$-cluster Hamiltonian conjugated by $U_{\text{g.f.}}^h(\theta)$ is also $2\pi$-periodic, which means that the model constructed in this way is parameterized by a circle $S^1$.

The unitary operator $U_{\text{g.f.}}^h(\theta)$ commutes with the $2\Rep(G_0) \boxtimes 2\Vect_{G_0}$ symmetry, which implies that the $S^1$-family obtained by the conjugation with $U_{\text{g.f.}}^h(\theta)$ preserves the symmetry.
To see this, we first recall that each local unitary $\left.u_h^{(i)}(\theta)\right|_{\text{g.f.}}$ is a polynomial of $L_{h^{-1}}^{(\partial i)}$ because $P_s[L_{h^{-1}}^{(\partial i)}]$ can be written as a polynomical
\begin{equation}
P_s[L_{h^{-1}}^{(\partial i)}] = \frac{1}{|G_0|} \sum_{n=0}^{\abs{G_0}-1} e^{-i\frac{2\pi s}{\abs{G_0}}n} (L_{h^{-1}}^{(\partial i)})^n.
\end{equation}
Thus, it suffices to show that $L_{h^{-1}}^{(\partial i)}$ commutes with the $2\Rep(G_0) \boxtimes 2\Vect_{G_0}$ symmetry.
The commutativity of $L_{h^{-1}}^{(\partial i)}$ and the $2\Vect_{G_0}$ symmetry is obvious because $L_{h^{-1}}^{(\partial i)}$ acts only on the gauge fields on the edges, while the $2\Vect_{G_0}$ symmetry acts only on the matter fields on the plaquettes.
Furthermore, since $L_{h^{-1}}^{(\partial i)}$ does not change the holonomy of the gauge fields (up to conjugation), it also commutes with the 1-form symmetry $\Rep(G_0)$ and hence commutes with its condensation completion $2\Rep(G_0)$.

\vspace*{\baselineskip}
\noindent{\bf Generalized Thouless pumps.}
Let us now discuss the generalized Thouless pump associated with the above $S^1$-parameterized family.
To this end, we consider the $\theta$-dependent textured Hamiltonian obtained by conjugating the $G_0$-cluster Hamiltonian~\eqref{eq: G-cluster ham} with the truncated unitary operator
\begin{equation}
U_{\text{g.f.}}^{h; x>0} \coloneq \prod_{i \in P_{>0}} \left.u_h^{(i)}(\theta)\right|_{\text{g.f.}}.
\label{eq: truncated Ugf G-cluster}
\end{equation}
Here, $P_{>0}$ is the set of plaquettes on the right half of the plane.\footnote{Here, for simplicity, we consider the model on an infinite plane.}
We note that the $\theta$-dependent textured Hamiltonian preserves the $2\Rep(G_0) \boxtimes 2\Vect_{G_0}$ symmetry for all $\theta \in [0, 2\pi]$.
By construction, the ground state of the textured Hamiltonian is given by
\begin{equation}
U_{\text{g.f.}}^{h; x>0}(\theta) \ket{\text{$G_0$-cluster}}.
\label{eq: textured G-cluster}
\end{equation}
At $\theta = 2\pi$, the truncated unitary operator~\eqref{eq: truncated Ugf G-cluster} reduces to a line operator along the $y$-axis, i.e.,
\begin{equation}
U_{\text{g.f.}}^{h; x>0}(2\pi) = \prod_{i \in P_{>0}} L_{h^{-1}}^{(\partial i)} = \prod_{e \in E|_{x=0}} \overleftarrow{X}_h^{(e)},
\end{equation}
where $E|_{x=0}$ is the set of edges on the line with $x=0$.
Thus, the ground state~\eqref{eq: textured G-cluster} at $\theta = 2\pi$ becomes
\begin{equation}
U_{\text{g.f.}}^{h; x>0}(2\pi) \ket{\text{$G_0$-cluster}} = \adjincludegraphics[scale=1,trim={10pt 10pt 10pt 10pt},valign = c]{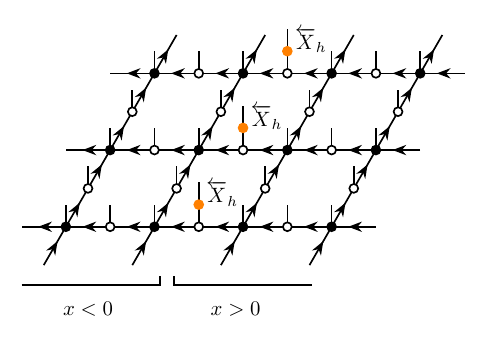},
\label{eq: G-cluster pump}
\end{equation}
where the black and white dots represent the local tensors~\eqref{eq: local tensor G-cluster} of the $G_0$-cluster state.
Since $h$ is in the center of $G_0^{\text{left}}$, there is no difference between the left and right multiplications of $h$, that is, $\overleftarrow{X}_h^{(e)} = \overrightarrow{X}_h^{(e)}$ for any edge $e \in E$.
Accordingly, $\overleftarrow{X}_h$ acting on the physical legs in \eqref{eq: G-cluster pump} can freely pass through the multiplication tensor (i.e., the white dot) as follows:
\begin{equation}
\adjincludegraphics[scale=1,trim={10pt 10pt 10pt 10pt},valign = c]{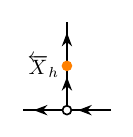} \;\;=\;\;
\adjincludegraphics[scale=1,trim={10pt 10pt 10pt 10pt},valign = c]{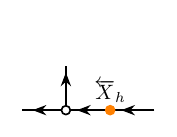} \;\;=\;\;
\adjincludegraphics[scale=1,trim={10pt 10pt 10pt 10pt},valign = c]{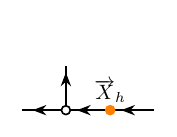} \;\;=\;\;
\adjincludegraphics[scale=1,trim={10pt 10pt 10pt 10pt},valign = c]{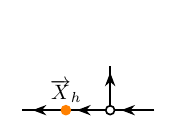}.
\end{equation}
Plugging this into \eqref{eq: G-cluster pump}, we find that the gapped excitation pumped by an adiabatic cycle is represented by the left multiplication of $h$ acting on the virtual degrees of freedom at the interface.
This result agrees with the general result obtained in Section~\ref{sec: Pumping of gapped interface modes}.

\section{Example 1: $2\Rep(G_0) \boxtimes 2\Vect_{G_0}$ SPT phases}
\label{sec: Gx2RepG SPT phases}
As our first examples, we consider 2+1d SPT phases with $2\Rep(G_0) \boxtimes 2\Vect_{G_0}$ symmetry, where $G_0$ is an arbitrary finite group.
We note that $2\Rep(G_0) \boxtimes 2\Vect_{G_0}$ is a group-theoretical fusion 2-category because
\begin{equation}
2\Rep(G_0) \boxtimes 2\Vect_{G_0} \cong \mathcal{C}(G_0 \times G_0; G_0^{\text{left}}),
\label{eq: 2Rep(G)x2VecG}
\end{equation}
where $G_0^{\text{left}} := \{(g, e) \mid g \in G_0\}$.
The examples that we will discuss below include the SPT phase realized by the $G_0$-cluster state~\eqref{eq: G cluster}.

\subsection{Classification}
\label{sec: Classification Gx2Rep(G)}
We first provide the classification of SPT phases with $2\Rep(G_0) \boxtimes 2\Vect_{G_0}$ symmetry.
These SPT phases are in one-to-one correspondence with fiber 2-functors of $2\Rep(G_0) \boxtimes 2\Vect_{G_0}$.
Due to the monoidal equivalence~\eqref{eq: 2Rep(G)x2VecG}, fiber 2-functors of $2\Rep(G_0) \boxtimes 2\Vect_{G_0}$ are labeled by pairs $(K, \lambda)$, where $K$ is a complement of $G_0^{\text{left}}$ in $G_0 \times G_0$, and $\lambda \in Z^3(K, \mathrm{U}(1))$ is a 3-cocycle on $K$.
See Section~\ref{sec: Fiber 2-functors} for the classification of fiber 2-functors of a general group-theoretical fusion 2-category.
By definition, the complement $K$ satisfies
\begin{equation}
G_0^{\text{left}}K = G_0 \times G_0, \qquad
G_0^{\text{left}} \cap K = \{(e, e)\}.
\label{eq: complement of Gleft}
\end{equation}
The first equality implies that $K$ must contain an element $(\bullet, g)$ for any $g \in G_0$, where $\bullet$ is an element of $G_0$.
Furthermore, the second equality implies that the element of the form $(\bullet, g) \in K$ must be unique for every $g \in G_0$.\footnote{If $K$ contains two elements $(g_1, g)$ and $(g_2, g)$ with $g_1 \neq g_2$, it also contains $(g_1, g)^{-1} \cdot (g_2, g) = (g_1^{-1}g_2, e)$ because $K$ is a group. This implies that $G_0^{\text{left}} \cap K$ contains $(g_1^{-1}g_2, e) \neq (e, e)$, which contradicts the second equality of \eqref{eq: complement of Gleft}.}
Therefore, the complement $K$ must be of the form
\begin{equation}
K = G_0^{\text{diag}}(f) := \{(f(g), g) \mid g \in G_0\},
\label{eq: Gright x}
\end{equation}
where $f: G_0 \to G_0$ is a map.
Moreover, since $K$ is a subgroup, $f$ must satisfy
\begin{equation}
(f(g), g) \cdot (f(h), h) \in K, \qquad \forall g, h \in G_0.
\end{equation}
This condition implies that $f: G_0 \to G_0$ must be a group homomorphism.
Conversely, it is clear that $G_0^{\text{diag}}(f)$ is a complement of $G_0^{\text{left}}$ if $f$ is a group homomorphism of $G_0$.
Therefore, a fiber 2-functor of $2\Rep(G_0) \boxtimes 2\Vect_{G_0}$ is labeled by the pair $(f, \lambda)$, where $f$ is a group homomorphism of $G_0$, and $\lambda$ is a 3-cocycle on $G_0^{\text{diag}}(f)$.
In what follows, we will focus on fiber 2-functors with trivial $\lambda$.
In this case, each fiber 2-functor is labeled only by a group homomorphism $f: G_0 \to G_0$.

The fiber 2-functors labeled by $f$ and $f^{\prime}$ are isomorphic to each other if and only if there exists an element $(g_{\ell}, g_r) \in G_0 \times G_0$ such that
\begin{equation}
G_0^{\text{diag}}(f^{\prime}) = (g_{\ell}, g_r) G_0^{\text{diag}}(f) (g_{\ell}, g_r)^{-1}.
\end{equation}
Since $G_0^{\text{diag}}(f)$ is a complement of $G_0^{\text{left}}$, we can take $(g_{\ell}, g_r)$ to be an element of $G_0^{\text{left}}$ without loss of generality.
Namely, we can choose $g_r = e$.
In this case, the above condition reduces to
\begin{equation}
\forall g \in G_0,~ \exists h \in G_0 ~ \text{ s.t. } ~ (f^{\prime}(g), g) = (g_{\ell}f(h)g_{\ell}^{-1}, h).
\end{equation}
This condition is satisfied if and only if 
\begin{equation}
f^{\prime} = \text{Ad}_{g_{\ell}} \circ f,
\label{eq: adjoint}
\end{equation}
where $\text{Ad}_{g_{\ell}}: G_0 \to G_0$ denotes the adjoint action of $g_{\ell} \in G_0$.
Thus, the fiber 2-functors labeled by $f$ and $f^{\prime}$ are isomorphic to each other if and only if there exists $g_{\ell} \in G_0$ that satisfies \eqref{eq: adjoint}.
This shows that the isomorphism classes of fiber 2-functors with trivial $\lambda$ are classified by
\begin{equation}
\End(G_0) / \text{Inn}(G_0),
\end{equation}
where $\End(G_0)$ denotes the set of group endomorphisms of $G_0$, and $\text{Inn}(G_0)$ is the group of inner automorphisms of $G_0$.

\subsection{Generalized cluster states}
\label{sec: Generalized cluster states Gx2Rep(G)}
The SPT phases labeled by fiber 2-functors with trivial $\lambda$ are realized in the generalized cluster models defined in Section~\ref{sec: 2+1d generalized cluster states with non-invertible symmetries}.
In the case of $2\Rep(G_0) \boxtimes 2\Vect_{G_0}$ symmetry, the input data of the model are 
\begin{equation}
G = G_0 \times G_0, \quad
H = G_0^{\text{left}}, \quad
K = G_0^{\text{diag}}(f).
\label{eq: G cluster f input}
\end{equation}
The generalized cluster model for the above input data is obtained by gauging the subgroup symmetry $H$ in a $G$-symmetric model that spontaneously breaks the symmetry $G$ down to $K$.
In what follows, we will write down the models both before and after the gauging.

\vspace*{\baselineskip}
\noindent{\bf $G/K$-SSB model.}
The state space of the $G$-symmetric model before the gauging is
\begin{equation}
\mathcal{H} = \bigotimes_{i \in P} \mathbb{C}^{|G_0 \times G_0|},
\end{equation}
where $P$ is the set of all plaquettes on a square lattice.
The on-site state space on each plaquette $i \in P$ is spanned by $\{\ket{g_i, g_i^{\prime}}_i \mid g_i, g^{\prime}_i \in G_0\}$.
The Hamiltonian of the model is 
\begin{equation}
H = -\sum_{i \in P} \hat{\mathsf{h}}_i - \sum_{[ij] \in E} \hat{\mathsf{h}}_{ij},
\label{eq: G0xG0 ham}
\end{equation}
where $\hat{\mathsf{h}}_i$ and $\hat{\mathsf{h}}_{ij}$ are defined by
\begin{align}
\hat{\mathsf{h}}_i \ket{g_i, g_i^{\prime}}_i &= \frac{1}{|G_0|} \sum_{g \in G_0} \ket{g_i f(g), g_i^{\prime}g}, \\
\hat{\mathsf{h}}_{ij} \ket{g_i, g_i^{\prime}}_i \otimes \ket{g_j, g_j^{\prime}}_j &= \delta_{g_i^{-1}g_j, f(g_i^{\prime})^{-1}f(g_j^{\prime})} \ket{g_i, g_i^{\prime}}_i \otimes \ket{g_j, g_j^{\prime}}_j.
\end{align}
The symmetry of this model is generated by
\begin{equation}
U_g^{(L)} = \bigotimes_{i \in P} \overrightarrow{X}_g^{(i; L)}, \qquad
U_g^{(R)} = \bigotimes_{i \in P} \overrightarrow{X}_g^{(i; R)}, \qquad \forall g \in G_0,
\label{eq: G0xG0 sym op}
\end{equation}
where the on-site operators $\overrightarrow{X}_g^{(i; L)}$ and $\overrightarrow{X}_g^{(i; R)}$ are defined by
\begin{equation}
\overrightarrow{X}_g^{(i; L)} \ket{g_i, g_i^{\prime}}_i = \ket{gg_i, g_i^{\prime}}, \qquad
\overrightarrow{X}_g^{(i; R)} \ket{g_i, g_i^{\prime}}_i = \ket{g_i, gg_i^{\prime}}.
\end{equation} 
The Hamiltonian~\eqref{eq: G0xG0 ham} has $|G_0|$-fold degenerate ground states
\begin{equation}
\ket{\text{GS}; g} = \bigotimes_{i \in P} \left(\frac{1}{\sqrt{|G_0|}} \sum_{g_i \in G_0} \ket{gf(g_i), g_i}_i\right), \qquad \forall g \in G.
\end{equation}
The symmetry operators~\eqref{eq: G0xG0 sym op} acts on these ground states as
\begin{equation}
U_g^{(L)} \ket{\text{GS}; g^{\prime}} = \ket{\text{GS}; gg^{\prime}}, \qquad
U_g^{(R)} \ket{\text{GS}; g^{\prime}} = \ket{\text{GS}; g^{\prime}f(g)^{-1}}.
\end{equation}
We note that $\ket{\text{GS}; g}$ is invariant under the action of $U_{gf(g_r)g^{-1}}^{(L)}U_{g_r}^{(R)}$ for all $g_r \in G_0$.
Thus, the unbroken symmetry of $\ket{\text{GS}; g}$ is 
\begin{equation}
\{(gf(g_r)g^{-1}, g_r) \mid g_r \in G_0\} = (g, e) \cdot G_0^{\text{diag}}(f) \cdot (g, e,)^{-1}.
\end{equation}
In particular, the unbroken symmetry of $\ket{\text{GS}; e}$ is $K = G_0^{\text{diag}}(f)$.

\vspace*{\baselineskip}
\noindent{\bf Generalized cluster model.}
The generalized cluster model with $2\Rep(G_0) \boxtimes 2\Vect_{G_0}$ symmetry is obtained by gauging the subgroup symmetry $H = G_0^{\text{left}}$ in the above model.
To describe the model, we choose a gauge so that the matter fields after the gauge fixing take values in
\begin{equation}
S_{H \backslash G} = G_0^{\text{right}} := \{(e, g) \mid g \in G_0\}.
\end{equation}
The state space of the gauged model after the gauge fixing is given by
\begin{equation}
\mathcal{H}_{\text{g.f.}} = \left(\bigotimes_{i \in P} \mathbb{C}^{|G_0^{\text{right}}|}\right) \otimes \hat{\pi}_{\text{flat}}\left(\bigotimes_{[ij] \in E} \mathbb{C}^{|G_0^{\text{left}}|}\right),
\label{eq: state space Gx2Rep(G)}
\end{equation}
where $\hat{\pi}_{\text{flat}}$ is the projector that imposes the flatness condition on the gauge fields.
The matter field on plaquette $i \in P$ and the gauge field on edge $[ij] \in E$ are labeled by $(e, g_i) \in G_0^{\text{right}}$ and $(g_{ij}, e) \in G_0^{\text{left}}$, respectively.
The corresponding states are simply denoted by $\ket{g_i}_i$ and $\ket{g_{ij}}_{ij}$.
The basis states of the above state space will be denoted by
\begin{equation}
\ket{\{g_i; g_{ij}\}} := \left(\bigotimes_{i \in P} \ket{g_i}_i\right) \otimes \hat{\pi}_{\text{flat}} \left(\bigotimes_{[ij] \in E} \ket{g_{ij}}_{ij}\right).
\label{eq: basis 2RepG x G}
\end{equation}
The Hamiltonian of the model is
\begin{equation}\label{eq: G cluster f ham}
H_{\text{g.f.}} = - \sum_{i \in P} \hat{\mathsf{h}}_i^{\text{g.f.}} - \sum_{[ij] \in E} \hat{\mathsf{h}}_{ij}^{\text{g.f.}},
\end{equation}
where $\hat{\mathsf{h}}_i^{\text{g.f.}}$ and $\hat{\mathsf{h}}_{ij}^{\text{g.f.}}$ are defined by
\begin{align}
\hat{\mathsf{h}}_i^{\text{g.f.}} \Ket{\adjincludegraphics[valign = c, trim={10, 10, 10, 10}]{tikz/out/G_cluster_ham1.pdf}} &= \frac{1}{|G_0|} \sum_{g \in G_0} \Ket{\adjincludegraphics[valign = c, trim={10, 10, 10, 10}]{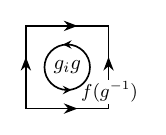}}, 
\label{eq: G cluster f ham1}\\
\hat{\mathsf{h}}_{ij}^{\text{g.f.}} \ket{g_i}_i \otimes \ket{g_{ij}}_{ij} \otimes \ket{g_j}_j &= \delta_{g_{ij}, f(g_i^{-1}g_j)} \ket{g_i}_i \otimes \ket{g_{ij}}_{ij} \otimes \ket{g_j}_j.
\label{eq: G cluster f ham2}
\end{align}
The loop on the right-hand side of \eqref{eq: G cluster f ham1} represents the loop operator acting on the boundary of the plaquette $i \in P$.\footnote{See \eqref{eq: loop op} for the general definition of a loop operator.}
The above Hamiltonian has a unique ground state on a square lattice with periodic boundary conditions.
The ground state can be written explicitly as
\begin{equation}
\ket{\text{$G_0$-cluster}(f)} := \frac{1}{|G_0|^{|P|/2}} \sum_{\{g_i \in G_0\}} \ket{\{g_i; f(g_i^{-1}g_j)\}},
\label{eq: G-cluster x}
\end{equation}
where $|P|$ is the number of plaquettes.
A tensor network representation of this ground state is
\begin{equation}
\ket{\text{$G_0$-cluster}(f)} = \adjincludegraphics[valign = c, trim={10, 10, 10, 10}]{tikz/out/cluster_PEPS.pdf},
\label{eq: G cluster f PEPS}
\end{equation}
where the virtual bonds take values in $G_0$, and the local tensors represented by the black and white dots are defined by
\begin{equation}
\adjincludegraphics[valign = c, trim={10, 10, 10, 10}]{tikz/out/copy.pdf} = \frac{1}{\sqrt{|G_0|}} \prod_{i = 1, 2, 3, 4} \delta_{g_i, f(g)}, \qquad
\adjincludegraphics[valign = c, trim={10, 10, 10, 10}]{tikz/out/multiplication.pdf} = \delta_{g, g_1^{-1}g_2}.
\label{eq: G cluster f tensor}
\end{equation}
We note that $\ket{\text{$G_0$-cluster}(f)}$ reduces to the $G_0$-cluster state~\eqref{eq: G-cluster state} when $f = \id$.
On the other hand, $\ket{\text{$G_0$-cluster}(f)}$ reduces to the trivial product state when $f(g) = e$ for all $g \in G_0$.

\subsection{Symmetry operators}
\label{sec: Symmetry operators Gx2Rep(G)}
The generalized cluster model in the previous subsection has $2\Rep(G_0) \boxtimes 2\Vect_{G_0}$ symmetry by construction.
In what follows, we will write down the symmetry operators for this symmetry.
We will also compute the action tensors for the symmetry operators acting on the generalized cluster state~\eqref{eq: G-cluster x}.
See Section~\ref{sec: Non-invertible symmetries of 2+1d generalized cluster states} for the symmetry of the generalized cluster model obtained from more general input data.

\vspace*{\baselineskip}
\noindent{\bf 1-form symmetry.}
The 1-form symmetry operators of the model correspond to 1-endomorphisms of the unit object $I \in 2\Rep(G_0) \boxtimes 2\Vect_{G_0}$.
These 1-morphisms form a symmetric fusion 1-category
\begin{equation}
\End_{2\Rep(G_0) \boxtimes 2\Vect_{G_0}}(I) \cong \Rep(G_0).
\end{equation}
The symmetry operator labeled by $\rho \in \Rep(G_0)$ is the Wilson line $W_{\rho}$.
The action of $W_{\rho}$ supported on an oriented loop $\gamma$ on the dual lattice is given by
\begin{equation}
W_{\rho}(\gamma) \ket{\{g_i; g_{ij}\}} = \text{tr}_{V_{\rho}}(\rho(\text{hol}_{\gamma}(\{g_{ij}\}))) \ket{\{g_i; g_{ij}\}},
\label{eq: Wilson line 2RepG x G}
\end{equation}
where $V_{\rho}$ is the representation space of $\rho$, and $\text{hol}_{\gamma}(\{g_{ij}\})$ is the holonomy of the gauge fields $\{g_{ij} \mid [ij] \in E\}$ along the loop $\gamma$.

\vspace{\baselineskip}
\noindent{\bf 0-form symmetry.}
The 0-form symmetry operators are labeled by objects of $2\Rep(G_0) \boxtimes 2\Vect_{G_0}$.
Any 0-form symmetry operator, up to condensation, can be written as
\begin{equation}
\mathsf{C} U_g, \qquad \forall g \in G_0,
\label{eq: CUg}
\end{equation}
where $\mathsf{C} := \mathsf{D}_{G_0^{\text{left}}} \overline{\mathsf{D}}_{G_0^{\text{left}}}$ is the symmetry operator corresponding to a simple object of $2\Rep(G_0)$, while $U_g$ is the symmetry operator corresponding to a simple object of $2\Vect_{G_0}$.
Here, $\mathsf{D}_{G_0^{\text{left}}}$ and $\overline{\mathsf{D}}_{G_0^{\text{left}}}$ denote the gauging and ungauging operators defined by
\begin{equation}
\mathsf{D}_{G_0^{\text{left}}} \ket{\{g_i, g_i^{\prime}\}} = \ket{\{g_i^{\prime}; g_i^{-1}g_j\}}, \qquad
\overline{\mathsf{D}}_{G_0^{\text{left}}} = \mathsf{D}_{G_0^{\text{left}}}^{\dagger}.
\end{equation}
The actions of $\mathsf{C}$ and $U_g$ are then given by
\begin{equation}
\mathsf{C} \ket{\{g_i; g_{ij}\}} = |G_0| \delta_{\text{hol}}(\{g_{ij}\}) \ket{\{g_i; g_{ij}\}}, \qquad
U_g \ket{\{g_i; g_{ij}\}} = \ket{\{gg_i; g_{ij}\}},
\label{eq: C and Ug}
\end{equation}
where $\delta_{\text{hol}}(\{g_{ij}\})$ is equal to $1$ if the holonomy of the gauge field is trivial and $0$ otherwise.
We note that $\mathsf{C}U_g$ can be also written as $\mathsf{D}_{G_0^{\text{left}}} U_g^{(R)} \overline{\mathsf{D}}_{G_0^{\text{left}}}$, where $U_g^{(R)}$ is the symmetry operator~\eqref{eq: G0xG0 sym op} of the ungauged model.

\vspace*{\baselineskip}
\noindent{\bf Tensor network representation.}
The symmetry operator~\eqref{eq: CUg} can be represented by a tensor network
\begin{equation}
\mathsf{C}U_g = \adjincludegraphics[valign = c, trim={10, 10, 10, 10}]{tikz/out/symmetry_PEPO_cluster.pdf},
\label{eq: CUg PEPO}
\end{equation}
where the virtual bonds take values in $G_0$, and the local tensors represented by the filled and unfilled green dots are defined by
\begin{equation}
\adjincludegraphics[valign = c, trim={10, 10, 10, 10}]{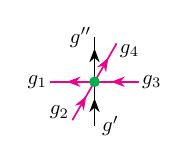} = \delta_{g^{\prime \prime}, gg^{\prime}} \prod_{i, j = 1, 2, 3, 4} \delta_{g_i, g_j}, \qquad
\adjincludegraphics[valign = c, trim={10, 10, 10, 10}]{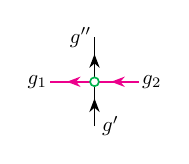} = \delta_{g^{\prime}, g_1^{-1}g_2} \delta_{g^{\prime \prime}, g^{\prime}}.
\label{eq: CUg local tensors}
\end{equation}
The other 0-form symmetry operators can be obtained by condensing line operators on \eqref{eq: CUg PEPO} as in \eqref{eq: sym op condensation}.
We refer the reader to Section~\ref{sec: Tensor network representations of symmetry operators} for more details on the symmetry operators.

\vspace*{\baselineskip}
\noindent{\bf Action tensors.}
The action of the symmetry operator $\mathsf{C}U_g$ on the local tensors of the generalized cluster state $\ket{\text{$G_0$-cluster}(f)}$ can be computed as
\begin{equation}
\adjincludegraphics[valign = c, trim={10, 10, 10, 10}]{tikz/out/local_action1.pdf} ~ = \sum_{h \in G_0} \adjincludegraphics[valign = c, trim={10, 10, 10, 10}]{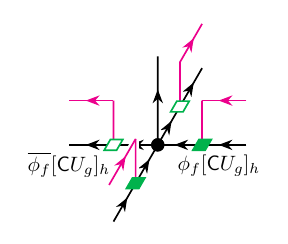}, \qquad
\adjincludegraphics[valign = c, trim={10, 10, 10, 10}]{tikz/out/local_action3.pdf} ~ = \sum_{h \in G_0} \adjincludegraphics[valign = c, trim={10, 10, 10, 10}]{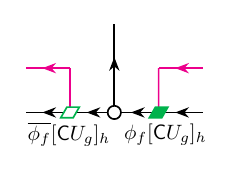},
\end{equation}
where the action tensors $\phi_f[\mathsf{C}U_g]_h$ and $\overline{\phi_f}[\mathsf{C}U_g]_h$ are given by
\begin{equation}
\adjincludegraphics[valign = c, trim={10, 10, 10, 10}]{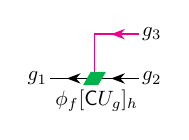} = \adjincludegraphics[valign = c, trim={10, 10, 10, 10}]{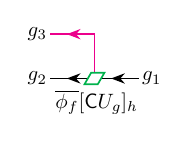} = \delta_{g_1g_3^{-1}, h} \delta_{g_2g_3^{-1}, f(g^{-1})h}.
\label{eq: action tensors G cluster f}
\end{equation}
These action tensors are obtained by applying the general formula \eqref{eq: action tensors} to the current example.\footnote{Using the notations in Section~\ref{sec: Non-invertible symmetries of 2+1d generalized cluster states}, we can write $\mathsf{C}U_g = \mathsf{D}[V_{G_0^{\text{left}}}^{(e, g)}]$ and $\phi_f[\mathsf{C}U_g]_h = \phi_{G_0^{\text{diag}}(f)}[V_{G_0^{\text{left}}}^{(e, g)}]_{(h, e)}$.}
One can also check the above equality by a direct computation using~\eqref{eq: G cluster f tensor} and \eqref{eq: CUg local tensors}.

\subsection{Interfaces: the case of $G_0 = \mathbb{Z}_2$}
\label{sec: Interfaces Gx2Rep(G)}
Let us consider the interfaces of the generalized cluster states with $2\Rep(G_0) \boxtimes 2\Vect_{G_0}$ symmetry.
For simplicity, we will focus on the case of $G_0 = \mathbb{Z}_2$.
In this case, the symmetry category is
\begin{equation}
2\Rep(\mathbb{Z}_2) \boxtimes 2\Vect_{\mathbb{Z}_2},
\end{equation}
i.e., the product of a $\mathbb{Z}_2$ 0-form symmetry and (the condensation completion of) a $\mathbb{Z}_2$ 1-form symmetry.
The generalized cluster states with this symmetry are in one-to-one correspondence with group endomorphisms of $\mathbb{Z}_2$.
The group $\mathbb{Z}_2$ has only two group endomorphisms $f^{(0)}$ and $f^{(1)}$, which are defined by
\begin{equation}
f^{(0)}(g) := e, \qquad
f^{(1)}(g) := g, \qquad \forall g \in \mathbb{Z}_2.
\end{equation}
The complements $\mathbb{Z}_2^{\text{right}}(f)$ corresponding to $f = f^{(0)}$ and $f = f^{(1)}$ are given by
\begin{equation}
\mathbb{Z}_2^{\text{right}}(f^{(0)}) := \mathbb{Z}_2^{\text{right}}, \qquad
\mathbb{Z}_2^{\text{right}}(f^{(1)}) := \mathbb{Z}_2^{\text{diag}}.
\label{eq: complements 2RepZ2 x Z2}
\end{equation}
In what follows, we will consider the interface of the generalized cluster states labeled by $f^{(m)}$ and $f^{(m^{\prime})}$ for each pair of $m, m^{\prime} \in \{0, 1\}$.

\vspace*{\baselineskip}
\noindent{\bf The models.}
Let us first write down the generalized cluster models and their ground states for both choices of the group endomorphism $f$.
The state space is independent of $f$ and is given by
\begin{equation}
\mathcal{H}_{\text{g.f.}} = \left(\bigotimes_{i \in P} \mathbb{C}^2\right) \otimes \hat{\pi}_{\text{flat}} \left(\bigotimes_{[ij] \in E} \mathbb{C}^2\right).
\end{equation}
Namely, we have a single qubit on each plaquette and on each edge.
The Hamiltonians for $f = f^{(0)}$ and $f = f^{(1)}$ are given by\footnote{We recall that we choose $S_{H \backslash G}$ to be $G_0^{\text{right}} = \mathbb{Z}_2^{\text{right}}$.}
\begin{align}
H_{\text{g.f.}}^{(0)} &= -\sum_{i \in P} \frac{1}{2}(1+X^{(i)}) - \sum_{[ij] \in E} \frac{1}{2}(1+Z^{(ij)}), \\
H_{\text{g.f.}}^{(1)} &= -\sum_{i \in P} \frac{1}{2}(1+X^{(i)} \prod_{[ij] \in \partial i}X^{(ij)}) - \sum_{[ij] \in E} \frac{1}{2}(1+Z^{(i)}Z^{(ij)}Z^{(j)}),
\end{align}
where $X^{(i)}$ and $Z^{(i)}$ are the Pauli operators acting on the qubit on the plaquette $i \in P$, and $X^{(ij)}$ and $Z^{(ij)}$ are those acting on the qubit on the edge $[ij] \in E$.
We note that $H^{(0)}_{\text{g.f.}}$ is the trivial Hamiltonian, while $H^{(1)}_{\text{g.f.}}$ is the cluster Hamiltonian.
The ground states of these Hamiltonians are
\begin{align}
\ket{\text{$\mathbb{Z}_2$-cluster}(f^{(0)})} := \ket{\text{trivial}} = \left(\bigotimes_{i \in P} \frac{1}{\sqrt{2}}\sum_{g_i \in \mathbb{Z}_2} \ket{g_i}_i\right) \otimes \left(\bigotimes_{[ij] \in E} \ket{e}_{ij}\right) , \\
\ket{\text{$\mathbb{Z}_2$-cluster}(f^{(1)})} := \ket{\text{$\mathbb{Z}_2$-cluster}} = \frac{1}{2^{|P|/2}} \sum_{\{g_i \in \mathbb{Z}_2\}} \ket{\{g_i; g_i^{-1}g_j\}}.
\end{align}

\vspace*{\baselineskip}
\noindent{\bf Category theoretical description of interfaces.}
As discussed in Section \ref{sec: Category theoretical description}, the symmetry at the interface of the generalized cluster states is generally described by the strip 2-algebra~\eqref{eq: strip 2-algebra}.
The strip 2-algebra for the interface of the SPT phases labeled by the complements $K_1$ and $K_2$ is denoted by $\mathcal{C}_{K_1, K_2}$.
In the case of $2\Rep(\mathbb{Z}_2) \boxtimes 2\Vect_{\mathbb{Z}_2}$ SPT phases, $K_1$ and $K_2$ are either $\mathbb{Z}_2^{\text{right}}$ or $\mathbb{Z}_2^{\text{diag}}$ as shown in \eqref{eq: complements 2RepZ2 x Z2}.
Based on the results in Section \ref{sec: Interface symmetries CT}, one can compute the strip 2-algebra $\mathcal{C}_{K_1, K_2}$ for each choice of $K_1$ and $K_2$ as follows:
\begin{align}
\mathcal{C}_{\mathbb{Z}_2^{\text{right}}, \mathbb{Z}_2^{\text{right}}} &\cong \mathcal{C}_{\mathbb{Z}_2^{\text{diag}}, \mathbb{Z}_2^{\text{diag}}} \cong \Vect_{\mathbb{Z}_2} \oplus \Vect_{\mathbb{Z}_2}, 
\label{eq: strip Z2 1}\\
\mathcal{C}_{\mathbb{Z}_2^{\text{right}}, \mathbb{Z}_2^{\text{diag}}} &\cong \mathcal{C}_{\mathbb{Z}_2^{\text{diag}}, \mathbb{Z}_2^{\text{right}}} \cong \Mat_2(\Vect).
\label{eq: strip Z2 2}
\end{align}
Namely, the symmetry at the interface of the same SPT phase is $\Vect_{\mathbb{Z}_2} \oplus \Vect_{\mathbb{Z}_2}$, while the symmetry at the interface of different SPT phases is $\Mat_2(\Vect)$.
The former should be regarded as the categorical description of a non-anomalous $\mathbb{Z}_2^{[0]} \times \mathbb{Z}_2^{[1]}$ symmetry in 1+1d, where the superscript denotes the form degree.
On the other hand, the latter should be regarded as the categorical description of a $\mathbb{Z}_2^{[0]} \times \mathbb{Z}_2^{[1]}$ symmetry with a mixed anomaly in 1+1d.

\vspace*{\baselineskip}
\noindent{\bf Tensor network description of interfaces.}
Let us now provide a tensor network description of the above interface symmetries.
For simplicity, as in Section~\ref{sec: Tensor network description}, we will focus on the interfaces that can be represented by tensor networks of the form~\eqref{eq: interface state}.
To obtain the symmetry operators at the interface, we first write down the action tensors.
Due to \eqref{eq: action tensors G cluster f}, the action tensors for the symmetry operator $\mathsf{C}U_g$ acting on $\ket{\text{$\mathbb{Z}_2$-cluster}(f^{(m)})}$ are given by
\begin{equation}
\adjincludegraphics[valign = c, trim={10, 10, 10, 10}]{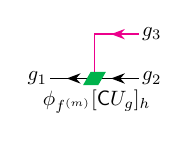} = \adjincludegraphics[valign = c, trim={10, 10, 10, 10}]{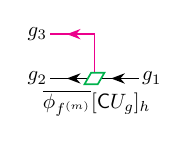} = \delta_{g_1g_3, h} \delta_{g_2g_3, g^mh},
\end{equation}
where $g, h, g_1, g_2, g_3 \in \mathbb{Z}_2$.
Using these action tensors, we can construct the symmetry operators acting on the interface of $\ket{\text{$\mathbb{Z}_2$-cluster}(f^{(m)})}$ and $\ket{\text{$\mathbb{Z}_2$-cluster}(f^{(m^{\prime})})}$ as
\begin{equation}
\hat{\mathsf{L}}(\mathsf{C}U_g; h, h^{\prime}) := \bigotimes_{i: \text{sites}} \hat{\mathcal{O}}_i(\mathsf{C}U_g; h, h^{\prime}),
\label{eq: interface sym op 2RepZ2 x Z2}
\end{equation}
where the on-site operator $\hat{\mathcal{O}}_i(\mathsf{C}U_g; h_1, h_2)$ is defined by
\begin{equation}
\hat{\mathcal{O}}_i(\mathsf{C}U_g; h, h^{\prime}) := \adjincludegraphics[valign = c, trim={10, 10, 10, 10}]{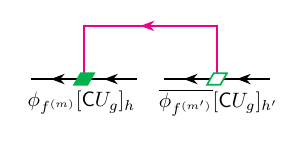} = 
\frac{1+\text{sgn}(hh^{\prime})Z_{\ell}^{(i)}Z_r^{(i)}}{2} (X_{\ell}^{(i)})^{m\delta_{g, \eta}} (X_r^{(i)})^{m^{\prime}\delta_{g, \eta}}.
\label{eq: onsite 2RepZ2 x Z2}
\end{equation}
Here, $\eta$ denotes the generator of $\mathbb{Z}_2$, and $\text{sgn}$ denotes the sign representation of $\mathbb{Z}_2$.
The operator on the right-hand side acts on the middle two legs in the middle diagram.\footnote{The small gap in the middle of the diagram corresponds to the interface.}
More specifically, $X_{\ell}^{(i)}$ and $Z_{\ell}^{(i)}$ are the Pauli operators acting on the left leg, while $X_r^{(i)}$ and $Z_r^{(i)}$ are those acting on the right leg.
By plugging \eqref{eq: onsite 2RepZ2 x Z2} into \eqref{eq: interface sym op 2RepZ2 x Z2}, we find
\begin{equation}
\hat{\mathsf{L}}(\mathsf{C}U_g; h, h^{\prime}) = P_{\text{sgn}(hh^{\prime})} X_{\ell}^{m\delta_{g, \eta}} X_r^{m^{\prime}\delta_{g, \eta}},
\label{eq: interface sym op 2RepZ2 x Z2 2}
\end{equation}
where $P_{\pm} := \bigotimes_{i: \text{sites}} \frac{1}{2}(1 \pm Z_{\ell}^{(i)}Z_r^{(i)})$, $X_{\ell} := \bigotimes_{i: \text{sites}} X_{\ell}^{(i)}$, and $X_r := \bigotimes_{i: \text{sites}} X_r^{(i)}$.

Equation~\eqref{eq: interface sym op 2RepZ2 x Z2 2} shows that the set of the symmetry operators at the interface of $\ket{\text{$\mathbb{Z}_2$-cluster}(f^{(m)})}$ and $\ket{\text{$\mathbb{Z}_2$-cluster}(f^{(m^{\prime})})}$ is given by
\begin{equation}
\{P_+, \qquad P_-, \qquad P_+ X_{\ell}^m X_r^{m^{\prime}}, \qquad P_- X_{\ell}^m X_r^{m^{\prime}}\}.
\label{eq: all interface sym op Z2}
\end{equation}
Let us compute the algebra of these symmetry operators for each pair of $m, m^{\prime} \in \{0, 1\}$.
\begin{itemize}
\item When $(m, m^{\prime}) = (0, 0)$, the set of the symmetry operators reduces to
\begin{equation}
\left\{P_+, \quad P_- \right\}.
\end{equation}
The algebra of these symmetry operators can be computed as
\begin{equation}
P_+P_+ = P_+, \quad P_-P_-=P_-, \quad P_+P_-=P_-P_+=0.
\label{eq: sym alg C 00 Z2}
\end{equation}
This suggests that the symmetry operators at the interface form a multifusion category $\Vect \oplus \Vect$, whose simple objects correspond to $P_+$ and $P_-$.
The action of the symmetry category~\eqref{eq: strip Z2 1} is thus implemented via a tensor functor $\Vect_{\mathbb{Z}_2} \oplus \Vect_{\mathbb{Z}_2} \to \Vect \oplus \Vect$.\footnote{We note that the symmetry action is unfaithful. This is because we focused on the interface of the simple form~\eqref{eq: interface state}. The symmetry action can become faithful if we consider more general interfaces.}
\item When $(m, m^{\prime}) = (0, 1)$, the set of the symmetry operators is
\begin{equation}
\left\{P_+, \quad P_-, \quad P_+X_r, \quad P_-X_r \right\}.
\end{equation}
The algebra of these symmetry operators can be computed as
\begin{equation}
(P_+X_r)P_+ = (P_-X_r)P_- = 0, \quad (P_+X_r)P_- = P_+X_r, \quad (P_-X_r)P_+ = P_-X_r.
\label{eq: sym alg C 01 Z2}
\end{equation}
The other products can be computed by using \eqref{eq: sym alg C 00 Z2} and $X_r^2 = 1$.
The above equation suggests that the symmetry operators at the interface form a multifusion category $\Mat_2(\Vect)$, which agrees with \eqref{eq: strip Z2 2}.
The simple objects of the diagonal components correspond to $P_+$ and $P_-$, while those of the off-diagonal components correspond to $P_+X_r$ and $P_-X_r$.
\item When $(m, m^{\prime}) = (1, 0)$, the set of the symmetry operators is
\begin{equation}
\left\{P_+, \quad P_-, \quad P_+X_{\ell}, \quad P_-X_{\ell}\right\}.
\end{equation}
The algebra of these symmetry operators is isomorphic to \eqref{eq: sym alg C 01 Z2}.
\item When $(m, m^{\prime}) = (1, 1)$, the set of the symmetry operators is
\begin{equation}
\left\{P_+, \quad P_-, \quad P_+X_{\ell}X_r, \quad P_-X_{\ell}X_r\right\}.
\end{equation}
The algebra of these symmetry operators can be computed as
\begin{equation}
(P_+X_{\ell}X_r) P_- = (P_-X_{\ell}X_r)P_+ = 0, \quad (P_+X_{\ell}X_r) P_+ = P_+X_{\ell}X_r, \quad (P_-X_{\ell}X_r) P_- = P_-X_{\ell}X_r.
\label{eq: sym alg C 11 Z2}
\end{equation}
The other products can be computed by using \eqref{eq: sym alg C 00 Z2} and $X_{\ell}^2 = X_r^2 = 1$.
The above equation suggests that the symmetry operators at the interface form $\Vect_{\mathbb{Z}_2} \oplus \Vect_{\mathbb{Z}_2}$, which agrees with \eqref{eq: strip Z2 1}.
The simple objects of the first component are $P_+$ and $P_+X_{\ell}X_r$, while those of the second component are $P_-$ and $P_-X_{\ell}X_r$.
\end{itemize}

\vspace*{\baselineskip}
\noindent{\bf Interface modes.}
Let us finally discuss the non-degenerate interface modes between $\ket{\text{$\mathbb{Z}_2$-cluster}(f^{(m)})}$ and $\ket{\text{$\mathbb{Z}_2$-cluster}(f^{(m^{\prime})})}$ for each pair of $m, m^{\prime} \in \{0, 1\}$.

\begin{itemize}
\item When $(m, m^{\prime}) = (0, 0)$, the interface symmetry is generated by $P_+$ and $P_-$, which are the projectors onto the subspaces with $Z_{\ell}^{(i)}Z_r^{(i)} = +1$ and $Z_{\ell}^{(i)}Z_r^{(i)} = -1$ for all $i$, respectively.
Since they are orthogonal projectors, we can divide the Hilbert space of interface modes into two sectors labeled by the eigenvalues of $Z_{\ell}^{(i)}Z_r^{(i)}$.
In the sector with $Z_{\ell}^{(i)}Z_r^{(i)} = +1$, we have infinitely many interface modes given by the tensor product states
\begin{equation}
\bigotimes_{i: \text{sites}} (\alpha_{i}\ket{e,e}_{i} + \beta_{i}\ket{\eta,\eta}_{i}),
\end{equation}
where $\eta$ is the generator of $\mathbb{Z}_2$ and $\alpha_i, \beta_i \in \mathbb{C}$.
However, these modes are continuously connected to each other by varying $\alpha_i$ and $\beta_i$ without breaking the symmetry.
Thus, we think of them as equivalent interface modes.
As a representative of this sector, we can choose 
\begin{equation}
       \ket{\psi(e)} = \bigotimes_{i: \text{sites}} \left(\sum_{g_{\ell}\in \mathbb{Z}_{2}}\ket{g_{\ell},g_{\ell}}_{i}\right).
\end{equation}
Similarly, in the sector with $Z_{\ell}^{(i)}Z_r^{(i)} = -1$, we also have infinitely many interface modes given by
\begin{equation}
\bigotimes_{i: \text{sites}} (\alpha_{i}\ket{e,\eta}_{i} + \beta_{i}\ket{\eta,e}_{i}),
\end{equation}
where $\alpha_i, \beta_i \in \mathbb{C}$.
Again, these modes are continuously connected to each other, and as a representative of this sector, we can choose 
\begin{equation}
       \ket{\psi(\eta)} = \bigotimes_{i: \text{sites}} \left(\sum_{g_{\ell}\in \mathbb{Z}_{2}}\ket{g_{\ell},\eta g_{\ell}}_{i}\right).
\end{equation} 
The symmetry operators $P_+$ and $P_-$ act on the above interface modes as
\begin{equation}
P_+ \ket{\psi(e)} = \ket{\psi(e)}, \qquad P_-\ket{\psi(\eta)} = \ket{\psi(\eta)}, \qquad
P_- \ket{\psi(e)} = P_+ \ket{\psi(\eta)} = 0,
\label{eq: interface sym action 00}
\end{equation}
which shows that $\ket{\psi(e)}$ and $\ket{\psi(\eta)}$ are non-degenerate.

From a category theoretical point of view, non-degenerate interface modes are classified by the fiber functors of the fusion category generated by the symmetry operators at the interface.
The symmetry operators $P_+$ and $P_-$ generate a multifusion category $\Vect \oplus \Vect$ and it has two fiber functors $F_{\pm}:\Vect \oplus \Vect \to \Vect$.
Here, $F_{+}$ and $F_{-}$ are projections onto the first and second components, respectively.
The symmetry action~\eqref{eq: interface sym action 00} implies that the two non-degenerate interface modes $\ket{\psi(e)}$ and $\ket{\psi(\eta)}$ correspond to $F_{+}$ and $F_{-}$, respectively.

\item When $(m, m^{\prime}) = (1, 1)$, the interface symmetry is generated by $P_{\pm}$ and $P_{\pm}X_{\ell}X_{r}$, where $P_{\pm}$ is the projector onto the subspaces with $Z_{\ell}^{(i)}Z_r^{(i)} = \pm 1$ for all $i$.
Since $P_+$ and $P_-$ are orthogonal projectors, we can divide the Hilbert space of interface modes into two sectors labeled by the eigenvalues of $Z_{\ell}^{(i)}Z_r^{(i)}$.
The operator $P_{\pm}X_{\ell}X_{r}$ acts as $X_{\ell}X_{r}$ within the sector with $Z_{\ell}^{(i)}Z_r^{(i)} = \pm1$, while it acts as zero on the other sector.
In the sector with $Z_{\ell}^{(i)}Z_r^{(i)} = +1$, there are infinitely many interface modes given by the tensor product states
\begin{equation}
       \bigotimes_{i: \text{sites}} \alpha_{i}(\ket{e,e}_{i} + (-1)^{n_{i}}\ket{\eta,\eta}_{i}),
\label{eq: interface mode 11}
\end{equation}
where $\alpha_{i} \in \mathbb{C}$ and $n_i \in \{0, 1\}$.
Contrary to the previous case, we cannot continuously connect the modes with $n_{i}=0$ and $n_{i}=1$ by only varying $\alpha_i$.
However, by a more general continuous deformation, we can change $n_i$'s in pairs without breaking the symmetry.
Indeed, for any pair $(i_1, i_2)$, we can change $n_{i_1}$ and $n_{i_2}$ simultaneously by applying $Z_{r}^{(i_1)}Z_{r}^{(i_2)}$, which can be continuously deformed into the identity operator via
\begin{equation}
e^{i\frac{\theta}{4}(1-Z_{r}^{(i_1)}\otimes Z_{r}^{(i_2)})}, \qquad 0 \leq \theta \leq 2\pi.
\end{equation}
Since this operator commutes with $P_{\pm}$ and $P_{\pm}X_{\ell}X_{r}$, the deformation does not break the symmetry.
By repeating this procedure, we can deform any state of the form~\eqref{eq: interface mode 11} to the one with at most one site with $n_i = 1$.\footnote{We note that the intermediate state of this deformation is not of the form~\eqref{eq: interface mode 11}.}
The remaining site with $n_i = 1$ cannot be removed by the above procedure.
This is interpreted as a point-like excitation on the interface.
This excitation can be moved freely by continuous deformations, and hence it is topological up to continuous deformation.
Therefore, the interface modes \eqref{eq: interface mode 11} are all equivalent up to topological point-like excitation.
As a representative of this sector, we can choose 
\begin{equation}
       \ket{\psi(e)} = \bigotimes_{i: \text{sites}} \left(\sum_{g_{\ell}\in \mathbb{Z}_{2}}\ket{g_{\ell},g_{\ell}}_{i}\right).
\end{equation}

Similarly, in the sector with $Z_{\ell}^{(i)}Z_r^{(i)} = -1$, there are also infinitely many interface modes given by
\begin{equation}
       \bigotimes_{i: \text{sites}} \alpha_{i}(\ket{e,\eta}_{i} + (-1)^{n_{i}}\ket{\eta,e}_{i}),
\end{equation}
where $\alpha_{i} \in \mathbb{C}$ and $n_i \in \{0, 1\}$.
For the same reason as above, these interface modes are all equivalent up to point-like excitation.
As a representative of this sector, we can choose 
\begin{equation}
       \ket{\psi(\eta)} = \bigotimes_{i: \text{sites}} \left(\sum_{g_{\ell}\in \mathbb{Z}_{2}}\ket{g_{\ell},\eta g_{\ell}}_{i}\right).
\end{equation}
The symmetry operators $P_{\pm}$ and $P_{\pm}X_{\ell}X_{r}$ act on the above interface modes as
\begin{equation}
\begin{aligned}
P_{+} \ket{\psi(e)} &= P_{+}X_{\ell}X_{r} \ket{\psi(e)} = \ket{\psi(e)}, \qquad
P_{-} \ket{\psi(\eta)} = P_{-}X_{\ell}X_{r} \ket{\psi(\eta)} = \ket{\psi(\eta)}, \\
P_{-} \ket{\psi(e)} &= P_{-}X_{\ell}X_{r} \ket{\psi(e)} = P_{+} \ket{\psi(\eta)} = P_{+}X_{\ell}X_{r} \ket{\psi(\eta)} = 0,
\end{aligned}
\label{eq: interface sym action 11}
\end{equation}
which shows that $\ket{\psi(e)}$ and $\ket{\psi(\eta)}$ are non-degenerate.

From a category theoretical point of view, non-degenerate interface modes are classified by the fiber functors of the fusion category generated by the symmetry operators at the interface.
The symmetry operators $P_{\pm}$ and $P_{\pm}X_{\ell}X_{r}$ generate a multifusion category $\Vect_{\mathbb{Z}_{2}} \oplus \Vect_{\mathbb{Z}_{2}}$ and it has two fiber functors $F_{\pm}:\Vect_{\mathbb{Z}_{2}} \oplus \Vect_{\mathbb{Z}_{2}} \to \Vect$.
Here, $F_{+}$ and $F_{-}$ are the unique fiber functor of the first and second $\Vect_{\mathbb{Z}_{2}}$, respectively.
The symmetry action~\eqref{eq: interface sym action 11} implies that the two non-degenerate interface modes $\ket{\psi(e)}$ and $\ket{\psi(\eta)}$ correspond to $F_{+}$ and $F_{-}$, respectively.

Moreover, the topological point-like excitations on the non-degenerate interface corresponding to $F_{\pm}$ are classified by natural endomorphisms of $F_{\pm}$.
Given that both $F_+$ and $F_-$ are the unique fiber functor of $\Vect_{\mathbb{Z}_2}$, the natural endomorphisms of $F_{\pm}$ are in one-to-one correspondence with objects of the dual category $(\Vect_{\mathbb{Z}_2})_{\Vect}^* \cong \Rep(\mathbb{Z}_2)$.
This category has two simple objects, i.e., the trivial representation and the sign representation of $\mathbb{Z}_2$.
Since both of these objects are invertible, the topological point-like excitations on the non-degenerate interface are classified by $\mathbb{Z}_2$.
The above discussion suggests that the non-trivial point-like excitation $n_i = 1$ corresponds to the sign representation, while the trivial point-like excitation $n_i = 0$ corresponds to the trivial representation.

\item When $(m, m^{\prime}) = (0, 1) \text{ or } (1, 0)$, we cannot find any non-degenerate interface modes.
This is consistent with the fact that the multifusion category $\Mat_2(\Vect)$ generated by the symmetry operators at the interface does not have any fiber functors.

\end{itemize}

\subsection{Parameterized families}
\label{sec: Parameterized families Gx2Rep(G)}
In this subsection, we consider $S^1$-parameterized families of $2\Rep(G_0) \boxtimes 2\Vect_{G_0}$-symmetric states in the same phase as $\ket{\text{$G_0$-cluster}(f)}$.

\vspace*{\baselineskip}
\noindent{\bf Classification.}
Due to the general classification mentioned in Section~\ref{sec: Parameterized families}, the $S^1$-parameterized families are expected to be classified by
\begin{equation}
\{(h, [\xi]) \mid h \in H \text{ s.t. } K= hKh^{-1}, ~ [\xi] \in H^2(K, \mathrm{U}(1))\},
\end{equation}
where $H = G_0^{\text{left}}$ and $K=G_0^{\text{diag}}(f)$.
In what follows, we will only consider $S^1$-parameterized families with trivial $[\xi]$.
The condition $K=hKh^{-1}$ for $h=(g, e) \in G_0^{\text{left}}$ means that for any $(f(g^{\prime}), g^{\prime}) \in G_0^{\text{diag}}(f)$, there exists $(f(g^{\prime \prime}), g^{\prime \prime}) \in G_0^{\text{diag}}(f)$ such that
\begin{equation}
(g, e) (f(g^{\prime}), g^{\prime}) (g, e)^{-1} = (f(g^{\prime \prime}), g^{\prime \prime}).
\end{equation}
This condition is equivalent to
\begin{equation}
gf(g^{\prime})g^{-1} = f(g^{\prime}), \qquad \forall g^{\prime} \in G_0.
\end{equation}
Therefore, $h=(g,e)$ satisfies $K=hKh^{-1}$ if and only if $g$ commutes with the image of $f$.
In other words, the condition $K=hKh^{-1}$ holds if and only is $g$ is in the centralizer of the image of $f$ in $G_0$, i.e.,
\begin{equation}
g \in C_{G_0}(\mathrm{Im}(f)).
\label{eq: centralizer of Imf}
\end{equation}
In particular, if $f=\mathrm{id}$, the condition holds for $g \in Z(G_0)$.
On the other hand, if $f$ is the trivial homomorphism, the condition holds for every $g \in G_{0}$.
Below, we will simply write $h = (g,e) \in H$ as $g$.

\vspace*{\baselineskip}
\noindent{\bf $S^1$-parameterized families.}
For each $g \in C_{G_0}(\mathrm{Im}(f))$, we can construct an $S^1$-parameterized family of $2\Rep(G_0) \boxtimes 2\Vect_{G_0}$-symmetric models by conjugating the Hamiltonian~\eqref{eq: G cluster f ham} with the following $\theta$-dependent unitary operator:
\begin{equation}
U_{\text{g.f.}}^g(\theta) \coloneq \prod_{i \in P} \left.u_g^{(i)}(\theta)\right|_{\text{g.f.}}, \qquad
\left.u_g^{(i)}(\theta)\right|_{\text{g.f.}} \coloneq \exp\left( \frac{\theta}{2\pi} \log L_{g^{-1}}^{(\partial i)} \right),
\label{eq: G cluster f U}
\end{equation}
Here, $\theta$ runs from $0$ to $2\pi$ and $L_{g^{-1}}^{(\partial i)}$ denotes the loop operator defined as in \eqref{eq: loop op}.
We choose the branch of the logarithm as
\begin{equation}
       \log L_{g^{-1}}^{(\partial i)} = \sum_{s=0}^{\abs{G_0} - 1} i \frac{2\pi s}{\abs{G_0}} P_s[L_{g^{-1}}^{(\partial i)}],
\end{equation}
where $P_s[L_{g^{-1}}^{(\partial i)}]$ is the projection onto the eigenspace of $L_{g^{-1}}^{(\partial i)}$ with eigenvalue $e^{i\frac{2\pi s}{\abs{G_0}}}$.
We note that $\left.u_g^{(i)}(\theta)\right|_{\text{g.f.}}$ interpolates between the identity operator and the loop operator $L_{g^{-1}}^{(\partial i)}$.
By a direct computation, we can show that $U_{\text{g.f.}}^g(\theta)$ at $\theta = 2\pi$ commutes with the Hamiltonian~\eqref{eq: G cluster f ham} due to \eqref{eq: centralizer of Imf}.
Therefore, conjugating \eqref{eq: G cluster f ham} with $U_{\text{g.f.}}^g(\theta)$ gives us a $2\pi$-periodic Hamiltonian.
This $2\pi$-periodic Hamiltonian preserves the $2\Rep(G_0) \boxtimes 2\Vect_{G_0}$ symmetry because $U_{\text{g.f.}}^g(\theta)$ commutes with the symmetry operators for all $\theta \in [0, 2\pi]$: see Section~\ref{sec: Example: parameterized families of the G-cluster states} for more details.

\vspace*{\baselineskip}
\noindent{\bf Generalized Thouless pumps.}
We can argue the non-triviality of the above $S^1$-parameterized families by considering the corresponding generalized Thouless pumps.
We will not study this here because the discussion is almost parallel to the case of the $G_0$-clutser state detailed in Section~\ref{sec: Example: parameterized families of the G-cluster states}.

\section{Example 2: Tambara-Yamagami SPT phases}
\label{sec: Tambara-Yamagami SPT phases}
As another class of examples, we consider 2+1d SPT phases with Tambara-Yamagami symmetry $2\TY(A, 1)$, where $A$ is a finite abelian group.
As we reviewed in Section~\ref{sec: Example: Tambara-Yamagami fusion 2-categories}, $2\TY(A, 1)$ is a group-theoretical fusion 2-category defined by
\begin{equation}
2\TY(A, 1) := \mathcal{C}(A \wr \mathbb{Z}_2; A_L),
\label{eq: TY def}
\end{equation}
where $A \wr \mathbb{Z}_2 := (A \times A) \rtimes \mathbb{Z}_2$ is the wreath product of $A$ by $\mathbb{Z}_2$, and $A_L$ is the subgroup~\eqref{eq: AL} of $A \wr \mathbb{Z}_2$.
We recall that $2\TY(A, 1)$ is a $\mathbb{Z}_2$-graded fusion 2-category
\begin{equation}
2\TY(A, 1) \cong 2\Rep(A) \boxtimes 2\Vect_A \oplus 2\Vect,
\label{eq: TY decomposition}
\end{equation}
where $2\Rep(A) \boxtimes 2\Vect_A$ is the trivially graded component.
Throughout this subsection, the elements of $\mathbb{Z}_2$ will be denoted additively, i.e., $\mathbb{Z}_2 = \{0, 1\}$, while the elements of $A$ will be denoted multiplicatively.

\subsection{Classification}
\label{sec: Classification TY}
We first provide the classification of SPT phases with Tambara-Yamagami symmetry $2\TY(A, 1)$.
The SPT phases with $2\TY(A, 1)$ symmetry are in one-to-one correspondence with fiber 2-functors of $2\TY(A, 1)$.
Due to \eqref{eq: TY def}, fiber 2-functors of $2\TY(A, 1)$ are labeled by pairs $(K, \lambda)$, where $K$ is a complement of $A_L$ in $A \wr \mathbb{Z}_2$, and $\lambda \in Z^3(K, \mathrm{U}(1))$ is a 3-cocycle on $K$.
We refer the reader to Section~\ref{sec: Fiber 2-functors} for the classification of fiber 2-functors of general group-theoretical fusion 2-categories.

As we will see below, a complement $K$ of $A_L$ in $A \wr \mathbb{Z}_2$ can always be written as
\begin{equation}
K = K(f) := \{(f_n(a), a, n) \mid a \in A, n \in \mathbb{Z}_2\},
\label{eq: TY complement}
\end{equation}
where $f_n: A \to A$ for $n \in \mathbb{Z}_2$ is a map that satisfies the following conditions:
\begin{itemize}
\item $f_0: A \to A$ is an involutive group homomorphism
\item $f_1(e) \in A$ is a fixed point of $f_0$, i.e., $f_0(f_1(e)) = f_1(e)$
\item $f_1: A \to A$ is determined by $f_1(a) = f_0(a) f_1(e)$ for all $a \in A$
\end{itemize}
We note that $2\TY(A, 1)$ always admits a fiber 2-functor because the identity map $f_0 = f_1 =\id$ satisfies the above conditions.
In what follows, we will focus on fiber 2-functors with trivial $\lambda$.
In this case, each fiber 2-functor is labeled only by a complement $K(f)$.

The fiber 2-functors labeled by $K(f)$ and $K(f^{\prime})$ are isomorphic to each other if and only if there exists $a \in A$ such that
\begin{equation}
f^{\prime}_0 = f_0, \quad f^{\prime}_1(e) = af_0(a) f_1(e).
\label{eq: isom fiber 2-functor}
\end{equation}
A proof will be provided shortly.
Therefore, fiber 2-functors with trivial $\lambda$ are classified by the set
\begin{equation}
\coprod_{f_0 \in \Aut(A)[2]} A^{f_0}/(A^{f_0})_0,
\end{equation}
where $\Aut(A)[2]$ is the 2-torsion subgroup of the automorphism group $\Aut(A)$, $A^{f_0} := \{a \in A \mid f_0(a) = a\}$ is the group of fixed points of $f_0$, and $(A^{f_0})_0 :=\{af_0(a) \mid a \in A\}$ is a subgroup of $A^{f_0}$.

Now, we show that any complement $K$ of $A_L$ in $A \wr \mathbb{Z}_2$ can be written as~\eqref{eq: TY complement}.
First of all, since a complement $K$ satisfies $A_L K = A \wr \mathbb{Z}_2$ by definition, it must contain an element of the form $(\bullet, a, n)$ for any $a \in A$ and $n \in \mathbb{Z}_2$, where $\bullet$ denotes some element of $A$.
In addition, since $K$ satisfies $A_L \cap K = \{(e, e, 0)\} \in A \wr \mathbb{Z}_2$, the element $(\bullet, a, n) \in K$ must be unique for every $a \in A$ and $n \in \mathbb{Z}_2$.
Thus, there exists a map $f: A \times \mathbb{Z}_2 \to A$ such that $K=\{(f(a, n), a, n) \mid a \in A, n \in \mathbb{Z}_2\}$.
If we define $f_n(a) := f(a, n)$, we find $K = K(f)$.
The set $K(f)$ is a complement of $A_L$ if and only if it is a subgroup of $A \wr \mathbb{Z}_2$, i.e., if and only
\begin{equation}
(f_n(a), a, n) \cdot (f_m(b), b, m) \in K(f), \qquad \forall a, b \in A, ~ \forall n, m \in \mathbb{Z}_2.
\label{eq: K subgroup}
\end{equation}
Let us look into the above condition for each choice of $n, m \in \mathbb{Z}_2$.
\begin{itemize}
\item When $(n, m) = (0, 0)$, the left-hand side of \eqref{eq: K subgroup} is equal to $(f_0(a)f_0(b), ab, 0)$.
This element is in $K(f)$ if and only if $f_0$ satisfies 
\begin{equation}
f_0(a) f_0(b) = f_0(ab).
\label{eq: (n, m) = (0, 0)}
\end{equation}
Thus, we find that $f_0: A \to A$ must be a group homomorphism.
\item When $(n, m) = (0, 1)$, the left-hand side of \eqref{eq: K subgroup} is equal to $(f_0(a)f_1(b), ab, 1)$.
This element is in $K(f)$ if and only if $f_0$ and $f_1$ satisfy
\begin{equation}
f_0(a)f_1(b) = f_1(ab).
\label{eq: (n, m) = (0, 1) 1}
\end{equation}
If we choose $b = e$, this condition reduces to
\begin{equation}
f_0(a) f_1(e) = f_1(a).
\label{eq: (n, m) = (0, 1) 2}
\end{equation}
Conversely, if equation~\eqref{eq: (n, m) = (0, 1) 2} holds, then equation~\eqref{eq: (n, m) = (0, 1) 1} also holds because $f_0$ is a group homomorphism due to \eqref{eq: (n, m) = (0, 0)}.
Thus, the condition~\eqref{eq: K subgroup} for $(n, m) = (0, 1)$ reduces to~\eqref{eq: (n, m) = (0, 1) 2} for all $a \in A$.
\item When $(n, m) = (1, 0)$, the left-hand side of \eqref{eq: K subgroup} is equal to $(f_1(a)b, af_0(b), 1)$.
This element is in $K(f)$ if and only if $f_0$ and $f_1$ satisfy
\begin{equation}
f_1(a)b = f_1(af_0(b)).
\end{equation}
Due to \eqref{eq: (n, m) = (0, 0)} and \eqref{eq: (n, m) = (0, 1) 2}, the above equation reduces to
\begin{equation}
f_0(f_0(b)) = b.
\label{eq: (n, m) = (1, 0)}
\end{equation}
Thus, we find that $f_0: A \to A$ must be an involution.
\item When $(n, m) = (1, 1)$, the left-hand side of \eqref{eq: K subgroup} is equal to $(f_1(a)b, af_1(b), 0)$.
This element is in $K(f)$ if and only if $f_0$ and $f_1$ satisfy
\begin{equation}
f_1(a)b = f_0(af_1(b)).
\end{equation}
Due to \eqref{eq: (n, m) = (0, 0)}, \eqref{eq: (n, m) = (0, 1) 2}, and \eqref{eq: (n, m) = (1, 0)}, the above equation reduces to
\begin{equation}
f_0(f_1(e)) = f_1(e).
\end{equation}
Thus, we find that $f_1(e)$ must be a fixed point of $f_0$.
\end{itemize}
This shows that $K(f)$ is a complement of $A_L$ in $A \wr \mathbb{Z}_2$ if and only if the three conditions below~\eqref{eq: TY complement} are satisfied.

Due to the general result reviewed in Section~\ref{sec: Fiber 2-functors}, the complements $K(f)$ and $K(f^{\prime})$ correspond to isomorphic fiber 2-functors if and only if there exists $g \in A \wr \mathbb{Z}_2$ such that
\begin{equation}
K(f^{\prime}) = g K(f) g^{-1}.
\end{equation}
Since $K(f)$ is a complement of $A_L$, we can take $g$ to be an element of $A_L$ without loss of generality.
For $g = (a, e, 0) \in A_L$, the above condition is equivalent to\footnote{Equation \eqref{eq: equivalence of K} is equivalent to $gK(f)g^{-1} \subset K(f^{\prime})$. On the other hand, $gK(f)g^{-1}$ and $K(f^{\prime})$ have the same order. Thus, the relation $gK(f)g^{-1} \subset K(f^{\prime})$ implies $gK(f)g^{-1} = K(f^{\prime})$.}
\begin{equation}
(a, e, 0) \cdot (f_n(b), b, n) \cdot (a, e, 0)^{-1} \in K(f^{\prime}), \qquad \forall b \in A, ~ \forall n \in \mathbb{Z}_2.
\label{eq: equivalence of K}
\end{equation}
Let us look into the above condition for each choice of $n \in \mathbb{Z}_2$.
\begin{itemize}
\item When $n = 0$, the left-hand side of \eqref{eq: equivalence of K} is equal to $(f_0(b), b, 0)$.
This element is in $K(f^{\prime})$ if and only if $f_0$ and $f_0^{\prime}$ satisfy
\begin{equation}
f_0(b) = f_0^{\prime}(b).
\label{eq: x0 = x0 prime}
\end{equation}
Thus, the condition~\eqref{eq: equivalence of K} for $n = 0$ reduces to the equality $f_0 = f_0^{\prime}$.
\item When $n = 1$, the left-hand side of \eqref{eq: equivalence of K} is equal to $(af_1(b), ba^{-1}, 1)$.
This element is in $K(f^{\prime})$ if and only if $f_1$ and $f_1^{\prime}$ satisfy
\begin{equation}
af_1(b) = f_1^{\prime}(ba^{-1}).
\end{equation}
Due to \eqref{eq: (n, m) = (0, 0)}, \eqref{eq: (n, m) = (0, 1) 2} and \eqref{eq: x0 = x0 prime}, the above equation reduces to
\begin{equation}
f_1^{\prime}(e) = a f_0(a) f_1(e).
\end{equation}
\end{itemize}
This shows that $K(f)$ and $K(f^{\prime})$ correspond to isomorphic fiber 2-functors if and only if there exists $a \in A$ such that the condition~\eqref{eq: isom fiber 2-functor} is satisfied.

\subsection{Generalized cluster states}
\label{sec: Generalized cluster states TY}
The SPT phases labeled by fiber 2-functors with trivial $\lambda$ are realized in the generalized cluster models defined in Section~\ref{sec: 2+1d generalized cluster states with non-invertible symmetries}.
In the case of the Tambara-Yamagami symmetry $2\TY(A, 1)$, the input data of the model are given by
\begin{equation}
G = A \wr \mathbb{Z}_2, \quad
H = A_L, \quad
K = K(f).
\label{eq: TY input}
\end{equation}
The generalized cluster model for the above input data is obtained by gauging the subgroup symmetry $H$ in a $G$-symmetric model that spontaneously breaks the symmetry $G$ down to $K$.
In what follows, we will write down the models both before and after the gauging.

\vspace*{\baselineskip}
\noindent{\bf $G/K$-SSB model.}
The state space of the $G$-symmetric model before the gauging is
\begin{equation}
\mathcal{H} = \bigotimes_{i \in P} \mathbb{C}^{|A \wr \mathbb{Z}_2|}.
\end{equation}
The on-site state space on each plaquette $i \in P$ is spanned by $\{\ket{a_i, b_i, n_i} \mid a_i,b_i \in A, n_i \in \mathbb{Z}_2\}$.
The Hamiltonian of the model is
\begin{equation}
H = -\sum_{i \in P} \hat{\mathsf{h}}_i - \sum_{[ij] \in E} \hat{\mathsf{h}}_{ij},
\label{eq: A wr Z2 ham}
\end{equation}
where $\hat{\mathsf{h}}_i$ and $\hat{\mathsf{h}}_{ij}$ are defined by
\begin{align}
\hat{\mathsf{h}}_i \ket{a_i, b_i, n_i}_i &= \frac{1}{2|A|} \sum_{a \in A} \sum_{n \in \mathbb{Z}_2} \ket{(a_i, b_i, n_i) \cdot (f_n(a), a, n)}_i, \\
\hat{\mathsf{h}}_{ij} \ket{a_i, b_i, n_i}_i \otimes\ket{a_j, b_j, n_j}_j &= \delta_{(a_i, b_i, n_i)^{-1} \cdot (a_j, b_j, n_j) \in K(f)} \ket{a_i, b_i, n_i}_i \otimes\ket{a_j, b_j, n_j}_j.
\end{align}
Here, $(a, b, n) \cdot (a^{\prime}, b^{\prime}, n^{\prime})$ denotes the group multiplication in $A \wr \mathbb{Z}_2$, which is defined by~\eqref{eq: wreath prod}.
The symmetry of the above model is generated by
\begin{equation}
U_a^{(L)} := \bigotimes_{i \in P} X_a^{(i; L)}, \qquad
U_a^{(R)} := \bigotimes_{i \in P} X_a^{(i; R)}, \qquad
U_{\text{SWAP}} := \bigotimes_{i \in P} \sigma_x^{(i)} \text{SWAP}_A^{(i)},
\label{eq: A wr Z2 sym op}
\end{equation}
where the on-site operators $X_a^{(i; L)}$, $X_a^{(i; R)}$, $\sigma_x^{(i)}$, and $\text{SWAP}_A^{(i)}$ are defined by
\begin{equation}
\begin{aligned}
X_a^{(i; L)} \ket{a_i, b_i, n_i}_i &= \ket{aa_i, b_i, n_i}_i, &
X_a^{(i; R)} \ket{a_i, b_i, n_i}_i &= \ket{a_i, ab_i, n_i}_i, \\
\sigma_x^{(i)} \ket{a_i, b_i, n_i}_i &= \ket{a_i, b_i, n_i + 1}_i, &
\text{SWAP}_A^{(i)} \ket{a_i, b_i, n_i}_i &= \ket{b_i, a_i, n_i}_i.
\end{aligned}
\end{equation}
The Hamiltonian~\eqref{eq: A wr Z2 ham} has $|A|$-fold degenerate ground states
\begin{equation}
\ket{\text{GS}; a} = \bigotimes_{i \in P} \left( \frac{1}{\sqrt{2|A|}} \sum_{a_i \in A} \sum_{n_i \in \mathbb{Z}_2} \ket{af_{n_i}(a_i), a_i, n_i}_i \right), \qquad \forall a \in A.
\end{equation}
The symmetry operators \eqref{eq: A wr Z2 sym op} act on these ground states as
\begin{equation}
U_a^{(L)} \ket{\text{GS}; a^{\prime}} = \ket{\text{GS}; aa^{\prime}}, \quad
U_a^{(R)} \ket{\text{GS}; a^{\prime}} = \ket{\text{GS}; f_0(a)^{-1}a^{\prime}}, \quad
U_{\text{SWAP}} \ket{\text{GS}; a^{\prime}} = \ket{\text{GS}; f_1(a^{\prime})^{-1}}.
\end{equation}
We note that $\ket{\text{GS}; a}$ is invariant under the action of $U_{(af_0(a))^n f_n(a_r)}^{(L)} U_{a_r}^{(R)} U_{\text{SWAP}}^n$ for all $a_r \in A$ and $n \in \mathbb{Z}_2$.
Thus, the unbroken symmetry of $\ket{\text{GS}; a}$ is
\begin{equation}
\{((af_0(a))^n f_n(a_r), a_r, n) \mid a_r \in A, n \in \mathbb{Z}_2\} = (a, e, 0) \cdot K(f) \cdot (a, e, 0)^{-1}.
\end{equation}
In particular, the unbroken symmetry of $\ket{\text{GS}; e}$ is $K = K(f)$.

\vspace*{\baselineskip}
\noindent{\bf Generalized cluster model.}
The generalized cluster model with Tambara-Yamagami symmetry $2\TY(A, 1)$ is obtained by gauging the subgroup symmetry $H = A_L$ in the above model.
To describe the model, we choose a gauge so that the matter fields after the gauge fixing take values in\footnote{This choice of a gauge guarantees that the symmetry operators of the gauged model can be written as tensor networks with small bond dimensions. We refer the reader to Section~\ref{sec: Symmetry operators} for more details on this point. Another choice of a gauge will be discussed in Appendix~\ref{sec: Tambara-Yamagami cluster model in another gauge}.}
\begin{equation}
S_{H \backslash G} = A^{\text{diag}} \times \mathbb{Z}_2 := \{(a, a, n) \mid a \in A, n \in \mathbb{Z}_2\}.
\label{eq: TY SHG}
\end{equation}
The state space of the gauged model after the gauge fixing is 
\begin{equation}
\mathcal{H}_{\text{g.f.}} = \left(\bigotimes_{i \in P} \mathbb{C}^{|A^{\text{diag}} \times \mathbb{Z}_2|}\right) \otimes \hat{\pi}_{\text{flat}} \left(\bigotimes_{[ij] \in E} \mathbb{C}^{|A_L|}\right),
\label{eq: state space TY}
\end{equation}
where $\hat{\pi}_{\text{flat}}$ is the projector that imposes the flatness condition on the gauge fields.
The matter field on plaquette $i \in P$ and the gauge field on edge $[ij] \in E$ are labeled by $(a_i, a_i, n_i) \in A^{\text{diag}} \times \mathbb{Z}_2$ and $(a_{ij}, e, 0) \in A_L$.
The corresponding states are simply written as $\ket{a_i, n_i}_i$ and $\ket{a_{ij}}_{ij}$, respectively.
The basis states of~\eqref{eq: state space TY} will be denoted by
\begin{equation}
\ket{\{a_i, n_i; a_{ij}\}} := \left(\bigotimes_{i \in P} \ket{a_i, n_i}_i\right) \otimes \hat{\pi}_{\text{flat}} \left(\bigotimes_{[ij] \in E} \ket{a_{ij}}_{ij}\right).
\end{equation}
The Hamiltonian of the model is
\begin{equation}
H_{\text{g.f.}} = - \sum_{i \in P} \hat{\mathsf{h}}_i^{\text{g.f.}} - \sum_{[ij] \in E} \hat{\mathsf{h}}_{ij}^{\text{g.f.}},
\label{eq: TY ham}
\end{equation}
where $\hat{\mathsf{h}}_i^{\text{g.f.}}$ and $\hat{\mathsf{h}}_{ij}^{\text{g.f.}}$ are defined by
\begin{align}
\hat{\mathsf{h}}_i^{\text{g.f.}} \Ket{\adjincludegraphics[valign = c, trim={10, 10, 10, 10}]{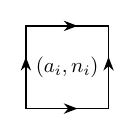}} &= \frac{1}{2|A|} \sum_{a \in A} \sum_{n \in \mathbb{Z}_2} \Ket{\adjincludegraphics[valign = c, trim={10, 10, 10, 10}]{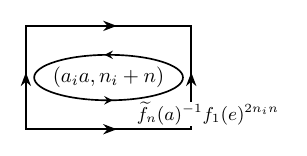}},
\label{eq: TY cluster ham p} \\
\hat{\mathsf{h}}_{ij}^{\text{g.f.}} \ket{a_i, n_i}_i \otimes \ket{a_{ij}}_{ij} \otimes \ket{a_j, n_j}_j &= \delta_{a_{ij}, \widetilde{f}_{n_i}(a_i)^{-1} \widetilde{f}_{n_j}(a_j)} \ket{a_i, n_i}_i \otimes \ket{a_{ij}}_{ij} \otimes \ket{a_j, n_j}_j.
\label{eq: TY cluster ham e}
\end{align}
Here, the map $\widetilde{f}_n: A \to A$ for $n \in \mathbb{Z}_2$ is defined by
\begin{equation}
\widetilde{f}_n(a) := f_n(a)a^{-1}.
\end{equation}
The above Hamiltonian has a unique ground state on a square lattice with periodic boundary conditions.
The ground state can be written explicitly as
\begin{equation}
\ket{\text{2TY-cluster}(A; f)} := \frac{1}{(2|A|)^{|P|/2}} \sum_{\{a_i \in A\}} \sum_{\{n_i \in \mathbb{Z}_2\}} \ket{\{a_i, n_i; \widetilde{f}_{n_i}(a_i)^{-1}\widetilde{f}_{n_j}(a_j)\}}.
\label{eq: TY cluster}
\end{equation}
A tensor network representation of the above ground state is
\begin{equation}
\ket{\text{2TY-cluster}(A; f)} = \adjincludegraphics[valign = c, trim={10, 10, 10, 10}]{tikz/out/cluster_PEPS.pdf},
\end{equation}
where the local tensors represented by the black and white dots are defined by
\begin{equation}
\adjincludegraphics[valign = c, trim={10, 10, 10, 10}]{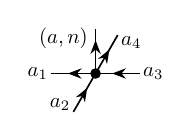} = \frac{1}{\sqrt{2|A|}} \prod_{i = 1, 2, 3, 4} \delta_{a_i, \widetilde{f}_{n}(a)}, \qquad
\adjincludegraphics[valign = c, trim={10, 10, 10, 10}]{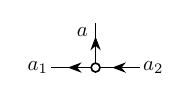} = \delta_{a, a_1^{-1}a_2}.
\label{eq: local tensor TY}
\end{equation}

\subsection{Symmetry operators}
\label{sec: Symmetry operators TY}
In this subsection, we write down the symmetry operators of the Tambara-Yamagami cluster model defined above.
We will also compute the action tensors for the symmetry operators acting on the Tambara-Yamagami cluster state~\eqref{eq: TY cluster}.

\vspace*{\baselineskip}
\noindent{\bf 1-form symmetry.}
The 1-form symmetry operators of the model are labeled by 1-endomorphisms of the unit object $I \in 2\TY(A, 1)$.
The equivalence~\eqref{eq: TY decomposition} implies that these 1-morphisms form a symmetric fusion 1-category
\begin{equation}
\End_{2\TY(A, 1)}(I) \cong \Rep(A).
\end{equation}
The symmetry operator corresponding to a simple object $\widehat{a} \in \Rep(A)$ is the Wilson line for $\widehat{a}$.
The Wilson line $W_{\widehat{a}}(\gamma)$ supported on an oriented loop $\gamma$ on the dual lattice is given by
\begin{equation}
W_{\widehat{a}}(\gamma) = \prod_{[ij] \in E_{\gamma}} (Z_{\widehat{a}}^{(ij)})^{s_{ij}},
\end{equation}
where $E_{\gamma}$ is the set of the edges that intersect $\gamma$, and $s_{ij}$ is either $+1$ or $-1$ depending on the relative orientation of $\gamma$ and $[ij]$, cf. \eqref{eq: sij}.

\vspace*{\baselineskip}
\noindent{\bf 0-form symmetry.}
The 0-form symmetry operators are labeled by objects of $2\TY(A, 1)$.
The equivalence~\eqref{eq: TY decomposition} implies that any 0-form symmetry operator, up to condensation, can be written as
\begin{equation}
\mathsf{C}U_a \quad \text{or} \quad \mathsf{D}, \qquad \forall a \in A,
\label{eq: sym op TY}
\end{equation}
where $\mathsf{C}$, $U_a$, and $\mathsf{D}$ are the symmetry operators corresponding to simple objects of $2\Rep(A)$, $2\Vect_A$, and $2\Vect$ in \eqref{eq: TY decomposition}.
Each operator in \eqref{eq: sym op TY} is given by\footnote{The expression of the invertible symmetry operator $U_a$ generally depends on a gauge choice. Nevertheless, the product $\mathsf{C}U_a$ has a gauge-independent expression $\mathsf{C}U_a = \mathsf{D}_{A_L} \left(\bigotimes_{i \in P} X_a^{(i; R)}\right) \overline{\mathsf{D}}_{A_L}$. We note that the action of $\mathsf{C}U_a$ still depends on a gauge choice because $\mathsf{D}_{A_L}$ and $\overline{\mathsf{D}}_{A_L}$ do. Similarly, the action of $\mathsf{D}$ also depends on a gauge.}
\begin{equation}
\mathsf{C} := \mathsf{D}_{A_L} \overline{\mathsf{D}}_{A_L}, \qquad
U_a := \bigotimes_{i \in P} X_a^{(i)}, \qquad
\mathsf{D} := \mathsf{D}_{A_L} \left(\bigotimes_{i \in P} \sigma_x^{(i)}\text{SWAP}_{A}^{(i)}\right) \overline{\mathsf{D}}_{A_L}.
\label{eq: C Ua D}
\end{equation}
Here, $\mathsf{D}_{A_L}$ and $\overline{\mathsf{D}}_{A_L}$ denote the gauging and ungauging operators defined by
\begin{equation}
\mathsf{D}_{A_L} \ket{\{a_i, b_i, n_i\}} = \ket{\{b_i, n_i; b_i a_i^{-1} a_jb_j^{-1}\}}, \qquad
\overline{\mathsf{D}}_{A_L} = \mathsf{\mathsf{D}}_{A_L}^{\dagger}.
\label{eq: TY gauging op}
\end{equation}
The actions of the above symmetry operators can then be computed explicitly as
\begin{equation}
\begin{aligned}
\mathsf{C}U_a\ket{\{a_i, n_i; a_{ij}\}} &= |A|\delta_{\text{hol}}(\{a_{ij}\}) \ket{\{aa_i, n_i; a_{ij}\}}, \\
\mathsf{D}\ket{\{a_i, n_i; a_{ij}\}} &= \sum_{\{b_i \in A\}} \left(\prod_{[ij] \in E} \delta_{a_{ij}, b_i^{-1}b_j}\right) \ket{\{a_ib_i, n_i+1; a_{ij}^{-1}\}},
\end{aligned}
\label{eq: actions of CUa and D}
\end{equation}
where $\delta_{\text{hol}}(\{a_{ij}\})$ is equal to $1$ if the holonomy of the gauge field is trivial and $0$ otherwise.

\vspace*{\baselineskip}
\noindent{\bf Tensor network representation.}
The above symmetry operators can be represented by the tensor networks of the form
\begin{equation}
\mathsf{C}U_a ~\text{or}~ \mathsf{D} = \adjincludegraphics[valign = c, trim={10, 10, 10, 10}]{tikz/out/symmetry_PEPO_cluster.pdf},
\end{equation}
where the virtual bonds take values in $A$.
The local tensors of $\mathsf{C}U_a$ are given by
\begin{equation}
\adjincludegraphics[valign = c, trim={10, 10, 10, 10}]{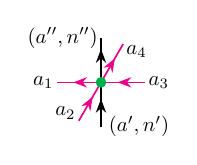} = \delta_{a^{\prime \prime}, aa^{\prime}} \delta_{n^{\prime \prime}, n^{\prime}} \prod_{i, j = 1, 2, 3, 4} \delta_{a_i, a_j}, \quad
\adjincludegraphics[valign = c, trim={10, 10, 10, 10}]{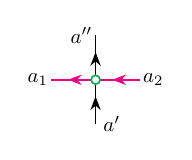} = \delta_{a^{\prime}, a_1^{-1} a_2} \delta_{a^{\prime \prime}, a^{\prime}}.
\end{equation}
On the other hand, the local tensors of $\mathsf{D}$ are given by
\begin{equation}
\adjincludegraphics[valign = c, trim={10, 10, 10, 10}]{tikz/out/TY_PEPO_green.pdf} = \delta_{a^{\prime \prime}, a_1a^{\prime}} \delta_{n^{\prime \prime}, n^{\prime}+1} \prod_{i, j = 1, 2, 3, 4} \delta_{a_i, a_j}, \quad
\adjincludegraphics[valign = c, trim={10, 10, 10, 10}]{tikz/out/TY_PEPO_white.pdf} = \delta_{a^{\prime}, a_1^{-1}a_2} \delta_{a^{\prime \prime}, (a^{\prime})^{-1}}.
\end{equation}
The other symmetry operators can be obtained by condensing line operators on $\mathsf{C}U_a$ as in \eqref{eq: sym op condensation}.
When $A=\mathbb{Z}_2$, the tensor network representation of $\mathsf{D}$ has been obtained in \cite{Cao:2025qhg}.

\vspace*{\baselineskip}
\noindent{\bf Action tensors.}
The actions of the symmetry operators $\mathsf{C}U_a$ and $\mathsf{D}$ on the local tensors of the Tambara-Yamagami cluster state~\eqref{eq: TY cluster} can be computed as
\begin{equation}
\adjincludegraphics[valign = c, trim={10, 10, 10, 10}]{tikz/out/local_action1.pdf} ~ = \sum_{b \in A} \adjincludegraphics[valign = c, trim={10, 10, 10, 10}]{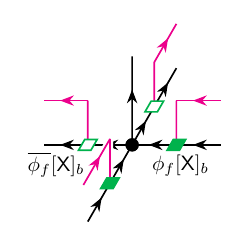}, \qquad
\adjincludegraphics[valign = c, trim={10, 10, 10, 10}]{tikz/out/local_action3.pdf} ~ = \sum_{b \in A} \adjincludegraphics[valign = c, trim={10, 10, 10, 10}]{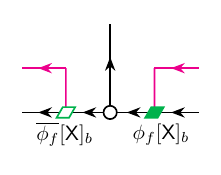},
\end{equation}
where the action tensors $\phi_f[\mathsf{X}]_b$ and $\overline{\phi_f}[\mathsf{X}]_b$ for $\mathsf{X} = \mathsf{C}U_a$ and $\mathsf{X} = \mathsf{D}$ are given by
\begin{align}
\adjincludegraphics[valign = c, trim={10, 10, 10, 10}]{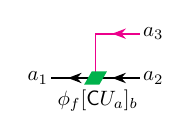} &= \adjincludegraphics[valign = c, trim={10, 10, 10, 10}]{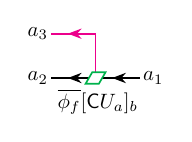} = \delta_{a_1a_3^{-1}, ba^{-1}} \delta_{a_2a_3^{-1}, bf_0(a^{-1})},
\label{eq: action (e, a, 0)} \\ 
\adjincludegraphics[valign = c, trim={10, 10, 10, 10}]{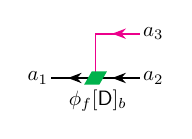} &= \adjincludegraphics[valign = c, trim={10, 10, 10, 10}]{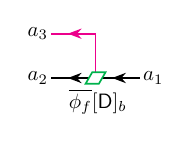} = \delta_{a_1a_3, b} \delta_{f_1(a_3a_2^{-1}), b}.
\label{eq: action (e, e, 1)}
\end{align}
These action tensors are obtained by applying the general formula~\eqref{eq: action tensors} to the current example.\footnote{Using the notations in Section~\ref{sec: Non-invertible symmetries of 2+1d generalized cluster states}, we can write $\mathsf{C}U_a = \mathsf{D}[V_{A_L}^{(e, a, 0)}]$, $\mathsf{D} = \mathsf{D}[V_{A_L}^{(e, e, 1)}]$, $\phi_f[\mathsf{C}U_a]_b = \phi_{K(f)}[V_{A_L}^{(e, a, 0)}]_{(b, e, 0)}$, and $\phi_f[\mathsf{D}]_b = \phi_{K(f)}[V_{A_L}^{(e, e, 1)}]_{(b, e, 0)}$.}

\subsection{Interfaces: the case of $A=\mathbb{Z}_2$}
\label{sec: Interfaces TY}
Let us consider the interfaces between the generalized cluster states with $2\TY(A, 1)$ symmetry.
For simplicity, we will focus on the case of $A = \mathbb{Z}_2$.
In this case, the symmetry category is \cite{Decoppet:2023bay, Bhardwaj:2022maz, Bartsch:2022ytj, Bartsch:2022mpm}
\begin{equation}
2\TY(\mathbb{Z}_2, 1) \cong 2\Rep((\mathbb{Z}_2^{[1]} \times \mathbb{Z}_2^{[1]}) \rtimes \mathbb{Z}_2^{[0]}).
\end{equation}
The generalized cluster states with this symmetry are classified by the equivalence classes of pairs of maps $(f_0, f_1)$ that satisfy the following conditions:
\begin{itemize}
\item $f_0: \mathbb{Z}_2 \to \mathbb{Z}_2$ is an involutive group homomorphism
\item $f_1(e) \in \mathbb{Z}_2$ is a fixed point of $f_0$, i.e., $f_0(f_1(e)) = f_1(e)$
\item $f_1: \mathbb{Z}_2 \to \mathbb{Z}_2$ is determined by $f_1(a) = f_0(a)f_1(e)$ for all $a \in \mathbb{Z}_2$.
\end{itemize}
The above conditions have only two solutions $f^{[0]}$ and $f^{[1]}$, which are defined by
\begin{equation}
f^{[0]}_n(a) = a, \qquad f_n^{[1]}(a) = a \eta^{n}.
\end{equation}
Here, $\eta$ denotes the generator of $A = \mathbb{Z}_2$.\footnote{We recall that the elements of $A$ are denoted multiplicatively, i.e., $\eta^2 = e$.}
The corresponding complements of $A_L \cong \mathbb{Z}_2$ in $A \wr \mathbb{Z}_2 \cong D_8$ are 
\begin{align}
K(f^{[0]}) &= \{(a, a, n) \mid a \in \mathbb{Z}_2, n \in \mathbb{Z}_2\} = \mathbb{Z}_2^{\text{diag}} \times \mathbb{Z}_2,
\label{eq: K f0}\\
K(f^{[1]}) &= \{(a\eta^n, a, n) \mid a \in \mathbb{Z}_2, n \in \mathbb{Z}_2\} \cong \mathbb{Z}_4.
\label{eq: K f1}
\end{align}
The above solutions are not equivalent to each other because there does not exist $a \in A$ that satisfies \eqref{eq: isom fiber 2-functor}.
Therefore, the corresponding generalized cluster states are in different SPT phases.\footnote{Lattice models for these two SPT phases have already been constructed in \cite{Choi:2024rjm}. One of these models is the ordinary $\mathbb{Z}_2$-cluster model, which realizes the SPT phase corresponding to $K = \mathbb{Z}_2^{\text{diag}} \times \mathbb{Z}_2$. The other model is called the cluster$^{\prime}$ model in \cite{Choi:2024rjm}, which is slightly different from our generalized cluster model. As we will see in Appendix~\ref{sec: cluster prime model}, the cluster$^{\prime}$ model realizes the SPT phase corresponding to $K = \mathbb{Z}_4$. It was shown in \cite{Furukawa:2025flp} that the interface of these models must be degenerate, which is consistent with our results. We also mention that Ref.~\cite{Cao:2025qhg} studied an interface of $2\TY(\mathbb{Z}_2, 1)$ SPT phases with different cocycles $\lambda$, which we do not discuss in this paper.}
In what follows, we will consider the interface between these two SPT phases, as well as the self-interface of each of them.

\vspace*{\baselineskip}
\noindent{\bf The models.}
Let us first write down the generalized cluster models and their ground states for $K = K(f^{[0]})$ and $K = K(f^{[1]})$.
To this end, as in Section~\ref{sec: Generalized cluster states TY}, we choose a gauge so that the matter fields take values in
\begin{equation}
S_{H \backslash G} = \mathbb{Z}_2^{\text{diag}} \times \mathbb{Z}_2.
\label{eq: Z2 TY gauge}
\end{equation}
The state space of the model for both $K = K(f^{[0]})$ and $K = K(f^{[1]})$ is given by
\begin{equation}
\mathcal{H}_{\text{g.f.}} = \left(\bigotimes_{i \in P} \mathbb{C}^{|\mathbb{Z}_2 \times \mathbb{Z}_2|}\right) \otimes \hat{\pi}_{\text{flat}}\left(\bigotimes_{[ij] \in E} \mathbb{C}^{|\mathbb{Z}_2|}\right).
\end{equation}
We note that each plaquette $i \in P$ has a pair of qubits $(a_i, n_i) \in \mathbb{Z}_2 \times \mathbb{Z}_2$, while each edge $[ij] \in E$ has a single qubit $a_{ij} \in \mathbb{Z}_2$.
The Hamiltonians for $K = K(f^{[0]})$ and $K = K({f^{[1]}})$ are given by
\begin{equation}
\begin{aligned}
H_{\text{g.f.}}^{[0]} &= -\sum_{i \in P} \frac{1}{4}(1+X^{(i)})(1+\sigma_x^{(i)}) - \sum_{[ij] \in E} \frac{1}{2}(1+Z^{(ij)}), \\
H_{\text{g.f.}}^{[1]} &= -\sum_{i \in P} \frac{1}{4}(1+X^{(i)})(1+\sigma_x^{(i)}\prod_{[ij] \in \partial i} X^{(ij)}) - \sum_{[ij] \in E} \frac{1}{2}(1+\sigma_z^{(i)}Z^{(ij)}\sigma_z^{(j)}).
\end{aligned}
\end{equation}
Here, $X^{(i)}$ and $Z^{(i)}$ are the Pauli operators acting on the qubit $a_i$ on the plaquette $i$, $\sigma_x^{(i)}$ and $\sigma_z^{(i)}$ are those acting on the qubit $n_i$ on the plaquette $i$, and $X^{(ij)}$ and $Z^{(ij)}$ are those acting on the qubit $a_{ij}$ on the edge $[ij]$.
The ground states of the above Hamiltonians are
\begin{equation}
\begin{aligned}
\ket{\text{$2\TY$-cluster}(\mathbb{Z}_2; f^{[0]})} &= \frac{1}{2^{|P|}} \sum_{\{a_i \in \mathbb{Z}_2\}} \sum_{\{n_i \in \mathbb{Z}_2\}} \ket{\{a_i, n_i; e\}}, \\
\ket{\text{$2\TY$-cluster}(\mathbb{Z}_2; f^{[1]})} &= \frac{1}{2^{|P|}} \sum_{\{a_i \in \mathbb{Z}_2\}} \sum_{\{n_i \in \mathbb{Z}_2\}} \ket{\{a_i, n_i; \eta^{n_i+n_j}\}}.
\end{aligned}
\end{equation}
We note that $\ket{\text{$2\TY$-cluster}(\mathbb{Z}_2; f^{[0]})}$ is the trivial product state,\footnote{This is because we chose a gauge so that $S_{H \backslash G} = K(f^{[0]})$. See Appendix~\ref{sec: Tambara-Yamagami cluster model in another gauge} for the models in another gauge.} while $\ket{\text{$2\TY$-cluster}(\mathbb{Z}_2; f^{[1]})}$ is the tensor product of the $\mathbb{Z}_2$-cluster state and the trivial product state.

\vspace*{\baselineskip}
\noindent{\bf Category theoretical description of interfaces.}
As discussed in Section \ref{sec: Category theoretical description}, the symmetry at the interface of the generalized cluster states is generally described by the strip 2-algebra~\eqref{eq: strip 2-algebra}.
The strip 2-algebra for the interface of the SPT phases labeled by the complements $K_1$ and $K_2$ is denoted by $\mathcal{C}_{K_1, K_2}$.
In the case of $2\TY(\mathbb{Z}_2, 1)$ SPT phases, $K_1$ and $K_2$ are either $K(f^{[0]}) = \mathbb{Z}_2^{\text{diag}} \times \mathbb{Z}_2$ or $K(f^{[1]}) = \mathbb{Z}_4$.
Based on the results in Section~\ref{sec: Interface symmetries CT}, we can compute $\mathcal{C}_{K_1, K_2}$ for each pair of $K_1$ and $K_2$ as follows:
\begin{align}
\mathcal{C}_{K(f^{[0]}), K(f^{[0]})} &\cong \Vect_{\mathbb{Z}_2 \times \mathbb{Z}_2} \oplus \Vect_{\mathbb{Z}_2 \times \mathbb{Z}_2}, \label{eq: C 00}\\
\mathcal{C}_{K(f^{[0]}), K(f^{[1]})} &\cong \mathcal{C}_{K(f^{[1]}), K(f^{[0]})} \cong \Vect_{\mathbb{Z}_2} \boxtimes \Mat_2(\Vect), \label{eq: C 01}\\
\mathcal{C}_{K(f^{[1]}), K(f^{[1]})} &\cong \Vect_{\mathbb{Z}_4} \oplus \Vect_{\mathbb{Z}_4}. \label{eq: C 11}
\end{align}
We note that $\mathcal{C}_{K(f^{[0]}), K(f^{[0]})} \not\cong \mathcal{C}_{K(f^{[1]}), K(f^{[1]})}$, which implies that the symmetry at the self-interface depends on the phase.
This is in contrast to the case of SPT phases with invertible symmetries.

\vspace*{\baselineskip}
\noindent{\bf Tensor network description of interfaces.}
We now provide the tensor network description of the above interface symmetries.
For simplicity, we will focus on the interfaces represented by the tensor networks of the form \eqref{eq: interface state}.
To obtain the symmetry operators at the interface, we first write down the action tensors.
Due to \eqref{eq: action (e, a, 0)} and \eqref{eq: action (e, e, 1)}, the action tensors for the symmetry operators $\mathsf{C}U_a$ and $\mathsf{D}$ acting on $\ket{\text{$2\TY$-cluster}(\mathbb{Z}_2; f^{[m]})}$ are given by
\begin{equation}
\adjincludegraphics[valign = c, trim={10, 10, 10, 10}]{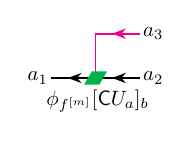} = \adjincludegraphics[valign = c, trim={10, 10, 10, 10}]{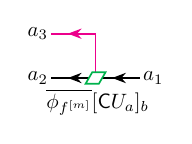} = \delta_{a_1, a_2} \delta_{a_3, a_1ab}, \quad
\adjincludegraphics[valign = c, trim={10, 10, 10, 10}]{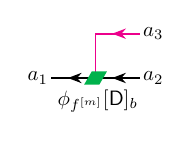} = \adjincludegraphics[valign = c, trim={10, 10, 10, 10}]{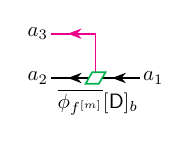} = \delta_{a_1, a_2\eta^m} \delta_{a_3, a_1b}.
\end{equation}
Using these action tensors, we can construct the symmetry operators acting on the interface of $\ket{\text{$2\TY$-cluster}(\mathbb{Z}_2; f^{[m]})}$ and $\ket{\text{$2\TY$-cluster}(\mathbb{Z}_2; f^{[m^{\prime}]})}$ as
\begin{equation}
\hat{\mathsf{L}}(\mathsf{C}U_a; b, b^{\prime}) = \bigotimes_{i: \text{sites}} \hat{\mathcal{O}}_i(\mathsf{C}U_a; b, b^{\prime}), \qquad
\hat{\mathsf{L}}(\mathsf{D}; b, b^{\prime}) = \bigotimes_{i: \text{sites}} \hat{\mathcal{O}}_i(\mathsf{D}; b, b^{\prime}),
\label{eq: TY interface sym op}
\end{equation}
where the on-site operators $\hat{\mathcal{O}}_i(\mathsf{C}U_a; b, b^{\prime})$ and $\hat{\mathcal{O}}_i(\mathsf{D}; b, b^{\prime})$ are defined by
\begin{align}
\hat{\mathcal{O}}_i(\mathsf{C}U_a; b, b^{\prime}) &= \adjincludegraphics[valign = c, trim={10, 10, 10, 10}]{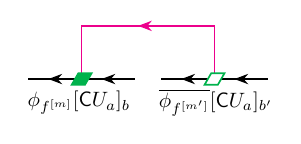} = \frac{1+\text{sgn}(bb^{\prime})Z^{(i)}_{\ell}Z^{(i)}_r}{2}, 
\label{eq: TYZ2 interface onsite1} \\
\hat{\mathcal{O}}_i(\mathsf{D}; b, b^{\prime}) &= \adjincludegraphics[valign = c, trim={10, 10, 10, 10}]{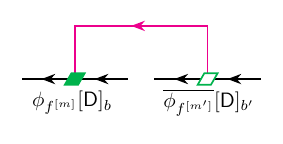} = \frac{1+\text{sgn}(bb^{\prime})Z^{(i)}_{\ell}Z^{(i)}_r}{2}(X^{(i)}_{\ell})^{m}(X^{(i)}_r)^{m^{\prime}}.
\label{eq: TYZ2 interface onsite2}
\end{align}
The operators on the right-hand side act on the middle two legs of the middle diagram.\footnote{The small gap in the middle represents the interface.}
More specifically, $X^{(i)}_{\ell}$ and $Z^{(i)}_{\ell}$ are the Pauli operators acting on the left leg, while $X^{(i)}_r$ and $Z^{(i)}_r$ are those acting on the right leg.
The sign $\text{sgn}(bb^{\prime}) = \pm 1$ is the sign representation of $\mathbb{Z}_2$.
By plugging~\eqref{eq: TYZ2 interface onsite1} and \eqref{eq: TYZ2 interface onsite2} into \eqref{eq: TY interface sym op}, we obtain
\begin{equation}
\hat{\mathsf{L}}(\mathsf{C}U_a; b, b^{\prime}) = P_{\text{sgn}(bb^{\prime})}, \qquad
\hat{\mathsf{L}}(\mathsf{D}; b, b^{\prime}) = P_{\text{sgn}(bb^{\prime})} X_{\ell}^m X_r^{m^{\prime}},
\label{eq: TY interface sym op2}
\end{equation}
where $P_{\pm} := \bigotimes_{i: \text{sites}}\frac{1}{2}(1 \pm Z_{\ell}^{(i)}Z_r^{(i)})$, $X_{\ell} := \bigotimes_{i: \text{sites}} X_{\ell}^{(i)}$, and $X_r := \bigotimes_{i: \text{sites}} X_r^{(i)}$.

Equation~\eqref{eq: TY interface sym op2} shows that the set of the symmetry operators acting on the interface of $\ket{\text{$2\TY$-cluster}(\mathbb{Z}_2; f^{[m]})}$ and $\ket{\text{$2\TY$-cluster}(\mathbb{Z}_2; f^{[m^{\prime}]})}$ is given by
\begin{equation}
\left\{P_+,\quad P_-,\quad P_+X_{\ell}^{m}X_r^{m^{\prime}},\quad P_-X_{\ell}^{m}X_r^{m^{\prime}}\right\}.
\end{equation}
This set agrees with the set of symmetry operators~\eqref{eq: all interface sym op Z2} at the interface of $\ket{\text{$\mathbb{Z}_2$-cluster}(f^{(m)})}$ and $\ket{\text{$\mathbb{Z}_2$-cluster}(f^{(m^{\prime})})}$.
The algebra of these symmetry operators is already computed in Section~\ref{sec: Interfaces Gx2Rep(G)}.
Nevertheless, for completeness, let us describe it again for each pair of $m, m^{\prime} \in \{0, 1\}$.
\begin{itemize}
\item When $(m, m^{\prime}) = (0, 0)$, the set of the symmetry operators reduces to
\begin{equation}
\left\{P_+, \quad P_- \right\}.
\end{equation}
The algebra of these symmetry operators is given by \eqref{eq: sym alg C 00 Z2}, which suggests that the symmetry operators form a multifusion category $\Vect \oplus \Vect$.
The action of the symmetry category~\eqref{eq: C 00} is thus implemented via a tensor functor $\Vect_{\mathbb{Z}_2 \times \mathbb{Z}_2} \oplus \Vect_{\mathbb{Z}_2 \times \mathbb{Z}_2} \to \Vect \oplus \Vect$.
\item When $(m, m^{\prime}) = (0, 1)$, the set of the symmetry operators is
\begin{equation}
\left\{P_+, \quad P_-, \quad P_+X_r, \quad P_-X_r \right\}.
\label{eq: sym alg C01 TY}
\end{equation}
The algebra of these symmetry operators is given by \eqref{eq: sym alg C 01 Z2}, which suggests that the symmetry operators form a multifusion category $\Mat_2(\Vect)$.
The action of the symmetry category~\eqref{eq: C 01} is thus implemented via a tensor functor $\Vect_{\mathbb{Z}_2} \boxtimes \Mat_2(\Vect) \to \Mat_2(\Vect)$.
\item When $(m, m^{\prime}) = (1, 0)$, the set of symmetry operators is
\begin{equation}
\left\{P_+, \quad P_-, \quad P_+X_{\ell}, \quad P_-X_{\ell}\right\}.
\end{equation}
The algebra of these symmetry operators is isomorphic to that of \eqref{eq: sym alg C01 TY}.
\item When $(m, m^{\prime}) = (1, 1)$, the set of the symmetry operators is
\begin{equation}
\left\{P_+, \quad P_-, \quad P_+X_{\ell}X_r, \quad P_-X_{\ell}X_r\right\}.
\end{equation}
The algebra of these symmetry operators is given by \eqref{eq: sym alg C 11 Z2}, which suggests that the symmetry operators form a multifusion category $\Vect_{\mathbb{Z}_2} \boxtimes \Vect_{\mathbb{Z}_2}$.
The action of the symmetry category~\eqref{eq: C 11} is thus implemented via a tensor functor $\Vect_{\mathbb{Z}_4} \oplus \Vect_{\mathbb{Z}_4} \to \Vect_{\mathbb{Z}_2} \oplus \Vect_{\mathbb{Z}_2}$.
\end{itemize}

\subsection{Parameterized families}
\label{sec: Parameterized families TY}
In this subsection, we consider $S^1$-parameterized families of $2\TY(A, 1)$-symmetric states in the same phase as $\ket{\text{$2\TY$-cluster}(A; f)}$.

\vspace*{\baselineskip}
\noindent{\bf Classification.}
Due to the general classification mentioned in Section~\ref{sec: Parameterized families}, the $S^1$-parameterized families are expected to be classified by
\begin{equation}
\{(h, [\xi]) \mid h \in H \text{ s.t. } K= hKh^{-1}, ~ [\xi] \in H^2(K, \mathrm{U}(1))\},
\end{equation}
where $H = A_L$ and $K=K(f)$.
In what follows, we will only consider $S^1$-parameterized families with trivial $[\xi]$.
The condition $K=hKh^{-1}$ for $h=(b,e,0) \in A_L$ means that for any $(f_{n}(a), a, n) \in K(f)$, there exists $(f_{n^{\prime}}(a'), a', n') \in K(f)$ such that
\begin{equation}
       (b,e,0)(f_{n}(a), a, n)(b,e,0)^{-1} = (f_{n^{\prime}}(a'), a', n').
\end{equation}
When $n=0$, this condition produces no constraint on $b\in A$.
When $n=1$, the above condition is equivalent to
\begin{equation}
       (bf_{1}(a), ab^{-1}, 1) = (f_{1}(ab^{-1}), ab^{-1}, 1)\Longleftrightarrow f_{0}(b) = b^{-1}.
\end{equation}
Therefore, $h = (b,e,0)$ satisfies $K=hKh^{-1}$ if and only if $f_{0}(b)=b^{-1}$.
Below, we will simply write $h = (b,e,0)\in H$ as $b$ and $(f_{n}(a),a,n)\in K(f)$ as $(a,n)$.

\vspace*{\baselineskip}
\noindent{\bf $S^1$-parameterized families.}
For each $b \in A$ satisfying $f_0(b) = b^{-1}$, we can construct an $S^1$-parameterized family of $2\TY(A, 1)$-symmetric models by conjugating the Hamiltonian~\eqref{eq: TY ham} with the following $\theta$-dependent unitary operator:
\begin{equation}
U_{\text{g.f.}}^b(\theta) \coloneq \prod_{i \in P} \left.u_b^{(i)}(\theta)\right|_{\text{g.f.}}, \qquad
\left.u_b^{(i)}(\theta)\right|_{\text{g.f.}} \coloneq \exp\left( \frac{\theta}{2\pi} \log\left.\overleftarrow{X}_b^{(i)}\right|_{\text{g.f.}} \right),
\label{eq: TY cluster U}
\end{equation}
Here, $\theta$ runs from $0$ to $2\pi$, and $\left.\overleftarrow{X}_b^{(i)}\right|_{\text{g.f.}}$ is defined by\footnote{This is an example of the dual of a domain wall creation operator, which we introduced in Section~\ref{sec: Parameterized families}. See \eqref{eq: dual of Xhi 2} for the general expression of the dual of a domain wall creation operator.}
\begin{align}
\left.\overleftarrow{X}_{b}^{(i)}\right|_{\text{g.f.}} &= L^{(\partial i)}_{b^{-1}} \otimes P_{0}^{(i)} + L^{(\partial i)}_{b} \otimes X_{b}^{(i)} P_{1}^{(i)},
\label{eq: Xbi gf}
\end{align}
where $P_{n}^{(i)}$ for $n = 0, 1$ is the projection onto the subspace with $n_i = n$, i.e., $P_n^{(i)} \coloneq \frac{1}{2}(1 + (-1)^n \sigma_z^{(i)})$.
We choose the branch of the logarithm as 
\begin{equation}
       \log \left.\overleftarrow{X}_b^{(i)}\right|_{\text{g.f.}} = \sum_{s=0}^{\abs{A} - 1} i \frac{2\pi s}{\abs{A}} P_s[\left.\overleftarrow{X}_b^{(i)}\right|_{\text{g.f.}}],
\end{equation}
where $P_s[\left.\overleftarrow{X}_{b}^{(i)}\right|_{\text{g.f.}}]$ is the projection onto the eigenspace of $\left.\overleftarrow{X}_{b}^{(i)}\right|_{\text{g.f.}}$ with eigenvalue $e^{i\frac{2\pi s}{\abs{A}}}$.
We note that $u_b^{(i)}(\theta)$ interpolates between the identity operator and $\left.\overleftarrow{X}_{b}^{(i)}\right|_{\text{g.f.}}$.
In particular, at $\theta = 2\pi$, the unitary operator in \eqref{eq: TY cluster U} reduces to
\begin{equation}
U_{\text{g.f.}}^b(2\pi) = \prod_{i \in P} \left.\overleftarrow{X}_b^{(i)}\right|_{\text{g.f.}}.
\end{equation}
By a direct computation, we can show that $U_{\text{g.f.}}^b(2\pi)$ commutes with the Hamiltonian~\eqref{eq: TY ham} due to the condition $f_0(b) = b^{-1}$.
Therefore, conjugating \eqref{eq: TY ham} with $U_{\text{g.f.}}^b(\theta)$ gives us a $2\pi$-periodic Hamiltonian.
This $2\pi$-periodic Hamiltonian preserves the $2\TY(A, 1)$ symmetry because $U_{\text{g.f.}}^b(\theta)$ commutes with the symmetry operators for all $\theta \in [0, 2\pi]$.

\vspace*{\baselineskip}
\noindent{\bf Generalized Thouless pumps.}
Let us discuss the generalized Thouless pump associated with the above $S^1$-parameterized family.
To this end, we consider a $\theta$-dependent textured Hamiltonian on an infinite plane, which is obtained by conjugating the Hamiltonian~\eqref{eq: TY ham} by the truncated unitary operator
\begin{equation}
U_{\text{g.f.}}^{b; x>0}(\theta) \coloneq \prod_{i \in P_{>0}} \left.u_b^{(i)}(\theta)\right|_{\text{g.f.}}.
\end{equation}
Here, $P_{>0}$ denotes the set of plaquettes on the right half-plane.
The textured Hamiltonian commutes with the $2\TY(A, 1)$ symmetry for any $\theta \in [0, 2\pi]$ because each local unitary $\left.u_b^{(i)}(\theta)\right|_{\text{g.f.}}$ commutes with this symmetry.
A continuous family of the unique ground states of the textured Hamiltonians is simply given by
\begin{equation}
U_{\text{g.f.}}^{b; x>0}(\theta) \ket{\text{$2\TY$-cluster}(A; f)}.
\end{equation}
In particular, at $\theta = 2\pi$, this ground state reduces to
\begin{equation}
\left(\prod_{i \in P_{>0}} \left.\overleftarrow{X}_b^{(i)}\right|_{\text{g.f.}} \right) \ket{\text{$2\TY$-cluster}(A; f)},
\label{eq: TY pump}
\end{equation}

To identify the pumped excitation at $\theta = 2\pi$, we compute the action of $\left.\overleftarrow{X}_b^{(i)}\right|_{\text{g.f.}}$ for each $i \in P_{>0}$ by using the tensor networks.
To this end, we first recall that the local tensors of $\ket{\text{$2\TY$-cluster}(A; f)}$ are given by 
\begin{equation}
\adjincludegraphics[valign = c, trim={10, 10, 10, 10}]{tikz/out/copy_TY.pdf} = \frac{1}{\sqrt{2|A|}} \prod_{i = 1, 2, 3, 4} \delta_{a_i, \widetilde{f}_{n}(a)}, \qquad
\adjincludegraphics[valign = c, trim={10, 10, 10, 10}]{tikz/out/multiplication_TY.pdf} = \delta_{a, a_1^{-1}a_2},
\end{equation}
where $\widetilde{f}_{0}(a) = f_{0}(a)a^{-1}$.
When the qubit on the plaquette $i$ is in the state with $n_i=0$, the operator $\left.\overleftarrow{X}_b^{(i)}\right|_{\text{g.f.}}$ acts only on the gauge field as the loop operator $L_{b^{-1}}^{(\partial i)}$.
This action can be pushed onto the virtual legs as
\begin{equation}
\begin{aligned}
       \adjincludegraphics[scale=1,trim={10pt 10pt 10pt 10pt},valign = c]{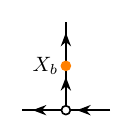}
       \;\;=\;\;
       \adjincludegraphics[scale=1,trim={10pt 10pt 10pt 10pt},valign = c]{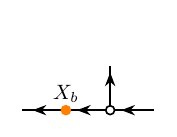}
       \;\;=\;\;
       \adjincludegraphics[scale=1,trim={10pt 10pt 10pt 10pt},valign = c]{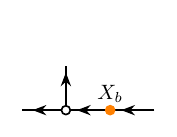},
       \\ 
       \adjincludegraphics[scale=1,trim={10pt 10pt 10pt 10pt},valign = c]{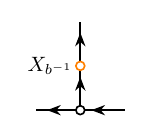}
       \;\;=\;\;
       \adjincludegraphics[scale=1,trim={10pt 10pt 10pt 10pt},valign = c]{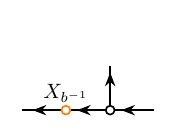}
       \;\;=\;\;
       \adjincludegraphics[scale=1,trim={10pt 10pt 10pt 10pt},valign = c]{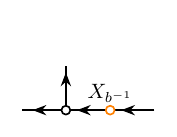}.
\end{aligned}
\label{eq: Xb fractionalization}
\end{equation}
On the other hand, when the qubit on the plaquette $i$ is in the state with $n_i = 1$, the operator $\left.\overleftarrow{X}_b^{(i)}\right|_{\text{g.f.}}$ acts on both the gauge field and the matter field.
The action on the gauge field is given by $L_{b}^{(\partial i)}$, which can be pushed onto the virtual legs as in \eqref{eq: Xb fractionalization}.
The action on the matter field is given by $X_b$, which can be pushed onto the virtual legs as
\begin{equation}
       \adjincludegraphics[scale=1,trim={10pt 10pt 10pt 10pt},valign = c]{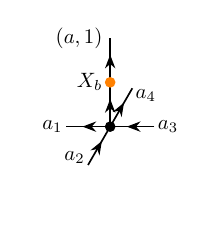}
       = \frac{1}{\sqrt{2|A|}} \prod_{i = 1, 2, 3, 4} \delta_{a_i, \widetilde{f}_{1}(a)b^{-2}}
       =\;\;
       \adjincludegraphics[scale=1,trim={10pt 10pt 10pt 10pt},valign = c]{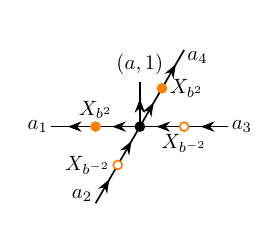}.
\end{equation}
Combining these actions, we obtain
\begin{equation}
\left.\overleftarrow{X}_b^{(i)}\right|_{\text{g.f.}} \adjincludegraphics[scale=1,trim={10pt 10pt 10pt 10pt},valign = c]{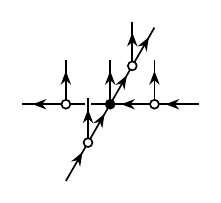}
\;\;=\;\;
\adjincludegraphics[scale=1,trim={10pt 10pt 10pt 10pt},valign = c]{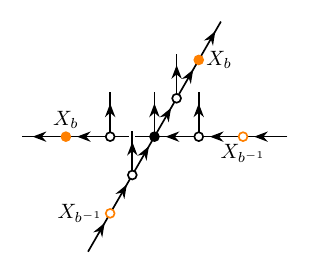}.
\end{equation}
We note that this equation holds for both $n_i = 0$ and $n_i = 1$.
By plugging this into \eqref{eq: TY pump}, we find that the pumped excitation is given by the tensor product of $X_b$'s on the virtual bonds along $x = 0$.
This agrees with the general result obtained in Section~\ref{sec: Parameterized families}.

\begin{acknowledgments}
KI is supported in part by the EPSRC Open Fellowship EP/X01276X/1 and by the Leverhulme-Peierls Fellowship funded by the Leverhulme Trust.
S.O. is supported by the European Union's Horizon 2020 research and innovation programme through grant no. 863476 (ERC-CoG SEQUAM). 
S.O. is also supported by JSPS KAKENHI Grant Number 24K00522.
\end{acknowledgments}

\appendix

\section{Remarks on gauge choice}
\label{sec: Remarks on the choice of representatives}
The generalized cluster model was defined by gauging a subgroup symmetry $H$ in a $G$-symmetric model and then fixing the gauge.
The Hamiltonian and the symmetry operators of the model depend on the gauge choice.
In this appendix, we discuss this dependency.

We first recall that the gauge fixing is specified by the choice of representatives of right $H$-cosets in $G$.
As in the main text, we denote the set of representatives by $S_{H \backslash G}$.
As discussed in Section~\ref{sec: Generalized cluster models}, the state space $\mathcal{H}_{\text{g.f.}}$ and the Hamiltonian $H_{\text{g.f.}}$ after the gauge fixing are given by
\begin{equation}
\mathcal{H}_{\text{g.f.}} = \left(\bigotimes_{i \in P} \mathbb{C}^{|S_{H \backslash G}|}\right) \otimes \hat{\pi}_{\text{flat}} \left(\bigotimes_{[ij] \in E} \mathbb{C}^{|H|}\right), \qquad
H_{\text{g.f.}} = \mathcal{U}_{\text{g.f.}} H_{\text{gauged}} \mathcal{U}_{\text{g.f.}}^{-1},
\label{eq: gauge fixed ss ham}
\end{equation}
where $H_{\text{gauged}}$ is the gauged Hamiltonian~\eqref{eq: gauged Ham} before the gauge fixing, and $\mathcal{U}_{\text{g.f.}}: \mathcal{H}_{\text{gauged}} \to \mathcal{H}_{\text{g.f.}}$ is an isomorphism~\eqref{eq: gauge fixing} between the state spaces before and after the gauge fixing.

Now, let us consider changing the gauge, i.e., we choose another set of representatives of right $H$-cosets in $G$.
The new set of representatives is denoted by $S^{\prime}_{H \backslash G}$.
Due to~\eqref{eq: gauge fixed ss ham}, the Hamiltonian $H^{\prime}_{\text{g.f.}}$ for this gauge is related to that for the previous gauge as
\begin{equation}
H^{\prime}_{\text{g.f.}} = \mathcal{U}^{\prime}_{\text{g.f.}} H_{\text{gauged}} (\mathcal{U}^{\prime}_{\text{g.f.}})^{-1} = \mathcal{U}^{\prime}_{\text{g.f.}} \mathcal{U}_{\text{g.f.}}^{-1} H_{\text{g.f.}} (\mathcal{U}^{\prime}_{\text{g.f.}} \mathcal{U}_{\text{g.f.}}^{-1})^{-1},
\label{eq: gauge changing ham}
\end{equation}
where $\mathcal{U}^{\prime}_{\text{g.f.}}$ is the gauge-fixing isomorphism~\eqref{eq: gauge fixing} for the new gauge.
The above equation shows that the Hamiltonians for different gauges are related by the gauge-changing operator $\mathcal{U}^{\prime}_{\text{g.f.}}\mathcal{U}_{\text{g.f.}}^{-1}$.
Using~\eqref{eq: gauge fixing}, we can compute the action of the gauge-changing operator explicitly as
\begin{equation}
\mathcal{U}^{\prime}_{\text{g.f.}}\mathcal{U}_{\text{g.f.}}^{-1} \hat{\pi}_{\text{flat}} \ket{\{r_i; h_{ij}\}} = \hat{\pi}_{\text{flat}} \ket{\{r_i^{\prime}; \delta r_i h_{ij} \delta r_j^{-1}\}},
\label{eq: gauge changing op}
\end{equation}
where $r_i \in S_{H \backslash G}$ and $r_i^{\prime} \in S^{\prime}_{H \backslash G}$ are the representatives of the same right $H$-coset in $G$, and $\delta r_i \in H$ is the unique element of $H$ that satisfies
$r_i^{\prime} = \delta r_i r_i$.
Equation~\eqref{eq: gauge changing ham} implies that the ground state of $H^{\prime}_{\text{g.f.}}$ is related to that of $H_{\text{g.f.}}$ as
\begin{equation}
\ket{\text{cluster}(G; H; K)}^{\prime} = \mathcal{U}^{\prime}_{\text{g.f.}}\mathcal{U}_{\text{g.f.}}^{-1} \ket{\text{cluster}(G; H; K)}.
\end{equation}
The symmetry operators are also conjugated by $\mathcal{U}^{\prime}_{\text{g.f.}}\mathcal{U}_{\text{g.f.}}^{-1}$ upon changing the gauge.
We emphasize that the symmetry category does not depend on the gauge choice, although the symmetry operators do.

\vspace*{\baselineskip}
\noindent{\bf Example.}
As an example, let us consider the gauge-changing operator for
\begin{equation}
G = G_0 \times G_0, \qquad
H = G_0^{\text{left}}, \qquad
S_{H \backslash G} = G_0^{\text{right}}, \qquad
S^{\prime}_{H \backslash G} = G_0^{\text{diag}},
\end{equation}
where $G_0$ is an arbitrary finite group.
As in the main text, the elements $(g, e) \in G_0^{\text{left}}$, $(e, g) \in G_0^{\text{right}}$, and $(g, g) \in G_0^{\text{diag}}$ will be simply denoted by $g$.
Due to~\eqref{eq: gauge changing op}, the action of the gauge-changing operator in this case can be computed as
\begin{equation}
\mathcal{U}^{\prime}_{\text{g.f.}} \mathcal{U}_{\text{g.f.}}^{-1} \hat{\pi}_{\text{flat}} \ket{\{g_i; g_{ij}\}} = \hat{\pi}_{\text{flat}} \ket{\{g_i; g_i g_{ij} g_j^{-1}\}}.
\label{eq: gauge changing op G0}
\end{equation}
We note that this operator maps the $G_0$-cluster state~\eqref{eq: G cluster} to the trivial product state.
A 1+1d analogue of such an operator was discussed in \cite{Fechisin:2023odt}.

The symmetry of the gauged model is described by $2\Rep(G_0) \boxtimes 2\Vect_{G_0}$ irrespective of the gauge choice.
Nevertheless, the symmetry operators generally depend on the gauge choice.
For example, the invertible symmetry operators for the gauge choices $S_{H \backslash G}$ and $S^{\prime}_{H \backslash G}$ are given by
\begin{equation}
U_g = \bigotimes_{i \in P} \overrightarrow{X}_g^{(i)}, \qquad 
U^{\prime}_g = \mathcal{U}^{\prime}_{\text{g.f.}} \mathcal{U}_{\text{g.f.}}^{-1} U_g (\mathcal{U}^{\prime}_{\text{g.f.}} \mathcal{U}_{\text{g.f.}}^{-1})^{-1} = \left(\bigotimes_{i \in P} \overrightarrow{X}_g^{(i)}\right) \otimes \left(\bigotimes_{[ij] \in E} \overrightarrow{X}_g^{(ij)} \overleftarrow{X}_{g^{-1}}^{(ij)}\right),
\end{equation}
where $\overrightarrow{X}_g$ and $\overleftarrow{X}_{g^{-1}}$ are the left multiplication of $g$ and the right multiplication of $g^{-1}$, respectively.
The above equation shows that $U_g = U^{\prime}_g$ if and only if $G_0$ is abelian.
On the other hand, the non-invertible condensation operator~$\mathsf{C}$ defined by~\eqref{eq: C and Ug} commutes with the gauge-changing operator~\eqref{eq: gauge changing op G0} for any $G_0$, and hence it does not depend on the gauge choice.

When $G_0$ is abelian, the gauge-changing operator $\mathcal{U}_{\text{g.f.}}^{\prime} \mathcal{U}_{\text{g.f.}}^{-1}$ is a symmetric disentangler for the $G_0$-cluster state.\footnote{On the other hand, when $G_0$ is non-abelian, $\mathcal{U}^{\prime}_{\text{g.f.}} \mathcal{U}_{\text{g.f.}}^{-1}$ is not a symmetric disentagler for the $G_0$-cluster state because it does not commute with the symmetry operators. When $G_0$ is non-abelian, it is natural to expect that there is no symmetric disentangler for the $G_0$-cluster state because the symmetry category at the self-interface of the $G_0$-cluster state is not equivalent to that of the trivial product state. A similar argument for the non-existence of symmetric (dis)entanglers in 1+1d is found in \cite{You:2025uxo}.}
In particular, this operator changes the SPT phase: that is, the SPT phase realized in the gauged model depends on the gauge choice.
More specifically, in the gauge specified by $S_{H \backslash G} = G_0^{\text{right}}$, the model with $K = G_0^{\text{right}}$ realizes the trivial phase,\footnote{The trivial phase is well-defined because the symmetry is invertible (up to condensation) when $G_0$ is abelian.} while the model with $K = G_0^{\text{diag}}$ realizes a non-trivial SPT phase.
On the other hand, in the gauge specified by $S_{H \backslash G}^{\prime} = G_0^{\text{diag}}$, the model with $K = G_0^{\text{right}}$ realizes a non-trivial SPT phase, while the model with $K = G_0^{\text{diag}}$ realizes the trivial phase.

\section{Condensation surfaces}
\label{sec: Condensation}
In this appendix, we consider the tensor network representations of condensation surfaces of the group-theoretical fusion 2-category $\mathcal{C}(G; H)$.
We will also discuss the condensation of line operators on some condensation surfaces.

\subsection{Tensor network representations}
\label{sec: DVHe as a condensation surface}
The condensation surfaces are obtained by condensing line operators on the identity surface.
The condensed lines form a $\Delta$-separable symmetric Frobenius algebra in the endomorphism 1-category
\begin{equation}
\End_{\mathcal{C}(G; H)}(I) \cong \Rep(H),
\end{equation}
where $I$ denotes the unit object of $\mathcal{C}(G; H)$.
The surface operator obtained by condensing $A \in \Rep(H)$ will be denoted by $\mathsf{D}[\id; A]$.
We note that $A$ is a semisimple $H$-equivariant algebra.

The condensation surface $\mathsf{D}[\id; A]$ acting on the state space \eqref{eq: state space gf} can be represented by the following tensor network \cite{Delcamp:2023kew, Vancraeynest-DeCuiper:2025wkh}:
\begin{equation}
\mathsf{D}[\id; A] = \adjincludegraphics[valign = c, trim={10, 10, 10, 10}]{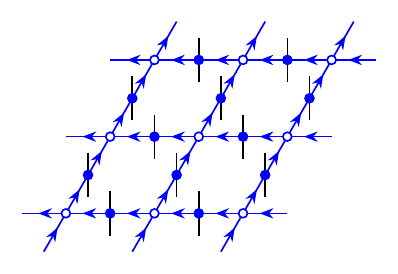}.
\label{eq: condensation TN}
\end{equation}
Here, the virtual Hilbert space is $A$, and the local tensors represented by the filled and unfilled dots are defined by
\begin{equation}
\adjincludegraphics[valign = c, trim={10, 10, 10, 10}]{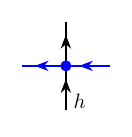} = \adjincludegraphics[valign = c, trim={10, 10, 10, 10}]{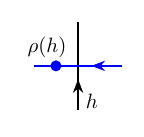}, \qquad
\adjincludegraphics[valign = c, trim={10, 10, 10, 10}]{tikz/out/condensation_tensor5.pdf} = \adjincludegraphics[valign = c, trim={10, 10, 10, 10}]{tikz/out/condensation_tensor6.pdf},
\label{eq: condensation local tensor}
\end{equation}
where $m: A \otimes A \to A$ and $\Delta: A \to A \otimes A$ are the multiplication and comultiplication of $A$, and $\rho: H \to \Aut(A)$ is the action of $H$ on $A$.
The physical legs for the matter fields are omitted in \eqref{eq: condensation TN} because $\mathsf{D}[\id; A]$ acts only on the gauge fields.

When $A \in \Rep(H)$ is the dual group algebra $\mathbb{C}[H]^*$ equipped with the regular $H$-action, the condensation surface $\mathsf{D}[\id; A]$ agrees with the symmetry operator $\mathsf{D}[V_H^e]$, i.e., we have
\begin{equation}
\mathsf{D}[\id; \mathbb{C}[H]^*] = \mathsf{D}[V_H^e].
\label{eq: full condensation surface}
\end{equation}
Here, we recall that $\mathsf{D}[V_H^e] = \mathsf{D}_H \overline{\mathsf{D}}_H$, where $\mathsf{D}_H$ and $\overline{\mathsf{D}}_H$ are the gauging operator~\eqref{eq: D gf} and the ungauging operator~\eqref{eq: D gf bar}.
To show \eqref{eq: full condensation surface}, let us first spell out the $H$-equivariant algebra structure on $\mathbb{C}[H]^*$.
\begin{itemize}
\item The multiplication $m$ and comultiplication $\Delta$ for $\mathbb{C}[H]^*$ are given by
\begin{equation}
m(v^{h_1} \otimes v^{h_2}) = \delta_{h_1, h_2} v^{h_1}, \qquad
\Delta(v^{h_1}) = v^{h_1} \otimes v^{h_1},
\label{eq: m Delta CH*}
\end{equation} 
where $\{v^h \mid h \in H\}$ is a basis of $\mathbb{C}[H]^*$.
\item The $H$-action $\rho: H \to \Aut(\mathbb{C}[H]^*)$ is given by
\begin{equation}
\rho(h)(v^{h_1}) = v^{h_1h^{-1}},
\end{equation}
with respect to which $m$ and $\Delta$ in \eqref{eq: m Delta CH*} are equivariant.
\end{itemize}
For the above choice of $A \in \Rep(H)$, the local tensors defined by \eqref{eq: condensation local tensor} reduce to
\begin{equation}
\adjincludegraphics[valign = c, trim={10, 10, 10, 10}]{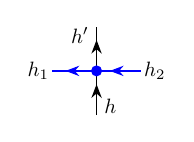} = \delta_{h, h_1^{-1}h_2} \delta_{h^{\prime}, h}, \qquad
\adjincludegraphics[valign = c, trim={10, 10, 10, 10}]{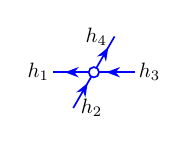} = \prod_{i, j = 1, 2, 3, 4} \delta_{h_i, h_j}.
\label{eq: H dual condensation}
\end{equation}
These tensors agree with the local tensors~\eqref{eq: unfilled} and \eqref{eq: filled} of the PEPO representation of $\mathsf{D}[V_H^g]$ with $g = e$.\footnote{The derivation of the local tensors \eqref{eq: filled} and \eqref{eq: unfilled} for $\mathsf{D}[V_H^g]$ with general $g \in G$ is based on the assumption that $\mathcal{C}(G; H)$ is non-anomalous. Nevertheless, when $g = e$, these expressions are applicable to any $\mathcal{C}(G; H)$.}
Thus, we conclude that $\mathsf{D}[\id; \mathbb{C}[H]^*] = \mathsf{D}[V_H^e]$, which shows \eqref{eq: full condensation surface}.

\subsection{Condensation on a condensation surface}
\label{sec: Condensation on a condensation surface}
One can recover the identity surface by condensing line operators on the condensation surfaces.
In what follows, we will show that condensing the line operators forming the group algebra $\mathbb{C}[H]$ on the condensation surface $\mathsf{D}[V_H^e]$ gives us back the identity surface.
Namely, we will show
\begin{equation}
\mathsf{D}[V_H^e; \mathbb{C}[H]] = \mathsf{D}[\id],
\label{eq: condensation back}
\end{equation}
where $\mathsf{D}[\id]$ is the identity operator acting on the state space~\eqref{eq: state space gf}.

To show \eqref{eq: condensation back}, we first recall that $\mathsf{D}[V_H^e; \mathbb{C}[H]]$ can be represented by the following PEPO (cf. \eqref{eq: sym op condensation reduced}):
\begin{equation}
\mathsf{D}[V_H^e; \mathbb{C}[H]] = \adjincludegraphics[valign = c, trim={10, 10, 10, 10}]{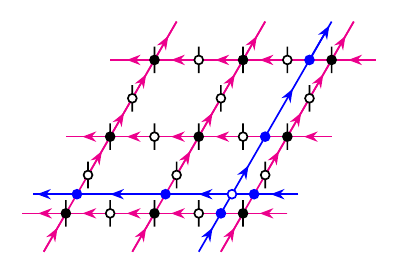}.
\label{eq: id PEPO}
\end{equation}
Here, all virtual bonds take values in $H$, and the local tensors in the above equation are defined by
\begin{align}
\adjincludegraphics[valign = c, trim={10, 10, 10, 10}]{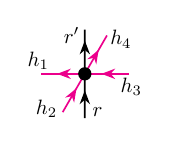} &= \delta_{r, r^{\prime}} \prod_{i, j = 1, 2, 3, 4} \delta_{h_i, h_j}, \qquad
\adjincludegraphics[valign = c, trim={10, 10, 10, 10}]{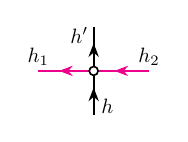} = \delta_{h, h_1^{-1}h_2} \delta_{h, h^{\prime}}, \label{eq: id PEPO tensor 1}\\
\adjincludegraphics[valign = c, trim={10, 10, 10, 10}]{tikz/out/full_condensation_tensor2.pdf} &= \frac{1}{|H|} \delta_{h_1h_4, h_2h_3}, \qquad
\adjincludegraphics[valign = c, trim={10, 10, 10, 10}]{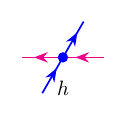}  = \adjincludegraphics[valign = c, trim={10, 10, 10, 10}]{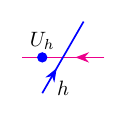} , \qquad
\adjincludegraphics[valign = c, trim={10, 10, 10, 10}]{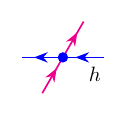}  = \adjincludegraphics[valign = c, trim={10, 10, 10, 10}]{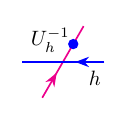}. \label{eq: id PEPO tensor 2}
\end{align}
We note that the PEPO in~\eqref{eq: id PEPO} reduces to $\mathsf{D}[\id; \mathbb{C}[H^*]]$ upon removing the additional blue lines. Indeed, the local tensors~\eqref{eq: id PEPO tensor 1} that do not involve the blue lines agree with the local tensors~\eqref{eq: H dual condensation} of $\mathsf{D}[\id; \mathbb{C}[H^*]]$.\footnote{We recall that the physical leg for the matter field is omitted in the second equation in~\eqref{eq: H dual condensation}.}
Due to \eqref{eq: id PEPO tensor 2}, the symmerty operator \eqref{eq: id PEPO} can be decomposed into a sum
\begin{equation}
\mathsf{D}[V_H^e; \mathbb{C}[H]] = \frac{1}{|H|} \sum_{h_x, h_y \in H} \delta_{h_xh_y, h_yh_x} \mathsf{D}[V_H^e; \mathbb{C}[H]]_{h_x, h_y},
\label{eq: id PEPO decomposition}
\end{equation}
where the summand on the right-hand side is defined by
\begin{equation}
\mathsf{D}[V_H^e; \mathbb{C}[H]]_{h_x, h_y} = \adjincludegraphics[valign = c, trim={10, 10, 10, 10}]{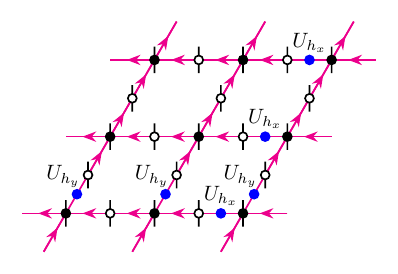}.
\label{eq: id PEPO summand}
\end{equation}
The action of $\mathsf{D}[V_H^e; \mathbb{C}[H]]_{h_x, h_y}$ on a state $\ket{\{r_i; h_{ij}\}}$ can be computed as
\begin{equation}
\mathsf{D}[V_H^e; \mathbb{C}[H]]_{h_x, h_y} \ket{\{r_i; h_{ij}\}} = \sum_{h \in H} \delta_{\text{hol}_{x; *}(\{h_{ij}\}), h^{-1}h_xh} \delta_{\text{hol}_{y; *}(\{h_{ij}\}), h^{-1}h_yh} \ket{\{r_i; h_{ij}\}},
\label{eq: id PEPO summand action}
\end{equation}
where $\text{hol}_{a; *}(\{h_{ij}\})$ for $a = x, y$ is the holonomy of the gauge fields along the $a$ direction with the base point $*$.
The above equation shows that the summand~\eqref{eq: id PEPO summand} is the projection to the subspace in which the gauge fields have the holonomy $h^{-1}h_xh$ and $h^{-1}h_yh$ along the $x$ and $y$ directions, where $h \in H$ is arbitrary. 
By substituting~\eqref{eq: id PEPO summand action} into \eqref{eq: id PEPO decomposition}, we find
\begin{equation}
\begin{aligned}
\mathsf{D}[V_H^e; \mathbb{C}[H]] \ket{\{r_i; h_{ij}\}} &= \frac{1}{|H|} \sum_{h_x, h_y, h \in H} \delta_{h_xh_y, h_yh_x} \delta_{\text{hol}_{x; *}(\{h_{ij}\}), h^{-1}h_xh} \delta_{\text{hol}_{y; *}(\{h_{ij}\}), h^{-1}h_yh} \ket{\{r_i; h_{ij}\}} \\
&= \frac{1}{|H|} \sum_{h \in H} \left(\sum_{h_x, h_y \in H} \delta_{h_xh_y, h_yh_x} \delta_{\text{hol}_{x; *}(\{h_{ij}\}), h_x} \delta_{\text{hol}_{y; *}(\{h_{ij}\}), h_y} \right) \ket{\{r_i; h_{ij}\}}\\
& = \ket{\{r_i; h_{ij}\}}.
\end{aligned}
\end{equation}
The last equality follows from $\sum_{h_x, h_y \in H} \delta_{h_xh_y, h_yh_x} \delta_{\text{hol}_{x; *}(\{h_{ij}\}), h_x} \delta_{\text{hol}_{y; *}(\{h_{ij}\}), h_y} = 1$.
The above equation implies that $\mathsf{D}[V_H^e]$ acts as the identity on the state space, which shows~\eqref{eq: condensation back}.\footnote{More specifically, the PEPO~\eqref{eq: id PEPO} acts as the projector $\hat{\pi}_{\text{flat}}$ on the vector space $\left(\bigotimes_{i \in P} \mathbb{C}^{|S_{H \backslash G}|}\right) \otimes \left(\bigotimes_{[ij] \in E} \mathbb{C}^{|H|}\right)$ and thus reduces to the identity operator on the state space~\eqref{eq: state space gf}.}
Intuitively, one may think of the insertion of the line operators in \eqref{eq: id PEPO} as summing over all possible holonomies of the gauge fields, which amounts to turning $\mathsf{D}[V_H^e]$ into the identity surface $\mathsf{D}[\id]$.

\section{Derivations}
\label{sec: Derivations}
In this appendix, we derive several equalities of tensor network operators.
We will follow the notations used in the main text.

\subsection{Fractionalized symmetries \eqref{eq: fractionalization}}
\label{sec: Derivation of fractionalized symmetry operators}

To derive the fractionalized symmetry operator~\eqref{eq: fractionalization}, we first compute the action of $\mathsf{D}[V_H^g]$ on the following local patch of the generalized cluster PEPS~\eqref{eq: generalized cluster PEPS}:
\begin{equation}
\adjincludegraphics[valign = c, trim={10, 10, 10, 10}]{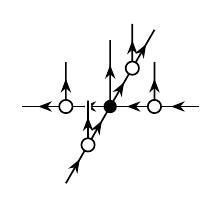}.
\label{eq: local patch}
\end{equation}
Since $\mathsf{D}[V_H^g]$ is given by the composition~\eqref{eq: 0-form} of three operators $\mathsf{D}_H$, $U_g$, and $\overline{\mathsf{D}}_H$, it suffices to compute the action of each of these operators consecutively.
We will later decompose the action of $\mathsf{D}[V_H^g]$ on~\eqref{eq: local patch} into the symmetry actions on the local tensors represented by the black and white dots.

\vspace*{\baselineskip}
\noindent{\bf Action of $\overline{\mathsf{D}}_H$.}
The action of the ungauging operator $\overline{\mathsf{D}}_H$ on the local patch~\eqref{eq: local patch} can be computed as
\begin{equation}
\adjincludegraphics[valign = c, trim={10, 10, 10, 10}]{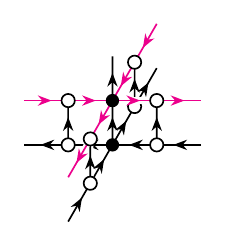} ~ = ~ \adjincludegraphics[valign = c, trim={10, 10, 10, 10}]{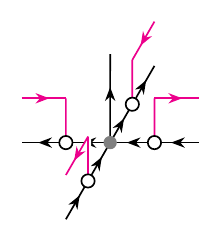},
\qquad \text{where }
\adjincludegraphics[valign = c, trim={10, 10, 10, 10}]{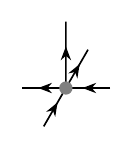} := \adjincludegraphics[valign = c, trim={10, 10, 10, 10}]{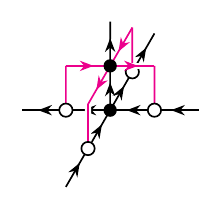}.
\label{eq: local action DH}
\end{equation}
The non-zero components of the gray tensor are given by
\begin{equation}
\adjincludegraphics[valign = c, trim={10, 10, 10, 10}]{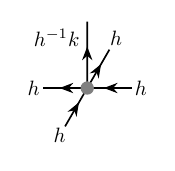} = \frac{1}{\sqrt{|K|}}, \qquad \forall h \in H, ~ \forall k \in K.
\end{equation}
We note that the PEPS generated by this tensor is the symmetric linear combination of the $G/K$-SSB ground states~\eqref{eq: GS G/K}.
Thus, equation~\eqref{eq: local action DH} is consistent with the global action~\eqref{eq: ungauging GS} of $\overline{\mathsf{D}}_H$ on the generalized cluster state.

\vspace*{\baselineskip}
\noindent{\bf Action of $U_g$.}
The action of the symmetry operator $U_g$ on the gray tensor in \eqref{eq: local action DH} can be computed as
\begin{equation}
\adjincludegraphics[valign = c, trim={10, 10, 10, 10}]{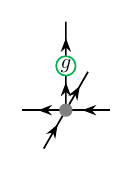} = \adjincludegraphics[valign = c, trim={10, 10, 10, 10}]{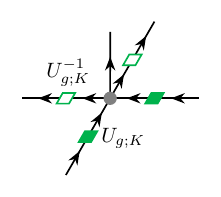},
\label{eq: local action Ug}
\end{equation}
where $U_{g; K}: H \to H$ is defined by the condition
\begin{equation}
U_{g; K}(h)gh^{-1} \in K.
\end{equation}
We note that $U_{g; K}(h)$ satisfying the above condition is unique for any $h \in H$ because $K$ is a complement of $H$ in $G$.
The map $U_{g; K}: H \to H$ is invertible (cf. footnote\ref{fn: UgK inverse}), and its inverse is denoted by $U_{g; K}^{-1}$.
Equation~\eqref{eq: local action Ug} makes it manifest that the PEPS generated by the gray tensor is $G$-symmetric.

\vspace*{\baselineskip}
\noindent{\bf Action of $\mathsf{D}_H$.}
The action of the gauging operator $\mathsf{D}_H$ on the gray tensor can be computed as
\begin{equation}
\adjincludegraphics[valign = c, trim={10, 10, 10, 10}]{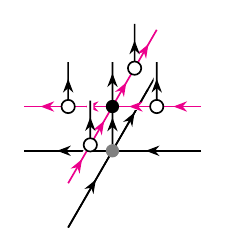} ~ = ~ \adjincludegraphics[valign = c, trim={10, 10, 10, 10}]{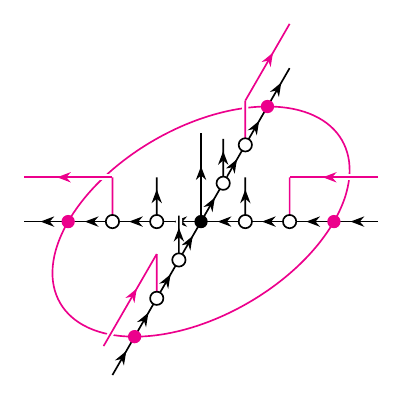}.
\label{eq: local action DH bar}
\end{equation}
The virtual bonds on the loop, as well as the other virtual bonds, have the bond dimension $|H|$, and the magenta dot on the right-hand side represents the copy tensor.

\vspace*{\baselineskip}
\noindent{\bf Action of $\mathsf{D}[V_H^g]$.}
Combining \eqref{eq: local action DH}, \eqref{eq: local action Ug}, and \eqref{eq: local action DH bar}, we find that the action of $\mathsf{D}[V_H^g]$ on the local patch \eqref{eq: local patch} is given by
\begin{equation}
\adjincludegraphics[valign = c, trim={10, 10, 10, 10}]{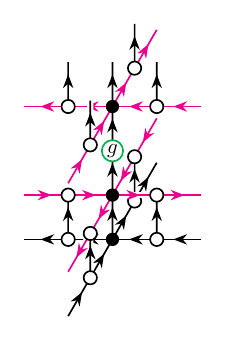} ~ = ~ \adjincludegraphics[valign = c, trim={10, 10, 10, 10}]{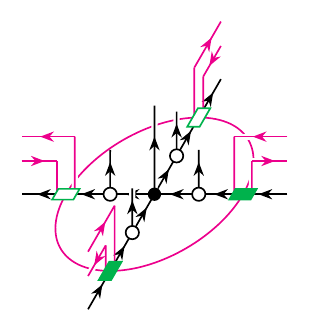}.
\label{eq: local action DVHg}
\end{equation}
The local tensors on the right-hand side are defined by
\begin{equation}
\adjincludegraphics[valign = c, trim={10, 10, 10, 10}]{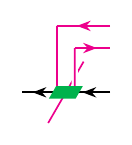} := \adjincludegraphics[valign = c, trim={10, 10, 10, 10}]{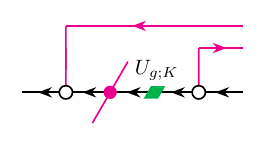}\,, \qquad \qquad
\adjincludegraphics[valign = c, trim={10, 10, 10, 10}]{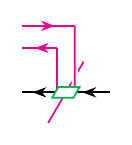} := \adjincludegraphics[valign = c, trim={10, 10, 10, 10}]{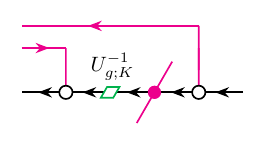}.
\end{equation}
Equation~\eqref{eq: local action DVHg} is not yet satisfactory in that it does not provide the symmetry action on each local tensor of the generalized cluster state.
In what follows, we obtain the action of $\mathsf{D}[V_H^g]$ on each local tensor by reducing the bond dimension of the symmetry PEPO as in~\eqref{eq: reduced PEPO}.

\vspace*{\baselineskip}
\noindent{\bf Reduction of bond dimension.}
As we discussed in Section~\ref{sec: Symmetry operators}, one can reduce the bond dimension of the symmetry PEPO by using the three-leg tensors in \eqref{eq: filled}.\footnote{We recall that $\mathcal{C}(G; H)$ is supposed to be non-anomalous, and hence the reduction of the bond dimension works.}
Reducing the bond dimensions on both sides of \eqref{eq: local action DVHg} leads to
\begin{equation}
\adjincludegraphics[valign = c, trim={10, 10, 10, 10}]{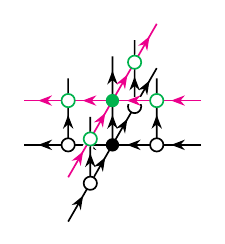} ~ = ~ \adjincludegraphics[valign = c, trim={10, 10, 10, 10}]{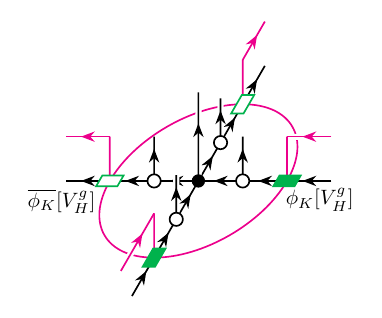}.
\end{equation}
The filled and unfilled green dots on the left-hand side are given by \eqref{eq: filled} and \eqref{eq: unfilled}, while $\phi_K[V_H^g]$ and $\overline{\phi_K}[V_H^g]$ on the right-hand side are defined by \eqref{eq: fractionalization MPO}.
When the physical legs on the edges are labeled by the unit element $e \in H$, the above equation reduces to
\begin{equation}
\adjincludegraphics[valign = c, trim={10, 10, 10, 10}]{tikz/out/local_action1.pdf} ~ = ~ \adjincludegraphics[valign = c, trim={10, 10, 10, 10}]{tikz/out/local_action2.pdf}.
\label{eq: sym action on black}
\end{equation}
Here, we used the following equations:
\begin{equation}
\adjincludegraphics[valign = c, trim={10, 10, 10, 10}]{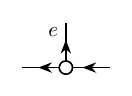} = \adjincludegraphics[valign = c, trim={10, 10, 10, 10}]{tikz/out/orthogonality2.pdf}, \qquad
\adjincludegraphics[valign = c, trim={10, 10, 10, 10}]{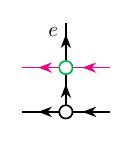} = \adjincludegraphics[valign = c, trim={10, 10, 10, 10}]{tikz/out/completeness2.pdf}.
\end{equation}
Equation~\eqref{eq: sym action on black} shows the action of $\mathsf{D}[V_H^g]$ on the local tensor represented by the black dot.
Similarly, the symmetry action on the local tensor represented by the white dot is given by
\begin{equation}
\adjincludegraphics[valign = c, trim={10, 10, 10, 10}]{tikz/out/local_action3.pdf} ~ = ~ \adjincludegraphics[valign = c, trim={10, 10, 10, 10}]{tikz/out/local_action4.pdf},
\label{eq: sym action on white}
\end{equation}
which follows from a direct computation as follows:
\begin{equation}
\begin{aligned}
\text{LHS}
&\overset{\eqref{eq: completeness}}{=} \adjincludegraphics[valign = c, trim={10, 10, 10, 10}]{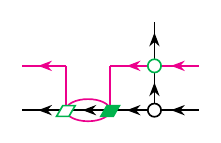}
\overset{\eqref{eq: unfilled}}{=} \adjincludegraphics[valign = c, trim={10, 10, 10, 10}]{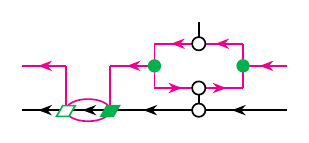}\\
& \overset{\eqref{eq: fractionalization MPO}}{=} \adjincludegraphics[valign = c, trim={10, 10, 10, 10}]{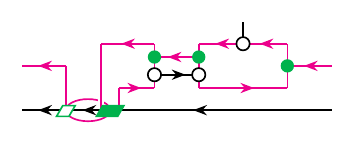}
= \adjincludegraphics[valign = c, trim={10, 10, 10, 10}]{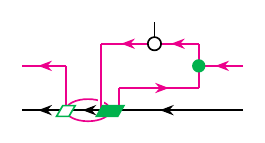}
 \overset{\eqref{eq: fractionalization MPO}}{=} \text{RHS}.
\end{aligned}
\end{equation}
In the first equality, we used the completeness relation~\eqref{eq: completeness}, which we will show in the next subsection.
Equations~\eqref{eq: sym action on black} and \eqref{eq: sym action on white} agree with \eqref{eq: fractionalization}, and hence this completes the derivation.

\subsection{Orthogonality and completeness \eqref{eq: orthogonality and completeness}}
\label{sec: Orthogonality and completeness relations}
In this subsection, we show the following orthogonality and completeness relations for the action tensors:
\begin{align}
\adjincludegraphics[valign = c, trim={10, 10, 10, 10}]{tikz/out/orthogonality.pdf} &= \delta_{h_1, h_2} \adjincludegraphics[valign = c, trim={10, 10, 10, 10}]{tikz/out/orthogonality2.pdf},
\label{eq: orthogonality} \\
\sum_{h \in H} \adjincludegraphics[valign = c, trim={10, 10, 10, 10}]{tikz/out/completeness.pdf} &= \adjincludegraphics[valign = c, trim={10, 10, 10, 10}]{tikz/out/completeness2.pdf}. 
\label{eq: completeness}
\end{align}
We recall that the action tensors $\phi_K[V_H^g]_h$ and $\overline{\phi_K}[V_H^g]_h$ are defined by \eqref{eq: action tensors}.

The orthogonality relation~\eqref{eq: orthogonality} can be verified by a direct computation as
\begin{equation}
\begin{aligned}
\text{LHS}
&= \adjincludegraphics[valign = c, trim={10, 10, 10, 10}]{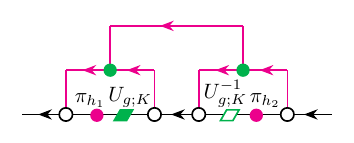}
= \adjincludegraphics[valign = c, trim={10, 10, 10, 10}]{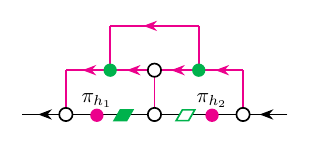} \\
&= \adjincludegraphics[valign = c, trim={10, 10, 10, 10}]{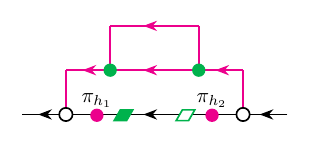}
= \delta_{h_1, h_2} \adjincludegraphics[valign = c, trim={10, 10, 10, 10}]{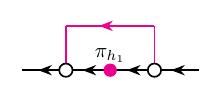}
= \text{RHS}.
\end{aligned}
\end{equation}
Here, the first equality on the second line follows from the fact that the last diagram on the first line is non-zero only when the middle vertical line is labeled by the unit element $e \in H$.

The completeness relation~\eqref{eq: completeness} automatically follows from the orthogonality relation~\eqref{eq: orthogonality}.
To see this, we note that \eqref{eq: orthogonality} and \eqref{eq: completeness} can be written respectively as
\begin{equation}
\adjincludegraphics[valign = c, trim={10, 10, 10, 10}]{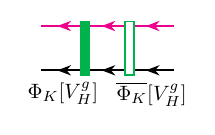} = \adjincludegraphics[valign = c, trim={10, 10, 10, 10}]{tikz/out/completeness2.pdf}, \qquad
\adjincludegraphics[valign = c, trim={10, 10, 10, 10}]{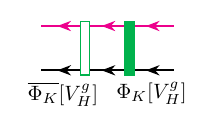} = \adjincludegraphics[valign = c, trim={10, 10, 10, 10}]{tikz/out/completeness2.pdf} \, ,
\label{eq: basis independent oc}
\end{equation}
where the four-leg tensors $\Phi_K[V_H^g]$ and $\overline{\Phi_K}[V_H^g]$ are defined by
\begin{equation}
\adjincludegraphics[valign = c, trim={10, 10, 10, 10}]{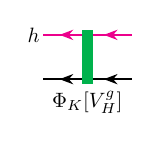} \coloneq \adjincludegraphics[valign = c, trim={10, 10, 10, 10}]{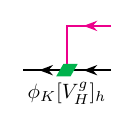}, \qquad
\adjincludegraphics[valign = c, trim={10, 10, 10, 10}]{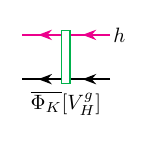} \coloneq \adjincludegraphics[valign = c, trim={10, 10, 10, 10}]{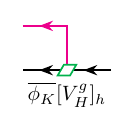} \,.
\label{eq: Phi four leg}
\end{equation}
The first equality (i.e., the orthogonality relation) of \eqref{eq: basis independent oc} shows that $\overline{\Phi_K}[V_H^g]$ is the right inverse of $\Phi_K[V_H^g] \in \End(\mathbb{C}[H]^{\otimes 2})$.
In general, the right inverse of a linear map $f \in \End(V)$ for a finite dimensional vector space $V$ is also the left inverse of $f$.
Therefore, $\overline{\Phi_K}[V_H^g]$ is also the left inverse of $\Phi_K[V_H^g]$.
This shows the second equality (i.e., the completeness relation) of \eqref{eq: basis independent oc}.

\subsection{Fusion rules \eqref{eq: interface sym fusion rules}}
\label{sec: Derivation of the fusion rules}
In this subsection, we derive the fusion rules~\eqref{eq: interface sym fusion rules} of the symmetry operators acting on the interface of the generalized cluster states.
Since the symmetry operators at the interface are given by the tensor product of on-site operators as in~\eqref{eq: interface symmetry op}, it suffices to compute the fusion rules of the on-site operators.
Using the tensor network representation~\eqref{eq: interface symmetry op TN}, we can compute the fusion rules of the on-site operators as follows:
\begin{equation}
\begin{aligned}
&\quad \hat{\mathcal{O}}_i(g; h_1, h_2) \hat{\mathcal{O}}_i(g^{\prime}; h_1^{\prime}, h_2^{\prime}) = \adjincludegraphics[valign = c, trim={10, 10, 10, 10}]{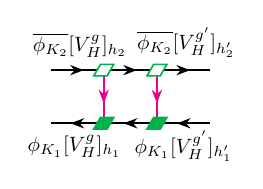} = \adjincludegraphics[valign = c, trim={10, 10, 10, 10}]{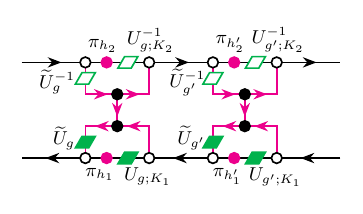}\\
&= \adjincludegraphics[valign = c, trim={10, 10, 10, 10}]{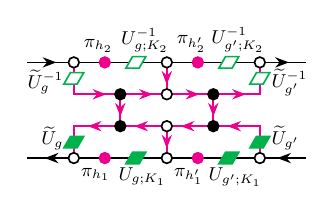}
= \delta_{h_1^{\prime \prime}, h_2^{\prime \prime}} \adjincludegraphics[valign = c, trim={10, 10, 10, 10}]{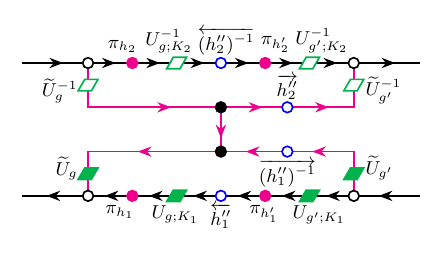}\\
&= \delta_{h_1^{\prime \prime}, h_2^{\prime \prime}} \adjincludegraphics[valign = c, trim={10, 10, 10, 10}]{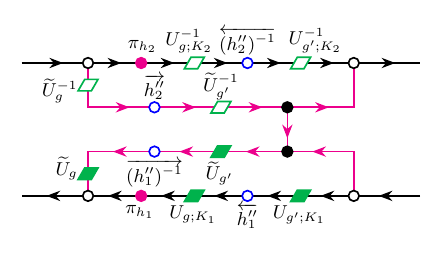}.
\end{aligned}
\label{eq: interface sym fusion derivation}
\end{equation}
Here, $U_{g; K}$ and $\widetilde{U}_g := \widetilde{U}_{g; S_{H \backslash G}}$ are defined by \eqref{eq: U and U tilde}.
The unlabeled black and white dots in the above diagrams represent the copy and multiplication tensors.
On the second line of \eqref{eq: interface sym fusion derivation}, we defined 
\begin{equation}
h_1^{\prime \prime} := (h_1^{\prime})^{-1} U^{-1}_{g; K_1}(h_1), \qquad
h_2^{\prime \prime} := (h_2^{\prime})^{-1} U^{-1}_{g; K_2}(h_2).
\end{equation}
The operators $\overrightarrow{h}$ and $\overleftarrow{h}$ denote the left multiplication and right multiplication of $h$.
Using the relations
\begin{equation}
\widetilde{U}_{g} \circ \overrightarrow{h} \circ \widetilde{U}_{g^{\prime}} = \widetilde{U}_{ghg^{\prime}}, \qquad
U_{g; K} \circ \overleftarrow{h} \circ U_{g^{\prime}; K} = U_{gh^{-1}g^{\prime}; K},
\end{equation}
we find that \eqref{eq: interface sym fusion derivation} reduces to
\begin{equation}
\hat{\mathcal{O}}_i(g; h_1, h_2) \hat{\mathcal{O}}_i(g^{\prime}; h_1^{\prime}, h_2^{\prime}) = \delta_{h_1^{\prime \prime}, h_2^{\prime \prime}} \adjincludegraphics[valign = c, trim={10, 10, 10, 10}]{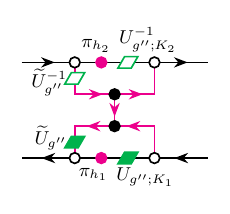}
= \delta_{h_1^{\prime \prime}, h_2^{\prime \prime}} \hat{\mathcal{O}}_i(g^{\prime \prime}; h_1, h_2),
\end{equation}
where $g^{\prime \prime} := g(h_1^{\prime \prime})^{-1}g^{\prime}$.
This equation implies the fusion rules~\eqref{eq: interface sym fusion rules}.

\subsection{Symmetry operators for isomorphic objects \eqref{eq: isomorphic sym op}}
\label{sec: Equality of symmetry operators}
In this subsection, we show \eqref{eq: isomorphic sym op}, i.e., for any $h_1, h_2, h_1^{\prime}, h_2^{\prime} \in H$ and $g, g^{\prime} \in G$, we show
\begin{equation}
(h_2h_1^{-1}, h_1gH) = (h_2^{\prime}(h_1^{\prime})^{-1}, h_1^{\prime}g^{\prime}H) ~\Rightarrow~ \hat{\mathsf{L}}(g; h_1, h_2) = \hat{\mathsf{L}}(g^{\prime}; h_1^{\prime}, h_2^{\prime}),
\end{equation}
where $\hat{\mathsf{L}}(g; h_1, h_2)$ is the symmetry operator~\eqref{eq: interface symmetry op} acting on the interface of the generalized cluster states.
To show the above equation, we choose $(g, h_1, h_2)$ and $(g^{\prime}, h_1^{\prime}, h_2^{\prime})$ so that they satisfy
\begin{equation}
h_2h_1^{-1} = h_2^{\prime}(h_1^{\prime})^{-1}, \qquad
h_1gH = h_1^{\prime} g^{\prime} H.
\label{eq: assumption}
\end{equation}
We note that this equation also implies $h_2gH = h_2^{\prime}g^{\prime}H$.
Due to \eqref{eq: on-site sym action}, the equality $\hat{\mathsf{L}}(g; h_1, h_2) = \hat{\mathsf{L}}(g^{\prime}; h_1^{\prime}, h_2^{\prime})$ reduces to
\begin{align}
U_{g; K_2}^{-1}(h_2) (U_{g; K_1}^{-1}(h_1))^{-1} &= U_{g^{\prime}; K_2}^{-1}(h_2^{\prime}) (U_{g^{\prime}; K_1}^{-1}(h_1^{\prime}))^{-1} \label{eq: first eq} \\
h_1 \widetilde{U}_{g; S_{H \backslash G}}((U_{g; K_1}^{-1}(h_1))^{-1}h_{\ell}) &= h_1^{\prime} \widetilde{U}_{g^{\prime}; S_{H \backslash G}}((U_{g^{\prime}; K_1}^{-1}(h_1^{\prime}))^{-1}h_{\ell}), \quad \forall h_{\ell} \in H, \label{eq: second eq}
\end{align}
where $U_{g; K}: H \to H$ and $\widetilde{U}_{g; S_{H \backslash G}}: H \to H$ are defined by \eqref{eq: U and U tilde}.
Thus, it suffices to show \eqref{eq: first eq} and \eqref{eq: second eq} under the assumption~\eqref{eq: assumption}.

We first show the first equality~\eqref{eq: first eq}.
To this end, we recall that due to the definition of $U_{g; K}$, there exist unique $k_1, k_1^{\prime} \in K_1$ and $k_2, k_2^{\prime} \in K_2$ such that
\begin{alignat}{2}
h_1g &= k_1 U_{g; K_1}^{-1}(h_1), & \qquad
h_2g &= k_2 U_{g; K_2}^{-1}(h_2), \label{eq: h1g h2g} \\
h_1^{\prime}g^{\prime} &= k_1^{\prime} U_{g^{\prime}; K_1}^{-1}(h_1^{\prime}), &
h_2^{\prime}g^{\prime} &= k_2^{\prime} U_{g^{\prime}; K_2}^{-1}(h_2^{\prime}). \label{eq: h1g h2g prime}
\end{alignat}
Therefore, the left-hand side and the right-hand side of \eqref{eq: first eq} can be written as
\begin{equation}
\text{LHS} = k_2^{-1}h_2h_1^{-1}k_1, \qquad
\text{RHS} = (k_2^{\prime})^{-1}h_2^{\prime}(h_1^{\prime})^{-1}k_1^{\prime}.
\label{eq: LHS RHS}
\end{equation}
On the other hand, since $h_1gH = h_1^{\prime}g^{\prime}H$ and $h_2gH = h_2^{\prime}g^{\prime}H$ by assumption, it follows that
\begin{equation}
k_1 = k_1^{\prime}, \qquad
k_2 = k_2^{\prime},
\label{eq: k = k prime}
\end{equation}
Here, we used the fact that $K_1$ and $K_2$ are complements of $H$ in $G$.
Assuming that $h_2h_1^{-1} = h_2^{\prime}(h_1^{\prime})^{-1}$ as in \eqref{eq: assumption}, equations \eqref{eq: LHS RHS} and \eqref{eq: k = k prime} lead to \eqref{eq: first eq}.

Next, we show the second equality \eqref{eq: second eq}.
To this end, we notice that the assumption $h_1gH = h_1^{\prime}g^{\prime}H$ implies that there exists a unique $h \in H$ such that
\begin{equation}
h_1^{\prime}g^{\prime} = h_1gh.
\label{eq: h1gh}
\end{equation}
Due to \eqref{eq: h1g h2g} and \eqref{eq: k = k prime}, the element $h \in H$ in \eqref{eq: h1gh} satisfies
\begin{equation}
U_{g^{\prime}; K_1}^{-1}(h_1^{\prime}) = U_{g; K_1}^{-1}(h_1)h.
\end{equation}
Using this relation, one can compute the right-hand side of \eqref{eq: second eq} as
\begin{equation}
\begin{aligned}
\text{RHS} &= h_1^{\prime} \widetilde{U}_{g^{\prime}; S_{H \backslash G}}(h^{-1}(U_{g; K_1}^{-1}(h_1))^{-1}h_{\ell}) = h_1^{\prime} \widetilde{U}_{g^{\prime}h^{-1}; S_{H \backslash G}}((U_{g; K_1}^{-1}(h_1))^{-1}h_{\ell}) \\
&= h_1^{\prime} \widetilde{U}_{(h_1^{\prime})^{-1}h_1g; S_{H \backslash G}}((U_{g; K_1}^{-1}(h_1))^{-1}h_{\ell}) = h_1 \widetilde{U}_{g; S_{H \backslash G}}((U_{g; K_1}^{-1}(h_1))^{-1}h_{\ell}) = \text{LHS}.
\end{aligned}
\end{equation}
Here, the second equality follows from $\widetilde{U}_{g^{\prime}; S_{H \backslash G}}(h^{-1} \bullet) = \widetilde{U}_{g^{\prime}h^{-1}; S_{H \backslash G}}(\bullet)$, and the last equality follows from $\widetilde{U}_{(h_1^{\prime})^{-1}h_1 g; S_{H \backslash G}}(\bullet) = (h_1^{\prime})^{-1}h_1 \widetilde{U}_{g; S_{H \backslash G}}(\bullet)$, where $\bullet$ denotes an arbitrary element of $H$.
The above equation shows \eqref{eq: second eq}.

\section{Uniqueness of the action tensors}
\label{sec: Uniqueness of action tensors}
In this appendix, we show that the action tensors in \eqref{eq: decomposed fractionalization} that satisfy the orthogonality and completeness relations \eqref{eq: orthogonality and completeness} are essentially unique.\footnote{As shown in Appendix~\ref{sec: Orthogonality and completeness relations}, the completeness relation automatically follows from the orthogonality relation.}
To this end, let us suppose that there exist two action tensors $\phi_K[V_H^g]$ and $\psi_K[V_H^g]$ that satisfy \eqref{eq: decomposed fractionalization}, i.e.,
\begin{align}
\adjincludegraphics[valign = c, trim={10, 10, 10, 10}]{tikz/out/local_action1.pdf} &= \sum_{h \in H} \adjincludegraphics[valign = c, trim={10, 10, 10, 10}]{tikz/out/local_action5.pdf} = \sum_{h \in H} \adjincludegraphics[valign = c, trim={10, 10, 10, 10}]{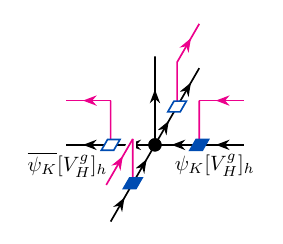} \,, \\
\adjincludegraphics[valign = c, trim={10, 10, 10, 10}]{tikz/out/local_action3.pdf} &= \sum_{h \in H} \adjincludegraphics[valign = c, trim={10, 10, 10, 10}]{tikz/out/local_action6.pdf} = \sum_{h \in H} \adjincludegraphics[valign = c, trim={10, 10, 10, 10}]{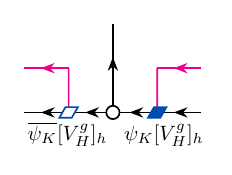} \,.
\end{align}
Due to the orthogonality relation for the action tensors, the above equations imply 
\begin{equation}
\adjincludegraphics[valign = c, trim={10, 10, 10, 10}]{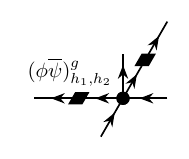} = \adjincludegraphics[valign = c, trim={10, 10, 10, 10}]{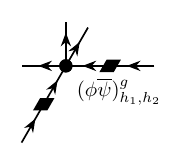} \,, \quad
\adjincludegraphics[valign = c, trim={10, 10, 10, 10}]{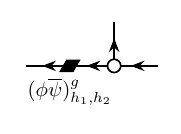} = \adjincludegraphics[valign = c, trim={10, 10, 10, 10}]{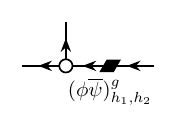} \,, \quad \forall h_1, h_2 \in H,
\end{equation}
where the two-leg tensor $(\phi \overline{\psi})^g_{h_1, h_2}$ is defined by
\begin{equation}
\adjincludegraphics[valign = c, trim={10, 10, 10, -10}]{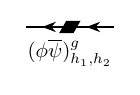} \coloneq \adjincludegraphics[valign = c, trim={10, 10, 10, 10}]{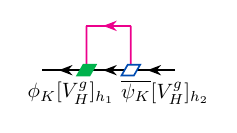}.
\label{eq: phi psi}
\end{equation}
In particular, the two-leg tensor $(\phi\overline{\psi})^g_{h_1, h_2}$ satisfies
\begin{equation}
\adjincludegraphics[valign = c, trim={10, 10, 10, -10}]{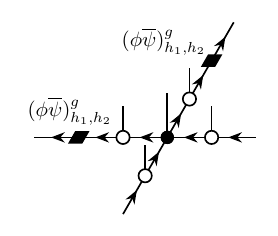} = \adjincludegraphics[valign = c, trim={10, 10, 10, 10}]{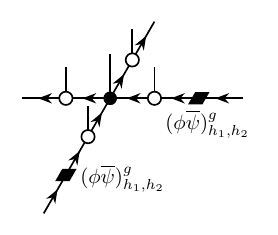} \,.
\label{eq: phi psi injective}
\end{equation}
We note that the composite tensor consisting of the middle black dot and the four white dots surrounding it is injective (as a linear map from the virtual bonds to the physical legs).\footnote{It immediately follows from \eqref{eq: local tensors} that the composite tensor in~\eqref{eq: phi psi injective} is surjective as a linear map from the physical legs to the virtual bonds. This in turn implies that it is injective as a linear map from the virtual bonds to the physical legs.}
Therefore, the above equation implies that $(\phi\overline{\psi})^g_{h_1, h_2}$ is a scalar multiple of the identity map, i.e.,
\begin{equation}
(\phi\overline{\psi})^g_{h_1, h_2} = f^g_{h_1, h_2} \id_{\mathbb{C}[H]}, \qquad f^g_{h_1, h_2} \in \mathbb{C}.
\end{equation}
Thus, by applying the completeness relation for the action tensors to \eqref{eq: phi psi}, we find that $\phi_K[V_H^g]$ and $\psi_K[V_H^g]$ are related by
\begin{equation}
\phi_K[V_H^g]_{h_1} = \sum_{h_2 \in H} f^g_{h_1, h_2} \psi_K[V_H^g]_{h_2}, \qquad
\overline{\psi_K}[V_H^g]_{h_2} = \sum_{h_1 \in H} f^g_{h_1, h_2} \overline{\phi_K}[V_H^g]_{h_1}.
\label{eq: phi = f psi}
\end{equation}
Similarly, by swapping the role of $\phi$ and $\psi$ in the above discussion, we can show that there exists a set of scalars $\overline{f}^g_{h_1, h_2} \in \mathbb{C}$ such that
\begin{equation}
\psi_K[V_H^g]_{h_2} = \sum_{h_1 \in H} \overline{f}^g_{h_2, h_1} \phi_K[V_H^g]_{h_1}, \qquad
\overline{\phi_K}[V_H^g]_{h_1} = \sum_{h_2 \in H} \overline{f}^g_{h_2, h_1} \overline{\psi_K}[V_H^g]_{h_2}.
\label{eq: psi = f phi}
\end{equation}
The orthogonality relation for the action tensors implies that $f^g_{h_1, h_2}$ and $\overline{f}^g_{h_2, h_1}$ are related by
\begin{equation}
\sum_{h_2 \in H} f^g_{h_1, h_2} \overline{f}^g_{h_2, h_1^{\prime}} = \delta_{h_1, h_1^{\prime}} ~\Leftrightarrow~ \overline{f}^g = (f^g)^{-1}.
\end{equation}
In particular, $f^g$ is invertible as an $|H| \times |H|$ matrix.
Equations \eqref{eq: phi = f psi} and \eqref{eq: psi = f phi} show that the action tensor $\phi_K[V_H^g]$ is unique up to the basis transformation $f^g$.
Using the four-leg tensors defined by~\eqref{eq: Phi four leg}, we can write \eqref{eq: phi = f psi} and \eqref{eq: psi = f phi} more concisely as
\begin{equation}
\adjincludegraphics[valign = c, trim={10, 10, 10, -10}]{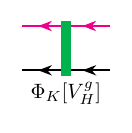} = \adjincludegraphics[valign = c, trim={10, 10, 10, -10}]{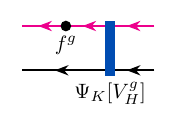} \,, \qquad
\adjincludegraphics[valign = c, trim={10, 10, 10, -10}]{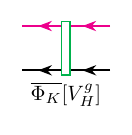} = \adjincludegraphics[valign = c, trim={10, 10, 10, -10}]{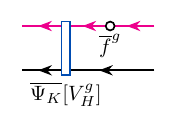} \,.
\end{equation}

\section{Multifusion categories}
\label{sec: Multifusion categories}
In this appendix, we review the basic properties of multifusion categories.

\subsection{Basic facts}
A multifusion category $\mathcal{D}$ is a finite semsimple $\mathbb{C}$-linear rigid monoidal category, whose unit object is not necessarily simple.
The simple decomposition of the unit object $I \in \mathcal{D}$ is given by
\begin{equation}
I = \bigoplus_{i = 1, 2, \cdots, n} I_i, \qquad n \in \mathbb{Z}_{+}.
\end{equation}
The simple summands $\{I_i \mid i = 1, 2, \cdots, n\}$ satisfy $I_i \otimes I_j \cong \delta_{i, j}I_i$.
The multifusion category $\mathcal{D}$ is called a fusion category when $I$ is simple, i.e., when $n = 1$.

As a semisimple category, $\mathcal{D}$ can be decomposed as
\begin{equation}
\mathcal{D} = \bigoplus_{i, j = 1, 2, \cdots, n} \mathcal{D}_{ij},
\label{eq: D decomposition}
\end{equation}
where $\mathcal{D}_{ij}$ is the full subcategory consisting of objects $X \in \mathcal{D}$ that satisfy $I_i \otimes X \cong X \otimes I_j \cong X$.
Any object $X_{ij} \in \mathcal{D}_{ij}$ is isomorphic to an object of the form $I_i \otimes X \otimes I_j$ for $X \in \mathcal{D}$.
The diagonal component $\mathcal{D}_{ii}$ is a fusion category, while the off-diagonal component $\mathcal{D}_{ij}$ is a $(\mathcal{D}_{ii}, \mathcal{D}_{jj})$-bimodule category, with the bimodule action given by the tensor product in $\mathcal{D}$.
The decomposition~\eqref{eq: D decomposition} implies that an object of $\mathcal{D}$ can be thought of as an $n \times n$ matrix whose $(i, j)$-component is an object of $\mathcal{D}_{ij}$.
The tensor product in $\mathcal{D}$ then corresponds to matrix multiplication.

A multifusion category $\mathcal{D}$ is said to be indecomposable if it cannot be decomposed into a direct sum of two non-zero multifusion categories.
Any component $\mathcal{D}_{ij}$ of an indecomposable multifusion category $\mathcal{D}$ is not empty due to \cite[Theorem 7.21.12]{EGNO2015}.

\subsection{Fiber functors}
\label{sec: Fiber functors of multifusion categories}
In this subsection, we show that an indecomposable multifusion category $\mathcal{D}$ is a fusion category if it admits a fiber functor.
To this end, let us suppose that $\mathcal{D}$ admits a fiber functor.
Equivalently, we suppose that $\mathcal{D}$ admits a left module category $\Vect$.
The left action of $\mathcal{D}$ on $\Vect$ is denoted by
\begin{equation}
\vartriangleright: \mathcal{D} \times \Vect \to \Vect.
\end{equation}
The unit object $I \in \mathcal{D}$ acts as the identity up to isomorphism, i.e., $I \vartriangleright M \cong M$ for any $M \in \Vect$.
In particular, for the unique simple object $\mathbb{C} \in \Vect$, we have 
\begin{equation}
I \vartriangleright \mathbb{C} = \bigoplus_{i = 1, 2, \cdots, n} I_i \vartriangleright \mathbb{C} \cong \mathbb{C}.
\end{equation}
This equation implies that there exists $i_0 \in \{1, 2, \cdots, n\}$ such that 
\begin{equation}
I_i \vartriangleright \mathbb{C} \cong \delta_{i, i_0} \mathbb{C}.
\end{equation}
This in turn implies that for any non-zero object $X_{ij} \in \mathcal{D}_{ij}$,\footnote{The category $\mathcal{D}_{ij}$ is not empty for any $i$ and $j$ because $\mathcal{D}$ is indecomposable.} we have
\begin{equation}
X_{ij} \vartriangleright \mathbb{C} \cong X_{ij} \vartriangleright (I_j \vartriangleright \mathbb{C}) \cong \delta_{j, i_0} X_{ij} \vartriangleright \mathbb{C}.
\label{eq: Xij C}
\end{equation}
On the other hand, for the dual object $\overline{X_{ij}} \in \mathcal{D}_{ji}$, we have
\begin{equation}
X_{ij} \vartriangleright (\overline{X_{ij}} \vartriangleright \mathbb{C}) \cong (X_{ij} \otimes \overline{X_{ij}}) \vartriangleright \mathbb{C} \supset I_i \vartriangleright \mathbb{C} \cong \delta_{i, i_0} \mathbb{C}.
\label{eq: X Xbar}
\end{equation}
If we choose $i = i_0$ in \eqref{eq: Xij C}, we find $X_{i_0 j} \vartriangleright (\overline{X_{i_0 j}} \vartriangleright \mathbb{C}) \supset \mathbb{C}$, which shows that $X_{i_0j} \vartriangleright \mathbb{C}$ is non-zero for any $j \in \{1, 2, \cdots, n\}$.
However, equation~\eqref{eq: Xij C} shows that $X_{i_0j} \vartriangleright \mathbb{C}$ can be non-zero only when $j = i_0$.
Therefore, it follows that $n$ must be one, which means that $\mathcal{D}$ must be fusion.

\section{Tambara-Yamagami cluster model in another gauge}
\label{sec: Tambara-Yamagami cluster model in another gauge}
In Section~\ref{sec: Generalized cluster states TY}, the generalized cluster model with Tambara-Yamagami symmetry $2\TY(A, 1)$ is defined in a gauge where the matter fields take values in $S_{H \backslash G} = A^{\text{diag}} \times \mathbb{Z}_2$.
In this appendix, we describe the model in a different gauge where the matter fields take values in\footnote{We note that $S^{\prime}_{H \backslash G}$ in \eqref{eq: TY another gauge} is not a complement of $A_L$ in $A \wr \mathbb{Z}_2$. Thus, there is no obvious way to reduce the bond dimensions of the symmetry PEPOs as in \ref{sec: Symmetry operators}.}
\begin{equation}
S^{\prime}_{H \backslash G} = A^{\text{right}} \times \mathbb{Z}_2 := \{(e, a, n) \mid a \in A, n \in \mathbb{Z}_2\}.
\label{eq: TY another gauge}
\end{equation}
We will also discuss the relation between the Tambara-Yamagami cluster model in this gauge and the ordinary $A$-cluster model.

\subsection{The model}
The state space of the Tambara-Yamagami cluster model in the above gauge is
\begin{equation}
\mathcal{H}_{\text{g.f.}} = \left(\bigotimes_{i \in P} \mathbb{C}^{|A^{\text{right}} \times \mathbb{Z}_2|}\right) \otimes \hat{\pi}_{\text{flat}}\left(\bigotimes_{[ij] \in E} \mathbb{C}^{|A_L|}\right).
\label{eq: TY state space another gauge}
\end{equation}
The matter field on plaquette $i \in P$ is labeled by $(e, a_i, n_i) \in A^{\text{right}} \times \mathbb{Z}_2$, while the matter field on edge $[ij] \in E$ is labeled by $(a_{ij}, e, 0) \in A_L$.
The corresponding states are denoted by $\ket{a_i, n_i}_i$ and $\ket{a_{ij}}_{ij}$, respectively.
The Hamiltonian of the model for $K = K(f)$ is
\begin{equation}
H_{\text{g.f.}} = -\sum_{i \in P} \hat{\mathsf{h}}_i^{\text{g.f.}} - \sum_{[ij] \in E} \hat{\mathsf{h}}_{ij}^{\text{g.f.}},
\label{eq: TY cluster ham another gauge}
\end{equation}
where $\hat{\mathsf{h}}_i$ and $\hat{\mathsf{h}}_{ij}$ are defined by
\begin{align}
\hat{\mathsf{h}}_i^{\text{g.f.}} \Ket{\adjincludegraphics[valign = c, trim={10, 10, 10, 10}]{tikz/out/TY_cluster_ham1.pdf}} &= \frac{1}{2|A|} \sum_{a \in A} \sum_{n \in \mathbb{Z}_2} \Ket{\adjincludegraphics[valign = c, trim={10, 10, 10, 10}]{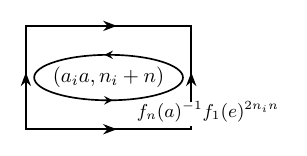}}, \\
\hat{\mathsf{h}}_{ij}^{\text{g.f.}} \ket{a_i, n_i}_i \otimes \ket{a_{ij}}_{ij} \otimes \ket{a_j, n_j}_j &= \delta_{a_{ij}, f_{n_i}(a_i)^{-1} f_{n_j}(a_j)} \ket{a_i, n_i}_i \otimes \ket{a_{ij}}_{ij} \otimes \ket{a_j, n_j}_j.
\end{align}
We note that the above Hamiltonian is obtained by replacing $\widetilde{f}$ with $f$ in \eqref{eq: TY cluster ham p} and \eqref{eq: TY cluster ham e}.
Accordingly, the ground state of the Hamiltonian~\eqref{eq: TY cluster ham another gauge} is also obtained by replacing $\widetilde{f}$ with $f$ in \eqref{eq: TY cluster}, i.e., 
\begin{equation}
\ket{\text{$2\TY$-cluster}(A; f)}^{\prime} = \frac{1}{(2|A|)^{|P|/2}} \sum_{\{a_i \in A\}} \sum_{\{n_i \in \mathbb{Z}_2\}} \ket{\{a_i, n_i; f_{n_i}(a_i)^{-1} f_{n_j}(a_j)\}}.
\end{equation}
The Hamiltonian~\eqref{eq: TY cluster ham another gauge} is related to \eqref{eq: TY ham} by the gauge-changing operator (cf. Appendix~\ref{sec: Remarks on the choice of representatives})
\begin{equation}
\mathcal{U}_{\text{g.f.}}^{\prime} \mathcal{U}_{\text{g.f.}}^{-1} \ket{a_i, n_i; a_{ij}} = \ket{a_i, n_i; a_i^{-1}a_{ij}a_j},
\end{equation}
which is the entangler for the $A$-cluster state.

\vspace*{\baselineskip}
\noindent{\bf Symmetry operators.}
The symmetry operators \eqref{eq: sym op TY} in a general gauge can be written as
\begin{equation}
\mathsf{C}U_a = \mathsf{D}_{A_L} U_a^{(R)} \overline{\mathsf{D}}_{A_L}, \qquad
\mathsf{D} = \mathsf{D}_{A_L} U_{\text{SWAP}} \overline{\mathsf{D}}_{A_L},
\label{eq: TY sym op general gauge}
\end{equation}
where $\mathsf{D}_{A_L}$ and $\overline{\mathsf{D}}_{A_L}$ are the gauging and ungauging operators for $A_L \subset A \wr \mathbb{Z}_2$, and $U_a^{(R)}$ and $U_{\text{SWAP}}$ are the symmetry operators~\eqref{eq: A wr Z2 sym op} of the ungauged model.
If we choose a gauge as in~\eqref{eq: TY another gauge}, the gauging and ungauging operators can be written explicitly as\footnote{We note that \eqref{eq: TY gauging op another gauge} differs from \eqref{eq: TY gauging op}, meaning that the gauging operator depends on a gauge choice.}
\begin{equation}
\mathsf{D}_{A_L} \ket{\{a_i, b_i, n_i\}} = \ket{\{b_i, n_i; a_i^{-1}a_j\}}, \qquad
\overline{\mathsf{D}}_{A_L} = \mathsf{D}_{A_L}^{\dagger}.
\label{eq: TY gauging op another gauge}
\end{equation}
The action of the symmetry operators~\eqref{eq: TY sym op general gauge} can then be computed as
\begin{align}
\mathsf{C}U_a \ket{\{a_i, n_i; a_{ij}\}} &= |A| \delta_{\text{hol}}(\{a_{ij}\}) \ket{\{aa_i, n_i; a_{ij}\}}, 
\label{eq: TY CUg another gauge}\\
\mathsf{D} \ket{\{a_i, n_i; a_{ij}\}} &= \sum_{\{b_i \in A\}} \left(\prod_{[ij] \in E} \delta_{a_{ij}, b_i^{-1}b_j}\right) \ket{b_i, n_i+1; a_i^{-1}a_j},
\label{eq: TY D another gauge}
\end{align}
where $\delta_{\text{hol}}(\{a_{ij}\})$ is one if the gauge fields have the trivial holonomy and zero otherwise.
The above symmetry operators can be expressed more concisely as
\begin{equation}
\mathsf{C} = \mathsf{D}_{A_L} \overline{\mathsf{D}}_{A_L}, \qquad
U_a = \bigotimes_{i \in P} X_a^{(i)}, \qquad
\mathsf{D} = \mathsf{D}_{A_R} \overline{\mathsf{D}}_{A_L} \left(\bigotimes_{i \in P} \sigma_x^{(i)}\right),
\label{eq: TY sym op another gauge}
\end{equation}
where $\mathsf{D}_{A_R}$ is the gauging operator for $A_R \subset A \wr \mathbb{Z}_2$, which is defined by
\begin{equation}
\mathsf{D}_{A_R}\ket{\{a_i, b_i, n_i\}} = \ket{\{a_i, n_i; b_i^{-1}b_j\}}.
\end{equation}
We note that the composition $\mathsf{D}_{A_R} \overline{\mathsf{D}}_{A_L}$ can be regarded as the gauging operator for the 2-group symmetry $A^{[0]} \times A^{[1]}$.

\vspace*{\baselineskip}
\noindent{\bf Example: $A = \mathbb{Z}_2$.}
When $A = \mathbb{Z}_2$, the state space \eqref{eq: TY state space another gauge} of the Tambara-Yamagami cluster model reduces to
\begin{equation}
\mathcal{H}_{\text{g.f.}} = \left(\bigotimes_{i \in P} \mathbb{C}^{|\mathbb{Z}_2 \times \mathbb{Z}_2|}\right) \otimes \hat{\pi}_{\text{flat}} \left(\bigotimes_{[ij] \in E} \mathbb{C}^{|\mathbb{Z}_2|}\right).
\end{equation}
That is, we have a pair of qubits on each plaquette and a single qubit on each edge.
If we choose a gauge as in \eqref{eq: TY another gauge}, the Hamiltonians for the two SPT phases corresponding to the complements \eqref{eq: K f0} and \eqref{eq: K f1} are given by
\begin{align}
H_{\text{g.f.}}^{[0]} &=  -\sum_{i \in P} \frac{1}{4}(1+X^{(i)}\prod_{[ij] \in \partial i} X^{(ij)})(1+\sigma_x^{(i)}) - \sum_{[ij] \in E} \frac{1}{2}(1+Z^{(i)}Z^{(ij)}Z^{(j)}), \\
H_{\text{g.f.}}^{[1]} &= -\sum_{i \in P} \frac{1}{4}(1+X^{(i)}\prod_{[ij] \in \partial i} X^{(ij)})(1+\sigma_x^{(i)}\prod_{[ij] \in \partial i} X^{(ij)}) - \sum_{[ij] \in E} \frac{1}{2}(1+\sigma_z^{(i)}Z^{(i)}Z^{(ij)}Z^{(j)}\sigma_z^{(j)}).
\end{align}
The ground states of these Hamiltonians can be written explicitly as
\begin{align}
\ket{\text{$2\TY$-cluster}(\mathbb{Z}_2; f^{[0]})}^{\prime} = \frac{1}{2^{|P|}} \sum_{\{a_i \in \mathbb{Z}_2\}} \sum_{\{n_i \in \mathbb{Z}_2\}} \ket{\{a_i, n_i; a_i^{-1}a_j\}}, \\
\ket{\text{$2\TY$-cluster}(\mathbb{Z}_2; f^{[1]})}^{\prime} = \frac{1}{2^{|P|}} \sum_{\{a_i \in \mathbb{Z}_2\}} \sum_{\{n_i \in \mathbb{Z}_2\}} \ket{\{a_i, n_i; a_i^{-1}a_j \eta^{n_i+n_j}\}},
\end{align}
where $\eta$ is the generator of $A = \mathbb{Z}_2$.
We note that neither of these states is the trivial product state, as opposed to the case of the gauge choice~\eqref{eq: Z2 TY gauge}.

\subsection{Relation to the $A$-cluster model}
If we choose a gauge as in \eqref{eq: TY another gauge}, the Tambara-Yamagami cluster model for $K = A^{\text{diag}} \times \mathbb{Z}_2$ becomes almost the same as the ordinary $A$-cluster model.
Indeed, in this case, the Hamiltonian~\eqref{eq: TY cluster ham another gauge} reduces to
\begin{equation}
H_{\text{g.f}} = -\sum_{i \in P} \left(\frac{1}{|A|}\sum_{a \in A} X_a^{(i)} L_{a^{-1}}^{(\partial i)}\right) \frac{1+\sigma_x^{(i)}}{2} - \sum_{[ij] \in E} \frac{1}{|A|} \sum_{\widehat{a} \in \widehat{A}} Z_{\widehat{a}}^{(i)} Z_{\widehat{a}}^{(ij)} Z_{\widehat{a}^{-1}}^{(j)}.
\label{eq: TY cluster vs A cluster}
\end{equation}
Here, $\widehat{A}$ is the Pontryagin dual of $A$, $\sigma_x$ is the Pauli-X operator acting on the qubit on a plaquette, $L_{a^{-1}}$ is the counterclockwise loop operator labeled by $a^{-1}$ (cf. \eqref{eq: loop op}), and $X_a$ and $Z_{\widehat{a}}$ are the generalized Pauli operators acting on the $A$-qudits as
\begin{equation}
X_a \ket{b} = \ket{ab} = \ket{ba}, \qquad
Z_{\widehat{a}} \ket{b} = \widehat{a}(b) \ket{b}.
\label{eq: A Pauli}
\end{equation}
The above Hamiltonian agrees with the Hamiltonian of the $A$-cluster model, except that the first term in \eqref{eq: TY cluster vs A cluster} has an additional projector $\frac{1}{2}(1+\sigma_x^{(i)})$ for all $i \in P$.
Thus, the ground state of this Hamiltonian is given by
\begin{equation}
\ket{\text{$A$-cluster}} \otimes \ket{\text{$\mathbb{Z}_2$-trivial}} = \frac{1}{(2|A|)^{|P|/2}} \sum_{\{a_i \in A\}} \sum_{\{n_i \in \mathbb{Z}_2\}} \ket{\{a_i, n_i; a_i^{-1}a_j\}},
\end{equation}
where $\ket{\text{$A$-cluster}} := \frac{1}{|A|^{|P|/2}} \sum_{\{a_i \in A\}} \ket{\{a_i; a_i^{-1}a_j\}}$ is the $A$-cluster state formed by the $A$-qudits on the plaquettes and edges, and $\ket{\text{$\mathbb{Z}_2$-trivial}} := \bigotimes_i \frac{1}{\sqrt{2}}\sum_{n_i \in \mathbb{Z}_2}\ket{n_i}_i$ is the trivial product state formed by the qubits on the plaquettes.

It is clear from \eqref{eq: TY cluster vs A cluster} that the model reduces to the $A$-cluster model if we project out the qubits on the plaquettes by the projector
\begin{equation}
\hat{\pi}_{\text{qubit}} := \bigotimes_{i \in P} \frac{1}{2}(1+ \sigma_x^{(i)}).
\label{eq: qubit projector}
\end{equation}
This projector commutes with the symmetry operators~\eqref{eq: TY CUg another gauge} and \eqref{eq: TY D another gauge}.
Therefore, the model after the projection, i.e., the $A$-cluster model, also has the Tambara-Yamagami symmetry $2\TY(A, 1)$.
This fact was originally pointed out in \cite{Choi:2024rjm} in the case of $A = \mathbb{Z}_2$.
The symmetry operator $\mathsf{D}$ on the projected subspace $\hat{\pi}_{\text{qubit}}\mathcal{H}_{\text{g.f.}} \cong \left(\bigotimes_{i \in P} \mathbb{C}^{|A|}\right) \otimes \hat{\pi}_{\text{flat}} \left( \bigotimes_{[ij] \in E} \mathbb{C}^{|A|}\right)$ reduces to
\begin{equation}
\left.\mathsf{D}\right|_{\hat{\pi}_{\text{qubit}}\mathcal{H}_{\text{g.f.}}} = \mathsf{D}_{A_R} \overline{\mathsf{D}}_{A_L},
\label{eq: projected D}
\end{equation}
whose action can be written as
\begin{equation}
\mathsf{D}_{A_R} \overline{\mathsf{D}}_{A_L} \ket{\{a_i; a_{ij}\}} = \sum_{\{b_i \in A\}} \left(\prod_{[ij] \in E} \delta_{a_{ij}, b_i^{-1}b_j}\right) \ket{\{b_i; a_i^{-1}a_j\}}.
\label{eq: DAR DAL}
\end{equation}
On the other hand, $\mathsf{C}U_a$ does not change under the projection.
When $A = \mathbb{Z}_2$, the operator~\eqref{eq: projected D} agrees with the symmetry operator constructed in \cite{Choi:2024rjm} (see Appendix~\ref{sec: Non-invertible swap operator} for more details).

\section{Lattice models for $2\TY(\mathbb{Z}_2, 1)$ SPT phases in \cite{Choi:2024rjm}}
\label{sec: cluster prime model}
Ref.\cite{Choi:2024rjm} constructed two lattice models for 2+1d SPT phases with Tambara-Yamagami symmetry $2\TY(\mathbb{Z}_2, 1)$.
One of them is the ordinary $\mathbb{Z}_2$-cluster model, while the other one is called the cluster$^{\prime}$ model in \cite{Choi:2024rjm}.\footnote{The cluster$^{\prime}$ model is called the $\beta$ model in \cite{Furukawa:2025flp}.}
In what follows, we will review these models and identify the SPT phases.
We will also discuss the relation between the non-invertible swap operator defined in \cite{Choi:2024rjm} and our symmetry operator~\eqref{eq: projected D}.
Throughout this appendix, the elements of $\mathbb{Z}_2$ will be denoted additively, i.e., $\mathbb{Z}_2 = \{0, 1\}$.

\subsection{Cluster model}
Let us first recall the $\mathbb{Z}_2$-cluster model on a square lattice with periodic boundary conditions.
The state space of the model is given by
\begin{equation}
\mathcal{H} = \left(\bigotimes_{i \in P} \mathbb{C}^2\right) \otimes \hat{\pi}_{\text{flat}} \left(\bigotimes_{[ij] \in P} \mathbb{C}^2\right),
\end{equation}
where $P$ is the set of the plaquettes, $E$ is the set of the edges, and $\hat{\pi}_{\text{flat}}$ is the projector that imposes the flatness condition on the qubits on the edges.\footnote{In \cite{Choi:2024rjm}, the flatness condition is not imposed at the level of the state space. As a result, the 1-form symmetry operators on the lattice become non-topological. In what follows, we impose the flatness condition so that the 1-form symmetry operators become topological.}
The Hamiltonian of the model is given by
\begin{equation}
H_{\text{cluster}} = -\sum_{i \in P} X^{(i)}\prod_{[ij] \in \partial i}X^{(ij)} - \sum_{[ij] \in E} Z^{(i)}Z^{(ij)}Z^{(j)},
\label{eq: cluster as TY SPT}
\end{equation}
where $X^{(i)}$ and $Z^{(i)}$ are the Pauli operators acting on the qubit on plaquette $i \in P$, while $X^{(ij)}$ and $Z^{(ij)}$ are those acting on the qubit on edge $[ij] \in E$.
The above Hamiltonian has a unique ground state given by
\begin{equation}
\ket{\text{cluster}} := \frac{1}{2^{|P|/2}} \sum_{\{n_i \in \mathbb{Z}_2\}} \ket{\{n_i; n_i+n_j \bmod 2\}},
\label{eq: cluster state as TY SPT}
\end{equation}
where $|P|$ is the number of plaquettes, and $\ket{\{n_i; n_{ij}\}} := \left(\bigotimes_{i \in P}\ket{n_i}_i\right) \otimes \left(\bigotimes_{[ij] \in E} \ket{n_{ij}}_{ij}\right)$ for $n_i, n_{ij} \in \mathbb{Z}_2$.
In what follows, $n_i + n_j \bmod 2$ will be written as $n_i + n_j$ when no confusion can arise.

The Hamiltonian~\eqref{eq: cluster as TY SPT} has a $\mathbb{Z}_2$ 0-form and a $\mathbb{Z}_2$ 1-form symmetries, which are generated by
\begin{equation}
X := \bigotimes_{i \in P} X^{(i)}, \qquad W(\gamma) := \bigotimes_{[ij] \in E_{\gamma}} Z^{(ij)},
\label{eq: cluster inv sym}
\end{equation}
where $\gamma$ is an arbitrary oriented loop on the dual lattice, and $E_{\gamma}$ is the set of the edges that intersect $\gamma$.
In addition to these symmetry operators, the Hamiltonian~\eqref{eq: cluster as TY SPT} also commutes with a non-invertible symmetry operator defined by \cite{Choi:2024rjm} (see also Appendix~\ref{sec: Non-invertible swap operator})
\begin{equation}
\mathsf{D} = \mathsf{D}_{\mathbb{Z}_2^{[0]}} \mathsf{D}_{\mathbb{Z}_2^{[1]}},
\label{eq: cluster non-inv sym}
\end{equation}
where $\mathsf{D}_{\mathbb{Z}_2^{[0]}}$ and $\mathsf{D}_{\mathbb{Z}_2^{[1]}}$ are the gauging operators for the $\mathbb{Z}_2$ 0-form and $\mathbb{Z}_2$ 1-form symmetries.
The action of $\mathsf{D}$ can be written explicitly as
\begin{equation}
\mathsf{D}\ket{\{n_i; n_{ij}\}} = \sum_{\{m_i \in \mathbb{Z}_2\}} \left(\prod_{[ij] \in E} \delta_{n_{ij}, m_i+m_j}\right) \ket{\{m_i; n_i+n_j\}}.
\end{equation}
The symmetry generated by \eqref{eq: cluster inv sym} and \eqref{eq: cluster non-inv sym} is described by the Tambara-Yamagami fusion 2-category
\begin{equation}
2\TY(\mathbb{Z}_2, 1) \cong 2\Rep((\mathbb{Z}_2^{[1]} \times \mathbb{Z}_2^{[1]}) \rtimes \mathbb{Z}_2^{[0]}).
\end{equation}
Thus, the cluster model can be regarded as a model of an SPT phase with Tambara-Yamagami symmetry. 

\vspace*{\baselineskip}
\noindent{\bf Gauging $\mathbb{Z}_2$ 1-form symmetry.}
Let us gauge the $\mathbb{Z}_2$ 1-form symmetry to identify the SPT phase realized in the above model.
The state space of the gauged model is given by
\begin{equation}
\widetilde{\mathcal{H}} = \bigotimes_{i \in P} (\mathbb{C}^2 \otimes \mathbb{C}^2).
\end{equation}
Namely, we have a pair of qubits on each plaquette.
The Hamiltonian of the gauged model is
\begin{equation}
\widetilde{H}_{\text{cluster}} = -\sum_{i \in P} X^{(i)}\widetilde{X}^{(i)} - \sum_{[ij] \in E} Z^{(i)}\widetilde{Z}^{(i)}Z^{(j)}\widetilde{Z}^{(j)},
\label{eq: gauged cluster}
\end{equation}
where $\widetilde{X}^{(i)}$ and $\widetilde{Z}^{(i)}$ are the Pauli operators acting on the second qubit on the plaquette $i \in P$.\footnote{The gauging of the $\mathbb{Z}_2$ 1-form symmetry effectively changes $\prod_{[ij] \in \partial i} X^{(ij)}$ and $Z^{(ij)}$ into $\widetilde{X}^{(i)}$ and $\widetilde{Z}^{(i)}\widetilde{Z}^{(j)}$.}
This model has an invertible 0-form symmetry generated by
\begin{equation}
X = \bigotimes_{i \in P} X^{(i)}, \qquad
\widetilde{X} = \bigotimes_{i \in P} \widetilde{X}^{(i)}, \qquad
S = \bigotimes_{i \in P} \mathrm{SWAP}^{(i)},
\label{eq: D8}
\end{equation}
where $\mathrm{SWAP}^{(i)}$ is defined by
\begin{equation}
\mathrm{SWAP}^{(i)} \ket{n_i, \widetilde{n}_i}_i = \ket{\widetilde{n}_i, n_i}_i.
\end{equation}
The algebra of the symmetry operators~\eqref{eq: D8} is
\begin{equation}
X^2 = \widetilde{X}^2 = S^2 = 1, \qquad
X\widetilde{X} = \widetilde{X}X, \qquad
XS = S\widetilde{X}.
\end{equation}
This implies that the symmetry group of the gauged model is $(\mathbb{Z}_2 \times \mathbb{Z}_2) \rtimes \mathbb{Z}_2 \cong D_8$, where the first two $\mathbb{Z}_2$'s are generated by $X$ and $\widetilde{X}$, while the third $\mathbb{Z}_2$ is generated by $S$.

The Hamiltonian~\eqref{eq: gauged cluster} has twofold degenerate ground states
\begin{equation}
\ket{\text{GS}; 1} := \bigotimes_{i \in P} \frac{1}{\sqrt{2}} (\ket{00}_i + \ket{11}_i), \qquad
\ket{\text{GS}; 2} := \bigotimes_{i \in P} \frac{1}{\sqrt{2}} (\ket{01}_i + \ket{10}_i).
\label{eq: gauged cluster GS}
\end{equation}
The symmetry operators~\eqref{eq: D8} act on these ground states as
\begin{equation}
X \ket{\text{GS}; 1} = \widetilde{X} \ket{\text{GS}; 1} = \ket{\text{GS}; 2}, \qquad
S \ket{\text{GS}; 1} = \ket{\text{GS}; 1}, \qquad
S \ket{\text{GS}; 2} = \ket{\text{GS}; 2}.
\label{eq: D8 sym action}
\end{equation}
Therefore, the symmetry $(\mathbb{Z}_2 \times \mathbb{Z}_2) \rtimes \mathbb{Z}_2$ is spontaneouly broken down to $\mathbb{Z}_2^{\text{diag}} \times \mathbb{Z}_2$ generated by $X\widetilde{X}$ and $S$.
This shows that the $2\TY(\mathbb{Z}_2, 1)$ SPT phase realized in the $\mathbb{Z}_2$-cluster model~\eqref{eq: cluster as TY SPT} corresponds to $K = \mathbb{Z}_2^{\text{diag}} \times \mathbb{Z}_2$.

\subsection{Cluster$^{\prime}$ model}
We now review the clutser$^{\prime}$ model in \cite{Choi:2024rjm}, which realizes another SPT phase with Tambara-Yamagami symmetry $2\TY(\mathbb{Z}_2, 1)$.
The state space of the model is again given by
\begin{equation}
\mathcal{H} = \left(\bigotimes_{i \in P} \mathbb{C}^2\right) \otimes \hat{\pi}_{\text{flat}} \left(\bigotimes_{[ij] \in E} \mathbb{C}^2\right).
\end{equation}
The Hamiltonian of the model is
\begin{equation}
H_{\text{cluster$^{\prime}$}} = \sum_{i \in P} X^{(i)}\prod_{[ij] \in \partial i}X^{(ij)} - \sum_{[ij] \in E} Y^{(i)}Z^{(ij)}Y^{(j)} + \Delta_{\text{cluster}^{\prime}},
\label{eq: cluster prime ham}
\end{equation}
where the last term $\Delta_{\text{cluster}^{\prime}}$ is defined by
\begin{equation}
\Delta_{\text{cluster}^{\prime}} = -\sum_{[ij] \in E} \left(X^{(i)} \prod_{[ii^{\prime}] \in \partial i}X^{(ii^{\prime})}\right) Y^{(i)}Z^{(ij)}Y^{(j)} \left(X^{(j)} \prod_{[jj^{\prime}] \in \partial j}X^{(jj^{\prime})}\right).
\end{equation}
We note that each local term of $\Delta_{\text{cluster}^{\prime}}$ is the product of the local terms in the first two sums in~\eqref{eq: cluster prime ham}.
This implies that $\Delta_{\text{cluster}^{\prime}}$ does not affect the ground state because the ground state is uniquely determined as the stabilizer of the first two terms.
As shown in \cite{Choi:2024rjm}, the Hamiltonian~\eqref{eq: cluster prime ham} commutes with the symmetry operators~\eqref{eq: cluster inv sym} and \eqref{eq: cluster non-inv sym}.
Thus, the cluster$^\prime$ model realizes an SPT phase with Tambara-Yamagami symmetry $2\TY(\mathbb{Z}_2, 1)$.

To find the ground state, we note that the cluster$^{\prime}$ Hamiltonian~\eqref{eq: cluster prime ham} is related to the cluster Hamiltonian~\eqref{eq: cluster as TY SPT} as
\begin{equation}
H_{\text{cluster}^{\prime}} = U (H_{\text{cluster}} + \Delta_{\text{cluster}}) U^{-1},
\label{eq: cluster prime conjugate}
\end{equation}
where $U$ and $\Delta_{\text{cluster}}$ are defined by
\begin{align}
U &= \bigotimes_{i \in P} Z^{(i)} e^{i\pi X^{(i)}/4},
\label{eq: on-site unitary}\\
\Delta_{\text{cluster}} &= -\sum_{[ij] \in E} \left(X^{(i)} \prod_{[ii^{\prime}] \in \partial i} X^{(ii^{\prime})}\right) Z^{(i)}Z^{(ij)}Z^{(j)} \left(X^{(j)} \prod_{[jj^{\prime}] \in \partial j}X^{(jj^{\prime})}\right).
\label{eq: Delta cluster}
\end{align}
Each local term of $\Delta_{\text{cluster}}$ is the product of the local terms of the cluster Hamiltonian~\eqref{eq: cluster as TY SPT}.
Hence, the ground state of $H_{\text{cluster}} + \Delta_{\text{cluster}}$ agrees with that of $H_{\text{cluster}}$, which is the cluster state~\eqref{eq: cluster state as TY SPT}.
Thus, due to \eqref{eq: cluster prime conjugate}, the ground state of $H_{\text{cluster}^{\prime}}$ is given by
\begin{equation}
\ket{\text{cluster}^{\prime}} := U \ket{\text{cluster}}.
\label{eq: cluster prime state}
\end{equation}
Since $U$ is the tensor product of on-site unitary operators, the ground state $\ket{\text{cluster}^{\prime}}$ has the same entanglement structure as the cluster state.
Nevertheless, as shown in \cite{Choi:2024rjm}, the cluster$^{\prime}$ state~\eqref{eq: cluster prime state} and the cluster state~\eqref{eq: cluster state as TY SPT} are in different SPT phases with $2\TY(\mathbb{Z}_2, 1)$ symmetry.

\vspace*{\baselineskip}
\noindent{\bf Gauging $\mathbb{Z}_2$ 1-form symmetry.}
Let us gauge the $\mathbb{Z}_2$ 1-form symmetry to identify the SPT phase realized in the cluster$^{\prime}$ model.
The state space of the gauge model is given by
\begin{equation}
\widetilde{\mathcal{H}} = \bigotimes_{i \in P} (\mathbb{C}^2 \otimes \mathbb{C}^2).
\end{equation}
The Hamiltonian of the gauge model is
\begin{equation}
\widetilde{H}_{\text{cluster}^{\prime}} = U(\widetilde{H}_{\text{cluster}} + \widetilde{\Delta}_{\text{cluster}})U^{-1},
\end{equation}
where $U$ is the unitary operator \eqref{eq: on-site unitary},\footnote{The conjugation by $U$ commutes with the gauging operation because $U$ does not act on the qubits on the edges.} $\widetilde{H}_{\text{cluster}}$ is the gauged cluster Hamiltonian~\eqref{eq: gauged cluster}, and $\widetilde{\Delta}_{\text{cluster}}$ is the gauged version of~\eqref{eq: Delta cluster}, which is given by
\begin{equation}
\widetilde{\Delta}_{\text{cluster}} = -\sum_{[ij] \in E} X^{(i)}\widetilde{X}^{(i)}Z^{(i)}\widetilde{Z}^{(i)}Z^{(j)}\widetilde{Z}^{(j)}X^{(j)}\widetilde{X}^{(j)}.
\end{equation}
The Hamiltonian $\widetilde{H}_{\text{cluster}} + \widetilde{\Delta}_{\text{cluster}}$ has the twofold degenerate ground states~\eqref{eq: gauged cluster GS}.
Therefore, the ground states of $\widetilde{H}_{\text{cluster}^{\prime}}$ are given by
\begin{equation}
\begin{aligned}
\ket{\text{GS}^{\prime}; 1} := U\ket{\text{GS}; 1} = \bigotimes_{i \in P} \frac{1}{\sqrt{2}} U^{(i)} (\ket{00}_i + \ket{11}_i), \\
\ket{\text{GS}^{\prime}; 2} := U\ket{\text{GS}; 2} = \bigotimes_{i \in P} \frac{1}{\sqrt{2}} U^{(i)} (\ket{01}_i + \ket{10}_i),
\end{aligned}
\label{eq: gauged cluster prime GS}
\end{equation}
where $U^{(i)} := Z^{(i)}e^{i\pi X^{(i)}/4}$ is the on-site unitary of $U$.
One can compute the action of the symmetry operators~\eqref{eq: D8} on the ground states~\eqref{eq: gauged cluster prime GS} by using the commutation relations
\begin{equation}
\begin{aligned}
(U^{(i)})^{-1}X^{(i)}U^{(i)} &= -X^{(i)}, \qquad
(U^{(i)})^{-1}\widetilde{X}^{(i)}U^{(i)} = \widetilde{X}^{(i)}, \\
(U^{(i)})^{-1} \mathrm{SWAP}^{(i)} U^{(i)} &= \frac{1}{2}(1-X^{(i)}\widetilde{X}^{(i)} -Z^{(i)}\widetilde{Y}^{(i)} - Y^{(i)}\widetilde{Z}^{(i)}).
\end{aligned}
\label{eq: commutation relation of U}
\end{equation}
Here, the last equality follows from the following expression of the SWAP operator:
\begin{equation}
\mathrm{SWAP}^{(i)} = \frac{1}{2}(1+X^{(i)}\widetilde{X}^{(i)}+Y^{(i)}\widetilde{Y}^{(i)}+Z^{(i)}\widetilde{Z}^{(i)}).
\end{equation}
Based on \eqref{eq: gauged cluster prime GS}, \eqref{eq: commutation relation of U}, and \eqref{eq: D8 sym action}, we can compute the symmetry action on the ground states as
\begin{align}
X\ket{\text{GS}^{\prime}; 1} &= U(U^{-1}XU) \ket{\text{GS}; 1} = (-1)^{|P|} \ket{\text{GS}^{\prime}; 2}, \\
\widetilde{X}\ket{\text{GS}^{\prime}; 1} &= U(U^{-1}\widetilde{X}U) \ket{\text{GS}; 1} = \ket{\text{GS}^{\prime}; 2}, \\
S\ket{\text{GS}^{\prime}; 1} &= U(U^{-1}SU) \ket{\text{GS}; 1} = (-i)^{|P|} \ket{\text{GS}^{\prime}; 2}.
\end{align}
This shows that the unbroken symmetry operators are $\{1, XS, \widetilde{X}S, X\widetilde{X}\}$, which form $\mathbb{Z}_4$ because
\begin{equation}
(XS)^2 = X\widetilde{X}, \qquad (XS)^3 = \widetilde{X}S, \qquad (XS)^4=1.
\end{equation}
Thus, the symmetry $(\mathbb{Z}_2 \times \mathbb{Z}_2)\rtimes \mathbb{Z}_2$ is spontaneously broken down to $\mathbb{Z}_4$, which implies that the $2\TY(\mathbb{Z}_2, 1)$ SPT phase realized in the cluster$^{\prime}$ model~\eqref{eq: cluster prime ham} corresponds to $K = \mathbb{Z}_4$.

\subsection{Non-invertible swap operator}
\label{sec: Non-invertible swap operator}
\begin{figure}[t]
\centering
\adjincludegraphics[trim={10, 10, 10, 10}]{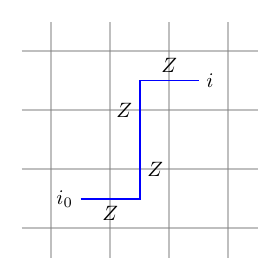}
\caption{The Wilson line $W^{(i_0i)}$ is the product of the Pauli-Z operators on the edges intersecting a path between $i_0$ and $i$. This operator does not depend on the choice of a path because we imposed the flatness condition on the state space.}
\label{fig: W i0i}
\end{figure}

In \cite{Choi:2024rjm}, it was shown that the Hamiltonians~\eqref{eq: cluster as TY SPT} and \eqref{eq: cluster prime ham} commute with the following operator called a non-invertible swap operator:
\begin{equation}
\mathsf{D}_{\text{SWAP}} = \frac{1}{2} \mathsf{C}_U \mathsf{C}_{\eta} \mathsf{S}.
\label{eq: non-invertible swap}
\end{equation}
Here, the operators on the right-hand side are defined by\footnote{The operator~\eqref{eq: non-invertible swap} is the Hermitian conjugate of the original non-invertible swap operator defined in \cite{Choi:2024rjm}. It was shown in \cite{Choi:2024rjm} that this operator is Hermitian.}
\begin{equation}
\mathsf{C}_U = 1 + \prod_{i \in P}X^{(i)}, \qquad
\mathsf{C}_{\eta} = \frac{1}{2^{|P|}} \sum_{\gamma: \text{loops}} W(\gamma), \qquad
\mathsf{S} = S_{i_0} \left(\prod_{i \in P \setminus \{i_0\}} S_i\right),
\label{eq: SCC}
\end{equation}
where $i_0 \in P$ is an arbitrary plaquette and $S_i$ is defined by
\begin{equation}
S_i = \frac{1}{2} \left( 1+G_i + \left(1 - G_i \right) W^{(i_0i)}Z^{(i)} \right), \qquad G_i = X^{(i)}\prod_{[ij] \in \partial i}X^{(ij)}.
\end{equation}
The operator $W^{(i_0i)}$ is the Wilson line on a path between $i_0$ and $i$ (cf. Figure~\ref{fig: W i0i}), and $W(\gamma)$ is the Wilson line on the loop $\gamma$.
The summation $\sum_{\gamma}$ in \eqref{eq: SCC} is taken over all (unoriented) loops on the dual lattice.
We note that $\mathsf{D}_{\text{SWAP}}$ does not depend on the choice of $i_0$ \cite{Choi:2024rjm}.
In what follows, we will show that the non-invertible swap operator~\eqref{eq: non-invertible swap} agrees with our symmetry operator~\eqref{eq: projected D}, i.e., 
\begin{equation}
\mathsf{D}_{\text{SWAP}} = \mathsf{D}_{(\mathbb{Z}_2)_R} \overline{\mathsf{D}}_{(\mathbb{Z}_2)_L},
\label{eq: non-invertible swap 2}
\end{equation}
where $\mathsf{D}_{(\mathbb{Z}_2)_R} = \mathsf{D}_{\mathbb{Z}_2^{[0]}}$ and $\overline{\mathsf{D}}_{(\mathbb{Z}_2)_L} = \mathsf{D}_{\mathbb{Z}_2^{[1]}}$ are the gauging operators for the $\mathbb{Z}_2$ 0-form and $\mathbb{Z}_2$ 1-form symmetries.

To show \eqref{eq: non-invertible swap 2}, we first compute the action of $\mathsf{S}$ on a state $\ket{\{n_i; n_{ij}\}}$.
To this end, we notice that the action of each $S_i$ for any $i \in P$ can be written as
\begin{equation}
S_i \ket{\{n_i; n_{ij}\}} = G_i^{n_{i_0i} + n_i} \ket{\{n_i; n_{ij}\}},
\label{eq: action of Si}
\end{equation}
where $n_{i_0 i}$ is the eigenvalue of $W^{(i_0i)}$, that is, $n_{i_0i} = \sum_{[jk]} n_{jk} \bmod 2$, where the summation is taken over all edges that intersect the path between $i_0$ and $i$.
In particular, we define $n_{i_0 i_0} = 0$.
Using~\eqref{eq: action of Si}, we can compute the action of $\mathsf{S}$ on $\ket{\{n_i; n_{ij}\}}$ as follows:
\begin{equation}
\begin{aligned}
\mathsf{S} \ket{\{n_i; n_{ij}\}} &= S_{i_0} \left(\prod_{i \in P \setminus \{i_0\}} G_i^{n_{i_0 i} + n_i}\right) \ket{\{n_i; n_{ij}\}}
= S_{i_0} G_{i_0}^{n_{i_0}} \left(\prod_{i \in P} G_i^{n_{i_0 i} + n_i} \right) \ket{\{n_i; n_{ij}\}} \\
&= S_{i_0} G_{i_0}^{n_{i_0}} \ket{\{n_{i_0 i}; n_i + n_j\}}
= \frac{1}{2} \left( 1+G_{i_0} + (1-G_{i_0}) (-1)^{n_{i_0}} \right) G_{i_0}^{n_{i_0}} \ket{\{n_{i_0 i}; n_i + n_j\}} \\
&= \ket{\{n_{i_0 i}; n_i + n_j\}}.
\end{aligned}
\end{equation}
The action of the non-invertible swap operator~\eqref{eq: non-invertible swap} can then be computed as
\begin{equation}
\begin{aligned}
\mathsf{D}_{\text{SWAP}} \ket{\{n_i; n_{ij}\}} &= \frac{1}{2} \mathsf{C}_U \mathsf{C}_{\eta} \ket{\{n_{i_0 i}; n_i + n_j\}} = \ket{\{n_{i_0 i}; n_i + n_j\}} + \ket{n_{i_0 i} +1; n_i + n_j\}} \\
&= \sum_{\{m_i \in \mathbb{Z}_2\}} \left(\prod_{[ij] \in E} \delta_{m_i+m_j, n_{ij}}\right) \ket{\{m_i; n_i+n_j\}}.
\end{aligned}
\label{eq: D SWAP}
\end{equation}
Here, the last equality follows from the fact that $\{m_i = n_{i_0 i} \mid i \in P\}$ and $\{m_i = n_{i_0 i}+1 \mid i \in P\}$ are the only solutions to the condition
\begin{equation}
m_i + m_j = n_{ij} \bmod 2, \qquad \forall [ij] \in E.
\end{equation}
Equations \eqref{eq: D SWAP} and \eqref{eq: DAR DAL} show that \eqref{eq: non-invertible swap 2} holds.

\bibliographystyle{ytphys}
\bibliography{bibliography}

\providecommand{\href}[2]{#2}\begingroup\raggedright\begin{thebibliography}{100}

\bibitem{Choi:2024rjm}
Y.~Choi, Y.~Sanghavi, S.-H. Shao, and Y.~Zheng, ``{Non-invertible and
  higher-form symmetries in 2+1d lattice gauge theories},''
  \href{http://dx.doi.org/10.21468/SciPostPhys.18.1.008}{{\em SciPost Phys.}
  {\bfseries 18} no.~1, (2025) 008},
  \href{http://arxiv.org/abs/2405.13105}{{\ttfamily arXiv:2405.13105
  [cond-mat.str-el]}}.

\bibitem{Gu_2009}
Z.-C. Gu and X.-G. Wen, ``{Tensor-entanglement-filtering renormalization
  approach and symmetry-protected topological order},''
  \href{http://dx.doi.org/10.1103/PhysRevB.80.155131}{{\em Physical Review B}
  {\bfseries 80} no.~15, (10, 2009) 155131},
  \href{http://arxiv.org/abs/0903.1069}{{\ttfamily arXiv:0903.1069
  [cond-mat.str-el]}}.

\bibitem{Pollmann:2009mhk}
F.~Pollmann, E.~Berg, A.~M. Turner, and M.~Oshikawa, ``{Symmetry protection of
  topological phases in one-dimensional quantum spin systems},''
  \href{http://dx.doi.org/10.1103/PhysRevB.85.075125}{{\em Phys. Rev. B}
  {\bfseries 85} no.~7, (2012) 075125},
  \href{http://arxiv.org/abs/0909.4059}{{\ttfamily arXiv:0909.4059
  [cond-mat.str-el]}}.

\bibitem{Pollmann_2010}
F.~Pollmann, A.~M. Turner, E.~Berg, and M.~Oshikawa, ``{Entanglement spectrum
  of a topological phase in one dimension},''
  \href{http://dx.doi.org/10.1103/PhysRevB.81.064439}{{\em Physical Review B}
  {\bfseries 81} no.~6, (2, 2010) 064439},
  \href{http://arxiv.org/abs/0910.1811}{{\ttfamily arXiv:0910.1811
  [cond-mat.str-el]}}.

\bibitem{Chen:2010zpc}
X.~Chen, Z.-C. Gu, and X.-G. Wen, ``{Classification of gapped symmetric phases
  in one-dimensional spin systems},''
  \href{http://dx.doi.org/10.1103/PhysRevB.83.035107}{{\em Phys. Rev. B}
  {\bfseries 83} no.~3, (2011) 035107},
  \href{http://arxiv.org/abs/1008.3745}{{\ttfamily arXiv:1008.3745
  [cond-mat.str-el]}}.

\bibitem{Chen_2011_complete}
X.~Chen, Z.-C. Gu, and X.-G. Wen, ``Complete classification of one-dimensional
  gapped quantum phases in interacting spin systems,''
  \href{http://dx.doi.org/10.1103/PhysRevB.84.235128}{{\em Physical Review B}
  {\bfseries 84} no.~23, (12, 2011) 235128},
  \href{http://arxiv.org/abs/1103.3323}{{\ttfamily arXiv:1103.3323
  [cond-mat.str-el]}}.

\bibitem{Chen:2011bcp}
X.~Chen, Z.-X. Liu, and X.-G. Wen, ``{Two-dimensional symmetry-protected
  topological orders and their protected gapless edge excitations},''
  \href{http://dx.doi.org/10.1103/PhysRevB.84.235141}{{\em Phys. Rev. B}
  {\bfseries 84} no.~23, (2011) 235141},
  \href{http://arxiv.org/abs/1106.4752}{{\ttfamily arXiv:1106.4752
  [cond-mat.str-el]}}.

\bibitem{Chen:2011pg}
X.~Chen, Z.-C. Gu, Z.-X. Liu, and X.-G. Wen, ``{Symmetry protected topological
  orders and the group cohomology of their symmetry group},''
  \href{http://dx.doi.org/10.1103/PhysRevB.87.155114}{{\em Phys. Rev. B}
  {\bfseries 87} no.~15, (2013) 155114},
  \href{http://arxiv.org/abs/1106.4772}{{\ttfamily arXiv:1106.4772
  [cond-mat.str-el]}}.

\bibitem{Schuch_2011}
N.~Schuch, D.~P\'{e}rez-Garc\'{i}a, and I.~Cirac, ``{Classifying quantum phases
  using matrix product states and projected entangled pair states},''
  \href{http://dx.doi.org/10.1103/PhysRevB.84.165139}{{\em Physical Review B}
  {\bfseries 84} no.~16, (10, 2011) 165139},
  \href{http://arxiv.org/abs/1010.3732}{{\ttfamily arXiv:1010.3732
  [cond-mat.str-el]}}.

\bibitem{Thouless83}
D.~J. Thouless, ``Quantization of particle transport,''
  \href{https://link.aps.org/doi/10.1103/PhysRevB.27.6083}{{\em Phys. Rev. B}
  {\bfseries 27} (May, 1983) 6083--6087}.

\bibitem{PhysRevB.82.115120}
J.~C.~Y. Teo and C.~L. Kane, ``Topological defects and gapless modes in
  insulators and superconductors,''
  \href{https://link.aps.org/doi/10.1103/PhysRevB.82.115120}{{\em Phys. Rev. B}
  {\bfseries 82} (Sep, 2010) 115120}.

\bibitem{Kitaev2011SCGP}
A.~Kitaev, ``Toward a topological classification of many-body quantum states
  with short-range entanglement,''.
  \url{http://scgp.stonybrook.edu/video_portal/video.php?id=336}. talk at
  Simons Center for Geometry and Physics.

\bibitem{Kitaev2013SCGP}
A.~Kitaev, ``On the classification of short-range entangled states,''.
  \url{http://scgp.stonybrook.edu/video_portal/video.php?id=2010}. talk at
  Simons Center for Geometry and Physics.

\bibitem{Kitaev2015IPAM}
A.~Kitaev, ``{Homotopy-theoretic approach to SPT phases in action: $Z_{16}$
  classification of three-dimensional superconductors},''.
  \url{https://www.ipam.ucla.edu/abstract/?tid=12389}. talk at Institute for
  Pure and Applied Mathematics.

\bibitem{Kapustin:2020mkl}
A.~Kapustin and L.~Spodyneiko, ``{Higher-dimensional generalizations of the
  Thouless charge pump},'' \href{http://arxiv.org/abs/2003.09519}{{\ttfamily
  arXiv:2003.09519 [cond-mat.str-el]}}.

\bibitem{PhysRevResearch.2.042024}
Y.~Kuno and Y.~Hatsugai, ``Interaction-induced topological charge pump,''
  \href{https://link.aps.org/doi/10.1103/PhysRevResearch.2.042024}{{\em Phys.
  Rev. Res.} {\bfseries 2} (Oct, 2020) 042024},
  \href{http://arxiv.org/abs/2007.11215}{{\ttfamily arXiv:2007.11215
  [cond-mat.quant-gas]}}.

\bibitem{Tantivasadakarn:2021wdv}
N.~Tantivasadakarn, R.~Thorngren, A.~Vishwanath, and R.~Verresen, ``{Pivot
  Hamiltonians as generators of symmetry and entanglement},''
  \href{http://dx.doi.org/10.21468/SciPostPhys.14.2.012}{{\em SciPost Phys.}
  {\bfseries 14} no.~2, (2023) 012},
  \href{http://arxiv.org/abs/2110.07599}{{\ttfamily arXiv:2110.07599
  [cond-mat.str-el]}}.

\bibitem{Shiozaki:2021weu}
K.~Shiozaki, ``{Adiabatic cycles of quantum spin systems},''
  \href{http://dx.doi.org/10.1103/PhysRevB.106.125108}{{\em Phys. Rev. B}
  {\bfseries 106} no.~12, (2022) 125108},
  \href{http://arxiv.org/abs/2110.10665}{{\ttfamily arXiv:2110.10665
  [cond-mat.str-el]}}.

\bibitem{Hermele2021CMSA}
M.~Hermele, ``Families of gapped systems and quantum pumps,''.
  \url{https://www.youtube.com/watch?v=wtaC0tqXZMU}. talk at Harvard CMSA.

\bibitem{Wen:2021gwc}
X.~Wen, M.~Qi, A.~Beaudry, J.~Moreno, M.~J. Pflaum, D.~Spiegel, A.~Vishwanath,
  and M.~Hermele, ``{Flow of higher Berry curvature and bulk-boundary
  correspondence in parametrized quantum systems},''
  \href{http://dx.doi.org/10.1103/PhysRevB.108.125147}{{\em Phys. Rev. B}
  {\bfseries 108} no.~12, (2023) 125147},
  \href{http://arxiv.org/abs/2112.07748}{{\ttfamily arXiv:2112.07748
  [cond-mat.str-el]}}.

\bibitem{Bachmann:2022bhx}
S.~Bachmann, W.~De~Roeck, M.~Fraas, and T.~Jappens, ``{A Classification of
  $G$-Charge Thouless Pumps in 1D Invertible States},''
  \href{http://dx.doi.org/10.1007/s00220-024-05010-w}{{\em Commun. Math. Phys.}
  {\bfseries 405} no.~7, (2024) 157},
  \href{http://arxiv.org/abs/2204.03763}{{\ttfamily arXiv:2204.03763
  [math-ph]}}.

\bibitem{Spodyneiko:2023vsw}
L.~Spodyneiko, ``{Hall conductivity pump},''
  \href{http://arxiv.org/abs/2309.14332}{{\ttfamily arXiv:2309.14332
  [cond-mat.mes-hall]}}.

\bibitem{Ohyama:2022cib}
S.~Ohyama, K.~Shiozaki, and M.~Sato, ``{Generalized Thouless pumps in
  (1+1)-dimensional interacting fermionic systems},''
  \href{http://dx.doi.org/10.1103/PhysRevB.106.165115}{{\em Phys. Rev. B}
  {\bfseries 106} no.~16, (2022) 165115},
  \href{http://arxiv.org/abs/2206.01110}{{\ttfamily arXiv:2206.01110
  [cond-mat.str-el]}}.

\bibitem{Inamura:2024jke}
K.~Inamura and S.~Ohyama, ``{1+1d SPT phases with fusion category symmetry:
  interface modes and non-abelian Thouless pump},''
  \href{http://arxiv.org/abs/2408.15960}{{\ttfamily arXiv:2408.15960
  [cond-mat.str-el]}}.

\bibitem{Jones:2025khc}
N.~G. Jones, R.~Thorngren, R.~Verresen, and A.~Prakash, ``{Charge pumps, pivot
  Hamiltonians, and symmetry-protected topological phases},''
  \href{http://dx.doi.org/10.1103/rtq1-pplf}{{\em Phys. Rev. B} {\bfseries 112}
  no.~16, (2025) 165123}, \href{http://arxiv.org/abs/2507.00995}{{\ttfamily
  arXiv:2507.00995 [cond-mat.str-el]}}.

\bibitem{Shiozaki:2025pyo}
K.~Shiozaki, ``{Equivariant Parameter Families of Spin Chains: A Discrete MPS
  Formulation},'' \href{http://arxiv.org/abs/2507.19932}{{\ttfamily
  arXiv:2507.19932 [quant-ph]}}.

\bibitem{Li:2025wes}
Y.~Li, M.~Dell'acqua, and A.~Mitra, ``{Classification of Thouless pumps with
  non-invertible symmetries and implications for Floquet phases},''
  \href{http://arxiv.org/abs/2510.01626}{{\ttfamily arXiv:2510.01626
  [cond-mat.str-el]}}.

\bibitem{Gaiotto:2014kfa}
D.~Gaiotto, A.~Kapustin, N.~Seiberg, and B.~Willett, ``{Generalized Global
  Symmetries},'' \href{http://dx.doi.org/10.1007/JHEP02(2015)172}{{\em JHEP}
  {\bfseries 02} (2015) 172}, \href{http://arxiv.org/abs/1412.5148}{{\ttfamily
  arXiv:1412.5148 [hep-th]}}.

\bibitem{Wigner1959}
E.~P. Wigner, {\em {Group Theory and its Application to the Quantum Mechanics
  of Atomic Spectra}}, vol.~5 of {\em Pure and Applied Physics}.
\newblock Academic Press, New York, 1959.
\newblock Originally published as
  \href{https://link.springer.com/book/10.1007/978-3-663-02555-9}{Gruppentheorie
  und ihre Anwendung auf die Quantenmechanik der Atomspektren} (1931).

\bibitem{KW1941}
H.~A. Kramers and G.~H. Wannier, ``{Statistics of the Two-Dimensional
  Ferromagnet. Part I},''
  \href{https://link.aps.org/doi/10.1103/PhysRev.60.252}{{\em Phys. Rev.}
  {\bfseries 60} (Aug, 1941) 252--262}.

\bibitem{Frohlich:2004ef}
J.~Frohlich, J.~Fuchs, I.~Runkel, and C.~Schweigert, ``{Kramers-Wannier duality
  from conformal defects},''
  \href{http://dx.doi.org/10.1103/PhysRevLett.93.070601}{{\em Phys. Rev. Lett.}
  {\bfseries 93} (2004) 070601},
  \href{http://arxiv.org/abs/cond-mat/0404051}{{\ttfamily
  arXiv:cond-mat/0404051}}.

\bibitem{Chang:2018iay}
C.-M. Chang, Y.-H. Lin, S.-H. Shao, Y.~Wang, and X.~Yin, ``{Topological Defect
  Lines and Renormalization Group Flows in Two Dimensions},''
  \href{http://dx.doi.org/10.1007/JHEP01(2019)026}{{\em JHEP} {\bfseries 01}
  (2019) 026}, \href{http://arxiv.org/abs/1802.04445}{{\ttfamily
  arXiv:1802.04445 [hep-th]}}.

\bibitem{Seiberg:2023cdc}
N.~Seiberg and S.-H. Shao, ``{Majorana chain and Ising model - (non-invertible)
  translations, anomalies, and emanant symmetries},''
  \href{http://dx.doi.org/10.21468/SciPostPhys.16.3.064}{{\em SciPost Phys.}
  {\bfseries 16} no.~3, (2024) 064},
  \href{http://arxiv.org/abs/2307.02534}{{\ttfamily arXiv:2307.02534
  [cond-mat.str-el]}}.

\bibitem{Seiberg:2024gek}
N.~Seiberg, S.~Seifnashri, and S.-H. Shao, ``{Non-invertible symmetries and
  LSM-type constraints on a tensor product Hilbert space},''
  \href{http://dx.doi.org/10.21468/SciPostPhys.16.6.154}{{\em SciPost Phys.}
  {\bfseries 16} (2024) 154}, \href{http://arxiv.org/abs/2401.12281}{{\ttfamily
  arXiv:2401.12281 [cond-mat.str-el]}}.

\bibitem{Cordova:2022ruw}
C.~C\'{o}rdova, T.~T. Dumitrescu, K.~Intriligator, and S.-H. Shao, ``{Snowmass
  White Paper: Generalized Symmetries in Quantum Field Theory and Beyond},'' in
  {\em {Snowmass 2021}}.
\newblock 5, 2022.
\newblock \href{http://arxiv.org/abs/2205.09545}{{\ttfamily arXiv:2205.09545
  [hep-th]}}.

\bibitem{McGreevy:2022oyu}
J.~McGreevy, ``{Generalized Symmetries in Condensed Matter},''
  \href{http://dx.doi.org/10.1146/annurev-conmatphys-040721-021029}{{\em Ann.
  Rev. Condensed Matter Phys.} {\bfseries 14} (2023) 57--82},
  \href{http://arxiv.org/abs/2204.03045}{{\ttfamily arXiv:2204.03045
  [cond-mat.str-el]}}.

\bibitem{Schafer-Nameki:2023jdn}
S.~Sch\"{a}fer-Nameki, ``{ICTP lectures on (non-)invertible generalized
  symmetries},'' \href{http://dx.doi.org/10.1016/j.physrep.2024.01.007}{{\em
  Phys. Rept.} {\bfseries 1063} (2024) 1--55},
  \href{http://arxiv.org/abs/2305.18296}{{\ttfamily arXiv:2305.18296
  [hep-th]}}.

\bibitem{Brennan:2023mmt}
T.~D. Brennan and S.~Hong, ``{Introduction to Generalized Global Symmetries in
  QFT and Particle Physics},''
  \href{http://arxiv.org/abs/2306.00912}{{\ttfamily arXiv:2306.00912
  [hep-ph]}}.

\bibitem{Bhardwaj:2023kri}
L.~Bhardwaj, L.~E. Bottini, L.~Fraser-Taliente, L.~Gladden, D.~S.~W. Gould,
  A.~Platschorre, and H.~Tillim, ``{Lectures on generalized symmetries},''
  \href{http://dx.doi.org/10.1016/j.physrep.2023.11.002}{{\em Phys. Rept.}
  {\bfseries 1051} (2024) 1--87},
  \href{http://arxiv.org/abs/2307.07547}{{\ttfamily arXiv:2307.07547
  [hep-th]}}.

\bibitem{Luo:2023ive}
R.~Luo, Q.-R. Wang, and Y.-N. Wang, ``{Lecture notes on generalized symmetries
  and applications},''
  \href{http://dx.doi.org/10.1016/j.physrep.2024.02.002}{{\em Phys. Rept.}
  {\bfseries 1065} (2024) 1--43},
  \href{http://arxiv.org/abs/2307.09215}{{\ttfamily arXiv:2307.09215
  [hep-th]}}.

\bibitem{Shao:2023gho}
S.-H. Shao, ``{What's Done Cannot Be Undone: TASI Lectures on Non-Invertible
  Symmetries},'' \href{http://arxiv.org/abs/2308.00747}{{\ttfamily
  arXiv:2308.00747 [hep-th]}}.

\bibitem{Carqueville:2023jhb}
N.~Carqueville, M.~D. Zotto, and I.~Runkel, ``Topological defects,''.
  \href{https://arxiv.org/abs/2311.02449}{arXiv:2311.02449 [math-ph]}.

\bibitem{Iqbal:2024pee}
N.~Iqbal, ``{Jena lectures on generalized global symmetries: principles and
  applications},'' \href{http://arxiv.org/abs/2407.20815}{{\ttfamily
  arXiv:2407.20815 [hep-th]}}.

\bibitem{Costa:2024wks}
D.~Costa, C.~C{\'o}rdova, M.~D. Zotto, D.~Freed, J.~G{\"o}dicke, A.~Hofer,
  D.~Jordan, D.~Morgante, R.~Moscrop, K.~Ohmori, E.~R. Garding, C.~Scheimbauer,
  and A.~{\v{S}}vraka, ``Simons lectures on categorical symmetries,''
  \href{http://arxiv.org/abs/2411.09082}{{\ttfamily arXiv:2411.09082
  [math-ph]}}.

\bibitem{Thorngren:2019iar}
R.~Thorngren and Y.~Wang, ``{Fusion category symmetry. Part I. Anomaly in-flow
  and gapped phases},'' \href{http://dx.doi.org/10.1007/JHEP04(2024)132}{{\em
  JHEP} {\bfseries 04} (2024) 132},
  \href{http://arxiv.org/abs/1912.02817}{{\ttfamily arXiv:1912.02817
  [hep-th]}}.

\bibitem{Inamura:2021wuo}
K.~Inamura, ``{Topological field theories and symmetry protected topological
  phases with fusion category symmetries},''
  \href{http://dx.doi.org/10.1007/JHEP05(2021)204}{{\em JHEP} {\bfseries 05}
  (2021) 204}, \href{http://arxiv.org/abs/2103.15588}{{\ttfamily
  arXiv:2103.15588 [cond-mat.str-el]}}.

\bibitem{Inamura:2021szw}
K.~Inamura, ``{On lattice models of gapped phases with fusion category
  symmetries},'' \href{http://dx.doi.org/10.1007/JHEP03(2022)036}{{\em JHEP}
  {\bfseries 03} (2022) 036}, \href{http://arxiv.org/abs/2110.12882}{{\ttfamily
  arXiv:2110.12882 [cond-mat.str-el]}}.

\bibitem{Garre-Rubio:2022uum}
J.~Garre-Rubio, L.~Lootens, and A.~Moln\'ar, ``{Classifying phases protected by
  matrix product operator symmetries using matrix product states},''
  \href{http://dx.doi.org/10.22331/q-2023-02-21-927}{{\em Quantum} {\bfseries
  7} (2023) 927}, \href{http://arxiv.org/abs/2203.12563}{{\ttfamily
  arXiv:2203.12563 [cond-mat.str-el]}}.

\bibitem{Fechisin:2023odt}
C.~Fechisin, N.~Tantivasadakarn, and V.~V. Albert, ``{Noninvertible
  Symmetry-Protected Topological Order in a Group-Based Cluster State},''
  \href{http://dx.doi.org/10.1103/PhysRevX.15.011058}{{\em Phys. Rev. X}
  {\bfseries 15} no.~1, (2025) 011058},
  \href{http://arxiv.org/abs/2312.09272}{{\ttfamily arXiv:2312.09272
  [cond-mat.str-el]}}.

\bibitem{Seifnashri:2024dsd}
S.~Seifnashri and S.-H. Shao, ``{Cluster State as a Noninvertible
  Symmetry-Protected Topological Phase},''
  \href{http://dx.doi.org/10.1103/PhysRevLett.133.116601}{{\em Phys. Rev.
  Lett.} {\bfseries 133} no.~11, (2024) 116601},
  \href{http://arxiv.org/abs/2404.01369}{{\ttfamily arXiv:2404.01369
  [cond-mat.str-el]}}.

\bibitem{Jia:2024bng}
Z.~Jia, ``{Generalized cluster states from Hopf algebras: non-invertible
  symmetry and Hopf tensor network representation},''
  \href{http://dx.doi.org/10.1007/JHEP09(2024)147}{{\em JHEP} {\bfseries 09}
  (2024) 147}, \href{http://arxiv.org/abs/2405.09277}{{\ttfamily
  arXiv:2405.09277 [quant-ph]}}.

\bibitem{Li:2024fhy}
Y.~Li and M.~Litvinov, ``{Non-invertible SPT, gauging and symmetry
  fractionalization},'' \href{http://arxiv.org/abs/2405.15951}{{\ttfamily
  arXiv:2405.15951 [cond-mat.str-el]}}.

\bibitem{Pace:2024acq}
S.~D. Pace, H.~T. Lam, and {\"O}.~M. Aksoy, ``{(SPT-)LSM theorems from
  projective non-invertible symmetries},''
  \href{http://dx.doi.org/10.21468/SciPostPhys.18.1.028}{{\em SciPost Phys.}
  {\bfseries 18} no.~1, (2025) 028},
  \href{http://arxiv.org/abs/2409.18113}{{\ttfamily arXiv:2409.18113
  [cond-mat.str-el]}}.

\bibitem{Meng:2024nxx}
C.~Meng, X.~Yang, T.~Lan, and Z.~Gu, ``{Non-invertible SPTs: an on-site
  realization of (1+1)d anomaly-free fusion category symmetry},''
  \href{http://arxiv.org/abs/2412.20546}{{\ttfamily arXiv:2412.20546
  [cond-mat.str-el]}}.

\bibitem{Cao:2025qhg}
W.~Cao, M.~Yamazaki, and L.~Li, ``{Duality viewpoint of noninvertible symmetry
  protected topological phases},''
  \href{http://arxiv.org/abs/2502.20435}{{\ttfamily arXiv:2502.20435
  [cond-mat.str-el]}}.

\bibitem{Aksoy:2025rmg}
{\"O}.~M. Aksoy and X.-G. Wen, ``{Phases with non-invertible symmetries in 1+1D
  -- symmetry protected topological orders as duality automorphisms},''
  \href{http://arxiv.org/abs/2503.21764}{{\ttfamily arXiv:2503.21764
  [cond-mat.str-el]}}.

\bibitem{Maeda:2025rxc}
J.~Maeda and T.~Oishi, ``{$N$-ality symmetry and SPT phases in (1+1)d},''
  \href{http://arxiv.org/abs/2504.20151}{{\ttfamily arXiv:2504.20151
  [hep-th]}}.

\bibitem{Lu:2025rwd}
D.-C. Lu, F.~Xu, and Y.-Z. You, ``{Strange correlator and string order
  parameter for non-invertible symmetry protected topological phases in
  1+1d},'' \href{http://arxiv.org/abs/2505.00673}{{\ttfamily arXiv:2505.00673
  [cond-mat.str-el]}}.

\bibitem{Furukawa:2025flp}
Y.~Furukawa, ``{Lattice models with subsystem/weak non-invertible
  symmetry-protected topological order},''
  \href{http://arxiv.org/abs/2505.11419}{{\ttfamily arXiv:2505.11419
  [cond-mat.str-el]}}.

\bibitem{ParayilMana:2025nxw}
A.~Parayil~Mana, Y.~Li, H.~Sukeno, and T.-C. Wei, ``{Higher-order topological
  phases protected by non-invertible and subsystem symmetries},''
  \href{http://arxiv.org/abs/2505.18119}{{\ttfamily arXiv:2505.18119
  [cond-mat.str-el]}}.

\bibitem{Inamura:2025cum}
K.~Inamura, S.-J. Huang, A.~Tiwari, and S.~Schafer-Nameki, ``{(2+1)d Lattice
  Models and Tensor Networks for Gapped Phases with Categorical Symmetry},''
  \href{http://arxiv.org/abs/2506.09177}{{\ttfamily arXiv:2506.09177
  [cond-mat.str-el]}}.

\bibitem{You:2025uxo}
M.~You, ``{Symmetric entanglers for non-invertible SPT phases},''
  \href{http://arxiv.org/abs/2509.04581}{{\ttfamily arXiv:2509.04581
  [cond-mat.str-el]}}.

\bibitem{Lu:2025yru}
D.-C. Lu and Z.~Sun, ``{Intrinsic NISPT Phases, igNISPT Phases, and Mixed
  Anomalies of Non-Invertible Symmetries},''
  \href{http://arxiv.org/abs/2511.01965}{{\ttfamily arXiv:2511.01965
  [hep-th]}}.

\bibitem{Cirac:2020obd}
J.~I. Cirac, D.~P\'erez-Garc\'ia, N.~Schuch, and F.~Verstraete, ``{Matrix
  product states and projected entangled pair states: Concepts, symmetries,
  theorems},'' \href{http://dx.doi.org/10.1103/RevModPhys.93.045003}{{\em Rev.
  Mod. Phys.} {\bfseries 93} no.~4, (2021) 045003},
  \href{http://arxiv.org/abs/2011.12127}{{\ttfamily arXiv:2011.12127
  [quant-ph]}}.

\bibitem{Molnar:2022nmh}
A.~Molnar, A.~R. de~Alarc\'on, J.~Garre-Rubio, N.~Schuch, J.~I. Cirac, and
  D.~P\'erez-Garc\'\i{}a, ``{Matrix product operator algebras I:
  representations of weak Hopf algebras and projected entangled pair states},''
  \href{http://arxiv.org/abs/2204.05940}{{\ttfamily arXiv:2204.05940
  [quant-ph]}}.

\bibitem{Lootens:2021tet}
L.~Lootens, C.~Delcamp, G.~Ortiz, and F.~Verstraete, ``{Dualities in
  One-Dimensional Quantum Lattice Models: Symmetric Hamiltonians and Matrix
  Product Operator Intertwiners},''
  \href{http://dx.doi.org/10.1103/PRXQuantum.4.020357}{{\em PRX Quantum}
  {\bfseries 4} no.~2, (2023) 020357},
  \href{http://arxiv.org/abs/2112.09091}{{\ttfamily arXiv:2112.09091
  [quant-ph]}}.

\bibitem{Lootens:2022avn}
L.~Lootens, C.~Delcamp, and F.~Verstraete, ``{Dualities in One-Dimensional
  Quantum Lattice Models: Topological Sectors},''
  \href{http://dx.doi.org/10.1103/PRXQuantum.5.010338}{{\em PRX Quantum}
  {\bfseries 5} no.~1, (2024) 010338},
  \href{http://arxiv.org/abs/2211.03777}{{\ttfamily arXiv:2211.03777
  [quant-ph]}}.

\bibitem{Gorantla:2024ocs}
P.~Gorantla, S.-H. Shao, and N.~Tantivasadakarn, ``{Tensor Networks for
  Noninvertible Symmetries in 3+1D and Beyond},''
  \href{http://dx.doi.org/10.1103/p32z-v884}{{\em Phys. Rev. X} {\bfseries 15}
  no.~4, (2025) 041006}, \href{http://arxiv.org/abs/2406.12978}{{\ttfamily
  arXiv:2406.12978 [quant-ph]}}.

\bibitem{Bhardwaj:2017xup}
L.~Bhardwaj and Y.~Tachikawa, ``{On finite symmetries and their gauging in two
  dimensions},'' \href{http://dx.doi.org/10.1007/JHEP03(2018)189}{{\em JHEP}
  {\bfseries 03} (2018) 189}, \href{http://arxiv.org/abs/1704.02330}{{\ttfamily
  arXiv:1704.02330 [hep-th]}}.

\bibitem{EGNO2015}
P.~Etingof, S.~Gelaki, D.~Nikshych, and V.~Ostrik,
  \href{http://dx.doi.org/10.1090/surv/205}{{\em {Tensor Categories}}},
  vol.~205 of {\em Mathematical Surveys and Monographs}.
\newblock American Mathematical Society, Providence, RI, 2015.
\newblock \url{http://www-math.mit.edu/~etingof/egnobookfinal.pdf}.

\bibitem{Douglas:2018qfz}
C.~L. Douglas and D.~J. Reutter, ``{Fusion 2-categories and a state-sum
  invariant for 4-manifolds},''
  \href{http://arxiv.org/abs/1812.11933}{{\ttfamily arXiv:1812.11933
  [math.QA]}}.

\bibitem{Decoppet:2024htz}
T.~D. D{\'e}coppet, P.~Huston, T.~Johnson-Freyd, D.~Nikshych, D.~Penneys,
  J.~Plavnik, D.~Reutter, and M.~Yu, ``{The Classification of Fusion
  2-Categories},'' \href{http://arxiv.org/abs/2411.05907}{{\ttfamily
  arXiv:2411.05907 [math.CT]}}.

\bibitem{Bhardwaj:2024qiv}
L.~Bhardwaj, D.~Pajer, S.~Schafer-Nameki, A.~Tiwari, A.~Warman, and J.~Wu,
  ``{Gapped phases in (2+1)d with non-invertible symmetries: Part I},''
  \href{http://dx.doi.org/10.21468/SciPostPhys.19.2.056}{{\em SciPost Phys.}
  {\bfseries 19} no.~2, (2025) 056},
  \href{http://arxiv.org/abs/2408.05266}{{\ttfamily arXiv:2408.05266
  [hep-th]}}.

\bibitem{Bullimore:2024khm}
M.~Bullimore and J.~J. Pearson, ``{Towards All Categorical Symmetries in 2+1
  Dimensions},'' \href{http://arxiv.org/abs/2408.13931}{{\ttfamily
  arXiv:2408.13931 [hep-th]}}.

\bibitem{Kaidi:2021xfk}
J.~Kaidi, K.~Ohmori, and Y.~Zheng, ``{Kramers-Wannier-like Duality Defects in
  (3+1)D Gauge Theories},''
  \href{http://dx.doi.org/10.1103/PhysRevLett.128.111601}{{\em Phys. Rev.
  Lett.} {\bfseries 128} no.~11, (2022) 111601},
  \href{http://arxiv.org/abs/2111.01141}{{\ttfamily arXiv:2111.01141
  [hep-th]}}.

\bibitem{Bhardwaj:2022lsg}
L.~Bhardwaj, S.~Sch\"{a}fer-Nameki, and J.~Wu, ``{Universal Non-Invertible
  Symmetries},'' \href{http://dx.doi.org/10.1002/prop.202200143}{{\em Fortsch.
  Phys.} {\bfseries 70} no.~11, (2022) 2200143},
  \href{http://arxiv.org/abs/2208.05973}{{\ttfamily arXiv:2208.05973
  [hep-th]}}.

\bibitem{Bhardwaj:2022maz}
L.~Bhardwaj, L.~E. Bottini, S.~Schafer-Nameki, and A.~Tiwari, ``{Non-invertible
  symmetry webs},'' \href{http://dx.doi.org/10.21468/SciPostPhys.15.4.160}{{\em
  SciPost Phys.} {\bfseries 15} no.~4, (2023) 160},
  \href{http://arxiv.org/abs/2212.06842}{{\ttfamily arXiv:2212.06842
  [hep-th]}}.

\bibitem{Bartsch:2022mpm}
T.~Bartsch, M.~Bullimore, A.~E.~V. Ferrari, and J.~Pearson, ``{Non-invertible
  symmetries and higher representation theory I},''
  \href{http://dx.doi.org/10.21468/SciPostPhys.17.1.015}{{\em SciPost Phys.}
  {\bfseries 17} no.~1, (2024) 015},
  \href{http://arxiv.org/abs/2208.05993}{{\ttfamily arXiv:2208.05993
  [hep-th]}}.

\bibitem{Bartsch:2022ytj}
T.~Bartsch, M.~Bullimore, A.~E.~V. Ferrari, and J.~Pearson, ``{Non-invertible
  symmetries and higher representation theory II},''
  \href{http://dx.doi.org/10.21468/SciPostPhys.17.2.067}{{\em SciPost Phys.}
  {\bfseries 17} no.~2, (2024) 067},
  \href{http://arxiv.org/abs/2212.07393}{{\ttfamily arXiv:2212.07393
  [hep-th]}}.

\bibitem{Delcamp:2023kew}
C.~Delcamp and A.~Tiwari, ``{Higher categorical symmetries and gauging in
  two-dimensional spin systems},''
  \href{http://dx.doi.org/10.21468/SciPostPhys.16.4.110}{{\em SciPost Phys.}
  {\bfseries 16} no.~4, (2024) 110},
  \href{http://arxiv.org/abs/2301.01259}{{\ttfamily arXiv:2301.01259
  [hep-th]}}.

\bibitem{Inamura:2023qzl}
K.~Inamura and K.~Ohmori, ``{Fusion Surface Models: 2+1d Lattice Models from
  Fusion 2-Categories},''
  \href{http://dx.doi.org/10.21468/SciPostPhys.16.6.143}{{\em SciPost Phys.}
  {\bfseries 16} (2024) 143}, \href{http://arxiv.org/abs/2305.05774}{{\ttfamily
  arXiv:2305.05774 [cond-mat.str-el]}}.

\bibitem{Hsin:2024aqb}
P.-S. Hsin, R.~Kobayashi, and C.~Zhang, ``{Fractionalization of coset
  non-invertible symmetry and exotic Hall conductance},''
  \href{http://dx.doi.org/10.21468/SciPostPhys.17.3.095}{{\em SciPost Phys.}
  {\bfseries 17} no.~3, (2024) 095},
  \href{http://arxiv.org/abs/2405.20401}{{\ttfamily arXiv:2405.20401
  [cond-mat.str-el]}}.

\bibitem{Cordova:2024jlk}
C.~Cordova, D.~B. Costa, and P.-S. Hsin, ``{Non-invertible symmetries in
  finite-group gauge theory},''
  \href{http://dx.doi.org/10.21468/SciPostPhys.18.1.019}{{\em SciPost Phys.}
  {\bfseries 18} no.~1, (2025) 019},
  \href{http://arxiv.org/abs/2407.07964}{{\ttfamily arXiv:2407.07964
  [cond-mat.str-el]}}.

\bibitem{Cordova:2024mqg}
C.~Cordova, D.~B. Costa, and P.-S. Hsin, ``{Non-Invertible Symmetries as
  Condensation Defects in Finite-Group Gauge Theories},''
  \href{http://arxiv.org/abs/2412.16681}{{\ttfamily arXiv:2412.16681
  [cond-mat.str-el]}}.

\bibitem{Bhardwaj:2025piv}
L.~Bhardwaj, S.~Schafer-Nameki, A.~Tiwari, and A.~Warman, ``{Gapped Phases in
  (2+1)d with Non-Invertible Symmetries: Part II},''
  \href{http://arxiv.org/abs/2502.20440}{{\ttfamily arXiv:2502.20440
  [hep-th]}}.

\bibitem{Eck:2025ldx}
L.~Eck, ``{Dualities between 2+1d fusion surface models from braided fusion
  categories},'' \href{http://dx.doi.org/10.21468/SciPostPhys.19.6.157}{{\em
  SciPost Phys.} {\bfseries 19} no.~6, (2025) 157},
  \href{http://arxiv.org/abs/2501.14722}{{\ttfamily arXiv:2501.14722
  [cond-mat.str-el]}}.

\bibitem{Vancraeynest-DeCuiper:2025wkh}
B.~Vancraeynest-De~Cuiper and C.~Delcamp, ``{Twisted gauging and topological
  sectors in (2+1)d Abelian lattice gauge theories},''
  \href{http://dx.doi.org/10.21468/SciPostPhys.19.2.054}{{\em SciPost Phys.}
  {\bfseries 19} no.~2, (2025) 054},
  \href{http://arxiv.org/abs/2501.16301}{{\ttfamily arXiv:2501.16301
  [cond-mat.str-el]}}.

\bibitem{Hsin:2025ria}
P.-S. Hsin, R.~Kobayashi, and C.~Zhang, ``{Anomalies of Coset Non-Invertible
  Symmetries},'' \href{http://arxiv.org/abs/2503.00105}{{\ttfamily
  arXiv:2503.00105 [cond-mat.str-el]}}.

\bibitem{KNBalasubramanian:2025vum}
M.~K.~N.~Balasubramanian, M.~Buican, C.~Delcamp, and R.~Radhakrishnan,
  ``{Gauging Non-Invertible Symmetries in (2+1)d Topological Orders},''
  \href{http://arxiv.org/abs/2507.01142}{{\ttfamily arXiv:2507.01142
  [hep-th]}}.

\bibitem{Decoppet:2023bay}
T.~D. D{\'e}coppet and M.~Yu, ``{Fiber 2-Functors and
  Tambara{\textendash}Yamagami Fusion 2-Categories},''
  \href{http://dx.doi.org/10.1007/s00220-025-05249-x}{{\em Commun. Math. Phys.}
  {\bfseries 406} no.~3, (2025) 64},
  \href{http://arxiv.org/abs/2306.08117}{{\ttfamily arXiv:2306.08117
  [math.CT]}}.

\bibitem{Cordova:2024iti}
C.~Cordova, N.~Holfester, and K.~Ohmori, ``{Representation theory of
  solitons},'' \href{http://dx.doi.org/10.1007/JHEP06(2025)001}{{\em JHEP}
  {\bfseries 06} (2025) 001}, \href{http://arxiv.org/abs/2408.11045}{{\ttfamily
  arXiv:2408.11045 [hep-th]}}.

\bibitem{Bridgeman:2022gdx}
J.~C. Bridgeman, L.~Lootens, and F.~Verstraete, ``{Invertible Bimodule
  Categories and Generalized Schur Orthogonality},''
  \href{http://dx.doi.org/10.1007/s00220-023-04781-y}{{\em Commun. Math. Phys.}
  {\bfseries 402} no.~3, (2023) 2691--2714},
  \href{http://arxiv.org/abs/2211.01947}{{\ttfamily arXiv:2211.01947
  [math.QA]}}.

\bibitem{Jia:2024rzr}
Z.~Jia, S.~Tan, and D.~Kaszlikowski, ``{Weak Hopf symmetry and tube algebra of
  the generalized multifusion string-net model},''
  \href{http://dx.doi.org/10.1007/JHEP07(2024)207}{{\em JHEP} {\bfseries 07}
  (2024) 207}, \href{http://arxiv.org/abs/2403.04446}{{\ttfamily
  arXiv:2403.04446 [hep-th]}}.

\bibitem{Choi:2024tri}
Y.~Choi, B.~C. Rayhaun, and Y.~Zheng, ``{Generalized Tube Algebras,
  Symmetry-Resolved Partition Functions, and Twisted Boundary States},''
  \href{http://arxiv.org/abs/2409.02159}{{\ttfamily arXiv:2409.02159
  [hep-th]}}.

\bibitem{Choi:2024wfm}
Y.~Choi, B.~C. Rayhaun, and Y.~Zheng, ``{Noninvertible Symmetry-Resolved
  Affleck-Ludwig-Cardy Formula and Entanglement Entropy from the Boundary Tube
  Algebra},'' \href{http://dx.doi.org/10.1103/PhysRevLett.133.251602}{{\em
  Phys. Rev. Lett.} {\bfseries 133} no.~25, (2024) 251602},
  \href{http://arxiv.org/abs/2409.02806}{{\ttfamily arXiv:2409.02806
  [hep-th]}}.

\bibitem{Jia:2024zdp}
Z.~Jia, ``{Weak Hopf non-invertible symmetry-protected topological spin liquid
  and lattice realization of (1+1)D symmetry topological field theory},''
  \href{http://arxiv.org/abs/2412.15336}{{\ttfamily arXiv:2412.15336
  [hep-th]}}.

\bibitem{Jia:2025yph}
Z.~Jia and S.~Tan, ``{Weak Hopf tube algebra for domain walls between 2d gapped
  phases of Turaev-Viro TQFTs},''
  \href{http://dx.doi.org/10.1007/JHEP11(2025)018}{{\em JHEP} {\bfseries 11}
  (2025) 018}, \href{http://arxiv.org/abs/2507.01515}{{\ttfamily
  arXiv:2507.01515 [hep-th]}}.

\bibitem{Choi:2023xjw}
Y.~Choi, B.~C. Rayhaun, Y.~Sanghavi, and S.-H. Shao, ``{Remarks on boundaries,
  anomalies, and noninvertible symmetries},''
  \href{http://dx.doi.org/10.1103/PhysRevD.108.125005}{{\em Phys. Rev. D}
  {\bfseries 108} no.~12, (2023) 125005},
  \href{http://arxiv.org/abs/2305.09713}{{\ttfamily arXiv:2305.09713
  [hep-th]}}.

\bibitem{Cordova:2024vsq}
C.~C\'{o}rdova, D.~Garc\'\i{}a-Sep\'ulveda, and N.~Holfester,
  ``{Particle-soliton degeneracies from spontaneously broken non-invertible
  symmetry},'' \href{http://dx.doi.org/10.1007/JHEP07(2024)154}{{\em JHEP}
  {\bfseries 07} (2024) 154}, \href{http://arxiv.org/abs/2403.08883}{{\ttfamily
  arXiv:2403.08883 [hep-th]}}.

\bibitem{Copetti:2024onh}
C.~Copetti, ``{Defect Charges, Gapped Boundary Conditions, and the Symmetry
  TFT},'' \href{http://arxiv.org/abs/2408.01490}{{\ttfamily arXiv:2408.01490
  [hep-th]}}.

\bibitem{Copetti:2024dcz}
C.~Copetti, L.~Cordova, and S.~Komatsu, ``{S-matrix bootstrap and
  non-invertible symmetries},''
  \href{http://dx.doi.org/10.1007/JHEP03(2025)204}{{\em JHEP} {\bfseries 03}
  (2025) 204}, \href{http://arxiv.org/abs/2408.13132}{{\ttfamily
  arXiv:2408.13132 [hep-th]}}.

\bibitem{Bhardwaj:2024igy}
L.~Bhardwaj, C.~Copetti, D.~Pajer, and S.~Schafer-Nameki, ``{Boundary
  SymTFT},'' \href{http://dx.doi.org/10.21468/SciPostPhys.19.2.061}{{\em
  SciPost Phys.} {\bfseries 19} no.~2, (2025) 061},
  \href{http://arxiv.org/abs/2409.02166}{{\ttfamily arXiv:2409.02166
  [hep-th]}}.

\bibitem{Cordova:2024nux}
C.~Cordova, D.~Garc{\'\i}a-Sep{\'u}lveda, and N.~Holfester, ``{Particle-Soliton
  Degeneracy in 2D Quantum Chromodynamics},''
  \href{http://arxiv.org/abs/2412.21153}{{\ttfamily arXiv:2412.21153
  [hep-th]}}.

\bibitem{Heymann:2024vvf}
J.~Heymann and T.~Quella, ``{Revisiting the symmetry-resolved entanglement for
  noninvertible symmetries in 1+1d conformal field theories},''
  \href{http://dx.doi.org/10.1103/lr47-yv3j}{{\em Phys. Rev. D} {\bfseries 112}
  no.~2, (2025) 025004}, \href{http://arxiv.org/abs/2409.02315}{{\ttfamily
  arXiv:2409.02315 [hep-th]}}.

\bibitem{AliAhmad:2025bnd}
S.~Ali~Ahmad, M.~S. Klinger, and Y.~Wang, ``{The Many Faces of Non-invertible
  Symmetries},'' \href{http://arxiv.org/abs/2509.18072}{{\ttfamily
  arXiv:2509.18072 [hep-th]}}.

\bibitem{Benini:2025lav}
F.~Benini, P.~Calabrese, M.~Fossati, A.~H. Singh, and M.~Venuti,
  ``{Entanglement Asymmetry for Higher and Noninvertible Symmetries},''
  \href{http://arxiv.org/abs/2509.16311}{{\ttfamily arXiv:2509.16311
  [hep-th]}}.

\bibitem{Kitaev:2011dxc}
A.~Kitaev and L.~Kong, ``{Models for Gapped Boundaries and Domain Walls},''
  \href{http://dx.doi.org/10.1007/s00220-012-1500-5}{{\em Commun. Math. Phys.}
  {\bfseries 313} no.~2, (2012) 351--373},
  \href{http://arxiv.org/abs/1104.5047}{{\ttfamily arXiv:1104.5047
  [cond-mat.str-el]}}.

\bibitem{Lan:2013wia}
T.~Lan and X.-G. Wen, ``{Topological quasiparticles and the holographic
  bulk-edge relation in (2+1) -dimensional string-net models},''
  \href{http://dx.doi.org/10.1103/PhysRevB.90.115119}{{\em Phys. Rev. B}
  {\bfseries 90} no.~11, (2014) 115119},
  \href{http://arxiv.org/abs/1311.1784}{{\ttfamily arXiv:1311.1784
  [cond-mat.str-el]}}.

\bibitem{Bridgeman:2018jdv}
J.~C. Bridgeman, D.~Barter, and C.~Jones, ``{Fusing Binary Interface Defects in
  Topological Phases: The $\operatorname{Vec}(\mathbb{Z}/p\mathbb{Z})$ case},''
  \href{http://dx.doi.org/10.1063/1.5095941}{{\em J. Math. Phys.} {\bfseries
  60} no.~12, (2019) 121701}, \href{http://arxiv.org/abs/1810.09469}{{\ttfamily
  arXiv:1810.09469 [math.QA]}}.

\bibitem{Bridgeman:2019axg}
J.~C. Bridgeman and D.~Barter, ``{Computing defects associated to bounded
  domain wall structures: the case},''
  \href{http://dx.doi.org/10.1088/1751-8121/ab7d60}{{\em J. Phys. A} {\bfseries
  53} no.~23, (2020) 235206}, \href{http://arxiv.org/abs/1901.08069}{{\ttfamily
  arXiv:1901.08069 [math.QA]}}.

\bibitem{Bridgeman:2019wyu}
J.~C. Bridgeman and D.~Barter, ``{Computing data for Levin-Wen with defects},''
  \href{http://dx.doi.org/10.22331/q-2020-06-04-277}{{\em Quantum} {\bfseries
  4} (2020) 277}, \href{http://arxiv.org/abs/1907.06692}{{\ttfamily
  arXiv:1907.06692 [quant-ph]}}.

\bibitem{Barter_2022}
D.~Barter, J.~Bridgeman, and R.~Wolf, ``{Computing associators of endomorphism
  fusion categories},''
  \href{http://dx.doi.org/10.21468/SciPostPhys.13.2.029}{{\em SciPost Physics}
  {\bfseries 13} no.~2, (2022) 029},
  \href{http://arxiv.org/abs/2110.03644}{{\ttfamily arXiv:2110.03644
  [math.QA]}}.

\bibitem{Gagliano:2025gwr}
F.~Gagliano, A.~Grigoletto, and K.~Ohmori, ``{Higher Representations and Quark
  Confinement},'' \href{http://arxiv.org/abs/2501.09069}{{\ttfamily
  arXiv:2501.09069 [hep-th]}}.

\bibitem{KV1994}
M.~M. Kapranov and V.~A. Voevodsky, {\em {2-categories and Zamolodchikov
  tetrahedra equations}}, vol.~56 of {\em Proc. Sympos. Pure Math.},
  p.~177{\textendash}259.
\newblock Amer. Math. Soc., Providence, RI, 1994.

\bibitem{gordon1995coherence}
R.~Gordon, A.~J. Power, and R.~Street, {\em {Coherence for tricategories}},
  vol.~558.
\newblock American Mathematical Soc., 1995.
\newblock \url{https://www.ams.org/books/memo/0558/}.

\bibitem{Gurski2007}
N.~Gurski, {\em {An algebraic theory of tricategories}}.
\newblock PhD thesis, University of Chicago, 2007.
\newblock
  \url{https://ncatlab.org/nlab/files/Gurski-AlgebraicTricategories.pdf}.

\bibitem{Kong:2014qka}
L.~Kong and X.-G. Wen, ``{Braided fusion categories, gravitational anomalies,
  and the mathematical framework for topological orders in any dimensions},''
  \href{http://arxiv.org/abs/1405.5858}{{\ttfamily arXiv:1405.5858
  [cond-mat.str-el]}}.

\bibitem{Gaiotto:2019xmp}
D.~Gaiotto and T.~Johnson-Freyd, ``{Condensations in higher categories},''
  \href{http://arxiv.org/abs/1905.09566}{{\ttfamily arXiv:1905.09566
  [math.CT]}}.

\bibitem{Johnson-Freyd:2020ivj}
T.~Johnson-Freyd and M.~Yu, ``{Fusion 2-categories With no Line Operators are
  Grouplike},'' \href{http://dx.doi.org/10.1017/S0004972721000095}{{\em Bull.
  Austral. Math. Soc.} {\bfseries 104} no.~3, (2021) 434--442},
  \href{http://arxiv.org/abs/2010.07950}{{\ttfamily arXiv:2010.07950
  [math.QA]}}.

\bibitem{Kong:2024ykr}
L.~Kong, Z.-H. Zhang, J.~Zhao, and H.~Zheng, ``{Higher condensation theory},''
  \href{http://arxiv.org/abs/2403.07813}{{\ttfamily arXiv:2403.07813
  [cond-mat.str-el]}}.

\bibitem{Decoppet2023Morita}
T.~D. D\'{e}coppet, ``{The Morita Theory of Fusion 2-Categories},''
  \href{http://dx.doi.org/10.21136/HS.2023.07}{{\em Higher Structures}
  {\bfseries 7} no.~1, (5, 2023) 234--292},
  \href{http://arxiv.org/abs/2208.08722}{{\ttfamily arXiv:2208.08722
  [math.CT]}}.

\bibitem{Greenough:0911.4979}
J.~Greenough, ``{Monoidal 2-structure of bimodule categories},'' {\em Journal
  of Algebra} {\bfseries 324} no.~8, (2010) 1818--1859,
  \href{http://arxiv.org/abs/0911.4979}{{\ttfamily arXiv:0911.4979 [math.QA]}}.

\bibitem{Tachikawa:2017gyf}
Y.~Tachikawa, ``{On gauging finite subgroups},''
  \href{http://dx.doi.org/10.21468/SciPostPhys.8.1.015}{{\em SciPost Phys.}
  {\bfseries 8} no.~1, (2020) 015},
  \href{http://arxiv.org/abs/1712.09542}{{\ttfamily arXiv:1712.09542
  [hep-th]}}.

\bibitem{ENO2010}
{P. Etingof, D. Nikshych, V. Ostrik, with an appendix by E. Meir}, ``{Fusion
  categories and homotopy theory},'' {\em Quantum topology} {\bfseries 1}
  no.~3, (2010) 209--273, \href{http://arxiv.org/abs/0909.3140}{{\ttfamily
  arXiv:0909.3140 [math.QA]}}.

\bibitem{Douglas:1406.4204}
C.~L. Douglas, C.~Schommer-Pries, and N.~Snyder, ``{The balanced tensor product
  of module categories},''
  \href{http://dx.doi.org/10.1215/21562261-2018-0006}{{\em Kyoto Journal of
  Mathematics} {\bfseries 59} no.~1, (Apr., 2019) 167--179},
  \href{http://arxiv.org/abs/1406.4204}{{\ttfamily arXiv:1406.4204 [math.QA]}}.

\bibitem{Delcamp:2021szr}
C.~Delcamp, ``{Tensor network approach to electromagnetic duality in (3+1)d
  topological gauge models},''
  \href{http://dx.doi.org/10.1007/JHEP08(2022)149}{{\em JHEP} {\bfseries 08}
  (2022) 149}, \href{http://arxiv.org/abs/2112.08324}{{\ttfamily
  arXiv:2112.08324 [cond-mat.str-el]}}.

\bibitem{Barkeshli:2022edm}
M.~Barkeshli, Y.-A. Chen, P.-S. Hsin, and R.~Kobayashi, ``{Higher-group
  symmetry in finite gauge theory and stabilizer codes},''
  \href{http://dx.doi.org/10.21468/SciPostPhys.16.4.089}{{\em SciPost Phys.}
  {\bfseries 16} no.~4, (2024) 089},
  \href{http://arxiv.org/abs/2211.11764}{{\ttfamily arXiv:2211.11764
  [cond-mat.str-el]}}.

\bibitem{Komargodski:2020mxz}
Z.~Komargodski, K.~Ohmori, K.~Roumpedakis, and S.~Seifnashri, ``{Symmetries and
  strings of adjoint QCD$_{2}$},''
  \href{http://dx.doi.org/10.1007/JHEP03(2021)103}{{\em JHEP} {\bfseries 03}
  (2021) 103}, \href{http://arxiv.org/abs/2008.07567}{{\ttfamily
  arXiv:2008.07567 [hep-th]}}.

\bibitem{Turaev:1992hq}
V.~G. Turaev and O.~Y. Viro, ``{State sum invariants of 3 manifolds and quantum
  6j symbols},'' \href{http://dx.doi.org/10.1016/0040-9383(92)90015-A}{{\em
  Topology} {\bfseries 31} (1992) 865--902}.

\bibitem{Barrett:1993ab}
J.~W. Barrett and B.~W. Westbury, ``{Invariants of piecewise linear three
  manifolds},'' \href{http://dx.doi.org/10.1090/S0002-9947-96-01660-1}{{\em
  Trans. Am. Math. Soc.} {\bfseries 348} (1996) 3997--4022},
  \href{http://arxiv.org/abs/hep-th/9311155}{{\ttfamily arXiv:hep-th/9311155}}.

\bibitem{Fuchs:2012dt}
J.~Fuchs, C.~Schweigert, and A.~Valentino, ``{Bicategories for boundary
  conditions and for surface defects in 3-d TFT},''
  \href{http://dx.doi.org/10.1007/s00220-013-1723-0}{{\em Commun. Math. Phys.}
  {\bfseries 321} (2013) 543--575},
  \href{http://arxiv.org/abs/1203.4568}{{\ttfamily arXiv:1203.4568 [hep-th]}}.

\bibitem{Meusburger:2022zul}
C.~Meusburger, ``{State sum models with defects based on spherical fusion
  categories},'' \href{http://dx.doi.org/10.1016/j.aim.2023.109177}{{\em Adv.
  Math.} {\bfseries 429} (2023) 109177},
  \href{http://arxiv.org/abs/2205.06874}{{\ttfamily arXiv:2205.06874
  [math.QA]}}.

\bibitem{TY1998}
D.~Tambara and S.~Yamagami, ``{Tensor Categories with Fusion Rules of
  Self-Duality for Finite Abelian Groups},''
  \href{http://www.sciencedirect.com/science/article/pii/S0021869398975585}{{\em
  Journal of Algebra} {\bfseries 209} no.~2, (1998) 692 -- 707}.

\bibitem{Haegeman:2014maa}
J.~Haegeman, K.~Van~Acoleyen, N.~Schuch, J.~I. Cirac, and F.~Verstraete,
  ``{Gauging quantum states: from global to local symmetries in many-body
  systems},'' \href{http://dx.doi.org/10.1103/PhysRevX.5.011024}{{\em Phys.
  Rev. X} {\bfseries 5} no.~1, (2015) 011024},
  \href{http://arxiv.org/abs/1407.1025}{{\ttfamily arXiv:1407.1025
  [quant-ph]}}.

\bibitem{Williamson:2017uzx}
D.~J. Williamson, N.~Bultinck, and F.~Verstraete, ``{Symmetry-enriched
  topological order in tensor networks: Defects, gauging and anyon
  condensation},'' \href{http://arxiv.org/abs/1711.07982}{{\ttfamily
  arXiv:1711.07982 [quant-ph]}}.

\bibitem{Tantivasadakarn:2022hgp}
N.~Tantivasadakarn, A.~Vishwanath, and R.~Verresen, ``{Hierarchy of Topological
  Order From Finite-Depth Unitaries, Measurement, and Feedforward},''
  \href{http://dx.doi.org/10.1103/PRXQuantum.4.020339}{{\em PRX Quantum}
  {\bfseries 4} no.~2, (2023) 020339},
  \href{http://arxiv.org/abs/2209.06202}{{\ttfamily arXiv:2209.06202
  [quant-ph]}}.

\bibitem{Brell_2015}
C.~G. Brell, ``Generalized cluster states based on finite groups,''
  \href{http://dx.doi.org/10.1088/1367-2630/17/2/023029}{{\em New Journal of
  Physics} {\bfseries 17} no.~2, (Feb., 2015) 023029},
  \href{http://arxiv.org/abs/1408.6237}{{\ttfamily arXiv:1408.6237
  [quant-ph]}}.

\bibitem{Ohyama:2026oay}
S.~Ohyama and K.~Inamura, ``{Parameterized families of 2+1d $G$-cluster
  states},'' \href{http://arxiv.org/abs/2601.08616}{{\ttfamily arXiv:2601.08616
  [cond-mat.str-el]}}.

\bibitem{Briegel:0004051}
H.~J. Briegel and R.~Raussendorf, ``{Persistent Entanglement in Arrays of
  Interacting Particles},''
  \href{https://link.aps.org/doi/10.1103/PhysRevLett.86.910}{{\em Phys. Rev.
  Lett.} {\bfseries 86} (Jan, 2001) 910--913},
  \href{http://arxiv.org/abs/quant-ph/0004051}{{\ttfamily
  arXiv:quant-ph/0004051 [quant-ph]}}.

\bibitem{PhysRevLett.86.5188}
R.~Raussendorf and H.~J. Briegel, ``{A One-Way Quantum Computer},''
  \href{https://link.aps.org/doi/10.1103/PhysRevLett.86.5188}{{\em Phys. Rev.
  Lett.} {\bfseries 86} (May, 2001) 5188--5191}.

\bibitem{Raussendorf:0301052}
R.~Raussendorf, D.~E. Browne, and H.~J. Briegel, ``{Measurement-based quantum
  computation on cluster states},''
  \href{https://link.aps.org/doi/10.1103/PhysRevA.68.022312}{{\em Phys. Rev. A}
  {\bfseries 68} (Aug, 2003) 022312},
  \href{http://arxiv.org/abs/quant-ph/0301052}{{\ttfamily
  arXiv:quant-ph/0301052 [quant-ph]}}.

\bibitem{Yoshida:2015cia}
B.~Yoshida, ``{Topological phases with generalized global symmetries},''
  \href{http://dx.doi.org/10.1103/PhysRevB.93.155131}{{\em Phys. Rev. B}
  {\bfseries 93} no.~15, (2016) 155131},
  \href{http://arxiv.org/abs/1508.03468}{{\ttfamily arXiv:1508.03468
  [cond-mat.str-el]}}.

\bibitem{Evans:2025msy}
D.~E. Evans and C.~Jones, ``{An operator algebraic approach to fusion category
  symmetry on the lattice},'' \href{http://arxiv.org/abs/2507.05185}{{\ttfamily
  arXiv:2507.05185 [math-ph]}}.

\bibitem{Seifnashri:2025vhf}
S.~Seifnashri and W.~Shirley, ``{Disentangling anomaly-free symmetries of
  quantum spin chains},'' \href{http://arxiv.org/abs/2503.09717}{{\ttfamily
  arXiv:2503.09717 [cond-mat.str-el]}}.

\bibitem{Feng:2025yge}
Y.~Feng, Y.-A. Chen, P.-S. Hsin, and R.~Kobayashi, ``{Onsiteability of
  Higher-Form Symmetries},'' \href{http://arxiv.org/abs/2510.23701}{{\ttfamily
  arXiv:2510.23701 [cond-mat.str-el]}}.

\bibitem{Bai:2025zze}
A.~Bai and Z.-H. Zhang, ``{On the Representation Categories of Weak Hopf
  Algebras Arising from Levin-Wen Models},''
  \href{http://arxiv.org/abs/2503.06731}{{\ttfamily arXiv:2503.06731
  [math.QA]}}.

\bibitem{FNW92}
M.~Fannes, B.~Nachtergaele, and R.~F. Werner, ``{Finitely Correlated States on
  Quantum Spin Chains},'' \href{http://dx.doi.org/10.1007/BF02099178}{{\em
  Commun. Math. Phys.} {\bfseries 144} (1992) 443--490}.

\bibitem{PhysRevB.73.094423}
F.~Verstraete and J.~I. Cirac, ``Matrix product states represent ground states
  faithfully,'' \href{https://link.aps.org/doi/10.1103/PhysRevB.73.094423}{{\em
  Phys. Rev. B} {\bfseries 73} (Mar, 2006) 094423},
  \href{http://arxiv.org/abs/cond-mat/0505140}{{\ttfamily
  arXiv:cond-mat/0505140 [cond-mat.str-el]}}.

\bibitem{Perez-Garcia:2006nqo}
D.~P\'erez-Garc\'ia, F.~Verstraete, M.~M. Wolf, and J.~I. Cirac, ``{Matrix
  product state representations},''
  \href{http://dx.doi.org/10.26421/QIC7.5-6-1}{{\em Quant. Inf. Comput.}
  {\bfseries 7} no.~5-6, (2007) 401--430},
  \href{http://arxiv.org/abs/quant-ph/0608197}{{\ttfamily
  arXiv:quant-ph/0608197}}.

\bibitem{Hastings:2007iok}
M.~B. Hastings, ``{An area law for one-dimensional quantum systems},''
  \href{http://dx.doi.org/10.1088/1742-5468/2007/08/P08024}{{\em J. Stat.
  Mech.} {\bfseries 0708} (2007) P08024},
  \href{http://arxiv.org/abs/0705.2024}{{\ttfamily arXiv:0705.2024
  [quant-ph]}}.

\bibitem{Huang:2021nvb}
T.-C. Huang, Y.-H. Lin, K.~Ohmori, Y.~Tachikawa, and M.~Tezuka, ``{Numerical
  Evidence for a Haagerup Conformal Field Theory},''
  \href{http://dx.doi.org/10.1103/PhysRevLett.128.231603}{{\em Phys. Rev.
  Lett.} {\bfseries 128} no.~23, (2022) 231603},
  \href{http://arxiv.org/abs/2110.03008}{{\ttfamily arXiv:2110.03008
  [cond-mat.stat-mech]}}.

\bibitem{Vanhove:2021zop}
R.~Vanhove, L.~Lootens, M.~Van~Damme, R.~Wolf, T.~J. Osborne, J.~Haegeman, and
  F.~Verstraete, ``{Critical Lattice Model for a Haagerup Conformal Field
  Theory},'' \href{http://dx.doi.org/10.1103/PhysRevLett.128.231602}{{\em Phys.
  Rev. Lett.} {\bfseries 128} no.~23, (2022) 231602},
  \href{http://arxiv.org/abs/2110.03532}{{\ttfamily arXiv:2110.03532
  [cond-mat.stat-mech]}}.

\bibitem{Liu:2022qwn}
Y.~Liu, Y.~Zou, and S.~Ryu, ``{Operator fusion from wave-function overlap:
  Universal finite-size corrections and application to the Haagerup model},''
  \href{http://dx.doi.org/10.1103/PhysRevB.107.155124}{{\em Phys. Rev. B}
  {\bfseries 107} no.~15, (2023) 155124},
  \href{http://arxiv.org/abs/2203.14992}{{\ttfamily arXiv:2203.14992
  [cond-mat.str-el]}}.

\bibitem{Hung:2025gcp}
L.-Y. Hung, K.~Ji, C.~Shen, Y.~Wan, and Y.~Zhao, ``{A 2D-CFT Factory: Critical
  Lattice Models from Competing Anyon Condensation Processes in
  SymTO/SymTFT},'' \href{http://arxiv.org/abs/2506.05324}{{\ttfamily
  arXiv:2506.05324 [cond-mat.str-el]}}.

\bibitem{Albert:2025umy}
J.~Albert, Y.~Honda, J.~Kaidi, and Y.~Zheng, ``{Haagerup Symmetry in
  $(E_8)_1$?},'' \href{http://arxiv.org/abs/2512.08225}{{\ttfamily
  arXiv:2512.08225 [hep-th]}}.

\bibitem{Bottini:2025hri}
L.~E. Bottini and S.~Schafer-Nameki, ``{Construction of a Gapless Phase with
  Haagerup Symmetry},''
  \href{http://dx.doi.org/10.1103/PhysRevLett.134.191602}{{\em Phys. Rev.
  Lett.} {\bfseries 134} no.~19, (2025) 191602},
  \href{http://arxiv.org/abs/2410.19040}{{\ttfamily arXiv:2410.19040
  [hep-th]}}.

\end{thebibliography}\endgroup

\end{document}